\documentclass[a4paper,11pt]{article}
\pdfoutput=1
\usepackage{jheppub}
\usepackage[utf8]{inputenc}

\usepackage{subcaption}
\usepackage{xcolor,colortbl}
\usepackage{mathtools}
\usepackage{ragged2e}
\usepackage{pdflscape}

\usepackage[font={small,it},labelfont=bf]{caption}  
\usepackage{todonotes}
\usepackage{booktabs}
\usepackage{paralist}
\usepackage{multirow}

\usepackage{tikz}
\usetikzlibrary{decorations.markings}
\usetikzlibrary{positioning}
\usetikzlibrary{arrows}
\usetikzlibrary{arrows.meta}
\usetikzlibrary{shapes}
\usetikzlibrary{chains}
\usetikzlibrary{arrows,fit,decorations.pathreplacing, shapes.misc,decorations.pathmorphing}
\tikzstyle{every picture}+=[remember picture]
\tikzstyle{na} = [baseline=-.5ex]

\tikzset{flavour/.style={regular polygon,regular polygon sides=4,inner sep=2.5pt, draw}}
\tikzset{gauge/.style={circle, draw,inner sep=2.5pt}}
\tikzset{hasse/.style={circle, fill,inner sep=2pt}}
\tikzset{sev/.style={inner sep=1mm,draw=none,fill=white,minimum size=4mm,circle, draw}}

\allowdisplaybreaks

\DeclareFontFamily{U}{rcjhbltx}{}
\DeclareFontShape{U}{rcjhbltx}{m}{n}{<->rcjhbltx}{}
\DeclareSymbolFont{hebrewletters}{U}{rcjhbltx}{m}{n}

\let\aleph\relax\let\beth\relax
\let\gimel\relax\let\daleth\relax

\DeclareMathSymbol{\aleph}{\mathord}{hebrewletters}{39}
\DeclareMathSymbol{\beth}{\mathord}{hebrewletters}{98}\let\bet\beth
\DeclareMathSymbol{\gimel}{\mathord}{hebrewletters}{103}
\DeclareMathSymbol{\daleth}{\mathord}{hebrewletters}{100}\let\dalet\daleth

\DeclareMathSymbol{\lamed}{\mathord}{hebrewletters}{108}
\DeclareMathSymbol{\mem}{\mathord}{hebrewletters}{109}\let\mim\mem
\DeclareMathSymbol{\ayin}{\mathord}{hebrewletters}{96}
\DeclareMathSymbol{\tsadi}{\mathord}{hebrewletters}{118}
\DeclareMathSymbol{\qof}{\mathord}{hebrewletters}{114}
\DeclareMathSymbol{\shin}{\mathord}{hebrewletters}{152}

\newcommand{\Ncal}{\mathcal{N}}
\newcommand{\C}{\mathbb{C}}

\newcommand{\HH}{\mathbb{H}}
\newcommand{\Z}{\mathbb{Z}}

\newcommand{\uo}{{ \mathrm{U}(1)}}
\newcommand{\su}{{{\rm SU}(2)}}
\newcommand{\urm}{{{\rm U}}}
\newcommand{\surm}{{{\rm SU}}}
\newcommand{\sorm}{{{\rm SO}}}
\newcommand{\orm}{{{\rm O}}}
\newcommand{\sprm}{{{\rm Sp}}}
\newcommand{\usprm}{{{\rm USp}}}

\newcommand{\sormL}{{\mathfrak{so}}}

\newcommand{\Coulomb}{\mathcal{C}}
\newcommand{\Higgs}{\mathcal{H}}

\newcommand{\orbit}{\mathcal{O}}
\newcommand{\clorbit}{\overline{\orbit}}

\newcommand{\NS}{\text{NS5}}
\newcommand{\Dp}{\text{D$p$}}
\newcommand{\Dpp}{\text{D($p{+}2$)}}
\newcommand{\Op}{\text{O$p$}}
\newcommand{\Opm}{\text{O$p{}^{-}$}}
\newcommand{\Opmt}{\text{$\widetilde{\text{O$p$}}^{-}$}}
\newcommand{\Opp}{\text{O$p{}^{+}$}}
\newcommand{\Oppt}{\text{$\widetilde{\text{O$p$}}^{+}$}}
\newcommand{\Oppm}{\text{O$p{}^{\pm}$}}
\newcommand{\Opmp}{\text{O$p{}^{\mp}$}}
\newcommand{\Oppmt}{\text{$\widetilde{\text{O$p$}}^{\pm}$}}
\newcommand{\Opmpt}{\text{$\widetilde{\text{O$p$}}^{\mp}$}}

\newcommand{\Dthree}{\text{D3}}
\newcommand{\Ot}{\text{O3}}

\newcommand{\Dfive}{\text{D5}}
\newcommand{\Of}{\text{O5}}
\newcommand{\Ofm}{\text{O5${}^{-}$}}
\newcommand{\Ofmt}{\text{$\widetilde{\text{O5}}^{-}$}}
\newcommand{\Ofp}{\text{O5${}^{+}$}}
\newcommand{\Ofpt}{\text{$\widetilde{\text{O5}}^{+}$}}

\newcommand{\Dseven}{\text{D7}}

\def\OPlus#1#2{
	\draw[dotted](#1)--(#2);
}

\def\OPlusTilde#1#2{
	\draw[dashed](#1)--(#2);
}

\def\OMinusTilde#1#2{
	\draw(#1)--(#2);
}

\def\Dbrane#1#2{
	\draw(#1)--(#2);
}

\def\MonoCut#1#2{
   \draw[dashed,red] (#1)--(#2);
}

\def\SevenB#1{
	\node[circle, draw, fill=white] at (#1){};
}
\def\DfiveOPlus#1#2{
	\draw(#2,0.1)--(#2+1,0.1);
	\draw(#2,-0.1)--(#2+1,-0.1);
	\draw[dotted] (#2,0)--(#2+1,0);
	\node[label=above:{{\scriptsize{#1}}}] at (#2+0.65,0) {};
}

\def\DfiveOPlusTilde#1#2{
	\draw(#2,0.1)--(#2+1,0.1);
	\draw(#2,-0.1)--(#2+1,-0.1);
	\draw[dashed] (#2,0)--(#2+1,0);
	\node[label=above:{{\scriptsize{#1}}}] at (#2+0.65,0) {};
}

\def\DfiveOMinus#1#2{
	\draw(#2,0.1)--(#2+1,0.1);
	\draw(#2,-0.1)--(#2+1,-0.1);
	\node[label=above:{{\scriptsize{#1}}}] at (#2+0.65,0) {};
}

\def\DfiveOMinusTilde#1#2{
	\draw(#2,0.1)--(#2+1,0.1);
	\draw(#2,-0.1)--(#2+1,-0.1);
	\draw (#2,0)--(#2+1,0);
	\node[label=above:{{\scriptsize{#1}}}] at (#2+0.65,0) {};
}
\newcommand{\dalg}{\text{D}}
\newcommand{\balg}{\text{B}}
\newcommand{\calg}{\text{C}}
\newcommand{\cc}[1]{$\scriptstyle{\calg_{#1}}$}
\newcommand{\bb}[1]{$\scriptstyle{\balg_{#1}}$}
\newcommand{\dd}[1]{$\scriptstyle{\dalg_{#1}}$}
\newcommand{\uu}[1]{$\scriptstyle{\urm_{#1}}$}

\newtheorem{myConj}{Conjecture}
\newtheorem{myObs}{Observation}

\tikzset{gauge/.style={inner sep=1mm,draw=none,fill=white,minimum size=2mm,circle, draw}}
\tikzset{gauger/.style={inner sep=1mm,draw=none,fill=red,minimum size=2mm,circle, draw}}
\tikzset{gaugeb/.style={inner sep=1mm,draw=none,fill=blue,minimum size=2mm,circle, draw}}

\preprint{Imperial/TP/20/AH/03}

\title{\boldmath{Magnetic Quivers from Brane Webs with O5 Planes}}

\author[\lamed]{Antoine Bourget}
\author[\lamed]{, Julius F.\ Grimminger}
\author[\lamed]{, Amihay Hanany}
\author[\beth]{, Marcus Sperling}
\author[\lamed]{, and Zhenghao Zhong}
\affiliation[\lamed]{Theoretical Physics Group, The Blackett Laboratory, 
Imperial College London, Prince Consort Road
London, SW7 2AZ, UK}
\affiliation[\beth]{Yau Mathematical Sciences Center, Tsinghua University, Haidian District, Beijing, 100084, China}
\emailAdd{a.bourget@imperial.ac.uk}
\emailAdd{julius.grimminger17@imperial.ac.uk}
\emailAdd{a.hanany@imperial.ac.uk}
\emailAdd{marcus.sperling@univie.ac.at}
\emailAdd{zhenghao.zhong14@imperial.ac.uk}

\abstract{
Magnetic quivers have led to significant progress in the understanding of gauge 
theories with 8 supercharges at UV fixed points. 
For a given low-energy gauge 
theory realised via a Type II brane construction, there exist magnetic quivers 
for the Higgs branches at finite and infinite gauge coupling. Comparing these 
moduli spaces allows one to study the 
non-perturbative effects when transitioning to the fixed point.
For 5d $\Ncal=1$ SQCD, 5-brane webs have 
been an important tool for deriving magnetic quivers. In this work, the 
emphasis is placed on 5-brane webs with orientifold 5-planes which give rise 
to 5d theories with orthogonal or symplectic gauge groups. 
For this set-up, the magnetic quiver prescription is derived and contrasted 
against a unitary magnetic quiver description extracted from an O$7^-$ construction. Further validation is achieved by a derivation of the 
associated Hasse diagrams.
An important class of families considered are the \emph{orthogonal exceptional 
$E_n$ families} ($-\infty < n \leq 8$), realised as infinite coupling 
Higgs branches of $\sprm(k)$ gauge theories with fundamental matter. 
In particular, the moduli spaces are realised 
by a novel type of magnetic quivers, called \emph{unitary-orthosymplectic 
quivers}.
}

\begin{document}

\maketitle

\section{Introduction}
\label{sec:introduction}
$5$-dimensional $\Ncal{=}1$ gauge theories are perturbatively non-renormalisable and can only meaningfully be defined as mass deformations of renormalisation group fixed points. Initially, these theories have been studied from various aspects: field theory \cite{Seiberg:1996bd,Morrison:1996xf,Intriligator:1997pq}, brane constructions \cite{Aharony:1997ju,Aharony:1997bh,DeWolfe:1999hj}, and geometry via M-theory backgrounds with Calabi-Yau singularities \cite{Douglas:1996xp}.

In this work, focus is placed on 5-brane webs in Type IIB superstring theory \cite{Aharony:1997ju,Aharony:1997bh,DeWolfe:1999hj} and generalisations that include orientifold planes \cite{Brunner:1997gk,Bergman:2015dpa,Zafrir:2015ftn,Hayashi:2015vhy}.
One advantage of these brane constructions is that they capture the dynamics of the corresponding 5d gauge theories and, simultaneously, their UV fixed points. Recent developments using brane webs include \cite{Bao:2011rc,Bergman:2014kza,Kim:2015jba,Hayashi:2015fsa,Gaiotto:2015una,Zafrir:2015rga,Hayashi:2015zka,Ohmori:2015tka,Zafrir:2016jpu,Hayashi:2016abm,Hayashi:2017btw,Hayashi:2018lyv,Cabrera:2018jxt,Hayashi:2019yxj,Hayashi:2019jvx}.
As known from the $\surm(2)$ example with $N_f <8$ flavours \cite{Seiberg:1996bd,Morrison:1996xf}, it is important to understand the enhancement of the global symmetry of these theories at the fixed point. Hence, this question has been studied via various techniques: for instance, superconformal indices \cite{Kim:2012gu,Rodriguez-Gomez:2013dpa,Bergman:2013koa,Bergman:2013ala,Taki:2013vka,Bergman:2013aca,Hwang:2014uwa,Zafrir:2014ywa,Bergman:2014kza,Zafrir:2015uaa,Hayashi:2015fsa,Yonekura:2015ksa}, Nekrasov partition functions and topological string partition functions \cite{Bao:2011rc,Bashkirov:2012re,Iqbal:2012xm,Bao:2013pwa,Hayashi:2013qwa,Hayashi:2014wfa,Mitev:2014jza,Kim:2014nqa,Hayashi:2015xla}.
The enhancement has been argued to be due to instanton operators \cite{Lambert:2014jna,Rodriguez-Gomez:2015xwa,Tachikawa:2015mha}, which create instanton particles in the UV superconformal field theory. Recall that in 5 dimensions, the instanton is a particle charged under the $\uo_I$ topological symmetry associated to the conserved current $\mathrm{Tr} \ast (F\wedge F)$.

Recently, there have been many works devoted to uncover further features \cite{Apruzzi:2019vpe,Apruzzi:2019opn,Apruzzi:2019enx,Apruzzi:2019kgb}; in particular, classifications of $5$d SCFTs \cite{Jefferson:2017ahm,Jefferson:2018irk,Bhardwaj:2019jtr} and 5d $\Ncal{=}1$ gauge theories \cite{Bhardwaj:2020gyu} have been proposed. 

An interesting question concerns the Higgs branch of the full vacuum moduli space: for the low-energy effective theory the Higgs branch is described by the hyper-K\"ahler quotient construction \cite{Hitchin:1986ea}; in contrast, for the Higgs branch $\Higgs_\infty$ at the fixed point the same is not true. The first studies \cite{Cremonesi:2015lsa,Ferlito:2017xdq} of $\Higgs_\infty$ indicated that in order to capture the geometric features of the moduli space at infinite gauge coupling, $3$d $\Ncal =4$ Coulomb branches of certain quiver gauge theories are useful.
This idea has been further developed and systematised in \cite{Cabrera:2018jxt}: for a given 5-brane web, where each external 5-brane ends on a 7-brane, \emph{magnetic quivers} can be derived such that the $3$d $\Ncal=4$ Coulomb branches thereof are equivalent geometric descriptions for the finite and infinite coupling $5$d $\Ncal=1$ Higgs branches.

So far, the magnetic quiver description has only been available for $5$d gauge theories with special unitary gauge groups. However, many interesting 5d dualities, see for instance \cite{Gaiotto:2015una,Hayashi:2015zka,Jefferson:2018irk,Bhardwaj:2020gyu}, are between unitary and orthogonal or symplectic gauge groups. Hence, it is natural to further develop the understanding of Higgs branches of $5$d theories with orthogonal or symplectic groups. Suitable 5-brane web realisations either contain \Of\ or O$7$ planes. In this work, the focus is placed on 5-brane webs in the presence of \Of\ orientifolds. For single gauge groups with fundamental matter, the field theory classification of theories with non-trivial interacting fixed point has been presented in \cite{Intriligator:1997pq}; the subsequent brane construction \cite{Brunner:1997gk} confirmed these results. However, more non-trivial fixed points have been proposed in \cite{Bergman:2015dpa}. 

Consider the Higgs branch moduli space in more detail. To begin with, the finite coupling, or classical, Higgs branches are conventionally treated by an F and D-term analysis. However, it is a known fact that $\sprm(k)$ theories with $N_f <2k$ fundamental flavours and $\orm(k)$ theories with $N_f \leq k-3$ fundamental flavours do not admit complete Higgsing. Consequently, their analysis is currently incomplete. For instance, the analogous behaviour exists for 5d $\Ncal=1$ $\surm(k)$ SQCD with $N_f < 2k$, which has only recently been addressed in \cite{Bourget:2019rtl}. In terms of brane webs, the quaternionic Higgs branch degrees of freedom can be counted by a decomposition into independent subwebs, as introduced in \cite{Aharony:1997bh} and demonstrated for 5-brane webs with \Of\ planes in \cite{Zafrir:2015ftn}. 
Moving on to the infinite coupling Higgs branches, the enhancement of the global symmetry has been studied via field theory \cite{Zafrir:2015uaa} and brane webs \cite{Bergman:2015dpa}. Moreover, the counting of additional new Higgs branch dimension at the fixed point has been demonstrated in \cite{Zafrir:2015ftn}. Hence, dimension and global symmetry of $\Higgs_\infty$ are known, but no geometrical description has been provided yet.

This is precisely the first aim of the present paper: to provide an improved description of finite and infinite coupling Higgs branches. The approach taken is known as \emph{magnetic quivers} \cite{Cabrera:2018jxt,Cabrera:2019izd,Cabrera:2019dob} (see also \cite{Hanany:1996ie}): in brief, a magnetic quiver $\mathsf{Q}$ is a combinatorial object that is derived from the Type II brane configuration with 8 supercharges, describing a given theory $\mathsf{T}$ in a certain phase $\mathcal{P}$. The Higgs branch of $\mathsf{T}$ in that phase equals the 3d $\Ncal=4 $ Coulomb branch of the magnetic quiver, meaning that the combinatorial data is taken as an input to derive a space of dressed monopole operators in the sense of \cite{Cremonesi:2013lqa}. Thus,
\begin{align}
    \Higgs \left(\text{phase $\mathcal{P}$ of theory $\mathsf{T}$}\right) = 
    \Coulomb\left( \text{magnetic quiver $\mathsf{Q}(\mathcal{P})$} \right)
\end{align}
holds as \emph{equality of moduli spaces}. To be more precise: the magnetic 
quivers compute a geometric space, called \emph{Higgs variety}. The Higgs branch 
chiral ring may contain nilpotent operators which makes the full Higgs branch a 
so called non-reduced scheme, called \emph{Higgs Scheme}. This problem is 
addressed for classical $4$d $\mathcal{N}=2$ SQCD in \cite{Bourget:2019rtl}. In 
the rest of the paper only the geometric parts of the moduli space are studied 
and the analysis of nilpotent elements is left for future work. The concept of 
magnetic quivers has proven itself useful in a variety of cases: for $6$d 
$\Ncal=(1,0)$ theories 
\cite{Mekareeya:2017jgc,Hanany:2018uhm,Hanany:2018vph,Cabrera:2019izd, 
Cabrera:2019dob}, 5d $\Ncal =1$ gauge theories 
\cite{Cremonesi:2015lsa,Ferlito:2017xdq,Cabrera:2018jxt}, and 4d Argyres-Douglas 
theories \cite{DelZotto:2014kka}. 

The second aim of the present paper is to derive the \emph{Hasse diagrams} for the finite and infinite coupling Higgs branches. Hasse diagrams for nilpotent orbits were studied in detail by Kraft and Procesi in \cite{kraft1980minimal,Kraft1982}. As described in \cite{Bourget:2019aer}, the Hasse diagram details the singularity structure of a Higgs branch understood as a symplectic singularity \cite{Beauville:2000sym}. For finite coupling Higgs branches, the Hasse diagram could, for example, be derived from the Higgs mechanism; however, the same method is not applicable at the fixed point. Alternatively, the Hasse diagram can also be derived from (i) the brane web via Kraft-Procesi transitions \cite{Cabrera:2016vvv,Cabrera:2017njm}, or (ii) from quiver subtraction \cite{Cabrera:2018ann} on the magnetic quivers. 

Once the combinatorial data of the magnetic quiver is known, the techniques developed allow to extract the following information:
\begin{compactenum}[(i)]
\item Dimension.
\item Global symmetry.
\item Representation content of the chiral ring via the Hilbert series.
\item Hasse diagram.
\item Comparison of dimensions, global symmetry, chiral ring representation content, and Hasse diagram between finite and infinite gauge coupling.
\end{compactenum}
Complementing these facts by approaches that rely on the \emph{same combinatorial input data}, like for instance \cite{Bullimore:2015lsa,Nakajima:2015txa,Braverman:2016wma}, one reaches the conclusion that the entire moduli space geometry is determined by this data.

\begin{table}[t]
      \hspace{-1.8cm}\scalebox{.95}{\begin{tabular}{c|c|c|c|c} \toprule
        Family    &  Theory SU  & Theory Sp & Magnetic quiver U & Magnetic quiver OSp   \\ \midrule 
        $E_8$ & \raisebox{-.5\height}{\begin{tikzpicture}
	\node (g1) [gauge,label=below:{$\mathrm{SU}(k{+}1)_{\pm \frac{1}{2}}$}] {};
	\node (g2) [flavour,above of=g1,label=above:{$2k{+}5$}] {};
	\draw (g1)--(g2);
	\end{tikzpicture}} & \raisebox{-.5\height}{\begin{tikzpicture}
	\node (g1) [gauge,label=below:{$\mathrm{Sp}(k)$}] {};
	\node (g2) [flavour,above of=g1,label=above:{$D_{2k+5}$}] {};
	\draw (g1)--(g2);
	\end{tikzpicture}} & \raisebox{-.5\height}{\begin{tikzpicture}
	\tikzset{node distance = .8cm};
	\node (g1) [gauge,label=below:{\footnotesize{\rotatebox[origin=c]{0}{$1$}}}] {};
	\node (g2) [right of=g1] {$\cdots$};
	\node (g3) [gauge,right of=g2,label=below:{\footnotesize{\rotatebox[origin=c]{-45}{$2k+4$}}}] {};
	\node (g4) [gauge,right of=g3,label=below:{\footnotesize{\rotatebox[origin=c]{-45}{$k+3$}}}] {};
	\node (g5) [gauge,right of=g4,label=below:{\footnotesize{\rotatebox[origin=c]{0}{$2$}}}] {};
	\node (g7) [gauge,above of=g3,label=left:{\footnotesize{$k+2$}}] {};
	\draw (g1)--(g2)--(g3)--(g4)--(g5) (g3)--(g7);
	\end{tikzpicture}} & \raisebox{-.5\height}{\begin{tikzpicture}
	\tikzset{node distance = .8cm};
	\node (g1) [gauge,label=below:{\dd{1}}] {};
	\node (g2) [gauge,right of=g1,label=below:{\cc{1}}] {};
	\node (g3) [right of=g2] {$\cdots$};
	\node (g4) [gauge,right of=g3,label=below:{\cc{k+2}}] {};
	\node (g5) [gauge,right of=g4,label=below:{\dd{k+3}}] {};
	\node (g6) [gauge,right of=g5,label=below:{\cc{k+2}}] {};
	\node (g7) [right of=g6] {$\cdots$};
	\node (g8) [gauge,right of=g7,label=below:{\cc{1}}] {};
	\node (g9) [gauge,right of=g8,label=below:{\dd{1}}] {};
	\node (g10) [gauge,above of=g5,label=left:{\cc{1}}] {};
	\draw (g1)--(g2) (g2)--(g3) (g3)--(g4) (g4)--(g5) (g5)--(g6) (g6)--(g7) (g7)--(g8) (g8)--(g9) (g5)--(g10);
	\end{tikzpicture}}  \\ 
           $E_7$ & \raisebox{-.5\height}{\begin{tikzpicture}
	\node (g1) [gauge,label=below:{$\mathrm{SU}(k{+}1)_{\pm 1}$}] {};
	\node (g2) [flavour,above of=g1,label=above:{$2k{+}4$}] {};
	\draw (g1)--(g2);
	\end{tikzpicture}} & \raisebox{-.5\height}{\begin{tikzpicture}
	\node (g1) [gauge,label=below:{$\mathrm{Sp}(k)$}] {};
	\node (g2) [flavour,above of=g1,label=above:{$D_{2k+4}$}] {};
	\draw (g1)--(g2);
	\end{tikzpicture}} & \raisebox{-.5\height}{\begin{tikzpicture}
	\tikzset{node distance = .8cm};
	\node (g1) [gauge,label=below:{\footnotesize{\rotatebox[origin=c]{0}{$1$}}}] {};
	\node (g2) [right of=g1] {$\cdots$};
	\node (g3) [gauge,right of=g2,label=below:{\footnotesize{\rotatebox[origin=c]{-45}{$2k+2$}}}] {};
	\node (g4) [gauge,right of=g3,label=below:{\footnotesize{\rotatebox[origin=c]{-45}{$k+2$}}}] {};
	\node (g5) [gauge,right of=g4,label=below:{\footnotesize{\rotatebox[origin=c]{0}{$2$}}}] {};
	\node (g6) [gauge,right of=g5,label=below:{\footnotesize{\rotatebox[origin=c]{0}{$1$}}}] {};
	\node (g7) [gauge,above of=g3,label=left:{\footnotesize{$k+1$}}] {};
	\draw (g1)--(g2)--(g3)--(g4)--(g5)--(g6) (g3)--(g7);
	\end{tikzpicture}} & \raisebox{-.5\height}{\begin{tikzpicture}
	\tikzset{node distance = .8cm};
	\node (g1) [gauge,label=below:{\dd{1}}] {};
	\node (g2) [gauge,right of=g1,label=below:{\cc{1}}] {};
	\node (g3) [right of=g2] {$\cdots$};
	\node (g4) [gauge,right of=g3,label=below:{\cc{k+1}}] {};
	\node (g5) [gauge,right of=g4,label=below:{\dd{k+2}}] {};
	\node (g6) [gauge,right of=g5,label=below:{\cc{k+1}}] {};
	\node (g7) [right of=g6] {$\cdots$};
	\node (g8) [gauge,right of=g7,label=below:{\cc{1}}] {};
	\node (g9) [gauge,right of=g8,label=below:{\dd{1}}] {};
	\node (g10) [gauge,above of=g5,label=left:{\cc{1}}] {};
	\node (g11) [gauge,above of=g10,label=left:{\uu{1}}] {};
	\draw (g1)--(g2) (g2)--(g3) (g3)--(g4) (g4)--(g5) (g5)--(g6) (g6)--(g7) (g7)--(g8) (g8)--(g9) (g5)--(g10) (g10)--(g11);
	\end{tikzpicture}}  \\ 
	$E_6$ & \raisebox{-.5\height}{\begin{tikzpicture}
	\node (g1) [gauge,label=below:{$\mathrm{SU}(k{+}1)_{\pm \frac{3}{2}}$}] {};
	\node (g2) [flavour,above of=g1,label=above:{$2k{+}3$}] {};
	\draw (g1)--(g2);
	\end{tikzpicture}} & \raisebox{-.5\height}{\begin{tikzpicture}
	\node (g1) [gauge,label=below:{$\mathrm{Sp}(k)$}] {};
	\node (g2) [flavour,above of=g1,label=above:{$D_{2k+3}$}] {};
	\draw (g1)--(g2);
	\end{tikzpicture}} & \raisebox{-.5\height}{\begin{tikzpicture}
	\tikzset{node distance = .8cm};
	\node (g1) [gauge,label=below:{\footnotesize{\rotatebox[origin=c]{0}{$1$}}}] {};
	\node (g2) [right of=g1] {$\cdots$};
	\node (g3) [gauge,right of=g2,label=below:{\footnotesize{\rotatebox[origin=c]{-45}{$2k+1$}}}] {};
	\node (g4) [gauge,right of=g3,label=below:{\footnotesize{\rotatebox[origin=c]{-45}{$k+1$}}}] {};
	\node (g5) [gauge,right of=g4,label=below:{\footnotesize{\rotatebox[origin=c]{0}{$1$}}}] {};
	\node (g7) [gauge,above of=g3,label=left:{\footnotesize{$k+1$}}] {};
	\node (g8) [gauge,right of=g7,label=right:{\footnotesize{$1$}}] {};
	\draw (g1)--(g2)--(g3)--(g4)--(g5) (g3)--(g7)--(g8);
	\end{tikzpicture}} & \raisebox{-.5\height}{\begin{tikzpicture}
	\tikzset{node distance = .8cm};
	\node (g1) [gauge,label=below:{\dd{1}}] {};
	\node (g2) [gauge,right of=g1,label=below:{\cc{1}}] {};
	\node (g3) [right of=g2] {$\cdots$};
	\node (g4) [gauge,right of=g3,label=below:{\dd{k+1}}] {};
	\node (g5) [gauge,right of=g4,label=below:{\cc{k+1}}] {};
	\node (g6) [gauge,right of=g5,label=below:{\dd{k+1}}] {};
	\node (g7) [right of=g6] {$\cdots$};
	\node (g8) [gauge,right of=g7,label=below:{\cc{1}}] {};
	\node (g9) [gauge,right of=g8,label=below:{\dd{1}}] {};
	\node (g10) [gauge,above of=g5,label=left:{\uu{1}}] {};
	\draw (g1)--(g2) (g2)--(g3) (g3)--(g4) (g4)--(g5) (g5)--(g6) (g6)--(g7) (g7)--(g8) (g8)--(g9) (g5)--(g10);
	\end{tikzpicture}} \\ 
		$E_5$ & \raisebox{-.5\height}{\begin{tikzpicture}
	\node (g1) [gauge,label=below:{$\mathrm{SU}(k{+}1)_{\pm 2}$}] {};
	\node (g2) [flavour,above of=g1,label=above:{$2k{+}2$}] {};
	\draw (g1)--(g2);
	\end{tikzpicture}} & \raisebox{-.5\height}{\begin{tikzpicture}
	\node (g1) [gauge,label=below:{$\mathrm{Sp}(k)$}] {};
	\node (g2) [flavour,above of=g1,label=above:{$D_{2k+2}$}] {};
	\draw (g1)--(g2);
	\end{tikzpicture}} & \raisebox{-.5\height}{\begin{tikzpicture}
	\tikzset{node distance = .8cm};
	\node (g1) [gauge,label=below:{\footnotesize{\rotatebox[origin=c]{0}{$1$}}}] {};
	\node (g2) [right of=g1] {$\cdots$};
	\node (g3) [gauge,right of=g2,label=below:{\footnotesize{\rotatebox[origin=c]{-45}{$2k$}}}] {};
	\node (g4) [gauge,right of=g3,label=below:{\footnotesize{\rotatebox[origin=c]{-45}{$k+1$}}}] {};
	\node (g5) [gauge,right of=g4,label=below:{\footnotesize{\rotatebox[origin=c]{0}{$1$}}}] {};
	\node (g7) [gauge,above of=g3,label=left:{\footnotesize{$k$}}] {};
	\node (g8) [gauge,above of=g4,label=right:{\footnotesize{$1$}}] {};
	\draw (g1)--(g2)--(g3)--(g4)--(g5) (g3)--(g7) (g4)--(g8);
	\end{tikzpicture}} & \raisebox{-.5\height}{\begin{tikzpicture}
	\tikzset{node distance = .8cm};
	\node (g1) [gauge,label=below:{\dd{1}}] {};
	\node (g2) [gauge,right of=g1,label=below:{\cc{1}}] {};
	\node (g3) [right of=g2] {$\cdots$};
	\node (g4) [gauge,right of=g3,label=below:{\cc{k}}] {};
	\node (g5) [gauge,right of=g4,label=below:{\dd{k+1}}] {};
	\node (g6) [gauge,right of=g5,label=below:{\cc{k}}] {};
	\node (g7) [right of=g6] {$\cdots$};
	\node (g8) [gauge,right of=g7,label=below:{\cc{1}}] {};
	\node (g9) [gauge,right of=g8,label=below:{\dd{1}}] {};
	\node (g10) [gauge,above of=g5,label=left:{\uu{1}}] {};
	\draw (g1)--(g2) (g2)--(g3) (g3)--(g4) (g4)--(g5) (g5)--(g6) (g6)--(g7) (g7)--(g8) (g8)--(g9) (g5)--(g10);
	\end{tikzpicture}}  \\  
		$\begin{matrix}E_{4-2l} \\ \scriptstyle{k\geq l\geq0} \end{matrix}$ & \raisebox{-.5\height}{\begin{tikzpicture}
	\node (g1) [gauge,label=below:{$\mathrm{SU}(k{+}1)_{\pm \left(\frac{5}{2} + l\right) } $}] {};
	\node (g2) [flavour,above of=g1,label=above:{$2k{-}2l{+}1$}] {};
	\draw (g1)--(g2);
	\end{tikzpicture}} & \raisebox{-.5\height}{\begin{tikzpicture}
	\node (g1) [gauge,label=below:{$\mathrm{Sp}(k)$}] {};
	\node (g2) [flavour,above of=g1,label=above:{$D_{2k{-}2l{+}1}$}] {};
	\draw (g1)--(g2);
	\end{tikzpicture}} & \raisebox{-.5\height}{\begin{tikzpicture}
	\tikzset{node distance = .8cm};
	\node (g1) [gauge,label=below:{\footnotesize{\rotatebox[origin=c]{0}{$1$}}}] {};
	\node (g2) [right of=g1] {$\cdots$};
	\node (g3) [gauge,right of=g2,label=below:{\footnotesize{\rotatebox[origin=c]{-45}{$2k{-}2l{-}1$}}}] {};
	\node (g4) [gauge,right of=g3,label=below:{\footnotesize{\rotatebox[origin=c]{-45}{$k{-}l$}}}] {};
	\node (g5) [gauge,right of=g4,label=below:{\footnotesize{\rotatebox[origin=c]{0}{$1$}}},label=above:{\footnotesize{\rotatebox[origin=c]{-45}{\hspace*{-.5cm}$l+1$}}}] {};
	\node (g7) [gauge,above of=g3,label=left:{\footnotesize{$k{-}l$}}] {};
	\node (g8) [gauge,above of=g4,label=above:{\footnotesize{$1$}}] {};
	\draw (g1)--(g2)--(g3)--(g4)--(g5)--(g8)--(g7)--(g3);
	\draw[double] (g5)--(g8);
	\end{tikzpicture}} &  
	\raisebox{-.5\height}{\begin{tikzpicture}
	\tikzset{node distance = .8cm};
	\node (g1) [gauge,label=below:{\dd{1}}] {};
	\node (g2) [gauge,right of=g1,label=below:{\cc{1}}] {};
	\node (g3) [right of=g2] {$\cdots$};
	\node (g4) [gauge,right of=g3,label=below:{\dd{k{-}l}}] {};
	\node (g5) [gauge,right of=g4,label=below:{\cc{k{-}l}}] {};
	\node (g6) [gauge,right of=g5,label=below:{\dd{k{-}l}}] {};
	\node (g7) [right of=g6] {$\cdots$};
	\node (g8) [gauge,right of=g7,label=below:{\cc{1}}] {};
	\node (g9) [gauge,right of=g8,label=below:{\dd{1}}] {};
	\node (g10) [gauge,above of=g5,label=left:{\uu{1}}] {};
	\draw (g1)--(g2) (g2)--(g3) (g3)--(g4) (g4)--(g5) (g5)--(g6) (g6)--(g7) (g7)--(g8) (g8)--(g9) (g5)--(g10);	\node (g11) [flavour,above of=g10,label=left:{$\scriptstyle{l+1}$}] {};
	\draw [line join=round,decorate, decoration={zigzag, segment length=4,amplitude=.9,post=lineto,post length=2pt}]  (g10) -- (g11);
	\end{tikzpicture}}\\  
		$\begin{matrix}E_{3-2l} \\ \scriptstyle{k\geq l\geq0} \end{matrix}$ & \raisebox{-.5\height}{\begin{tikzpicture}
	\node (g1) [gauge,label=below:{$\mathrm{SU}(k{+}1)_{\pm \left(3 + l\right) } $}] {};
	\node (g2) [flavour,above of=g1,label=above:{$2k{-}2l$}] {};
	\draw (g1)--(g2);
	\end{tikzpicture}} & \raisebox{-.5\height}{\begin{tikzpicture}
	\node (g1) [gauge,label=below:{$\mathrm{Sp}(k)$}] {};
	\node (g2) [flavour,above of=g1,label=above:{$D_{2k{-}2l}$}] {};
	\draw (g1)--(g2);
	\end{tikzpicture}} & \begin{tabular}{c}
	    \raisebox{-.5\height}{\begin{tikzpicture}
	\tikzset{node distance = .8cm};
	\node (g1) [gauge,label=below:{\footnotesize{\rotatebox[origin=c]{0}{$1$}}}] {};
	\node (g2) [right of=g1] {$\cdots$};
	\node (g3) [gauge,right of=g2,label=below:{\footnotesize{\rotatebox[origin=c]{-45}{$2k{-}2l{-}2$}}}] {};
	\node (g4) [gauge,right of=g3,label=below:{\footnotesize{\rotatebox[origin=c]{-45}{$k{-}l$}}}] {};
	\node (g5) [gauge,right of=g4,label=below:{\footnotesize{\rotatebox[origin=c]{0}{$1$}}},label=above:{\footnotesize{\rotatebox[origin=c]{-45}{\hspace*{-.5cm}$l+1$}}}] {};
	\node (g7) [gauge,above of=g3,label=left:{\footnotesize{$k{-}l{-}1$}}] {};
	\node (g8) [gauge,above of=g4,label=above:{\footnotesize{$1$}}] {};
	\draw (g1)--(g2)--(g3)--(g4)--(g5)--(g8)--(g4);
	\draw (g3)--(g7);
	\draw[double] (g5)--(g8);
	\end{tikzpicture}} \\
	   \raisebox{-.5\height}{\begin{tikzpicture}
	\tikzset{node distance = .8cm}; 
	\node (g1) [gauge,label=below:{\footnotesize{\rotatebox[origin=c]{0}{$1$}}}] {};
	\node (g2) [right of=g1] {$\cdots$};
	\node (g3) [gauge,right of=g2,label=below:{\footnotesize{\rotatebox[origin=c]{-45}{$2k{-}2l{-}2$}}}] {};
	\node (g4) [gauge,right of=g3,label=below:{\footnotesize{\rotatebox[origin=c]{-45}{$k{-}l$}}}] {};
	\node (g5) [gauge,right 
of=g4,label=below:{\footnotesize{$1$}},label=above:{\footnotesize{\hspace*{-.8cm}$2$}}] {};
	\node (g7) [gauge,above of=g3,label=left:{\footnotesize{$k{-}l{-}1$}}] {};
	\draw (g1)--(g2)--(g3)--(g4);
	\draw (g3)--(g7);
	\draw[double] (g4)--(g5);
	\end{tikzpicture}}
	\end{tabular}  &  
	\begin{tabular}{c}
	 	\raisebox{-.5\height}{\begin{tikzpicture}
	\tikzset{node distance = .8cm};
	\node (g1) [gauge,label=below:{\dd{1}}] {};
	\node (g2) [gauge,right of=g1,label=below:{\cc{1}}] {};
	\node (g3) [right of=g2] {$\cdots$};
	\node (g4) [gauge,right of=g3,label=below:{\cc{k{-}l{-}1}}] {};
	\node (g5) [gauge,right of=g4,label=below:{\dd{k{-}l}}] {};
	\node (g6) [gauge,right of=g5,label=below:{\cc{k{-}l{-}1}}] {};
	\node (g7) [right of=g6] {$\cdots$};
	\node (g8) [gauge,right of=g7,label=below:{\cc{1}}] {};
	\node (g9) [gauge,right of=g8,label=below:{\dd{1}}] {};
	\node (g10) [gauge,above of=g5,label=left:{\uu{1}}] {};
	\draw (g1)--(g2) (g2)--(g3) (g3)--(g4) (g4)--(g5) (g5)--(g6) (g6)--(g7) (g7)--(g8) (g8)--(g9) (g5)--(g10);	\node (g11) [flavour,above of=g10,label=left:{$\scriptstyle{l+1}$}] {};
	\draw [line join=round,decorate, decoration={zigzag, segment length=4,amplitude=.9,post=lineto,post length=2pt}]  (g10) -- (g11);
	\end{tikzpicture}}      \\   \begin{tabular}{c}
	 \vspace{.5cm}  \\   Cannot be read from brane diagram    \\ 
	\end{tabular}
	\end{tabular}
	\\ 
	\bottomrule 
         \end{tabular}}
         \caption{Magnetic quivers at infinite coupling. The $5$d $\Ncal=1$ 
duality between ``Theory SU" and ``Theory Sp" has been observed in 
\cite{Gaiotto:2015una}, also \cite{Hayashi:2015zka}. The wiggly link denotes 
a charge 2 hypermultiplet. The ``Magnetic quiver OSp" are derived in Section 
\ref{sec:Spk}, while ``Magnetic Quiver U" are subject of Section 
\ref{sec:O7_plane}. For $k=0$, the moduli spaces are free hypermultiplets 
transforming as spinors of the global symmetry. The ``Magnetic quiver OSp" for 
 $E_{8,7,6}$ can be obtained from class S 
\cite{Chacaltana:2011ze,Chacaltana:2012ch}.}
         \label{tab:Efamilies}
     \end{table}
     
\paragraph{Summary of main results.}
The methods developed in this paper yield a multitude of results and are summarised here for convenience.
\begin{compactitem}
\item Among the most interesting results are the magnetic quivers for $\sprm(k)$ theories with $0\leq N_f \leq 2k+5 $ fundamental flavours, summarised in Table \ref{tab:Efamilies}. Here, two different constructions, by \Ofp\ and O$7^-$ orientifolds, give rise to two different sets of magnetic quivers: unitary-orthosymplectic quivers and unitary quivers, respectively. 
The results are remarkable for various reasons: for the unitary magnetic quivers 
the global symmetry is straightforwardly evaluated, see Tables 
\ref{tab:CompareFiniteHasse}, \ref{tab:CompareFiniteHasse2}, and, for $k=1$, 
reproduce the known enhancement for $E_{N_f+1}$. Therefore, the cases $\sprm(k)$ 
with $N_f$ fundamentals are referred to as generalised exceptional families. For 
the unitary-orthosymplectic magnetic quivers, the global symmetry is less 
obvious from the quiver itself. However, a Hilbert series computation confirms 
that for each case both magnetic quivers have the same highest weight generating 
(HWG) function, see Table \ref{tab:CompareFiniteHWG}, which is a necessary 
condition that both quivers describe the same moduli space. The details of the 
computational challenges are a subject of a companion paper 
\cite{Bourget:2020xdz}.
\item Besides the two different magnetic quiver descriptions, the geometry of the Higgs branches is further detailed by the Hasse diagram. A non-trivial consistency check has been passed by verifying that both, the unitary and the unitary-orthosymplectic, magnetic quivers for the exceptional families lead to the same Hasse diagram, see Tables \ref{tab:CompareFiniteHasse} and \ref{tab:CompareFiniteHasse2}. The appearance of minimal nilpotent orbit closures $e_{n}$ of exceptional Lie algebras $E_n$ at the top of the infinite coupling Hasse diagram yields another reason for calling these families \emph{exceptional families}.

While the Hasse diagram derivation for the unitary quivers is straightforward from the results of \cite{Cabrera:2018ann,Bourget:2019aer}, the algorithm for orthosymplectic quivers is more subtle. Quiver subtraction with (framed) orthosymplectic quivers have been used in \cite{Hanany:2018uhm,Hanany:2019tji} and the first Hasse diagrams derived in this class have been presented in \cite{Cabrera:2019dob}, in the context of 6d theories.
\item Concerning orthogonal or symplectic gauge theories without 
complete Higgsing, the magnetic quivers provide \emph{predictions} on 
the finite and infinite coupling Higgs branches. 
\end{compactitem}

\begin{table}[t]
         \centering
         \vspace*{1cm}
         \hspace*{-1cm}
         \begin{tabular}{c|c|c|c|c} \toprule
            Family  & \begin{tabular}{c}
                 Dimension and  \\
                symmetry for $k>1$ \\ at finite coupling
            \end{tabular}  & \begin{tabular}{c}
                 Dimension and  \\
                symmetry for $k>1$  \\ at infinite coupling
            \end{tabular}   &     \begin{tabular}{c}
                Hasse diagram  \\
                finite coupling
            \end{tabular} &  \begin{tabular}{c}
                Hasse diagram  \\
                infinite coupling
            \end{tabular}   \\ \midrule
             $E_8$ & \begin{tabular}{c}
                 $2k^2+9k$  \\
                 $\mathfrak{so}(4k+10)$
             \end{tabular} & \begin{tabular}{c}
                 $2k^2+11k+16$  \\
                 $\mathfrak{so}(4k+12)$
             \end{tabular}  & \scalebox{.879}{\raisebox{-.5\height}{\begin{tikzpicture}
		\tikzstyle{hasse} = [circle, fill,inner sep=2pt];
		\node at (0,.3) {};\node at (0,-2.7) {};
		\node (1) [hasse] at (0,0) {};
		\node (2) [hasse] at (0,-.6) {};
		\node (3) [hasse] at (0,-1.2) {};
		\node (4) at (0,-1.4) {$\vdots$};
		\node (5) [hasse] at (0,-1.8) {};
		\node (6) [hasse] at (0,-2.4) {};
		\draw (1) edge [] node[label=right:$d_7$] {} (2);
		\draw (2) edge [] node[label=right:$d_{9}$] {} (3);
		\draw (5) edge [] node[label=right:$d_{2k+5}$] {} (6);
	         \end{tikzpicture}}} & \scalebox{.879}{\raisebox{-.5\height}{\begin{tikzpicture}
		\tikzstyle{hasse} = [circle, fill,inner sep=2pt];
		\node at (0,.3) {};\node at (0,-2.7) {};
		\node (1) [hasse] at (0,0) {};
		\node (2) [hasse] at (0,-.6) {};
		\node (3) [hasse] at (0,-1.2) {};
		\node (4) at (0,-1.4) {$\vdots$};
		\node (5) [hasse] at (0,-1.8) {};
		\node (6) [hasse] at (0,-2.4) {};
		\draw (1) edge [] node[label=right:$e_8$] {} (2);
		\draw (2) edge [] node[label=right:$d_{10}$] {} (3);
		\draw (5) edge [] node[label=right:$d_{2k+6}$] {} (6);
	         \end{tikzpicture}}}  \\  
             $E_7$ & \begin{tabular}{c}
                 $2k^2+7k$  \\
                 $\mathfrak{so}(4k+8)$
             \end{tabular} & \begin{tabular}{c}
                 $2k^2+7k+8$  \\
                 $\mathfrak{so}(4k+8) \oplus \mathfrak{su}(2)$
             \end{tabular} & \scalebox{.879}{\raisebox{-.5\height}{\begin{tikzpicture}
		\tikzstyle{hasse} = [circle, fill,inner sep=2pt];
		\node at (0,.3) {};\node at (0,-2.7) {};
		\node (1) [hasse] at (0,0) {};
		\node (2) [hasse] at (0,-.6) {};
		\node (3) [hasse] at (0,-1.2) {};
		\node (4) at (0,-1.4) {$\vdots$};
		\node (5) [hasse] at (0,-1.8) {};
		\node (6) [hasse] at (0,-2.4) {};
		\draw (1) edge [] node[label=right:$d_6$] {} (2);
		\draw (2) edge [] node[label=right:$d_{8}$] {} (3);
		\draw (5) edge [] node[label=right:$d_{2k+4}$] {} (6);
	         \end{tikzpicture}}}   & \scalebox{.879}{\raisebox{-.5\height}{\begin{tikzpicture}
		\tikzstyle{hasse} = [circle, fill,inner sep=2pt];
		\node at (0,.3) {};\node at (0,-2.9) {};
		\node (1) [hasse] at (0,0) {};
		\node (2) [hasse] at (0,-.6) {};
		\node (3) [hasse] at (0,-1.2) {};
		\node (3b) [hasse] at (0,-1.8) {};
		\node (4) at (0,-2) {$\vdots$};
		\node (5) [hasse] at (0,-2.4) {};
		\node (6) [hasse] at (0,-3) {};
		\draw (1) edge [] node[label=left:$e_7$] {} (2);
		\draw (2) edge [] node[label=left:$d_{8}$] {} (3);
		\draw (3) edge [] node[label=left:$d_{10}$] {} (3b);
		\draw (5) edge [] node[label=left:$d_{2k+4}$] {} (6);
		\node (7) [hasse] at (1,-.9) {};
		\node (8) [hasse] at (1,-1.5) {};
		\node (9) at (1,-1.7) {$\vdots$};
		\node (10) [hasse] at (1,-2.1) {};
		\node (11) [hasse] at (1,-2.7) {};
		\draw (1) edge [] node[label=right:$e_8$] {} (7);
		\draw (7) edge [] node[label=right:$d_{10}$] {} (8);
		\draw (10) edge [] node[label=right:$d_{2k+4}$] {} (11);
		\draw (7) edge [] node[label=below:$a_1$] {} (3);
		\draw (8) edge [] node[label=below:$a_1$] {} (3b);
		\draw (10) edge [] node[label=below:$a_1$] {} (5);
		\draw (11) edge [] node[label=below:$a_1$] {} (6);
	         \end{tikzpicture}}} \\ 
             $E_6$ & \begin{tabular}{c}
                 $2k^2+5k$  \\
                 $\mathfrak{so}(4k+6)$
             \end{tabular} & \begin{tabular}{c}
                 $2k^2+5k+4$  \\
                 $\mathfrak{so}(4k+6) \oplus \mathfrak{u}(1)$
             \end{tabular}  & \scalebox{.879}{\raisebox{-.5\height}{\begin{tikzpicture}
		\tikzstyle{hasse} = [circle, fill,inner sep=2pt];
		\node at (0,.3) {};\node at (0,-2.7) {};
		\node (1) [hasse] at (0,0) {};
		\node (2) [hasse] at (0,-.6) {};
		\node (3) [hasse] at (0,-1.2) {};
		\node (4) at (0,-1.4) {$\vdots$};
		\node (5) [hasse] at (0,-1.8) {};
		\node (6) [hasse] at (0,-2.4) {};
		\draw (1) edge [] node[label=right:$d_5$] {} (2);
		\draw (2) edge [] node[label=right:$d_{7}$] {} (3);
		\draw (5) edge [] node[label=right:$d_{2k+3}$] {} (6);
	         \end{tikzpicture}}}    & \scalebox{.879}{\raisebox{-.5\height}{\begin{tikzpicture}
		\tikzstyle{hasse} = [circle, fill,inner sep=2pt];
		\node at (0,.3) {};\node at (0,-2.7) {};
		\node (1) [hasse] at (0,0) {};
		\node (2) [hasse] at (0,-.6) {};
		\node (3) [hasse] at (0,-1.2) {};
		\node (4) at (0,-1.4) {$\vdots$};
		\node (5) [hasse] at (0,-1.8) {};
		\node (6) [hasse] at (0,-2.4) {};
		\draw (1) edge [] node[label=right:$e_6$] {} (2);
		\draw (2) edge [] node[label=right:$d_{7}$] {} (3);
		\draw (5) edge [] node[label=right:$d_{2k+3}$] {} (6);
	         \end{tikzpicture}}}
	           \\  
             $E_5$ & \begin{tabular}{c}
                 $2k^2+3k$ \\
                 $\mathfrak{so}(4k+4) $
             \end{tabular} & \begin{tabular}{c}
                 $2k^2+3k+2$ \\
                 $\mathfrak{so}(4k+4) \oplus \mathfrak{u}(1)$
             \end{tabular} & \scalebox{.879}{\raisebox{-.5\height}{\begin{tikzpicture}
		\tikzstyle{hasse} = [circle, fill,inner sep=2pt];
		\node at (0,.3) {};\node at (0,-2.7) {};
		\node (1) [hasse] at (0,0) {};
		\node (2) [hasse] at (0,-.6) {};
		\node (3) [hasse] at (0,-1.2) {};
		\node (4) at (0,-1.4) {$\vdots$};
		\node (5) [hasse] at (0,-1.8) {};
		\node (6) [hasse] at (0,-2.4) {};
		\draw (1) edge [] node[label=right:$d_4$] {} (2);
		\draw (2) edge [] node[label=right:$d_{6}$] {} (3);
		\draw (5) edge [] node[label=right:$d_{2k+2}$] {} (6);
	         \end{tikzpicture}}}     & \scalebox{.879}{\raisebox{-.5\height}{\begin{tikzpicture}
		\tikzstyle{hasse} = [circle, fill,inner sep=2pt];
		\node at (0,.3) {};\node at (0,-2.7) {};
		\node (1) [hasse] at (0,0) {};
		\node (2) [hasse] at (0,-.6) {};
		\node (3) [hasse] at (0,-1.2) {};
		\node (4) at (0,-1.4) {$\vdots$};
		\node (5) [hasse] at (0,-1.8) {};
		\node (6) [hasse] at (0,-2.4) {};
		\draw (1) edge [] node[label=right:$e_5$] {} (2);
		\draw (2) edge [] node[label=right:$d_{6}$] {} (3);
		\draw (5) edge [] node[label=right:$d_{2k+2}$] {} (6);
	         \end{tikzpicture}}}
	        \\ \bottomrule 
         \end{tabular}
         \caption{ Coulomb branch quaternionic dimension, global symmetry and Hasse diagram for the orthogonal exceptional families, both at finite and infinite gauge coupling. The Hasse diagrams are further detailed in Section \ref{sec:Hasse}. Note the following changes between finite and infinite coupling. The constant terms in the dimension formulae change in powers of two, as indicated by the spinor representations for the free hypermultiplets in the $k=0$ case. The global symmetry at finite coupling does not include the $U(1)_I$ factor since it is realized by the gaugino bilinear which lives in a nilpotent supermultiplet. At infinite coupling, it ceases to be nilpotent and becomes part of the geometry. As a consequence the rank of the global symmetry of the Higgs branch variety increases by one. The Hasse diagram is modified only in its top dimensional symplectic leaf for the $E_6$ and $E_5$ families, while the change is deeper for the $E_7$ and $E_8$ families. A recurring pattern for the $E_n$ families is the transformation of a $d_{n-1}$ transition into an $e_{n}$ transition. 
   } 
         \vspace*{1cm}
         \label{tab:CompareFiniteHasse}
     \end{table}

\begin{table}[t]
         \centering
         \vspace*{1.0cm}\hspace*{-1cm}\begin{tabular}{c|c|c|c|c} \toprule
            Family  & \begin{tabular}{c}
                 Dimension and  \\
                symmetry for $k>1$  \\ at finite coupling
            \end{tabular} & \begin{tabular}{c}
                 Dimension and  \\
                symmetry for $k>1$  \\ at infinite coupling
            \end{tabular} & \begin{tabular}{c}
                Hasse diagram  \\
                finite coupling
            \end{tabular}  &  \begin{tabular}{c}
                Hasse diagram  \\
                infinite coupling
            \end{tabular}  \\ \midrule
             $E_{4}$ & \begin{tabular}{c}
                 $2k^2+k$  \\
                 $\mathfrak{so}(4k+2)$
             \end{tabular}& \begin{tabular}{c}
                 $2k^2+k+1$  \\
                 $\mathfrak{so}(4k+2) \oplus \mathfrak{u}(1)$
             \end{tabular}  & \scalebox{.879}{\raisebox{-.5\height}{\begin{tikzpicture}
		\tikzstyle{hasse} = [circle, fill,inner sep=2pt];
		\node at (0,.3) {};\node at (0,-2.7) {};
		\node (1) [hasse] at (0,0) {};
		\node (2) [hasse] at (0,-.6) {};
		\node (3) [hasse] at (0,-1.2) {};
		\node (4) at (0,-1.4) {$\vdots$};
		\node (5) [hasse] at (0,-1.8) {};
		\node (6) [hasse] at (0,-2.4) {};
		\draw (1) edge [] node[label=right:$d_3$] {} (2);
		\draw (2) edge [] node[label=right:$d_{5}$] {} (3);
		\draw (5) edge [] node[label=right:$d_{2k+1}$] {} (6);
	         \end{tikzpicture}}}     & \scalebox{.879}{\raisebox{-.5\height}{\begin{tikzpicture}
		\tikzstyle{hasse} = [circle, fill,inner sep=2pt];
		\node at (0,.3) {};\node at (0,-2.7) {};
		\node (1) [hasse] at (0,0) {};
		\node (2) [hasse] at (0,-.6) {};
		\node (3) [hasse] at (0,-1.2) {};
		\node (4) at (0,-1.4) {$\vdots$};
		\node (5) [hasse] at (0,-1.8) {};
		\node (6) [hasse] at (0,-2.4) {};
		\draw (1) edge [] node[label=right:$e_4$] {} (2);
		\draw (2) edge [] node[label=right:$d_{5}$] {} (3);
		\draw (5) edge [] node[label=right:$d_{2k+1}$] {} (6);
	         \end{tikzpicture}}}
	           \\
             $E_3$ & \begin{tabular}{c}
                 $2k^2-k$  \\
                 $\mathfrak{so}(4k)$
             \end{tabular} & \begin{tabular}{c}
                 $2k^2-k+1$  \\
                 $\mathfrak{so}(4k) \oplus \mathfrak{u}(1)$
             \end{tabular} & \scalebox{.879}{\raisebox{-.5\height}{\begin{tikzpicture}
		\tikzstyle{hasse} = [circle, fill,inner sep=2pt];
		\node at (0,.3) {};
		\node at (0,-2.7) {};
		\node (1b) [hasse] at (-.6,0) {};
		\node (1) [hasse] at (0,0) {};
		\node (2) [hasse] at (0,-.6) {};
		\node (3) [hasse] at (0,-1.2) {};
		\node (4) at (0,-1.4) {$\vdots$};
		\node (5) [hasse] at (0,-1.8) {};
		\node (6) [hasse] at (0,-2.4) {};
		\draw (1) edge [] node[label=right:$a_1$] {} (2);
		\draw (2) edge [] node[label=right:$d_{4}$] {} (3);
		\draw (5) edge [] node[label=right:$d_{2k}$] {} (6);
		\draw (1b) edge [] node[label=left:$A_1$] {} (2);
	         \end{tikzpicture}}}   & \scalebox{.879}{\raisebox{-.5\height}{\begin{tikzpicture}
		\tikzstyle{hasse} = [circle, fill,inner sep=2pt];
		\node at (0,.3) {};
		\node at (0,-2.7) {};
		\node (1b) [hasse] at (-.6,0) {};
		\node (1) [hasse] at (0,.6) {};
		\node (2) [hasse] at (0,-.6) {};
		\node (3) [hasse] at (0,-1.2) {};
		\node (4) at (0,-1.4) {$\vdots$};
		\node (5) [hasse] at (0,-1.8) {};
		\node (6) [hasse] at (0,-2.4) {};
		\draw (1) edge [] node[label=right:$a_2$] {} (2);
		\draw (2) edge [] node[label=right:$d_{4}$] {} (3);
		\draw (5) edge [] node[label=right:$d_{2k}$] {} (6);
		\draw (1b) edge [] node[label=left:$A_1$] {} (2);
	         \end{tikzpicture}}} 
	           \\ 
	           \begin{tabular}{c}
	               $E_{4-2l}$ \\
	                $\scriptstyle{k\geq l > 0}$
	           \end{tabular} & \begin{tabular}{c}
                 $(k-l)(2k-2l+1)$  \\
                 $\mathfrak{so}(4k-4l+2)$
             \end{tabular}  & \begin{tabular}{c}
                 $(k-l)(2k-2l+1)+1$  \\
                 $\mathfrak{so}(4k-4l+2) \oplus \mathfrak{u}(1)$
             \end{tabular}   & \scalebox{.879}{\raisebox{-.5\height}{\begin{tikzpicture}
		\tikzstyle{hasse} = [circle, fill,inner sep=2pt];
		\node at (0,.3) {};\node at (0,-2.7) {};
		\node (1) at (0,0) {};
		\node (2) [hasse] at (0,-.6) {};
		\node (3) [hasse] at (0,-1.2) {};
		\node (4) at (0,-1.4) {$\vdots$};
		\node (5) [hasse] at (0,-1.8) {};
		\node (6) [hasse] at (0,-2.4) {};
		\draw (2) edge [] node[label=right:$d_{3}$] {} (3);
		\draw (5) edge [] node[label=right:$d_{2k-2l+1}$] {} (6);
	         \end{tikzpicture}}}    & \scalebox{.879}{\raisebox{-.5\height}{\begin{tikzpicture}
		\tikzstyle{hasse} = [circle, fill,inner sep=2pt];
		\node at (0,.3) {};\node at (0,-2.7) {};
		\node (1) [hasse] at (0,0) {};
		\node (2) [hasse] at (0,-.6) {};
		\node (3) [hasse] at (0,-1.2) {};
		\node (4) at (0,-1.4) {$\vdots$};
		\node (5) [hasse] at (0,-1.8) {};
		\node (6) [hasse] at (0,-2.4) {};
		\draw (1) edge [] node[label=right:$A_{l}$] {} (2);
		\draw (2) edge [] node[label=right:$d_{3}$] {} (3);
		\draw (5) edge [] node[label=right:$d_{2k-2l+1}$] {} (6);
	         \end{tikzpicture}}} 
	           \\ \begin{tabular}{c}
	               $E_{3-2l}$ \\
	                $\scriptstyle{k\geq l > 0}$
	           \end{tabular}   & \begin{tabular}{c}
                 $(k-l)(2k-2l-1)$\\
                 $\mathfrak{so}(4k-4l)$
             \end{tabular}  & \begin{tabular}{c}
                 $(k-l)(2k-2l-1)+1$\\
                 $\mathfrak{so}(4k-4l) \oplus \mathfrak{u}(1)$
             \end{tabular} & 
              \scalebox{.879}{\raisebox{-.5\height}{\begin{tikzpicture}
		\tikzstyle{hasse} = [circle, fill,inner sep=2pt];
		\node at (0,.3) {};
		\node at (0,-2.7) {};
		\node (0) at (0,.6) {};
		\node (1b) [hasse] at (-.6,0) {};
		\node (1) [hasse] at (0,0) {};
		\node (2) [hasse] at (0,-.6) {};
		\node (3) [hasse] at (0,-1.2) {};
		\node (4) at (0,-1.4) {$\vdots$};
		\node (5) [hasse] at (0,-1.8) {};
		\node (6) [hasse] at (0,-2.4) {};
		\draw (1) edge [] node[label=right:$A_1$] {} (2);
		\draw (2) edge [] node[label=right:$d_{4}$] {} (3);
		\draw (5) edge [] node[label=right:$d_{2k-2l}$] {} (6);
		\draw (1b) edge [] node[label=left:$A_1$] {} (2);
	         \end{tikzpicture}}}  & 
              \scalebox{.879}{\raisebox{-.5\height}{\begin{tikzpicture}
		\tikzstyle{hasse} = [circle, fill,inner sep=2pt];
		\node at (0,.3) {};
		\node at (0,-2.7) {};
		\node (0) [hasse] at (0,.6) {};
		\node (1b) [hasse] at (-.6,0) {};
		\node (1) [hasse] at (0,0) {};
		\node (2) [hasse] at (0,-.6) {};
		\node (3) [hasse] at (0,-1.2) {};
		\node (4) at (0,-1.4) {$\vdots$};
		\node (5) [hasse] at (0,-1.8) {};
		\node (6) [hasse] at (0,-2.4) {};
		\draw (0) edge [] node[label=right:$A_l$] {} (1);
		\draw (1) edge [] node[label=right:$A_1$] {} (2);
		\draw (2) edge [] node[label=right:$d_{4}$] {} (3);
		\draw (5) edge [] node[label=right:$d_{2k-2l}$] {} (6);
		\draw (1b) edge [] node[label=left:$A_1$] {} (2);
	         \end{tikzpicture}}}
	           \\  \bottomrule 
         \end{tabular}
         \caption{ Coulomb branch quaternionic dimension, global symmetry and Hasse diagram for the orthogonal exceptional families, both at finite and infinite gauge coupling. The Hasse diagrams are further detailed in Section \ref{sec:Hasse}. Note the following changes between finite and infinite coupling. The quaternionic dimension changes by one. The global symmetry at finite coupling does not include the $U(1)_I$ factor since it is realized by the gaugino bilinear which lives in a nilpotent supermultiplet. At infinite coupling, it ceases to be nilpotent and becomes part of the geometry. As a consequence the rank of the global symmetry of the Higgs branch variety increases by one. The Hasse diagrams show, in some cases, a bifurcation; as detailed in Section \ref{sec:Hasse}, the O5 brane construction studied in Sections \ref{sec:magQuiv_known_cases} and \ref{sec:Spk} only renders the part of the Hasse diagram depicted using vertical lines. The $\mathrm{O7}^-$ brane construction studied in Section \ref{sec:O7_plane} produces the entire Hasse diagrams. In the $E_3$ case, the bifurcation at the top of the finite coupling diagram could equivalently be denoted as $d_2$, and the bifurcation at the top of the infinite coupling diagram could equivalently be denoted as $e_3$, so that the modification at infinite coupling can be seen as $d_2 \rightarrow e_3$.    } 
         \label{tab:CompareFiniteHasse2}
     \end{table}

\begin{table}[t]
         \centering
         \vspace*{1cm}\hspace*{-2cm}\begin{tabular}{c|c|c} \toprule
           Family  & HWG  at finite coupling & HWG at infinite coupling \\  
\midrule 
         $E_8$ &     $\sum\limits_{i=1}^{k} \mu_{2i} t^{2i}$ & \begin{tabular}{c}
	       $\left( \sum\limits_{i=1}^{k+2} \mu_{2i} t^{2i} \right) + t^4  + \mu_{2k+6} (t^{k+1}+t^{k+3})$
	         \end{tabular}   \\    [7ex]
             $E_7$ &  $\sum\limits_{i=1}^{k} \mu_{2i} t^{2i}$ & 
	         \begin{tabular}{c}
	            $\left( \sum\limits_{i=1}^{k+1} \mu_{2i} t^{2i} \right) + \nu^2 t^2 + t^4 $ \\ $+ \nu \mu_{2k+4} (t^{k+1}+t^{k+3})   + \mu_{2k+4}^2 t^{2k+4} - \nu^2 \mu_{2k+4}^2  t^{2k+6}$
	         \end{tabular}  \\    [7ex]
             $E_6$ &  $\sum\limits_{i=1}^{k} \mu_{2i} t^{2i}$ &
	         \begin{tabular}{c}
	          $\left( \sum\limits_{i=1}^{k} \mu_{2i} t^{2i} \right) + t^2   + (\mu_{2k+2} q + \mu_{2k+3} q^{-1} )t^{k+1}$
	         \end{tabular}
	           \\  [7ex]
             $E_5$ &  $\sum\limits_{i=1}^{k} \mu_{2i} t^{2i}$  &
	        \begin{tabular}{c}
	           $\left( \sum\limits_{i=1}^{k} \mu_{2i} t^{2i} \right) + t^2  + ( q + q^{-1} )\mu_{2k+2} t^{k+1}$ 
	        \end{tabular}    \\ [7ex]
	 $E_{4}$  & 	         
	             $\left(\sum\limits_{i=1}^{k-1} \mu_{2i} t^{2i} \right) +\mu_{2k}\mu_{2k+1} t^{2k}$   
	            & 
	         \begin{tabular}{c}
	             $\left( \sum\limits_{i=1}^{k-1} \mu_{2i} t^{2i} \right) + t^2   + ( q\mu_{2k} + q^{-1}\mu_{2k+1} ) t^{k+1}  $ \\ $ +\mu_{2k}\mu_{2k+1} (t^{2k} {-} t^{2k+2})$ 
	         \end{tabular}
	           \\ [7ex]
             $E_3$ &  \begin{tabular}{c}
	              $ \left( \sum\limits_{i=1}^{k-1} \mu_{2i}t^{2i}\right) +(\mu_{2k-1}^2 + \mu_{2k}^2)t^{2k} $ \\ $ -\mu_{2k-1}^2\mu_{2k}^2t^{4k}$ 
	         \end{tabular}  &
 \begin{tabular}{c}
	              $\left(\sum\limits_{i=1}^{k-1} \mu_{2i}t^{2i}\right)+t^2   +(q+\dfrac{1}{q})\mu_{2k}t^{k+1} +\mu_{2k}^2t^{2k} $ \\ $ -\mu_{2k}^2t^{2k+2}$ 
	         \end{tabular}
	           \\  [7ex]
	           \begin{tabular}{c}
	               $E_{4-2l}$ \\
	               $\scriptstyle{k\geq l\geq0}$
	           \end{tabular} & $\left(\sum\limits_{i=1}^{k-l-1} \mu_{2i} t^{2i} \right) +\mu_{2k-2l}\mu_{2k-2l+1} t^{2k-2l}$   &
	         	         \begin{tabular}{c}
	              $\left( \sum\limits_{i=1}^{k-l-1} \mu_{2i} t^{2i} \right) + t^2     + ( q\mu_{2k-2l} + \frac{1}{q}\mu_{2k-2l+1} ) t^{k+1}  $ \\ $  +\mu_{2k-2l}\mu_{2k-2l+1} (t^{2k-2l} {-} t^{2k+2}) 
	             $ 
	         \end{tabular}
	           \\  [7ex] \begin{tabular}{c}
	               $E_{3-2l}$ \\
	               $\scriptstyle{k\geq l\geq0}$
	           \end{tabular}  & \begin{tabular}{c}
	              $ \left( \sum\limits_{i=1}^{k-l-1} \mu_{2i}t^{2i}\right) +(\mu_{2k-2l-1}^2 + \mu_{2k-2l}^2)t^{2k-2l}  $ \\ $   -\mu_{2k-2l-1}^2\mu_{2k-2l}^2t^{4k-2l}$ 
	         \end{tabular}   &  \begin{tabular}{c}
	              $\left(\sum\limits_{i=1}^{k-l-1} \mu_{2i}t^{2i}\right)+t^2   +(q+\dfrac{1}{q})\mu_{2k-2l}t^{k+1}  $ \\ $  +\mu_{2k-2l}^2 (t^{2k-2l} {-} t^{2k+2})$ 
	         \end{tabular}
	           \\  \bottomrule 
         \end{tabular} 
         \caption{Highest weight generating function (HWG) \cite{Hanany:2014dia,zhenghaoTropical} 
for the orthogonal exceptional families. The highest weight fugacities are 
assigned as follows: $\mu_i$ for $\mathfrak{so}$, $\nu$ for $\mathfrak{su}(2)$, 
and $q$ for $\mathfrak{u}(1)$. The global symmetry in each case can be read in 
Tables \ref{tab:CompareFiniteHasse} and \ref{tab:CompareFiniteHasse2}. Moreover, 
$t$ denotes the $\surm(2)_R$ fugacity along the 3d $\Ncal=4$ Coulomb branch. 
} 
         \vspace*{1cm}
         \label{tab:CompareFiniteHWG}
     \end{table}

\paragraph{Outline.}
The remainder of this paper is organised as follows: after reviewing the set-up, Section \ref{sec:magQuiv_known_cases} is devoted to a study of the Higgs branches for $\sprm(1)$ with $N_f<8$ flavours. In each case, the brane web  and the Higgs branch degrees of freedom are detailed. Building on that, the rules to read off the magnetic quivers are established.
Thereafter, the magnetic quiver proposal is applied to $5$d $\sprm(k)$ theories with fundamental matter in Section \ref{sec:Spk}; in each case, the 5-brane webs are detailed and the magnetic quivers are derived.
The results for $\sprm(k)$ theories are studied from an alternative O$7^-$ construction in Section \ref{sec:O7_plane}, and contrasted to the results of Section \ref{sec:Spk}. 
Having derived the magnetic quivers for finite and infinite coupling Higgs branch in a variety of cases, Section \ref{sec:Hasse} details the derivation of the associated Hasse diagrams.
In Section \ref{sec:quiver_theories}, a family of linear orthosymplectic quiver gauge theories is studied and the multitude of different infinite coupling phases is discussed. 
Lastly, Section \ref{sec:conclusion} provides a conclusion and outlook.
Appendix \ref{app:background} provides background material on general Type II 
brane configurations with 8 supercharges, 5-brane webs, and $3$d $\Ncal=4$ 
Coulomb branches, as well as a brief summary of \cite{Bourget:2020xdz}.
%
%
\section{Magnetic quivers and known examples}
\label{sec:magQuiv_known_cases}
\subsection{Set-up}
5-dimensional $\Ncal=1$ gauge theories can be constructed as low-energy 
effective theories for 5-brane webs in Type IIB superstring theory 
\cite{Aharony:1997ju,Aharony:1997bh,DeWolfe:1999hj}. The different branes occupy 
the space-time dimensions as summarised in Table \ref{tab:brane_setup}. Aiming 
for symplectic and orthogonal gauge symmetries, one may utilise orientifold 
planes 
\cite{Evans:1997hk,Landsteiner:1997vd}, in particular \Of\ planes 
\cite{Brunner:1997gk,Bergman:2015dpa,Zafrir:2015ftn,Hayashi:2015vhy} for the 
purposes of this work. As displayed in Table \ref{tab:orientifold}, the effect 
is twofold: the gauge symmetry along 5-brane parallel to the \Of\ is projected 
to an orthogonal or symplectic algebra; while the flavour symmetry originating 
from transversal 7-branes is also projected to ortho-symplectic algebras, but in 
the opposite way. 
Further details on brane webs are summarised in Appendix \ref{app:background}.

\begin{table}[t]
\begin{center}
\begin{tabular}{c|cccccccccc}
\toprule
Type IIB & $x^0$ & $x^1$ & $x^2$ & $x^3$ & $x^4$ & $x^5$ & $x^6$ & $x^7$ & $x^8$ & $x^9$\\
\midrule
\NS & $\times$ & $\times$ & $\times$ & $\times$ & $\times$ & $\times$ & & & &  \\
\Dfive & $\times$ & $\times$ & $\times$ & $\times$ & $\times$ & & $\times$ & & & \\
\Of & $\times$ & $\times$ & $\times$ & $\times$ & $\times$ & & $\times$ & & & \\
$(p,q)$ 5-brane & $\times$ & $\times$ & $\times$ & $\times$ & $\times$ & \multicolumn{2}{c}{angle $\alpha$} & & & \\
$[p,q]$ 7-brane & $\times$ & $\times$ & $\times$ & $\times$ & $\times$ & & & $\times$ & $\times$ & $\times$ \\
\bottomrule
\end{tabular}
\caption{Type IIB  5-brane web set-up: $\times$ indicates the space-time directions spanned by the various branes and the orientifold plane. A $(p,q)$ 5-brane is a line of slope $\tan(\alpha)=q \tau_2/(p+q\tau_1)$ in the $x^{5,6}$ plane where the axiodilaton is $\tau = \tau_1 + i \tau_2$. In this paper all the brane webs are drawn for the value $\tau = i$, so that $\tan(\alpha)=q/p$.}
\label{tab:brane_setup}
\end{center}
\end{table}

\begin{table}[t]
    \centering
    \begin{tabular}{c|c|c|c|c}
    \toprule
        Orientifold & Drawing & Gauge algebra & Flavour algebra & Charge \\ \midrule 
        \Opmt & \begin{tikzpicture}
    \OMinusTilde{0,0}{1,0}
    \end{tikzpicture} & $B_k$ & $C_n$ & $\frac{1}{2}-2^{p-5}$ \\
        \Opp & \begin{tikzpicture}
    \OPlus{0,0}{1,0}
    \end{tikzpicture} & $C_k$ & $D_n$ & $2^{p-5}$ \\
        \Oppt & \begin{tikzpicture}
    \OPlusTilde{0,0}{1,0}
    \end{tikzpicture} & $C_k$ & $D_n$ & $2^{p-5}$  \\
        \Opm & & $D_k$ & $C_n$  & $-2^{p-5}$ \\
        \bottomrule
    \end{tabular}
    \caption{Summary of orientifolds. The gauge algebra is indicated for a stack of $k$ half \Dp\ branes on a \Op\ orientifold plane. Whereas the flavour algebra is provided for a stack of $n$ half \Dpp\ branes perpendicular to orientifold.}
    \label{tab:orientifold}
\end{table}

\subsection{Magnetic quivers}
A 5-brane web, where every external $(p,q)$ 5-brane ends on a $[p,q]$ 7-brane, in the presence of an \Of\ plane has various phases: for instance, the pure Coulomb branch phase, the pure Higgs branch phase, as well as mixed phases. 
As customary, the brane web in the massive Coulomb branch phase is reached by 
placing gauge and flavour \Dfive\ branes in various positions along the $x^5$ 
direction, i.e.\ assigning VEVs to scalars in the gauge multiplets as well as 
assigning masses to the fundamental flavours. This phase is convenient for 
reading off the \emph{electric gauge theory} description via the suspension 
pattern of fundamental strings.

The pure Higgs branch phase at finite gauge coupling can be entered in two steps: 
Firstly, the flavour \Dfive\ branes are moved to coincide on the orientifold as 
well, i.e.\ all mass parameters are set to zero. 
Secondly, all gauge \Dfive\ branes are aligned on the \Of\ plane, i.e.\ the 
VEVs of the $5$d gauge multiplet are tuned to zero.
This results in a brane system which represents the origin of the Coulomb branch where Higgs branch directions can open up.
For this, it is important to recall that the half $[1,0]$ 7-branes, on which the flavour \Dfive s end, merge with their mirrors on the orientifold plane, such that the resulting physical 7-brane can be split along the \Of\ plane. The S-rule then determines how many branes are connected between each half 7-brane.
The Higgs branch degrees of freedom are realised by independent 5-brane subwebs that are free to move along the 7-branes in directions $x^7$, $x^8$, $x^9$. Each such subweb contributes one quaternionic dimension to the total Higgs branch. 

Besides the classical Higgs branch, meaning the Higgs branch of the low-energy 
effective gauge theory at finite gauge coupling, there exists also a Higgs 
branch at the RG-fixed point of the theory. For the purposes of this paper, the 
fixed point is always described by a $5$d $\Ncal=1$ SCFT and is further referred 
to as infinite (gauge) coupling phase. In terms of the 5-brane web, this point 
is reached when all external 5-branes meet at single point.
The changes between the Higgs branch at finite and infinite coupling have been 
first emphasised in 
\cite{Seiberg:1996bd} and further elaborated in \cite{Cremonesi:2015lsa,
Zafrir:2015ftn,Ferlito:2017xdq}. While \cite{Zafrir:2015ftn} computed the 
change in Higgs branch dimension via 5-brane webs, it has been demonstrated in 
\cite{Cremonesi:2015lsa,Ferlito:2017xdq} that infinite coupling Higgs branches 
can be described by Coulomb branches of $3$d $\Ncal=4$ gauge theories. Building 
on that, a systematic procedure on deriving the \emph{magnetic quiver} for 
any phase of a given 5-brane web has been developed in \cite{Cabrera:2018jxt}. 
The magnetic quiver proposal \cite[Conj.\ 1]{Cabrera:2018jxt} can be summarised 
as follows:
\begin{compactenum}[(i)]
    \item \textbf{Quiver:} Find all inequivalent maximal subdivisions $\{\mathcal{S}_i\}$ of a given 5-brane web which is suspended between 7-branes. Whether a decomposition is compatible with supersymmetry can be determined by the S-rule \cite{Hanany:1996ie}, see also \cite{Benini:2009gi}. Each subdivision $\mathcal{S}_i$ gives rise to one magnetic quiver with the following data:
    \item \textbf{Gauge groups:} The different subwebs in a given subdivision $\mathcal{S}_i$ are associated to magnetic gauge groups, wherein the number of identical subwebs determines the rank.
    \item \textbf{Matter:} The magnetic hypermultiplets  between two magnetic gauge groups are derived from the generalised intersection number of two different subwebs. This number is composed of intersection number of lines in the $x^{5,6}$ plane plus a modification if the subwebs end on common 7-branes. 
\end{compactenum}
Consequently, the Higgs branch associated to given phase of a 5-brane web is described via
\begin{align}
    \Higgs = \bigcup_{ \{\mathcal{S}_i\} } \Coulomb \left( \mathsf{Q} (\mathcal{S}_i)\right)
\end{align}
such that the moduli space is generically a union of symplectic singularities, one for each subdivison.

In order to derive magnetic quivers for 5-brane webs with \Of\  planes, the major conceptual challenge lies in the treatment of the orientifolds. In more detail, the T-dual set-up of \cite{Hanany:1996ie}, involving \Dthree-\Dfive-\NS\ brane configurations, can accommodate for \Ot\ planes \cite{Feng:2000eq} rather easily, because S-duality maps one \Ot\ into another \Ot\ plane. However, the same is not true for \Of\ planes, as  S-duality, for instance, relates an ON plane with a \Of\ plane \cite{Sen:1996na}, see also \cite{Hanany:2000fq}. Therefore, as emphasised in all magnetic quiver constructions \cite{Hanany:1996ie,Cabrera:2018jxt,Cabrera:2019izd,Cabrera:2019dob}, S-duality is neither correct nor necessary to change between electric and magnetic theory. Consequently, the proposal of \cite{Cabrera:2019dob} is inspired from the \Ot\ planes and involves \emph{magnetic orientifolds} as summarise in Table \ref{tab:magnetic_orientifold}. 
In other words, besides moving the 5-brane web into the Higgs branch phase, one also needs to convert the electric orientifolds into their magnetic counterparts in order to derive the magnetic quivers.

\begin{table}[t]
    \centering
    \begin{tabular}{c|c||c|c}
    \toprule
        \multirow{2}{*}{Orientifold} & Electric & Magnetic & Magnetic\\ 
         & algebra & orientifold & algebra \\\midrule 
        \Opp &  $C_k$ & \Opmt& $B_k$  \\
        \Opm & $D_k$ & \Opm & $D_k$  \\
        \Oppt & $C_k$ & \Oppt & $C_k$ \\
        \Opmt &  $B_k$ & \Opp & $C_k$ \\
        \bottomrule
    \end{tabular}
    \caption{Proposal for magnetic orientifolds \cite{Cabrera:2019dob}. A stack 
of $k$ physical \Dfive\ branes on top of a \Of\ plane, while being suspended 
between \NS\ branes, has a 5-dimensional low-energy description given by a gauge 
field for the electric algebra. In contrast, $k$ identical subwebs suspended 
between 7-branes on top of a magnetic orientifold \Of\ contribute to the 
magnetic quiver as gauge multiplet for the magnetic algebra.}
    \label{tab:magnetic_orientifold}
\end{table}

Before exploring potential predictions from magnetic quivers for 5-branes with 
orientifolds, one needs to derive a proposal that reproduces known cases. In 
Section \ref{sec:SU2}, the case of $5$d $\Ncal =1$ $\su\cong \sprm(1)$ gauge 
theory with $0\leq N_f <8 $ fundamental flavours is tested against the known 
symmetry enhancement at infinite coupling \cite{Seiberg:1996bd,Morrison:1996xf} 
and against the magnetic quiver construction for 5-brane webs without 
orientifolds \cite{Cabrera:2018jxt}. 
Thereafter, a precise proposal is formulated in Section \ref{sec:rules}.
One is then in the position to analyse various $5$d brane web constructions with this technique. In particular, Section \ref{sec:Spk} is focused on the general case of 5d $\Ncal =1$ $ \sprm(k)$ gauge theory with $N_f$ fundamental flavours. As a proof of concept, a class of 5d $\Ncal=1$ linear quiver gauge theories is analysed in Section \ref{sec:quiver_theories}.
%
%
\subsection{Single Sp(1) gauge group}
\label{sec:SU2}
In this section, a 5d $\Ncal=1$ $\sprm(1)$ gauge theory with $ 0\leq N_f <8$ fundamental flavours is considered. The 5-brane web is presented, from which the magnetic quiver is derived at finite and infinite gauge coupling.
A crucial consistency check is given by reproducing the enlargement of the 
Higgs branch from the minimal nilpotent orbit closure of $\sormL(2N_f)$ at 
finite coupling to the minimal nilpotent orbit closure of $E_{N_f+1}$ at the UV 
fixed point.
As detailed below, the finite coupling magnetic quivers agree with the results 
of \cite{Feng:2000eq}.
\subsubsection{\texorpdfstring{$E_8$}{E8}: Sp(1) with 7 flavours}
For the first case, consider $\sprm(1)$ with $7$ flavours. The brane web in the Coulomb branch phase with massive flavours is given by
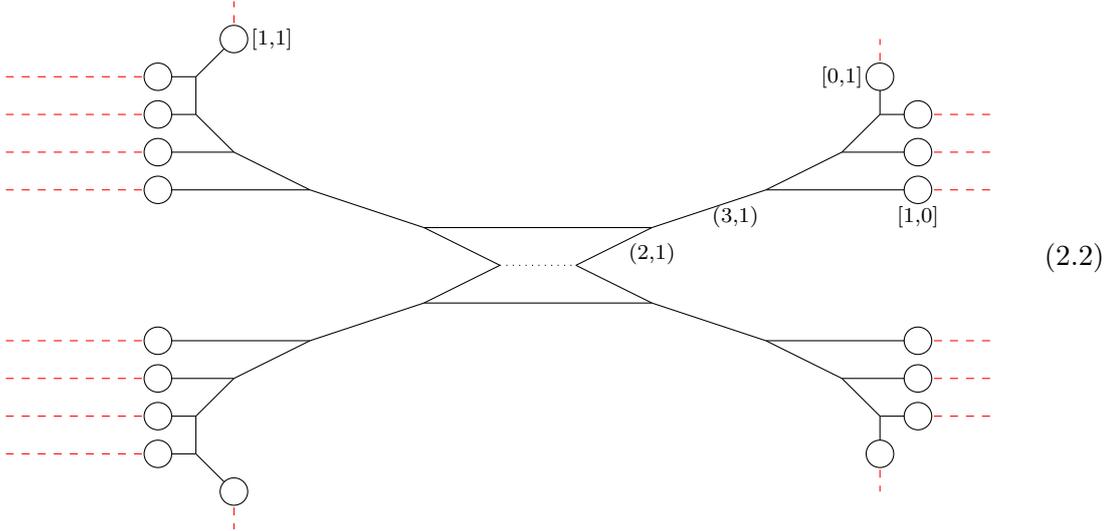
\begin{align}
\raisebox{-.5\height}{
    \begin{tikzpicture}
    \OPlus{0,0}{1,0}
    \Dbrane{0,0}{-1,0.5}
    \Dbrane{1,0}{2,0.5}
    \draw (2,0.15) node {$\scriptstyle{(2,1)}$};
    \Dbrane{-1,0.5}{2,0.5}
    \Dbrane{-1,0.5}{-2.5,1}
    \Dbrane{2,0.5}{3.5,1}
    \draw (3.1,0.65) node {$\scriptstyle{(3,1)}$};
    \Dbrane{-2.5,1}{-4.5,1}
    \Dbrane{3.5,1}{5.5,1}
    \draw (5.5,0.65) node {$\scriptstyle{[1,0]}$};
    \Dbrane{-2.5,1}{-3.5,1.5}
    \Dbrane{3.5,1}{4.5,1.5}
    \Dbrane{-3.5,1.5}{-4.5,1.5}
    \Dbrane{4.5,1.5}{5.5,1.5}
    \Dbrane{-3.5,1.5}{-4,2}
    \Dbrane{4.5,1.5}{5,2}
    \Dbrane{-4,2}{-4.5,2}
    \Dbrane{5,2}{5.5,2}
    \Dbrane{-4,2}{-4,2.5}
    \Dbrane{5,2}{5,2.5}
    \draw (4.5,2.5) node {$\scriptstyle{[0,1]}$};
    \Dbrane{-4,2.5}{-4.5,2.5}
    \Dbrane{-4,2.5}{-3.5,3}
    \draw (-3,3) node {$\scriptstyle{[1,1]}$};
    \MonoCut{-4.5,2.5}{-6.5,2.5}
    \MonoCut{-4.5,2}{-6.5,2}
    \MonoCut{-4.5,1.5}{-6.5,1.5}
    \MonoCut{-4.5,1}{-6.5,1}
    \MonoCut{5.5,1}{6.5,1}
    \MonoCut{5.5,1.5}{6.5,1.5}
    \MonoCut{5.5,2}{6.5,2}
    \MonoCut{-3.5,3}{-3.5,3.5}
    \MonoCut{5,2.5}{5,3}
    \SevenB{-3.5,3}
    \SevenB{5,2.5}
    \SevenB{-4.5,2.5}
    \SevenB{-4.5,2}
    \SevenB{-4.5,1.5}
    \SevenB{-4.5,1}
    \SevenB{5.5,2}
    \SevenB{5.5,1.5}
    \SevenB{5.5,1}
    \begin{scope}[yscale=-1,xscale=1]
    \Dbrane{0,0}{-1,0.5}
    \Dbrane{1,0}{2,0.5}
    \Dbrane{-1,0.5}{2,0.5}
    \Dbrane{-1,0.5}{-2.5,1}
    \Dbrane{2,0.5}{3.5,1}
    \Dbrane{-2.5,1}{-4.5,1}
    \Dbrane{3.5,1}{5.5,1}
    \Dbrane{-2.5,1}{-3.5,1.5}
    \Dbrane{3.5,1}{4.5,1.5}
    \Dbrane{-3.5,1.5}{-4.5,1.5}
    \Dbrane{4.5,1.5}{5.5,1.5}
    \Dbrane{-3.5,1.5}{-4,2}
    \Dbrane{4.5,1.5}{5,2}
    \Dbrane{-4,2}{-4.5,2}
    \Dbrane{5,2}{5.5,2}
    \Dbrane{-4,2}{-4,2.5}
    \Dbrane{5,2}{5,2.5}
    \Dbrane{-4,2.5}{-4.5,2.5}
    \Dbrane{-4,2.5}{-3.5,3}
    \MonoCut{-4.5,2.5}{-6.5,2.5}
    \MonoCut{-4.5,2}{-6.5,2}
    \MonoCut{-4.5,1.5}{-6.5,1.5}
    \MonoCut{-4.5,1}{-6.5,1}
    \MonoCut{5.5,1}{6.5,1}
    \MonoCut{5.5,1.5}{6.5,1.5}
    \MonoCut{5.5,2}{6.5,2}
    \MonoCut{-3.5,3}{-3.5,3.5}
    \MonoCut{5,2.5}{5,3}
    \SevenB{-3.5,3}
    \SevenB{5,2.5}
    \SevenB{-4.5,2.5}
    \SevenB{-4.5,2}
    \SevenB{-4.5,1.5}
    \SevenB{-4.5,1}
    \SevenB{5.5,2}
    \SevenB{5.5,1.5}
    \SevenB{5.5,1}
    \end{scope}
    \end{tikzpicture}
}
\label{eq:web_Sp1_7_Coulomb}
\end{align}
where the red dashed lines indicate the half monodromy cuts associated to each 
half 7-brane. The orientifold notation is summarised in Table 
\ref{tab:orientifold}. Moreover, the distance between the flavour 5-branes and 
the orientifold gives the bare mass.
The transition into the Higgs branch, where all masses are set to zero, is achieved by two steps: firstly, aligning the flavour 7-branes and the gauge $(1,0)$ 5-brane on the orientifold. Secondly, the half 7-branes merge with their mirror images such that the physical 7-branes can be split along the \Of\ plane.

The novelty, compared to the classical 5-brane web construction with $N_f<7$, is that one of the external non-flavour branes bends inside, i.e.\ the $(1,1)$ 5-brane. Consequently, there are two intermediate representations of the Higgs branch phase. On the one hand, the non-flavour branes are kept separated such that the brane web \eqref{eq:web_Sp1_7_Coulomb} turns into
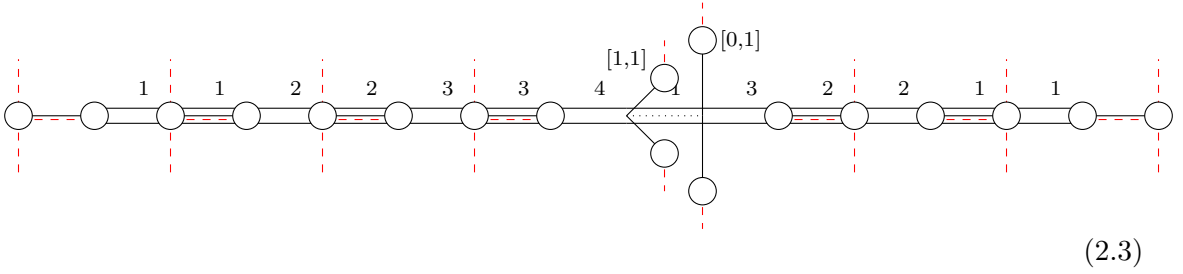
\begin{align}
\raisebox{-.5\height}{
    \begin{tikzpicture}
    \DfiveOPlus{1}{0}
    \DfiveOMinus{4}{-1}
    \DfiveOMinus{3}{1}
    \DfiveOMinusTilde{3}{-2}
    \MonoCut{-2,-0.05}{-1,-0.05}
    \DfiveOMinusTilde{2}{2}
    \MonoCut{2,-0.05}{3,-0.05}
    \DfiveOMinus{3}{-3}
    \DfiveOMinus{2}{3}
    \DfiveOMinusTilde{2}{-4}
    \MonoCut{-4,-0.05}{-3,-0.05}
    \DfiveOMinusTilde{1}{4}
    \MonoCut{4,-0.05}{5,-0.05}
    \DfiveOMinus{2}{-5}
    \DfiveOMinusTilde{1}{-6}
    \MonoCut{-6,-0.05}{-5,-0.05}
    \DfiveOMinus{1}{-7}
    \DfiveOMinus{1}{5}
    \OMinusTilde{6,0}{7,0}
    \MonoCut{6,-0.05}{7,-0.05}
    \OMinusTilde{-8,0}{-7,0}
    \MonoCut{-8,-0.05}{-7,-0.05}
    \Dbrane{1,0}{1,1}
    \draw (1.5,1) node {$\scriptstyle{[0,1]}$};
    \Dbrane{1,0}{1,-1}
    \Dbrane{0,0}{0.5,0.5}
    \draw (0,0.75) node {$\scriptstyle{[1,1]}$};
    \Dbrane{0,0}{0.5,-0.5}
    \MonoCut{-2,-0.75}{-2,0.75}
    \MonoCut{-4,-0.75}{-4,0.75}
    \MonoCut{-6,-0.75}{-6,0.75}
    \MonoCut{-8,-0.75}{-8,0.75}
    \MonoCut{3,-0.75}{3,0.75}
    \MonoCut{5,-0.75}{5,0.75}
    \MonoCut{7,-0.75}{7,0.75}
    \MonoCut{0.5,0.5}{0.5,1.0}
    \MonoCut{1,1}{1,1.5}
    \MonoCut{0.5,-0.5}{0.5,-1.0}
    \MonoCut{1,-1}{1,-1.5}
    \SevenB{0.5,0.5}
    \SevenB{1,1}
    \SevenB{0.5,-0.5}
    \SevenB{1,-1}
    \SevenB{-1,0}
    \SevenB{-2,0}
    \SevenB{-3,0}
    \SevenB{-4,0}
    \SevenB{-5,0}
    \SevenB{-6,0}
    \SevenB{-7,0}
    \SevenB{-8,0}
    \SevenB{2,0}
    \SevenB{3,0}
    \SevenB{4,0}
    \SevenB{5,0}
    \SevenB{6,0}
    \SevenB{7,0}
    \end{tikzpicture}
}
\label{eq:web_Sp1_7_Higgs1}
\end{align}
The numbers displayed above the branes count physical $(1,0)$ 5-branes 
on top of \Of\ planes, see Table \ref{tab:orientifold} for 
conventions.

On the other hand, the $(1,0)$ and $(1,1)$ 5-branes can cross such that the brane configuration becomes
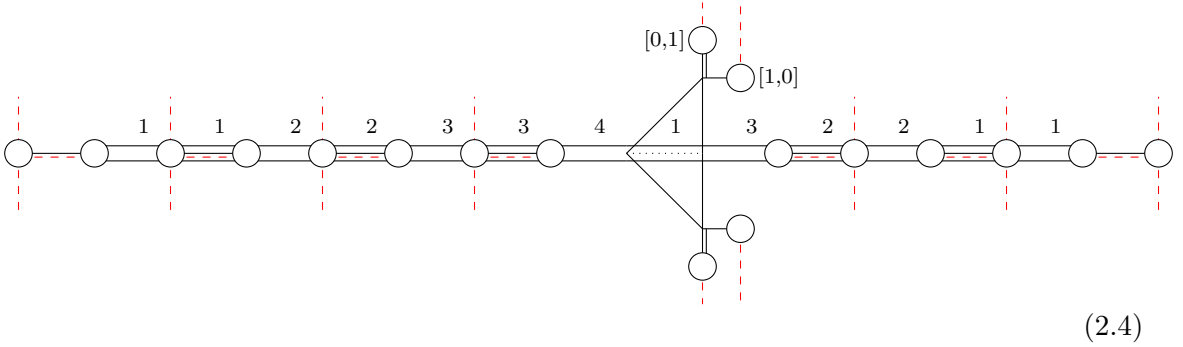
\begin{align}
 \raisebox{-.5\height}{
    \begin{tikzpicture}
    \DfiveOPlus{1}{0}
    \DfiveOMinus{4}{-1}
    \DfiveOMinus{3}{1}
    \DfiveOMinusTilde{3}{-2}
    \MonoCut{-2,-0.05}{-1,-0.05}
    \DfiveOMinusTilde{2}{2}
    \MonoCut{2,-0.05}{3,-0.05}
    \DfiveOMinus{3}{-3}
    \DfiveOMinus{2}{3}
    \DfiveOMinusTilde{2}{-4}
    \MonoCut{-4,-0.05}{-3,-0.05}
    \DfiveOMinusTilde{1}{4}
    \MonoCut{4,-0.05}{5,-0.05}
    \DfiveOMinus{2}{-5}
    \DfiveOMinusTilde{1}{-6}
    \MonoCut{-6,-0.05}{-5,-0.05}
    \DfiveOMinus{1}{-7}
    \DfiveOMinus{1}{5}
    \OMinusTilde{6,0}{7,0}
    \MonoCut{6,-0.05}{7,-0.05}
    \OMinusTilde{-8,0}{-7,0}
    \MonoCut{-8,-0.05}{-7,-0.05}
    \Dbrane{1,0}{1,1.5}
    \Dbrane{1.05,1}{1.05,1.5}
    \draw (0.5,1.5) node {$\scriptstyle{[0,1]}$};
    \Dbrane{1,0}{1,-1.5}
    \Dbrane{1.05,-1}{1.05,-1.5}
    \Dbrane{0,0}{1,1}
    \Dbrane{0,0}{1,-1}
    \Dbrane{1,1}{1.5,1}
    \draw (2,1) node {$\scriptstyle{[1,0]}$};
    \Dbrane{1,-1}{1.5,-1}
    \MonoCut{-2,-0.75}{-2,0.75}
    \MonoCut{-4,-0.75}{-4,0.75}
    \MonoCut{-6,-0.75}{-6,0.75}
    \MonoCut{-8,-0.75}{-8,0.75}
    \MonoCut{3,-0.75}{3,0.75}
    \MonoCut{5,-0.75}{5,0.75}
    \MonoCut{7,-0.75}{7,0.75}
    \MonoCut{1,1.5}{1,2}
    \MonoCut{1,-1.5}{1,-2}
    \MonoCut{1.5,1}{1.5,2}
    \MonoCut{1.5,-1}{1.5,-2}
    \SevenB{1,1.5}
    \SevenB{1.5,1}
    \SevenB{1,-1.5}
    \SevenB{1.5,-1}
    \SevenB{-1,0}
    \SevenB{-2,0}
    \SevenB{-3,0}
    \SevenB{-4,0}
    \SevenB{-5,0}
    \SevenB{-6,0}
    \SevenB{-7,0}
    \SevenB{-8,0}
    \SevenB{2,0}
    \SevenB{3,0}
    \SevenB{4,0}
    \SevenB{5,0}
    \SevenB{6,0}
    \SevenB{7,0}
    \end{tikzpicture}
}
\label{eq:web_Sp1_7_Higgs2}
\end{align}
where one accounts for brane creation, see Appendix \ref{app:background}.
Both, \eqref{eq:web_Sp1_7_Higgs1} and \eqref{eq:web_Sp1_7_Higgs2}, are useful starting points as shown below.
\paragraph{Finite coupling.}
To analyse the finite coupling Higgs branch, one considers \eqref{eq:web_Sp1_7_Higgs1} and notices that several branes are non-dynamical as they serve to connect the $(0,1)$ and $(1,1)$ 5-brane with several half 7-branes in order to render the configuration supersymmetric. The number of these branes can be minimised by transitioning the $(0,1)$ and $(1,1)$ through several half 7-branes such that brane annihilation leads to the following configuration:
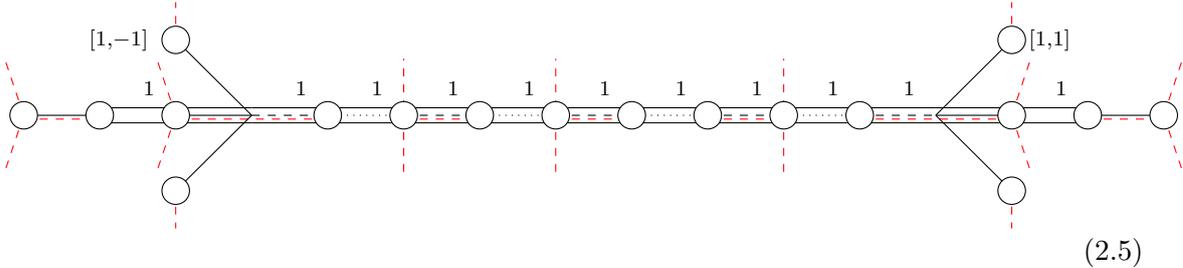
\begin{align}
 \raisebox{-.5\height}{
    \begin{tikzpicture}
    \DfiveOPlus{1}{0}
    \DfiveOPlusTilde{1}{-1}
    \DfiveOPlusTilde{1}{1}
    \DfiveOPlus{1}{-2}
    \MonoCut{-1,-0.05}{0,-0.05}
    \DfiveOPlus{1}{2}
    \MonoCut{1,-0.05}{2,-0.05}
    \DfiveOPlusTilde{1}{-3}
    \DfiveOPlusTilde{1}{3}
    \DfiveOPlus{1}{-4}
    \MonoCut{-3,-0.05}{-2,-0.05}
    \DfiveOMinusTilde{}{4}
    \MonoCut{3,-0.05}{5,-0.05}
    \DfiveOPlusTilde{1}{-5}
    \DfiveOMinusTilde{}{-6}
    \MonoCut{-6,-0.05}{-4,-0.05}
    \DfiveOMinus{1}{-7}
    \DfiveOMinus{1}{5}
    \OMinusTilde{6,0}{7,0}
    \MonoCut{6,-0.05}{7,-0.05}
    \OMinusTilde{-8,0}{-7,0}
    \MonoCut{-8,-0.05}{-7,-0.05}
    \Dbrane{4,0}{5,1}
    \draw (5.5,1) node {$\scriptstyle{[1,1]}$};
    \Dbrane{4,0}{5,-1}
    \Dbrane{-5,0}{-6,1}
    \draw (-6.75,1) node {$\scriptstyle{[1,-1]}$};
    \Dbrane{-5,0}{-6,-1}
    \MonoCut{-1,-0.75}{-1,0.75}
    \MonoCut{-3,-0.75}{-3,0.75}
    \MonoCut{-6,0}{-6-0.25,0.75}
    \MonoCut{-6,0}{-6-0.25,-0.75}
    \MonoCut{-8,0}{-8-0.25,0.75}
    \MonoCut{-8,0}{-8-0.25,-0.75}
    \MonoCut{2,-0.75}{2,0.75}
    \MonoCut{5,0}{5+0.25,0.75}
    \MonoCut{5,0}{5+0.25,-0.75}
    \MonoCut{7,0}{7+0.25,0.75}
    \MonoCut{7,0}{7+0.25,-0.75}
    \MonoCut{-6,1}{-6,1.5}
    \MonoCut{-6,-1}{-6,-1.5}
    \MonoCut{5,1}{5,1.5}
    \MonoCut{5,-1}{5,-1.5}
    \SevenB{-6,1}
    \SevenB{5,1}
    \SevenB{-6,-1}
    \SevenB{5,-1}
    \SevenB{0,0}
    \SevenB{-1,0}
    \SevenB{-2,0}
    \SevenB{-3,0}
    \SevenB{-4,0}
    \SevenB{-6,0}
    \SevenB{-7,0}
    \SevenB{-8,0}
    \SevenB{1,0}
    \SevenB{2,0}
    \SevenB{3,0}
    \SevenB{5,0}
    \SevenB{6,0}
    \SevenB{7,0}
    \end{tikzpicture}
}
\label{eq:web_Sp1_7_finite}
\end{align}
where the monodromy cuts have been chosen for unambiguous presentation. 

Starting from \eqref{eq:web_Sp1_7_finite}, the magnetic quiver is read off as 
follows: to begin with, the orientifolds are converted into magnetic 
orientifolds, see Table \ref{tab:magnetic_orientifold}. Then one needs to find 
all maximal subdivisions: in this case, there is only one. Between the half 
7-branes along the orientifold plane, all physical \Dfive\ branes constitute 
independent subwebs. Note that there are no two identical subwebs, because the 
\Dfive s are suspended between different 7-branes. Consequently, all magnetic 
gauge groups associated to these branes have rank one because each subweb 
contributes one quaternionic dimension to the Higgs branch. The character of the 
magnetic gauge multiplet is deduced from the magnetic orientifold between two 
adjacent $[1,0]$ 7-branes. 
The remaining parts that need to be understood are the $[1,\pm1]$ 5-branes on the left and right-hand-side. To be specific focus on the left-hand-side, one could transition the $(1,-1)$ 5-brane through one more half 7-brane to the left without brane creation or annihilation. Due to the effects of the monodromy line, one would end up with a $(2,-1)$ 5-brane, which is known to be a supersymmetric configuration by itself, see \eqref{eq:bending_O5-_O5+}. 
Therefore, the original $(1,-1)$ 5-brane is also supersymmetric, but, in 
contrast to all other 5-branes, the $(1,-1)$ 5-brane is not an independent 
subweb. In particular, this means that both the $(1,\pm1)$ branes are 
non-dynamical as these cannot be moved along the $x^{7,8,9}$ directions. As such 
these pieces of the brane web contribute as magnetic $\balg_0$ 
flavour nodes only. Analogous to \cite[Eq.\ (2.66)]{Cabrera:2019dob}, the 
$\balg_0$ denotes a single half-hypermultiplet for a $\calg_k$ magnetic gauge 
node.
As a remark, for reading off the magnetic gauge algebra, it is convenient to have the free consistent configuration on top of \Ofpt\ and \Ofmt\ planes, because the magnetic orientifolds of Table \ref{tab:magnetic_orientifold} imply an unambiguous magnetic algebra in contrast to a configuration on top of \Ofp\ and \Ofm\ planes.

This discussion reveals the magnetic gauge multiplets and background 
gauge multiplets. In addition, there is matter derived from the generalised 
intersection number \cite{Cabrera:2018jxt}. To be explicit, consider two 
adjacent \Dfive\ branes connecting a common half 7-brane from the left and 
right-hand-side, respectively. Then the intersection number between the two 
\Dfive s is zero, but there is a positive contribution because they end on the 
common 7-brane from opposite sides. The generalised intersection number turns 
out to be one. Alternatively, one can simply consider \Dthree\ branes suspended 
between the two subwebs and recognise the bifundamental hypermultiplet  between 
the gauge groups, see also \cite{Cabrera:2019dob}. Lastly, consider the 
$(1,\pm1)$ branes together with the closest \Dfive\ brane. Here, one computes 
the intersection number between a $(1,\pm1)$ and a $(1,0)$ 5-brane, which is 
simply one. Since there are no common 7-branes, the generalised intersection 
number is one too, indicating a simple magnetic fundamental hypermultiplet 
between the flavour and gauge node.

Collecting all the pieces, the magnetic quiver becomes
\begin{align}
 \raisebox{-.5\height}{
 	\begin{tikzpicture}
 	\tikzset{node distance = 1cm};
	\tikzstyle{gauge} = [circle, draw,inner sep=2.5pt];
	\tikzstyle{flavour} = [regular polygon,regular polygon sides=4,inner 
sep=2.5pt, draw];
	\node (g1) [gauge,label=below:{\dd{1}}] {};
	\node (g2) [gauge,right of=g1,label=below:{\cc{1}}] {};
	\node (g3) [gauge,right of=g2,label=below:{\bb{1}}] {};
	\node (g4) [gauge,right of=g3,label=below:{\cc{1}}] {};
	\node (g5) [gauge,right of=g4,label=below:{\bb{1}}] {};
	\node (g6) [gauge,right of=g5,label=below:{\cc{1}}] {};
	\node (g7) [gauge,right of=g6,label=below:{\bb{1}}] {};
	\node (g8) [gauge,right of=g7,label=below:{\cc{1}}] {};
	\node (g9) [gauge,right of=g8,label=below:{\bb{1}}] {};
	\node (g10) [gauge,right of=g9,label=below:{\cc{1}}] {};
	\node (g11) [gauge,right of=g10,label=below:{\dd{1}}] {};
	\node (f2) [flavour,above of=g2,label=above:{\bb{0}}] {};
	\node (f10) [flavour,above of=g10,label=above:{\bb{0}}] {};
	\draw (g1)--(g2) (g2)--(g3) (g3)--(g4) (g4)--(g5) (g5)--(g6) (g6)--(g7) (g7)--(g8) (g8)--(g9) (g9)--(g10) (g10)--(g11) (g2)--(f2) (g10)--(f10);
	\end{tikzpicture}
	} 
\label{eq:magQuiv_Sp1_7_finite}
\end{align}
such that Coulomb branch dimension and global symmetry, see Appendix \ref{app:Coulomb_branch}, are computed to be
\begin{subequations}
\label{eq:results_Sp1_7_finite}
\begin{align}
\dim_{\HH} \Coulomb \eqref{eq:magQuiv_Sp1_7_finite} = 11
\: , \qquad 
G = \sorm(14) \,.
\end{align}
Upon choosing the magnetic gauge groups suitably, the full Coulomb branch 
geometry is known \cite{Benini:2010uu,Chacaltana:2012zy,Cabrera:2017ucb}
\begin{align}
    \Coulomb\eqref{eq:magQuiv_Sp1_7_finite} = \clorbit_{\dalg_7}^{\min } \,,
\end{align}
\end{subequations}
i.e.\ the closure of the minimal nilpotent orbit of $\sormL(14)$.
The details of the assignment of the magnetic gauge group are the subject of the 
companion paper \cite{Bourget:2020xdz}, see also Appendix 
\ref{app:companion}.
The properties \eqref{eq:results_Sp1_7_finite} match the classical Higgs branch 
as well as the non-abelian part of the global symmetry.

The alert reader will notice that the $\uo_I$ symmetry is missing from the 
classical global symmetry computed from the magnetic quiver. This is because this global symmetry is associated with the gaugino bilinear. Being a nilpotent operator in the chiral ring, the gaugino bilinear does not enter the geometric part of the classical moduli space, called \emph{Higgs variety} in \cite{Bourget:2019rtl}. The current status of magnetic quivers is not sensitive to nilpotent elements, i.e.\ it is does not compute the whole \emph{Higgs Scheme}. If nilpotent operators are present in the chiral ring, the Higgs branch is a non-reduced scheme, and the magnetic quiver computes its geometric reduction. See \cite{Cremonesi:2015lsa} for a discussion of the gaugino biliniear and \cite{Bourget:2019rtl} for a detailed analysis of the difference between the Higgs variety and the Higgs scheme.
\paragraph{Infinite coupling.}
For the brane web at infinite coupling, the configuration \eqref{eq:web_Sp1_7_Higgs2} is convenient. Here, moving the infinite coupling is realised by moving the half $[1,0]$ 7-brane, which is away from the orientifold, onto the \Of\ plane. Then it merges with its mirror image and the resulting physical 7-brane can be split along the orientifold. As a consequence, the brane web becomes
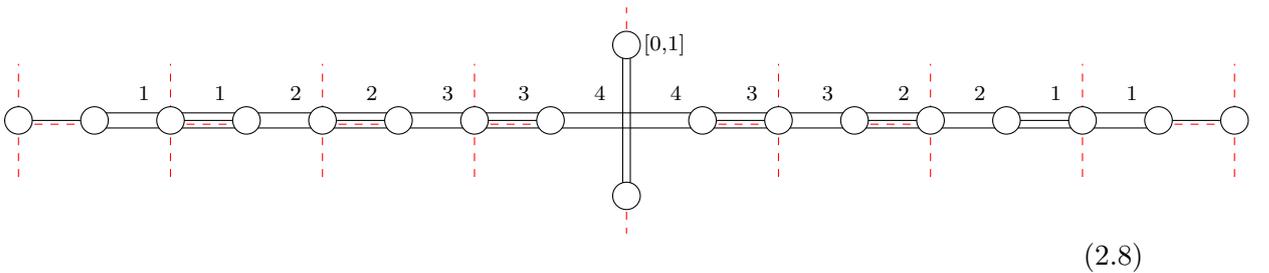
\begin{align}
 \raisebox{-.5\height}{
    \begin{tikzpicture}
    \DfiveOMinus{4}{0}
    \DfiveOMinus{4}{-1}
    \DfiveOMinusTilde{3}{1}
    \MonoCut{1,-0.05}{2,-0.05}
    \DfiveOMinus{3}{2}
    \DfiveOMinusTilde{3}{-2}
    \MonoCut{-2,-0.05}{-1,-0.05}
    \DfiveOMinusTilde{2}{3}
    \MonoCut{3,-0.05}{4,-0.05}
    \DfiveOMinus{3}{-3}
    \DfiveOMinus{2}{4}
    \DfiveOMinusTilde{2}{-4}
    \MonoCut{-4,-0.05}{-3,-0.05}
    \DfiveOMinusTilde{1}{5}
    \DfiveOMinus{2}{-5}
    \DfiveOMinusTilde{1}{-6}
    \MonoCut{-6,-0.05}{-5,-0.05}
    \DfiveOMinus{1}{-7}
    \DfiveOMinus{1}{6}
    \OMinusTilde{7,0}{8,0}
    \MonoCut{7,-0.05}{8,-0.05}
    \OMinusTilde{-8,0}{-7,0}
    \MonoCut{-8,-0.05}{-7,-0.05}
    \Dbrane{0.05,-1}{0.05,1}
    \draw (0.5,1) node {$\scriptstyle{[0,1]}$};
    \Dbrane{-0.05,-1}{-0.05,1}
    \MonoCut{-2,-0.75}{-2,0.75}
    \MonoCut{-4,-0.75}{-4,0.75}
    \MonoCut{-6,-0.75}{-6,0.75}
    \MonoCut{-8,-0.75}{-8,0.75}
    \MonoCut{2,-0.75}{2,0.75}
    \MonoCut{4,-0.75}{4,0.75}
    \MonoCut{6,-0.75}{6,0.75}
    \MonoCut{8,-0.75}{8,0.75}
    \MonoCut{0,1}{0,1.5}
    \MonoCut{0,-1}{0,-1.5}
    \SevenB{0,1}
    \SevenB{0,-1}
    \SevenB{-1,0}
    \SevenB{-2,0}
    \SevenB{-3,0}
    \SevenB{-4,0}
    \SevenB{-5,0}
    \SevenB{-6,0}
    \SevenB{-7,0}
    \SevenB{-8,0}
    \SevenB{1,0}
    \SevenB{2,0}
    \SevenB{3,0}
    \SevenB{4,0}
    \SevenB{5,0}
    \SevenB{6,0}
    \SevenB{7,0}
    \SevenB{8,0}
    \end{tikzpicture}
}
\label{eq:web_Sp1_7_infinite}
\end{align}
and, again, the numbers denote physical \Dfive\ branes in between half 7-branes.
The magnetic quiver is derived as follows: inspecting \eqref{eq:web_Sp1_7_infinite} shows that there is only one maximal subdivision. Therein, all subwebs are given either by \Dfive\ branes suspended between adjacent 7-branes along the orientifold or by \NS\ branes suspended between a half 7-brane and its mirror image away from the \Of\ plane. 
There are two new features compared to the finite coupling case: firstly, there 
are multiple copies of identical subwebs. The number of physical \Dfive\ branes 
between two adjacent half 7-branes coincides with the number of identical webs, 
and, therefore, yields the rank of the magnetic gauge group after transition to 
the magnetic orientifold, see Table \ref{tab:magnetic_orientifold}. Secondly, 
the \NS\ branes also form one subweb, but the orientifold projections now acts 
differently. Since the magnetic orientifold for the \NS s is a \Ofm\ plane, the 
\NS\ branes contribute a $\calg_1$ magnetic gauge group.

The determination of the matter content proceeds as above: the stacks of 
adjacent \Dfive\ brane subwebs have generalised intersection number of one; 
while the \NS\ branes and the \Dfive\ branes in the central segment have 
intersection number one with no additional corrections from common 7-branes. 
Therefore, all magnetic gauge nodes are connected by bifundamental magnetic 
hypermultilets. Alternatively, the same conclusion is reached by inspecting the 
suspension pattern of \Dthree\ branes, analogous to the 6d setting of 
\cite{Cabrera:2019dob}.

Consequently, the magnetic quiver associated to \eqref{eq:web_Sp1_7_infinite} is 
given by
\begin{align}
 \raisebox{-.5\height}{
 	\begin{tikzpicture}
 	\tikzset{node distance = 1cm};
	\tikzstyle{gauge} = [circle, draw,inner sep=2.5pt];
	\tikzstyle{flavour} = [regular polygon,regular polygon sides=4,inner 
sep=2.5pt, draw];
	\node (g1) [gauge,label=below:{\dd{1}}] {};
	\node (g2) [gauge,right of=g1,label=below:{\cc{1}}] {};
	\node (g3) [gauge,right of=g2,label=below:{\dd{2}}] {};
	\node (g4) [gauge,right of=g3,label=below:{\cc{2}}] {};
	\node (g5) [gauge,right of=g4,label=below:{\dd{3}}] {};
	\node (g6) [gauge,right of=g5,label=below:{\cc{3}}] {};
	\node (g7) [gauge,right of=g6,label=below:{\dd{4}}] {};
	\node (g8) [gauge,right of=g7,label=below:{\cc{3}}] {};
	\node (g9) [gauge,right of=g8,label=below:{\dd{3}}] {};
	\node (g10) [gauge,right of=g9,label=below:{\cc{2}}] {};
	\node (g11) [gauge,right of=g10,label=below:{\dd{2}}] {};
	\node (g12) [gauge,right of=g11,label=below:{\cc{1}}] {};
	\node (g13) [gauge,right of=g12,label=below:{\dd{1}}] {};
	\node (g14) [gauge,above of=g7,label=above:{\cc{1}}] {};
	\draw (g1)--(g2) (g2)--(g3) (g3)--(g4) (g4)--(g5) (g5)--(g6) (g6)--(g7) (g7)--(g8) (g8)--(g9) (g9)--(g10) (g10)--(g11) (g11)--(g12) (g12)--(g13) (g7)--(g14);
	\end{tikzpicture}
	} 
	\label{eq:magQuiv_Sp1_7_infinite}
\end{align}
with a Coulomb branch dimension of 
\begin{align}
    \dim_{\HH} \Coulomb \eqref{eq:magQuiv_Sp1_7_infinite} = 29 \,.
\end{align}
It has been concluded in \cite[Eq.\ (2.43)]{Hanany:2018uhm} and \cite{Zhenghao} 
that the Coulomb branch of \eqref{eq:web_Sp1_7_infinite} is the minimal 
nilpotent orbit of $E_8$. However, this result relies on clarifying what 
magnetic gauge group has to be assigned to \eqref{eq:magQuiv_Sp1_7_infinite}, 
which is elaborated on in a companion paper \cite{Bourget:2020xdz} see 
Appendix \ref{app:companion} for a summary).

Comparing to class $\mathcal{S}$ technology, one can decompose the 
star-shaped quiver \eqref{eq:magQuiv_Sp1_7_infinite} into three 
$D_4$ punctures associated with the following linear quivers: 
$T_{(1^8)}[\sorm(8)] $, $T_{(1^8)}[\sorm(8)] $, and $T_{(5,3)}[\sorm(8)] $. As 
demonstrated in \cite[Sec.\ 3.2.2]{Chacaltana:2011ze}, the fixture obtained via 
gluing gives rise to the so-called $E_8$ SCFT, which yields further validation 
to the approach of this paper.
\subsubsection{\texorpdfstring{$E_7$}{E7}: Sp(1) with 6 flavours}
Considering $\sprm(1)$ with $6$ fundamental flavours, the brane web in the Coulomb branch phase with massive flavours is given by
\begin{align}
\raisebox{-.5\height}{
    \begin{tikzpicture}
    \OPlus{0,0}{1,0}
    \Dbrane{0,0}{-1,0.5}
    \Dbrane{1,0}{2,0.5}
    \draw (2,0.15) node {$\scriptstyle{(2,1)}$};
    \Dbrane{-1,0.5}{2,0.5}
    \Dbrane{-1,0.5}{-2.5,1}
    \Dbrane{2,0.5}{3.5,1}
    \draw (3.1,0.65) node {$\scriptstyle{(3,1)}$};
    \Dbrane{-2.5,1}{-4.5,1}
    \Dbrane{3.5,1}{5.5,1}
    \draw (5.5,0.65) node {$\scriptstyle{[1,0]}$};
    \Dbrane{-2.5,1}{-3.5,1.5}
    \Dbrane{3.5,1}{4.5,1.5}
    \Dbrane{-3.5,1.5}{-4.5,1.5}
    \Dbrane{4.5,1.5}{5.5,1.5}
    \Dbrane{-3.5,1.5}{-4,2}
    \Dbrane{4.5,1.5}{5,2}
    \Dbrane{-4,2}{-4.5,2}
    \Dbrane{5,2}{5.5,2}
    \Dbrane{-4,2}{-4,2.5}
    \Dbrane{5,2}{5,2.5}
    \draw (4.5,2.5) node {$\scriptstyle{[0,1]}$};
    \MonoCut{-4.5,2}{-6.5,2}
    \MonoCut{-4.5,1.5}{-6.5,1.5}
    \MonoCut{-4.5,1}{-6.5,1}
    \MonoCut{5.5,1}{6.5,1}
    \MonoCut{5.5,1.5}{6.5,1.5}
    \MonoCut{5.5,2}{6.5,2}
    \MonoCut{-4,2.5}{-4,3}
    \MonoCut{5,2.5}{5,3}
    \SevenB{-4,2.5}
    \SevenB{5,2.5}
    \SevenB{-4.5,2}
    \SevenB{-4.5,1.5}
    \SevenB{-4.5,1}
    \SevenB{5.5,2}
    \SevenB{5.5,1.5}
    \SevenB{5.5,1}
    \begin{scope}[yscale=-1,xscale=1]
    \Dbrane{0,0}{-1,0.5}
    \Dbrane{1,0}{2,0.5}
    \Dbrane{-1,0.5}{2,0.5}
    \Dbrane{-1,0.5}{-2.5,1}
    \Dbrane{2,0.5}{3.5,1}
    \Dbrane{-2.5,1}{-4.5,1}
    \Dbrane{3.5,1}{5.5,1}
    \Dbrane{-2.5,1}{-3.5,1.5}
    \Dbrane{3.5,1}{4.5,1.5}
    \Dbrane{-3.5,1.5}{-4.5,1.5}
    \Dbrane{4.5,1.5}{5.5,1.5}
    \Dbrane{-3.5,1.5}{-4,2}
    \Dbrane{4.5,1.5}{5,2}
    \Dbrane{-4,2}{-4.5,2}
    \Dbrane{5,2}{5.5,2}
    \Dbrane{-4,2}{-4,2.5}
    \Dbrane{5,2}{5,2.5}
    \MonoCut{-4.5,2}{-6.5,2}
    \MonoCut{-4.5,1.5}{-6.5,1.5}
    \MonoCut{-4.5,1}{-6.5,1}
    \MonoCut{5.5,1}{6.5,1}
    \MonoCut{5.5,1.5}{6.5,1.5}
    \MonoCut{5.5,2}{6.5,2}
    \MonoCut{-4,2.5}{-4,3}
    \MonoCut{5,2.5}{5,3}
    \SevenB{-4,2.5}
    \SevenB{5,2.5}
    \SevenB{-4.5,2}
    \SevenB{-4.5,1.5}
    \SevenB{-4.5,1}
    \SevenB{5.5,2}
    \SevenB{5.5,1.5}
    \SevenB{5.5,1}
    \end{scope}
    \end{tikzpicture}
}
\label{eq:web_Sp1_6_Coulomb}
\end{align}
where the semi-infinite flavour 5-branes are set to end on half 7-branes.
Next, the Higgs branch phase is reached by moving the half $[1,0]$ 7-branes and the $(1,0)$ gauge 5-brane onto the \Of\ plane. The half 7-branes merge with their mirror images such that the resulting physical 7-brane can be split along the orientifold.  The brane web becomes 
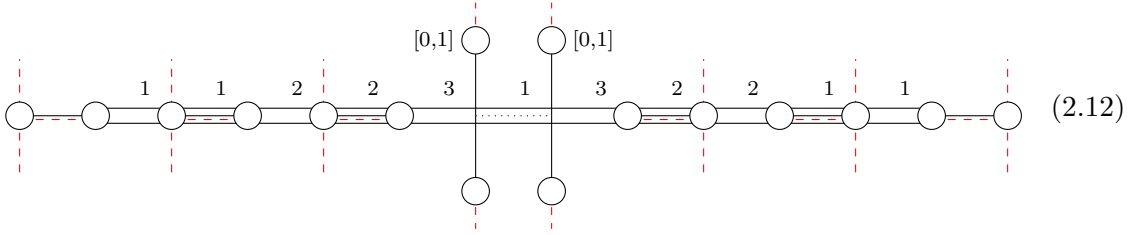
\begin{align}
\raisebox{-.5\height}{
    \begin{tikzpicture}
    \DfiveOPlus{1}{0}
    \DfiveOMinus{3}{-1}
    \DfiveOMinus{3}{1}
    \DfiveOMinusTilde{2}{-2}
    \MonoCut{-2,-0.05}{-1,-0.05}
    \DfiveOMinusTilde{2}{2}
    \MonoCut{2,-0.05}{3,-0.05}
    \DfiveOMinus{2}{-3}
    \DfiveOMinus{2}{3}
    \DfiveOMinusTilde{1}{-4}
    \MonoCut{-4,-0.05}{-3,-0.05}
    \DfiveOMinusTilde{1}{4}
    \MonoCut{4,-0.05}{5,-0.05}
    \DfiveOMinus{1}{-5}
    \DfiveOMinus{1}{5}
    \OMinusTilde{6,0}{7,0}
    \MonoCut{6,-0.05}{7,-0.05}
    \OMinusTilde{-6,0}{-5,0}
    \MonoCut{-6,-0.05}{-5,-0.05}
    \Dbrane{0,0}{0,1}
    \draw (-0.55,1) node {$\scriptstyle{[0,1]}$};
    \Dbrane{0,0}{0,-1}
    \Dbrane{1,0}{1,1}
    \draw (1.55,1) node {$\scriptstyle{[0,1]}$};
    \Dbrane{1,0}{1,-1}
    \MonoCut{-2,-0.75}{-2,0.75}
    \MonoCut{-4,-0.75}{-4,0.75}
    \MonoCut{-6,-0.75}{-6,0.75}
    \MonoCut{3,-0.75}{3,0.75}
    \MonoCut{5,-0.75}{5,0.75}
    \MonoCut{7,-0.75}{7,0.75}
    \MonoCut{0,1}{0,1.5}
    \MonoCut{0,-1}{0,-1.5}
    \MonoCut{1,1}{1,1.5}
    \MonoCut{1,-1}{1,-1.5}
    \SevenB{0,1}
    \SevenB{1,1}
    \SevenB{0,-1}
    \SevenB{1,-1}
    \SevenB{-1,0}
    \SevenB{-2,0}
    \SevenB{-3,0}
    \SevenB{-4,0}
    \SevenB{-5,0}
    \SevenB{-6,0}
    \SevenB{2,0}
    \SevenB{3,0}
    \SevenB{4,0}
    \SevenB{5,0}
    \SevenB{6,0}
    \SevenB{7,0}
    \end{tikzpicture}
}
\label{eq:web_Sp1_6_Higgs}
\end{align}
and the S-rule implies that the two $(0,1)$ 5-branes need to be connected to more than one 7-brane  via $(1,0)$ 5-branes.
\paragraph{Finite coupling.}
The non-dynamical parts of the brane web \eqref{eq:web_Sp1_6_Higgs} can be minimised by transitioning the $(0,1)$ 5-branes to the left or right, respectively. Accounting for brane annihilation leads to 
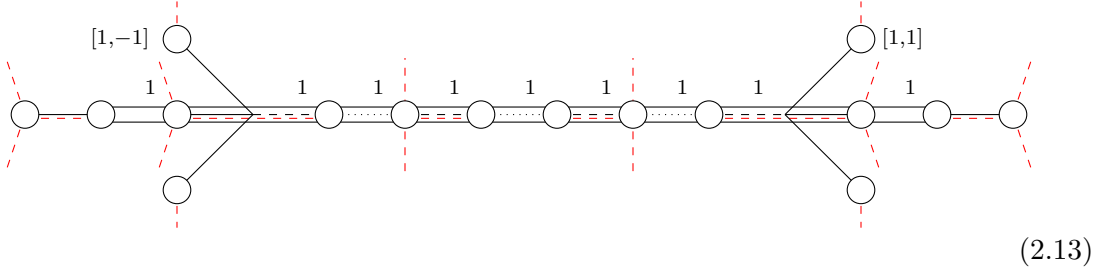
\begin{align}
\raisebox{-.5\height}{
    \begin{tikzpicture}
    \DfiveOPlus{1}{0}
    \DfiveOPlusTilde{1}{-1}
    \DfiveOPlusTilde{1}{1}
    \DfiveOPlus{1}{-2}
    \MonoCut{-1,-0.05}{0,-0.05}
    \DfiveOPlus{1}{2}
    \MonoCut{1,-0.05}{2,-0.05}
    \DfiveOPlusTilde{1}{-3}
    \DfiveOPlusTilde{1}{3}
    \DfiveOMinusTilde{}{-4}
    \MonoCut{-4,-0.05}{-2,-0.05}
    \DfiveOMinusTilde{}{4}
    \MonoCut{3,-0.05}{5,-0.05}
    \DfiveOMinus{1}{-5}
    \DfiveOMinus{1}{5}
    \OMinusTilde{6,0}{7,0}
    \MonoCut{6,-0.05}{7,-0.05}
    \OMinusTilde{-6,0}{-5,0}
    \MonoCut{-6,-0.05}{-5,-0.05}
    \Dbrane{-3,0}{-4,1}
    \draw (-4.75,1) node {$\scriptstyle{[1,-1]}$};
    \Dbrane{-3,0}{-4,-1}
    \Dbrane{4,0}{5,1}
    \draw (5.55,1) node {$\scriptstyle{[1,1]}$};
    \Dbrane{4,0}{5,-1}
    \MonoCut{-1,-0.75}{-1,0.75}
    \MonoCut{-4,0}{-4-0.25,0.75}
    \MonoCut{-4,0}{-4-0.25,-0.75}
    \MonoCut{-6,0}{-6-0.25,0.75}
    \MonoCut{-6,0}{-6-0.25,-0.75}
    \MonoCut{2,-0.75}{2,0.75}
    \MonoCut{5,0}{5+0.25,0.75}
    \MonoCut{5,0}{5+0.25,-0.75}
    \MonoCut{7,0}{7+0.25,0.75}
    \MonoCut{7,0}{7+0.25,-0.75}
    \MonoCut{-4,1}{-4,1.5}
    \MonoCut{-4,-1}{-4,-1.5}
    \MonoCut{5,1}{5,1.5}
    \MonoCut{5,-1}{5,-1.5}
    \SevenB{-4,1}
    \SevenB{5,1}
    \SevenB{-4,-1}
    \SevenB{5,-1}
    \SevenB{0,0}
    \SevenB{-1,0}
    \SevenB{-2,0}
    \SevenB{-4,0}
    \SevenB{-5,0}
    \SevenB{-6,0}
    \SevenB{1,0}
    \SevenB{2,0}
    \SevenB{3,0}
    \SevenB{5,0}
    \SevenB{6,0}
    \SevenB{7,0}
    \end{tikzpicture}
}
\label{eq:web_Sp1_6_finite}
\end{align}
where the monodromy cuts have been chosen for the sake of a clear presentation.
The only maximal subdivision is composed of subwebs that are \Dfive\ branes suspended between adjacent half 7-branes. As above, the $(1,\pm1)$ 5-branes between \Ofpt\ and \Ofmt\ are supersymmetric configurations by themselves. However, these branes are non-dynamical, as they are not part of any subweb in a maximal subdivision of \eqref{eq:web_Sp1_6_finite}. 
The magnetic quiver associated to \eqref{eq:web_Sp1_6_finite} is read off by the same logic as above; hence, one finds
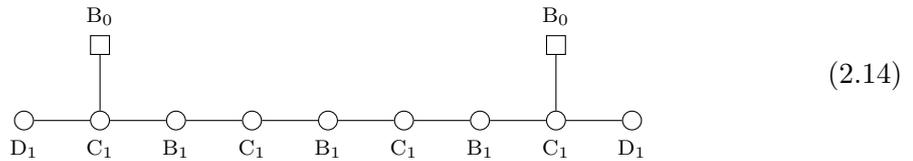
\begin{align}
 \raisebox{-.5\height}{
 	\begin{tikzpicture}
 	\tikzset{node distance = 1cm};
	\tikzstyle{gauge} = [circle, draw,inner sep=2.5pt];
	\tikzstyle{flavour} = [regular polygon,regular polygon sides=4,inner 
sep=2.5pt, draw];
	\node (g1) [gauge,label=below:{\dd{1}}] {};
	\node (g2) [gauge,right of=g1,label=below:{\cc{1}}] {};
	\node (g3) [gauge,right of=g2,label=below:{\bb{1}}] {};
	\node (g4) [gauge,right of=g3,label=below:{\cc{1}}] {};
	\node (g5) [gauge,right of=g4,label=below:{\bb{1}}] {};
	\node (g6) [gauge,right of=g5,label=below:{\cc{1}}] {};
	\node (g7) [gauge,right of=g6,label=below:{\bb{1}}] {};
	\node (g8) [gauge,right of=g7,label=below:{\cc{1}}] {};
	\node (g9) [gauge,right of=g8,label=below:{\dd{1}}] {};
	\node (f2) [flavour,above of=g2,label=above:{\bb{0}}] {};
	\node (f8) [flavour,above of=g8,label=above:{\bb{0}}] {};
	\draw (g1)--(g2) (g2)--(g3) (g3)--(g4) (g4)--(g5) (g5)--(g6) (g6)--(g7) (g7)--(g8) (g8)--(g9) (g2)--(f2) (g8)--(f8);
	\end{tikzpicture}
	} 
\label{eq:magQuiv_Sp1_6_finite}
\end{align}
and Coulomb branch dimension and global symmetry, see Appendix \ref{app:Coulomb_branch}, are given by
\begin{subequations}
\label{eq:results_Sp1_6_finite}
\begin{align}
    \dim_{\HH} \Coulomb \eqref{eq:magQuiv_Sp1_6_finite} = 9
    \; , \qquad 
    G = \sorm(12) \,.
\end{align}
Even more is true, because if the magnetic gauge groups for the quiver \eqref{eq:magQuiv_Sp1_6_finite} are chosen appropriately then it is known \cite{Benini:2010uu,Chacaltana:2012zy,Cabrera:2017ucb} that
\begin{align}
\Coulomb\eqref{eq:magQuiv_Sp1_6_finite} = \clorbit_{\dalg_6}^{\min } \,,
\end{align}
\end{subequations}
which denotes the closure of the minimal nilpotent orbit of $\sormL(12)$. 
Again, the correct choice of magnetic gauge group is subject of the companion 
paper \cite{Bourget:2020xdz}.
Compared to the classical Higgs branch and the non-abelian part of the global symmetry, the magnetic quiver matches these properties \eqref{eq:results_Sp1_6_finite} exactly.
\paragraph{Infinite coupling.}
Proceeding to infinite coupling means that the two $(0,1)$ 5-branes in \eqref{eq:web_Sp1_6_Higgs} coincide on the \Of\ plane. Since both branes end on half 7-branes, these can be vertically displaced slightly to a brane configuration of the form    
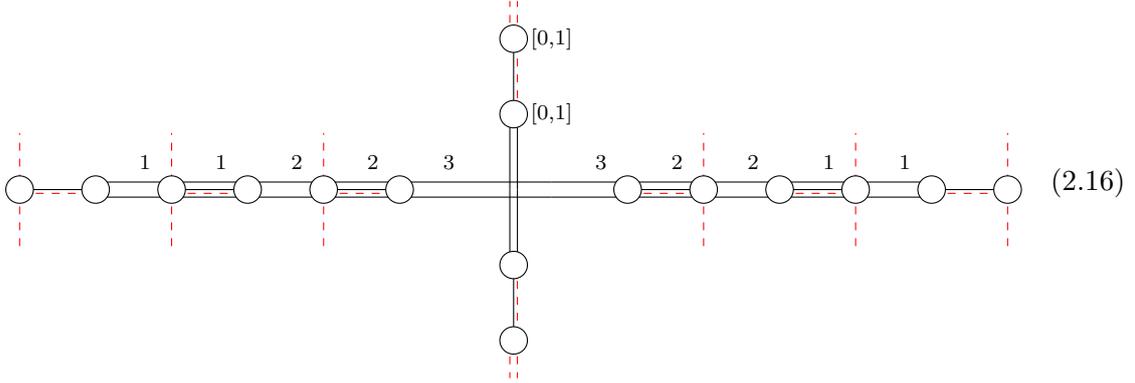
\begin{align}
\raisebox{-.5\height}{
    \begin{tikzpicture}
    \DfiveOMinus{}{0}
    \DfiveOMinus{3}{-1}
    \DfiveOMinus{3}{1}
    \DfiveOMinusTilde{2}{-2}
    \MonoCut{-2,-0.05}{-1,-0.05}
    \DfiveOMinusTilde{2}{2}
    \MonoCut{2,-0.05}{3,-0.05}
    \DfiveOMinus{2}{-3}
    \DfiveOMinus{2}{3}
    \DfiveOMinusTilde{1}{-4}
    \MonoCut{-4,-0.05}{-3,-0.05}
    \DfiveOMinusTilde{1}{4}
    \MonoCut{4,-0.05}{5,-0.05}
    \DfiveOMinus{1}{-5}
    \DfiveOMinus{1}{5}
    \OMinusTilde{6,0}{7,0}
    \MonoCut{6,-0.05}{7,-0.05}
    \OMinusTilde{-6,0}{-5,0}
    \MonoCut{-6,-0.05}{-5,-0.05}
    \Dbrane{0.45,0}{0.45,1}
    \Dbrane{0.55,0}{0.55,1}
    \draw (1,1) node {$\scriptstyle{[0,1]}$};
    \Dbrane{0.45,0}{0.45,-1}
    \Dbrane{0.55,0}{0.55,-1}
    \Dbrane{0.5,1}{0.5,2}
    \draw (1,2) node {$\scriptstyle{[0,1]}$};
    \Dbrane{0.5,-1}{0.5,-2}
    \MonoCut{-2,-0.75}{-2,0.75}
    \MonoCut{-4,-0.75}{-4,0.75}
    \MonoCut{-6,-0.75}{-6,0.75}
    \MonoCut{3,-0.75}{3,0.75}
    \MonoCut{5,-0.75}{5,0.75}
    \MonoCut{7,-0.75}{7,0.75}
    \MonoCut{0.55,1}{0.55,2}
    \MonoCut{0.55,-1}{0.55,-2}
    \MonoCut{0.45,2}{0.45,2.5}
    \MonoCut{0.55,2}{0.55,2.5}
    \MonoCut{0.45,-2}{0.45,-2.5}
    \MonoCut{0.55,-2}{0.55,-2.5}
    \SevenB{0.5,1}
    \SevenB{0.5,2}
    \SevenB{0.5,-1}
    \SevenB{0.5,-2}
    \SevenB{-1,0}
    \SevenB{-2,0}
    \SevenB{-3,0}
    \SevenB{-4,0}
    \SevenB{-5,0}
    \SevenB{-6,0}
    \SevenB{2,0}
    \SevenB{3,0}
    \SevenB{4,0}
    \SevenB{5,0}
    \SevenB{6,0}
    \SevenB{7,0}
    \end{tikzpicture}
}
\label{eq:web_Sp1_6_infinite}
\end{align}
Finding a maximal subdivision is similar to the cases above, the only new feature is the set of vertically aligned half 7-branes outside the \Of\ plane. The two (half) $(0,1)$ 5-branes that pass through the orientifold form two identical subwebs that are subjected to the projections of the magnetic orientifolds as discussed above. The $(0,1)$ 5-brane that is suspended between the two half $[0,1]$ 7-branes outside the orientifold is, in contrast, not affected by the orientifold projection and, consequently, induces a unitary magnetic gauge multiplet of rank one. 
The matter content between the magnetic $\uo$ and $\calg_1$ node is determined by the intersection property. The intersection number is zero, but the generalised intersection number receives a positive contribution because the corresponding subwebs end on opposite side on the common half 7-brane. Therefore, there exists a magnetic bifundamental hypermultiplet between these nodes.

Adding all the pieces, one then derives the following magnetic quiver:
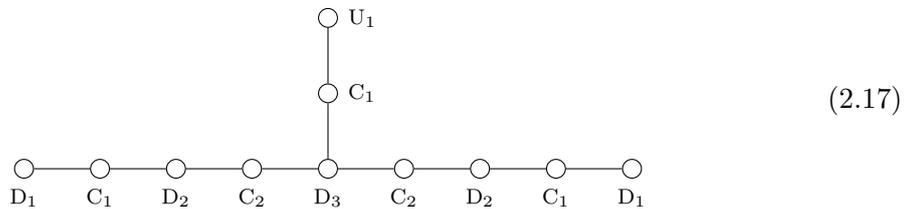
\begin{align}
 \raisebox{-.5\height}{
 	\begin{tikzpicture}
 	\tikzset{node distance = 1cm};
	\tikzstyle{gauge} = [circle, draw,inner sep=2.5pt];
	\tikzstyle{flavour} = [regular polygon,regular polygon sides=4,inner 
sep=2.5pt, draw];
	\node (g1) [gauge,label=below:{\dd{1}}] {};
	\node (g2) [gauge,right of=g1,label=below:{\cc{1}}] {};
	\node (g3) [gauge,right of=g2,label=below:{\dd{2}}] {};
	\node (g4) [gauge,right of=g3,label=below:{\cc{2}}] {};
	\node (g5) [gauge,right of=g4,label=below:{\dd{3}}] {};
	\node (g6) [gauge,right of=g5,label=below:{\cc{2}}] {};
	\node (g7) [gauge,right of=g6,label=below:{\dd{2}}] {};
	\node (g8) [gauge,right of=g7,label=below:{\cc{1}}] {};
	\node (g9) [gauge,right of=g8,label=below:{\dd{1}}] {};
	\node (g10) [gauge,above of=g5,label=right:{\cc{1}}] {};
	\node (g11) [gauge,above of=g10,label=right:{\uu{1}}] {};
	\draw (g1)--(g2) (g2)--(g3) (g3)--(g4) (g4)--(g5) (g5)--(g6) (g6)--(g7) (g7)--(g8) (g8)--(g9) (g5)--(g10) (g10)--(g11);
	\end{tikzpicture}
	} 
\label{eq:magQuiv_Sp1_6_infinite}
\end{align}
and the Coulomb branch dimension is computed to be
\begin{align}
    \dim_{\HH} \Coulomb \eqref{eq:magQuiv_Sp1_6_infinite} = 17 \,.
\end{align}
The physical choice of magnetic gauge groups for any orthosymplectic quiver is 
subject of the companion paper \cite{Bourget:2020xdz}; such that a Hilbert 
series 
analysis suggests that 
$\Coulomb\eqref{eq:magQuiv_Sp1_6_infinite}=\clorbit_{E_7}^{\min}$ holds.

Borrowing from class $\mathcal{S}$ literature, the 
star-shaped quiver \eqref{eq:magQuiv_Sp1_6_infinite} can be decomposed into 
three 
$D_3$ punctures corresponding to the linear quivers $T_{(1^6)}[\sorm(6)] $, 
$T_{(1^6)}[\sorm(6)] $, and
$T_{(3,1^3)}[\sorm(6)]$. As detailed in \cite[Sec.\ 
3.2.2]{Chacaltana:2011ze}, the resulting fixture gives rise to the so-called 
$E_7$ SCFT.
\subsubsection{\texorpdfstring{$E_6$}{E6}: Sp(1) with 5 flavours}
The 5-brane web for $\sprm(1)$ with 5 fundamental flavours in the Coulomb branch phase with masses for the flavours is given by
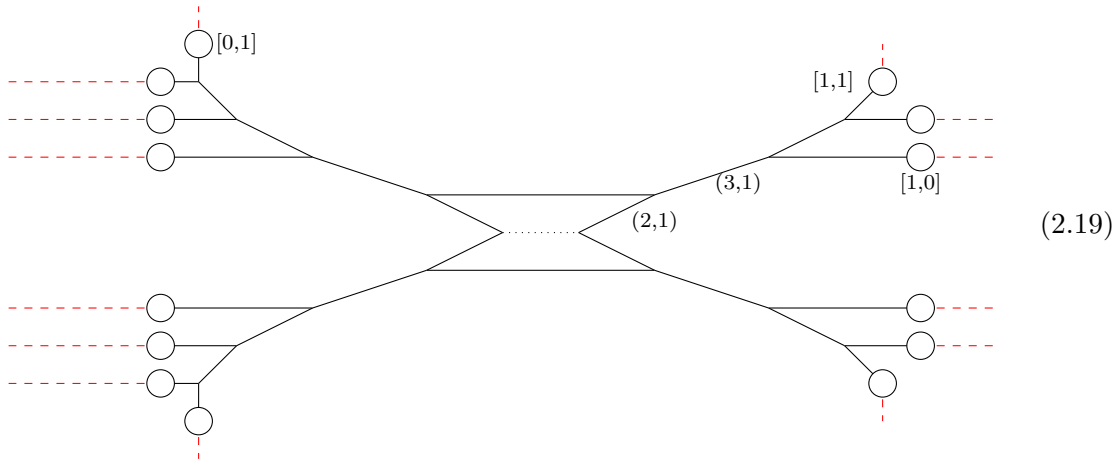
\begin{align}
\raisebox{-.5\height}{
    \begin{tikzpicture}
    \OPlus{0,0}{1,0}
    \Dbrane{0,0}{-1,0.5}
    \Dbrane{1,0}{2,0.5}
    \draw (2,0.15) node {$\scriptstyle{(2,1)}$};
    \Dbrane{-1,0.5}{2,0.5}
    \Dbrane{-1,0.5}{-2.5,1}
    \Dbrane{2,0.5}{3.5,1}
        \draw (3.1,0.65) node {$\scriptstyle{(3,1)}$};
    \Dbrane{-2.5,1}{-4.5,1}
    \Dbrane{3.5,1}{5.5,1}
    \draw (5.5,0.65) node {$\scriptstyle{[1,0]}$};
    \Dbrane{-2.5,1}{-3.5,1.5}
    \Dbrane{3.5,1}{4.5,1.5}
    \Dbrane{-3.5,1.5}{-4.5,1.5}
    \Dbrane{4.5,1.5}{5.5,1.5}
    \Dbrane{-3.5,1.5}{-4,2}
    \Dbrane{4.5,1.5}{5,2}
        \draw (4.35,2) node {$\scriptstyle{[1,1]}$};
    \Dbrane{-4,2}{-4.5,2}
    \Dbrane{-4,2}{-4,2.5}
    \draw (-3.5,2.5) node {$\scriptstyle{[0,1]}$};
    \MonoCut{-4.5,2}{-6.5,2}
    \MonoCut{-4.5,1.5}{-6.5,1.5}
    \MonoCut{-4.5,1}{-6.5,1}
    \MonoCut{5.5,1}{6.5,1}
    \MonoCut{5.5,1.5}{6.5,1.5}
    \MonoCut{-4,2.5}{-4,3}
    \MonoCut{5,2}{5,2.5}
    \SevenB{-4,2.5}
    \SevenB{5,2}
    \SevenB{-4.5,2}
    \SevenB{-4.5,1.5}
    \SevenB{5.5,1.5}
    \SevenB{-4.5,1}
    \SevenB{5.5,1}
    \begin{scope}[yscale=-1,xscale=1]
    \Dbrane{0,0}{-1,0.5}
    \Dbrane{1,0}{2,0.5}
    \Dbrane{-1,0.5}{2,0.5}
    \Dbrane{-1,0.5}{-2.5,1}
    \Dbrane{2,0.5}{3.5,1}
    \Dbrane{-2.5,1}{-4.5,1}
    \Dbrane{3.5,1}{5.5,1}
    \Dbrane{-2.5,1}{-3.5,1.5}
    \Dbrane{3.5,1}{4.5,1.5}
    \Dbrane{-3.5,1.5}{-4.5,1.5}
    \Dbrane{4.5,1.5}{5.5,1.5}
    \Dbrane{-3.5,1.5}{-4,2}
    \Dbrane{4.5,1.5}{5,2}
    \Dbrane{-4,2}{-4.5,2}
    \Dbrane{-4,2}{-4,2.5}
    \MonoCut{-4.5,2}{-6.5,2}
    \MonoCut{-4.5,1.5}{-6.5,1.5}
    \MonoCut{-4.5,1}{-6.5,1}
    \MonoCut{5.5,1}{6.5,1}
    \MonoCut{5.5,1.5}{6.5,1.5}
    \MonoCut{-4,2.5}{-4,3}
    \MonoCut{5,2}{5,2.5}
    \SevenB{-4,2.5}
    \SevenB{5,2}
    \SevenB{-4.5,2}
    \SevenB{-4.5,1.5}
    \SevenB{5.5,1.5}
    \SevenB{-4.5,1}
    \SevenB{5.5,1}
    \end{scope}
    \end{tikzpicture}
}
\label{eq:web_Sp1_5_Coulomb}
\end{align}
and the splitting into $3$ flavours on the left and $2$ flavours on the right-hand-side is a convenient choice. Subsequently moving to the origin of the 5d Coulomb branch is realised by moving the flavour half $[1,0]$ 7-branes and the gauge $(1,0)$ 5-brane onto the \Of\ plane. Once the half 7-branes merge with their respective mirror branes, the physical 7-branes can split on the orientifold leading to the following brane web in the Higgs branch phase: 
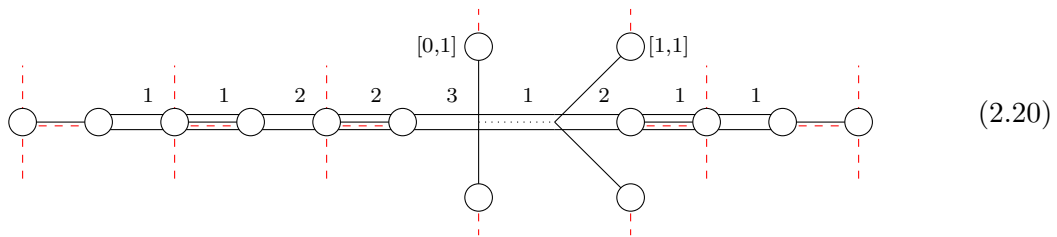
\begin{align}
\raisebox{-.5\height}{
    \begin{tikzpicture}
    \DfiveOPlus{1}{0}
    \DfiveOMinus{3}{-1}
    \DfiveOMinus{2}{1}
    \DfiveOMinusTilde{2}{-2}
    \MonoCut{-2,-0.05}{-1,-0.05}
    \DfiveOMinusTilde{1}{2}
    \MonoCut{2,-0.05}{3,-0.05}
    \DfiveOMinus{2}{-3}
    \DfiveOMinus{1}{3}
    \DfiveOMinusTilde{1}{-4}
    \MonoCut{-4,-0.05}{-3,-0.05}
    \DfiveOMinus{1}{-5}
    \OMinusTilde{4,0}{5,0}
    \MonoCut{4,-0.05}{5,-0.05}
    \OMinusTilde{-6,0}{-5,0}
    \MonoCut{-6,-0.05}{-5,-0.05}
    \Dbrane{0,0}{0,1}
    \draw (-0.55,1) node {$\scriptstyle{[0,1]}$};
    \Dbrane{0,0}{0,-1}
    \Dbrane{1,0}{2,1}
    \draw (2.5,1) node {$\scriptstyle{[1,1]}$};
    \Dbrane{1,0}{2,-1}
    \MonoCut{-2,-0.75}{-2,0.75}
    \MonoCut{-4,-0.75}{-4,0.75}
    \MonoCut{-6,-0.75}{-6,0.75}
    \MonoCut{3,-0.75}{3,0.75}
    \MonoCut{5,-0.75}{5,0.75}
    \MonoCut{0,1}{0,1.5}
    \MonoCut{0,-1}{0,-1.5}
    \MonoCut{2,1}{2,1.5}
    \MonoCut{2,-1}{2,-1.5}
    \SevenB{0,1}
    \SevenB{2,1}
    \SevenB{0,-1}
    \SevenB{2,-1}
    \SevenB{-1,0}
    \SevenB{-2,0}
    \SevenB{-3,0}
    \SevenB{-4,0}
    \SevenB{-5,0}
    \SevenB{-6,0}
    \SevenB{2,0}
    \SevenB{3,0}
    \SevenB{4,0}
    \SevenB{5,0}
    \end{tikzpicture}
}
\label{eq:web_Sp1_5_Higgs}
\end{align}
The $(1,1)$ 5-brane is connected via another $(1,0)$ 5-brane to a 7-brane on the right-hand-side, due to the S-rule. Moreover, the $(0,1)$ 5-brane has to be connected to several 7-branes on the left-hand-side in order to preserve supersymmetry.
\paragraph{Finite coupling.}
To analyse the finite coupling Higgs branch, one eliminates the non-dynamical 5-branes as much as possible by transitioning the $(0,1)$ and $(1,1)$ 5-brane through 7-branes on the left or right, respectively. Carefully considering brane annihilation reveals that the configuration \eqref{eq:web_Sp1_5_Higgs} becomes 
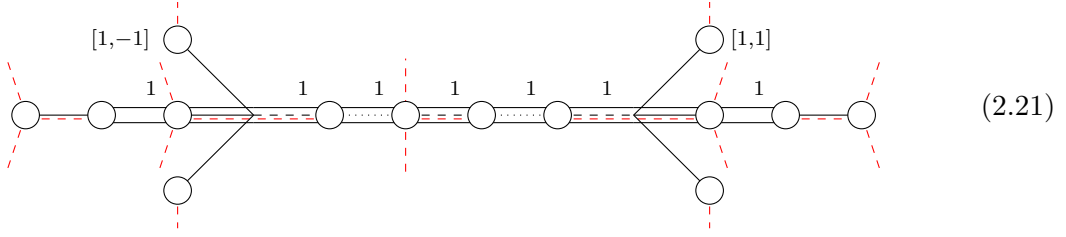
\begin{align}
 \raisebox{-.5\height}{
    \begin{tikzpicture}
    \DfiveOPlus{1}{0}
    \DfiveOPlusTilde{1}{-1}
    \DfiveOPlusTilde{1}{1}
    \DfiveOPlus{1}{-2}
    \MonoCut{-1,-0.05}{0,-0.05}
    \DfiveOMinusTilde{}{2}
    \MonoCut{1,-0.05}{3,-0.05}
    \DfiveOPlusTilde{1}{-3}
    \DfiveOMinus{1}{3}
    \DfiveOMinusTilde{}{-4}
    \MonoCut{-4,-0.05}{-2,-0.05}
    \DfiveOMinus{1}{-5}
    \OMinusTilde{4,0}{5,0}
    \MonoCut{4,-0.05}{5,-0.05}
    \OMinusTilde{-6,0}{-5,0}
    \MonoCut{-6,-0.05}{-5,-0.05}
    \Dbrane{-3,0}{-4,1}
    \draw (-4.75,1) node {$\scriptstyle{[1,-1]}$};
    \Dbrane{-3,0}{-4,-1}
    \Dbrane{2,0}{3,1}
    \draw (3.55,1) node {$\scriptstyle{[1,1]}$};
    \Dbrane{2,0}{3,-1}
    \MonoCut{-1,-0.75}{-1,0.75}
    \MonoCut{-4,0}{-4-0.25,0.75}
    \MonoCut{-4,0}{-4-0.25,-0.75}
    \MonoCut{-6,0}{-6-0.25,0.75}
    \MonoCut{-6,0}{-6-0.25,-0.75}
    \MonoCut{3,0}{3+0.25,0.75}
    \MonoCut{3,0}{3+0.25,-0.75}
    \MonoCut{5,0}{5+0.25,0.75}
    \MonoCut{5,0}{5+0.25,-0.75}
    \MonoCut{-4,1}{-4,1.5}
    \MonoCut{-4,-1}{-4,-1.5}
    \MonoCut{3,1}{3,1.5}
    \MonoCut{3,-1}{3,-1.5}
    \SevenB{-4,1}
    \SevenB{3,1}
    \SevenB{-4,-1}
    \SevenB{3,-1}
    \SevenB{0,0}
    \SevenB{-1,0}
    \SevenB{-2,0}
    \SevenB{-4,0}
    \SevenB{-5,0}
    \SevenB{-6,0}
    \SevenB{1,0}
    \SevenB{3,0}
    \SevenB{4,0}
    \SevenB{5,0}
    \end{tikzpicture}
}
\label{eq:web_Sp1_5_finite}
\end{align}
where the monodromy cuts have been chosen for an unambiguous presentation.
As above the $(1,\pm1)$ 5-branes between \Ofpt\ and \Ofmt\ are supersymmetric, 
but non-dynamical branes. Hence, these contribute as flavours to the magnetic 
quiver. The remaining brane web is subdivided into subwebs, where each $(1,0)$ 
5-brane ending on 7-branes is such a consistent subweb. Then, following Table 
\ref{tab:magnetic_orientifold}, the magnetic quiver associated to 
\eqref{eq:web_Sp1_5_finite} is given by
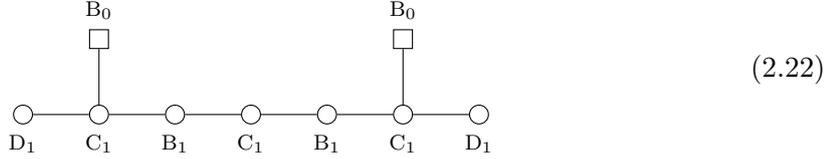
\begin{align}
 \raisebox{-.5\height}{
 	\begin{tikzpicture}
 	\tikzset{node distance = 1cm};
	\tikzstyle{gauge} = [circle, draw,inner sep=2.5pt];
	\tikzstyle{flavour} = [regular polygon,regular polygon sides=4,inner 
sep=2.5pt, draw];
	\node (g1) [gauge,label=below:{\dd{1}}] {};
	\node (g2) [gauge,right of=g1,label=below:{\cc{1}}] {};
	\node (g3) [gauge,right of=g2,label=below:{\bb{1}}] {};
	\node (g4) [gauge,right of=g3,label=below:{\cc{1}}] {};
	\node (g5) [gauge,right of=g4,label=below:{\bb{1}}] {};
	\node (g6) [gauge,right of=g5,label=below:{\cc{1}}] {};
	\node (g7) [gauge,right of=g6,label=below:{\dd{1}}] {};
	\node (f2) [flavour,above of=g2,label=above:{\bb{0}}] {};
	\node (f6) [flavour,above of=g6,label=above:{\bb{0}}] {};
	\draw (g1)--(g2) (g2)--(g3) (g3)--(g4) (g4)--(g5) (g5)--(g6) (g6)--(g7) (g2)--(f2) (g6)--(f6);
	\end{tikzpicture}
	} 
	\label{eq:magQuiv_Sp1_5_finite}
\end{align}
and the Coulomb branch dimension and global symmetry, see Appendix \ref{app:Coulomb_branch}, can be computed to be
\begin{subequations}
\label{eq:results_Sp1_5_finite}
\begin{align}
    \dim_{\HH} \Coulomb \eqref{eq:magQuiv_Sp1_5_finite} = 7
    \; ; \qquad 
    G = \sorm(10) \,.
\end{align}
Upon choosing the magnetic gauge groups for \eqref{eq:magQuiv_Sp1_5_finite} suitably, the entire moduli spaces is known \cite[Tab.\ 11]{Cabrera:2017ucb}, see also \cite{Benini:2010uu,Chacaltana:2012zy}, 
\begin{align}
    \Coulomb\eqref{eq:magQuiv_Sp1_5_finite}  = \clorbit_{\dalg_5}^{\min }
\end{align}
\end{subequations}
i.e.\ the closure of the minimal nilpotent orbit of $\sormL(10)$.
The rationale behind the choice of magnetic gauge group for orthosymplectic 
magnetic quivers is addressed in a companion paper \cite{Bourget:2020xdz}.
Consequently, the properties \eqref{eq:results_Sp1_5_finite} correctly reproduce the finite coupling Higgs branch as well as the non-abelian part of the global symmetry.
\paragraph{Infinite coupling.}
In order to transition to the Higgs branch phase at infinite coupling, the 
$(0,1)$ and $(1,1)$ 5-brane in \eqref{eq:web_Sp1_5_Higgs} have to become 
coincident on the \Of\ plane. However, as remarked above, there is a 
non-dynamical brane in \eqref{eq:web_Sp1_5_Higgs}, which can be eliminated by 
passing the $(0,1)$ and $(1,1)$ 5-brane through one half 7-brane on the 
left-hand-side. Accounting for brane-annihilation leads to
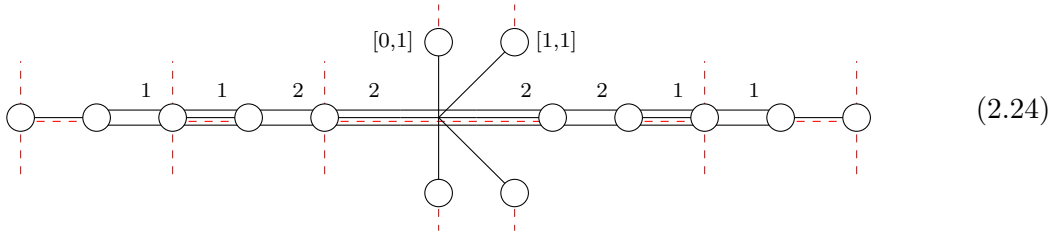
\begin{align}
\raisebox{-.5\height}{
    \begin{tikzpicture}
    \DfiveOMinusTilde{2}{0}
    \DfiveOMinusTilde{}{-1}
    \DfiveOMinus{2}{1}
    \DfiveOMinusTilde{2}{-2}
    \MonoCut{-2,-0.05}{1,-0.05}
    \DfiveOMinusTilde{1}{2}
    \MonoCut{2,-0.05}{3,-0.05}
    \DfiveOMinus{2}{-3}
    \DfiveOMinus{1}{3}
    \DfiveOMinusTilde{1}{-4}
    \MonoCut{-4,-0.05}{-3,-0.05}
    \DfiveOMinus{1}{-5}
    \OMinusTilde{4,0}{5,0}
    \MonoCut{4,-0.05}{5,-0.05}
    \OMinusTilde{-6,0}{-5,0}
    \MonoCut{-6,-0.05}{-5,-0.05}
    \Dbrane{-0.5,0}{-0.5,1}
    \draw (-1.1,1) node {$\scriptstyle{[0,1]}$};
    \Dbrane{-0.5,0}{-0.5,-1}
    \Dbrane{-0.5,0}{0.5,1}
    \draw (1.05,1) node {$\scriptstyle{[1,1]}$};
    \Dbrane{-0.5,0}{0.5,-1}
    \MonoCut{-2,-0.75}{-2,0.75}
    \MonoCut{-4,-0.75}{-4,0.75}
    \MonoCut{-6,-0.75}{-6,0.75}
    \MonoCut{3,-0.75}{3,0.75}
    \MonoCut{5,-0.75}{5,0.75}
    \MonoCut{-0.5,1}{-0.5,1.5}
    \MonoCut{-0.5,-1}{-0.5,-1.5}
    \MonoCut{0.5,1}{0.5,1.5}
    \MonoCut{0.5,-1}{0.5,-1.5}
    \SevenB{-0.5,1}
    \SevenB{0.5,1}
    \SevenB{-0.5,-1}
    \SevenB{0.5,-1}
    \SevenB{1,0}
    \SevenB{-2,0}
    \SevenB{-3,0}
    \SevenB{-4,0}
    \SevenB{-5,0}
    \SevenB{-6,0}
    \SevenB{2,0}
    \SevenB{3,0}
    \SevenB{4,0}
    \SevenB{5,0}
    \end{tikzpicture}
}
\label{eq:web_Sp1_5_infinite}
\end{align}
where the $(1,1)$ and $(0,1)$ 5-branes constitute an independent subweb in the 
maximal subdivision of \eqref{eq:web_Sp1_5_infinite}, see also Appendix 
\ref{app:background}. Again, this subweb passes through the \Of\ plane, but is mapped to itself, 
such that the magnetic 
degrees of freedom are not affected by the orientifold projection. The subweb 
contributes a unitary magnetic vector multiplet. 
The generalised intersection number between this subweb and the \Dfive\ branes 
in the central segment is computed to be one, simply from the intersection 
number without further corrections. Therefore, the $\uo$ node connects via a 
bifundamental magnetic hypermultiplet to another magnetic gauge node. In addition, since the $(0,1)$ and $(1,1)$ 5-brane have a non-trivial intersection before orientifolding, one may wonder if this implies additional matter. However, due to the identification of the 7-branes which leads to common 7-branes, the generalised intersection evaluates to zero and no matter multiplet arises. 

The remaining subwebs are, again, given by \Dfive\ branes between half 7-branes and represent no new challenge. Accounting for identical subwebs and transitioning to magnetic orientifolds, see Table \ref{tab:magnetic_orientifold}, leads to the following magnetic quiver:
\begin{align}
 \raisebox{-.5\height}{
 	\begin{tikzpicture}
 	\tikzset{node distance = 1cm};
	\tikzstyle{gauge} = [circle, draw,inner sep=2.5pt];
	\tikzstyle{flavour} = [regular polygon,regular polygon sides=4,inner 
sep=2.5pt, draw];
	\node (g1) [gauge,label=below:{\dd{1}}] {};
	\node (g2) [gauge,right of=g1,label=below:{\cc{1}}] {};
	\node (g3) [gauge,right of=g2,label=below:{\dd{2}}] {};
	\node (g4) [gauge,right of=g3,label=below:{\cc{2}}] {};
	\node (g5) [gauge,right of=g4,label=below:{\dd{2}}] {};
	\node (g6) [gauge,right of=g5,label=below:{\cc{1}}] {};
	\node (g7) [gauge,right of=g6,label=below:{\dd{1}}] {};
	\node (g8) [gauge,above of=g4,label=above:{\uu{1}}] {};
	\draw (g1)--(g2) (g2)--(g3) (g3)--(g4) (g4)--(g5) (g5)--(g6) (g6)--(g7) (g4)--(g8);
	\end{tikzpicture}
	} 
		\label{eq:magQuiv_Sp1_5_infinite}
\end{align}
The Coulomb branch dimension is computed to be
\begin{align}
    \dim_{\HH} \Coulomb \eqref{eq:magQuiv_Sp1_5_infinite}  =11 \,.
\end{align}
The identification of the magnetic gauge groups for 
\eqref{eq:magQuiv_Sp1_5_infinite} is addressed in the companion paper 
\cite{Bourget:2020xdz}, such that a Hilbert series computation indicates the 
stronger claim $\Coulomb \eqref{eq:magQuiv_Sp1_5_infinite} = 
\clorbit_{E_6}^{\min}$.

In terms of the class $\mathcal{S}$ dictionary, the star-shaped 
quiver \eqref{eq:magQuiv_Sp1_5_infinite} can be decomposed into one untwisted 
$A_3$ puncture corresponding to the linear quiver $T_{(3,1)}[\surm(4)]$ and two 
twisted $A_3$ punctures associated with the linear quiver 
$T_{(1^5)}[\usprm(4)]$. The resulting twisted $A_3$ fixture is referred to as 
$E_6$ SCFT \cite[Sec.\ A.1.5]{Chacaltana:2012ch}.
\subsubsection{\texorpdfstring{$E_5$}{E5}: Sp(1) with 4 flavours}
The 5-brane web for $\sprm(1)$ with $4$ massive flavours in the Coulomb 
branch phase reads 
\begin{align}
\raisebox{-.5\height}{
    \begin{tikzpicture}
    \OPlus{0,0}{1,0}
    \Dbrane{0,0}{-1,0.5}
    \Dbrane{1,0}{2,0.5}
    \draw (2,0.15) node {$\scriptstyle{(2,1)}$};
    \Dbrane{-1,0.5}{2,0.5}
    \Dbrane{-1,0.5}{-2.5,1}
    \Dbrane{2,0.5}{3.5,1}
    \draw (3,0.65) node {$\scriptstyle{(3,1)}$};
    \Dbrane{-2.5,1}{-4.5,1}
    \Dbrane{3.5,1}{5.5,1}
    \Dbrane{-2.5,1}{-3.5,1.5}
    \Dbrane{3.5,1}{4.5,1.5}
    \Dbrane{-3.5,1.5}{-4.5,1.5}
    \Dbrane{4.5,1.5}{5.5,1.5}
    \draw (5.5,0.65) node {$\scriptstyle{[1,0]}$};
    \Dbrane{-3.5,1.5}{-4,2}
    \Dbrane{4.5,1.5}{5,2}
    \draw (4.35,2) node {$\scriptstyle{[1,1]}$};
    \MonoCut{-4.5,1.5}{-5.5,1.5}
    \MonoCut{-4.5,1}{-5.5,1}
    \MonoCut{5.5,1}{6.5,1}
    \MonoCut{5.5,1.5}{6.5,1.5}
    \MonoCut{-4,2}{-4,2.5}
    \MonoCut{5,2}{5,2.5}
    \SevenB{-4,2}
    \SevenB{5,2}
    \SevenB{-4.5,1.5}
    \SevenB{5.5,1.5}
    \SevenB{-4.5,1}
    \SevenB{5.5,1}
    \begin{scope}[yscale=-1,xscale=1]
    \Dbrane{0,0}{-1,0.5}
    \Dbrane{1,0}{2,0.5}
    \Dbrane{-1,0.5}{2,0.5}
    \Dbrane{-1,0.5}{-2.5,1}
    \Dbrane{2,0.5}{3.5,1}
    \Dbrane{-2.5,1}{-4.5,1}
    \Dbrane{3.5,1}{5.5,1}
    \Dbrane{-2.5,1}{-3.5,1.5}
    \Dbrane{3.5,1}{4.5,1.5}
    \Dbrane{-3.5,1.5}{-4.5,1.5}
    \Dbrane{4.5,1.5}{5.5,1.5}
    \Dbrane{-3.5,1.5}{-4,2}
    \Dbrane{4.5,1.5}{5,2}
    \MonoCut{-4.5,1.5}{-5.5,1.5}
    \MonoCut{-4.5,1}{-5.5,1}
    \MonoCut{5.5,1}{6.5,1}
    \MonoCut{5.5,1.5}{6.5,1.5}
    \MonoCut{-4,2}{-4,2.5}
    \MonoCut{5,2}{5,2.5}
    \SevenB{-4,2}
    \SevenB{5,2}
    \SevenB{-4.5,1.5}
    \SevenB{5.5,1.5}
    \SevenB{-4.5,1}
    \SevenB{5.5,1}
    \end{scope}
    \end{tikzpicture}
}
\label{eq:web_Sp1_4_Coulomb}
\end{align}
    and the flavour branes have been chosen symmetrically for convenience.  
Transitioning to the Higgs branch phase, configuration \eqref{eq:web_Sp1_4_Coulomb} becomes
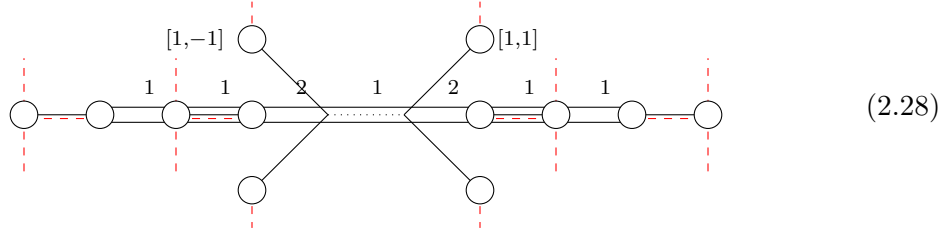
\begin{align}
\raisebox{-.5\height}{
    \begin{tikzpicture}
    \DfiveOPlus{1}{0}
    \DfiveOMinus{2}{-1}
    \DfiveOMinus{2}{1}
    \DfiveOMinusTilde{1}{-2}
    \MonoCut{-2,-0.05}{-1,-0.05}
    \DfiveOMinusTilde{1}{2}
    \MonoCut{2,-0.05}{3,-0.05}
    \DfiveOMinus{1}{-3}
    \DfiveOMinus{1}{3}
    \OMinusTilde{-3,0}{-4,0}
    \MonoCut{-3,-0.05}{-4,-0.05}
    \OMinusTilde{4,0}{5,0}
    \MonoCut{4,-0.05}{5,-0.05}
    \Dbrane{0,0}{-1,1}
    \draw (-1.75,1) node {$\scriptstyle{[1,-1]}$};
    \Dbrane{0,0}{-1,-1}
    \Dbrane{1,0}{2,1}
    \draw (2.5,1) node {$\scriptstyle{[1,1]}$};
    \Dbrane{1,0}{2,-1}
    \MonoCut{-2,-0.75}{-2,0.75}
    \MonoCut{-4,-0.75}{-4,0.75}
    \MonoCut{3,-0.75}{3,0.75}
    \MonoCut{5,-0.75}{5,0.75}
    \MonoCut{-1,1}{-1,1.5}
    \MonoCut{-1,-1}{-1,-1.5}
    \MonoCut{2,1}{2,1.5}
    \MonoCut{2,-1}{2,-1.5}
    \SevenB{-1,1}
    \SevenB{2,1}
    \SevenB{-1,-1}
    \SevenB{2,-1}
    \SevenB{-1,0}
    \SevenB{-2,0}
    \SevenB{-3,0}
    \SevenB{-4,0}
    \SevenB{2,0}
    \SevenB{3,0}
    \SevenB{4,0}
    \SevenB{5,0}
    \end{tikzpicture}
}
\label{eq:web_Sp1_4_Higgs}
\end{align}
As above, the $(1,\pm1)$ 5-branes each need to be connected to a $[1,0]$ 7-brane via a $(1,0)$ 5-brane for consistency.
\paragraph{Finite coupling.}
The finite coupling Higgs branch phase is most conveniently analysed when the 
$(1,\pm1)$ 5-branes in \eqref{eq:web_Sp1_4_Higgs} are passed through the 
7-branes on the left or right-hand-side, respectively. Accounting for brane 
annihilation, the resulting 5-brane web becomes
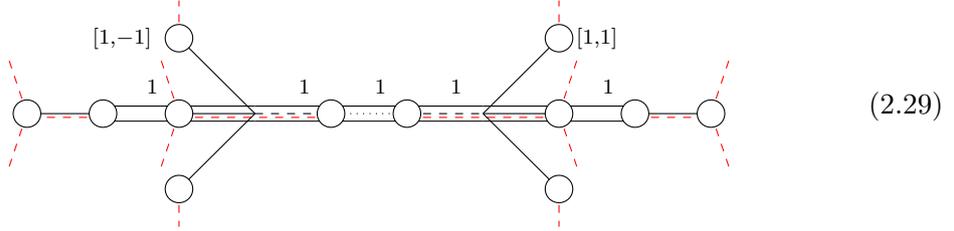
\begin{align}
\raisebox{-.5\height}{
    \begin{tikzpicture}
    \DfiveOPlus{1}{0}
    \DfiveOPlusTilde{1}{-1}
    \DfiveOPlusTilde{1}{1}
    \DfiveOMinusTilde{}{-2}
    \MonoCut{-2,-0.05}{0,-0.05}
    \DfiveOMinusTilde{}{2}
    \MonoCut{1,-0.05}{3,-0.05}
    \DfiveOMinus{1}{-3}
    \DfiveOMinus{1}{3}
    \OMinusTilde{-3,0}{-4,0}
    \MonoCut{-3,-0.05}{-4,-0.05}
    \OMinusTilde{4,0}{5,0}
    \MonoCut{4,-0.05}{5,-0.05}
    \Dbrane{-1,0}{-2,1}
    \Dbrane{-1,0}{-2,-1}
    \draw (-2.75,1) node {$\scriptstyle{[1,-1]}$};
    \Dbrane{2,0}{3,1}
    \draw (3.5,1) node {$\scriptstyle{[1,1]}$};
    \Dbrane{2,0}{3,-1}
    \MonoCut{-2,0}{-2-0.25,0.75}
    \MonoCut{-2,0}{-2-0.25,-0.75}
    \MonoCut{-4,0}{-4-0.25,0.75}
    \MonoCut{-4,0}{-4-0.25,-0.75}
    \MonoCut{3,0}{3+0.25,0.75}
    \MonoCut{3,0}{3+0.25,-0.75}
    \MonoCut{5,0}{5+0.25,0.75}
    \MonoCut{5,0}{5+0.25,-0.75}
    \MonoCut{-2,1}{-2,1.5}
    \MonoCut{-2,-1}{-2,-1.5}
    \MonoCut{3,1}{3,1.5}
    \MonoCut{3,-1}{3,-1.5}
    \SevenB{-2,1}
    \SevenB{3,1}
    \SevenB{-2,-1}
    \SevenB{3,-1}
    \SevenB{0,0}
    \SevenB{-2,0}
    \SevenB{-3,0}
    \SevenB{-4,0}
    \SevenB{1,0}
    \SevenB{3,0}
    \SevenB{4,0}
    \SevenB{5,0}
    \end{tikzpicture}
}
\label{eq:web_Sp1_4_finite}
\end{align}    
where the monodormy cuts have been slightly displaced for the sake of an unambiguous configuration.
Now, one can analyse the Higgs branch degrees of freedom. The $(1,\pm1)$ 5-branes between \Ofpt\ and \Ofmt\ are not freely-moving subwebs, and contribute only as magnetic flavours.
The $(1,0)$ 5-branes between the different 7-branes are all subwebs and induce magnetic gauge multiplets according to which orientifold they coincide with, see Table \ref{tab:magnetic_orientifold}. The associated magnetic quiver reads
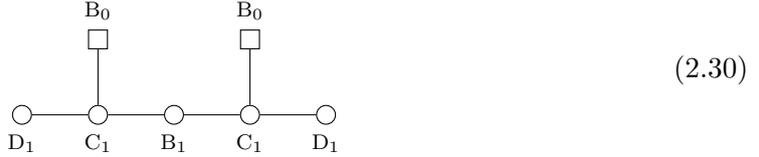
\begin{align}
 \raisebox{-.5\height}{
 	\begin{tikzpicture}
 	\tikzset{node distance = 1cm};
	\tikzstyle{gauge} = [circle, draw,inner sep=2.5pt];
	\tikzstyle{flavour} = [regular polygon,regular polygon sides=4,inner 
sep=2.5pt, draw];
	\node (g1) [gauge,label=below:{\dd{1}}] {};
	\node (g2) [gauge,right of=g1,label=below:{\cc{1}}] {};
	\node (g3) [gauge,right of=g2,label=below:{\bb{1}}] {};
	\node (g4) [gauge,right of=g3,label=below:{\cc{1}}] {};
	\node (g5) [gauge,right of=g4,label=below:{\dd{1}}] {};
	\node (f2) [flavour,above of=g2,label=above:{\bb{0}}] {};
	\node (f4) [flavour,above of=g4,label=above:{\bb{0}}] {};
	\draw (g1)--(g2) (g2)--(g3) (g3)--(g4) (g4)--(g5) (g2)--(f2) (g4)--(f4);
	\end{tikzpicture}
	} 
	\label{eq:magQuiv_Sp1_4_finite}
\end{align}
and Coulomb branch dimension and global symmetry, see Appendix \ref{app:Coulomb_branch}, are computed to be
\begin{subequations}
\label{eq:result_Sp1_4_finite}
\begin{align}
 \dim_{\HH}  \Coulomb \eqref{eq:magQuiv_Sp1_4_finite} = 5 \;, \qquad 
 G = \sorm(8) \,.
\end{align}
In fact, it is known \cite[Tab.\ 7]{Cabrera:2017ucb}, see also \cite{Benini:2010uu,Chacaltana:2012zy}, that upon choosing the magnetic gauge groups corresponding to the algebras in \eqref{eq:magQuiv_Sp1_4_finite} correctly,
\begin{align}
    \Coulomb\eqref{eq:magQuiv_Sp1_4_finite}  = \clorbit_{\dalg_4}^{\min }
\end{align}
\end{subequations}
meaning that \eqref{eq:magQuiv_Sp1_4_finite} describes the closure of the 
minimal nilpotent orbit of $\sormL(8)$. The reasoning behind choosing the 
magnetic gauge group from an orthosynmplectic magnetic quiver is subject of a 
companion paper \cite{Bourget:2020xdz}.
Consequently, the properties \eqref{eq:result_Sp1_4_finite} correctly match the classical Higgs branch as well as the non-abelian part of the global symmetry.
\paragraph{Infinite coupling.}
The brane configuration for infinite gauge coupling is reached by starting from \eqref{eq:web_Sp1_4_Higgs} and making the $(1,\pm1)$ 5-branes become coincident on the \Of\ plane.  In more detail
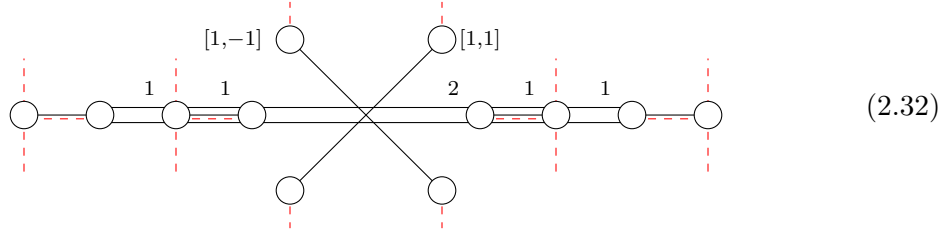
\begin{align}
\raisebox{-.5\height}{
    \begin{tikzpicture}
    \DfiveOMinus{}{-1}
    \DfiveOMinus{}{0}
    \DfiveOMinus{2}{1}
    \DfiveOMinusTilde{1}{-2}
    \MonoCut{-2,-0.05}{-1,-0.05}
    \DfiveOMinusTilde{1}{2}
    \MonoCut{2,-0.05}{3,-0.05}
    \DfiveOMinus{1}{-3}
    \DfiveOMinus{1}{3}
    \OMinusTilde{-3,0}{-4,0}
    \MonoCut{-3,-0.05}{-4,-0.05}
    \OMinusTilde{4,0}{5,0}
    \MonoCut{4,-0.05}{5,-0.05}
    \Dbrane{0.5,0}{-0.5,1}
    \draw (-1.25,1) node {$\scriptstyle{[1,-1]}$};
    \Dbrane{0.5,0}{-0.5,-1}
    \Dbrane{0.5,0}{1.5,1}
    \draw (2,1) node {$\scriptstyle{[1,1]}$};
    \Dbrane{0.5,0}{1.5,-1}
    \MonoCut{-2,-0.75}{-2,0.75}
    \MonoCut{-4,-0.75}{-4,0.75}
    \MonoCut{3,-0.75}{3,0.75}
    \MonoCut{5,-0.75}{5,0.75}
    \MonoCut{-0.5,1}{-0.5,1.5}
    \MonoCut{-0.5,-1}{-0.5,-1.5}
    \MonoCut{1.5,1}{1.5,1.5}
    \MonoCut{1.5,-1}{1.5,-1.5}
    \SevenB{-0.5,1}
    \SevenB{1.5,1}
    \SevenB{-0.5,-1}
    \SevenB{1.5,-1}
    \SevenB{-1,0}
    \SevenB{-2,0}
    \SevenB{-3,0}
    \SevenB{-4,0}
    \SevenB{2,0}
    \SevenB{3,0}
    \SevenB{4,0}
    \SevenB{5,0}
    \end{tikzpicture}
}
\label{eq:web_Sp1_4_infinite}
\end{align}
such that a new Higgs branch degree of freedom appears because the $(1,\pm1)$ 
5-branes form an independent subweb. In contrast to the subwebs formed by 
\Dfive\ branes suspended between $[1,0]$ 7-branes, the subweb given by the 
$(1,\pm1)$ branes passes through the \Of\ plane, but is mapped to 
itself under the orientfold action. Hence, the \Of\ plane does not affect the 
magnetic degrees of freedom of this subweb, such that the contribution is a 
unitary magnetic gauge node as in \cite{Cabrera:2018jxt}.
Without the \Of\ plane, the $(1,\pm1)$ 5-branes have a non-trivial 
intersection, but the orientifold identifies the 7-branes on which the 5-brane 
ends. Thus, the generalised intersection evaluates to zero because of the 
negative contributions from the common 7-brane, such that no new matter 
multiplet arises. 

Using the prescription of the magnetic orientifolds of Table 
\ref{tab:magnetic_orientifold} for the remaining subwebs of \Dfive\ branes and 
computing the intersection numbers, the magnetic quiver is read off to be
\begin{align}
 \raisebox{-.5\height}{
 	\begin{tikzpicture}
 	\tikzset{node distance = 1cm};
	\tikzstyle{gauge} = [circle, draw,inner sep=2.5pt];
	\tikzstyle{flavour} = [regular polygon,regular polygon sides=4,inner 
sep=2.5pt, draw];
	\node (g1) [gauge,label=below:{\dd{1}}] {};
	\node (g2) [gauge,right of=g1,label=below:{\cc{1}}] {};
	\node (g3) [gauge,right of=g2,label=below:{\dd{2}}] {};
	\node (g4) [gauge,right of=g3,label=below:{\cc{1}}] {};
	\node (g5) [gauge,right of=g4,label=below:{\dd{1}}] {};
	\node (g6) [gauge,above of=g3,label=above:{\uu{1}}] {};
	\draw (g1)--(g2) (g2)--(g3) (g3)--(g4) (g4)--(g5) (g3)--(g6);
	\end{tikzpicture}
	} 
	\label{eq:magQuiv_Sp1_4_infinite}
\end{align}
with a Coulomb branch of dimension
\begin{align}
    \dim_{\HH} \Coulomb \eqref{eq:magQuiv_Sp1_4_infinite}=7 \,.
\end{align}
In a companion paper \cite{Bourget:2020xdz}, the choice of magnetic gauge group 
for 
\eqref{eq:magQuiv_Sp1_4_infinite} will be clarified, which then indicates that $\Coulomb \eqref{eq:magQuiv_Sp1_4_infinite} = \clorbit_{E_5}^{\min}$ from a Hilbert series analysis.
\subsubsection{\texorpdfstring{$E_4$}{E4}: Sp(1) with 3 flavours}
The 5-brane web for $\sprm(1)$ with $3$ fundamental flavours can be chosen to be
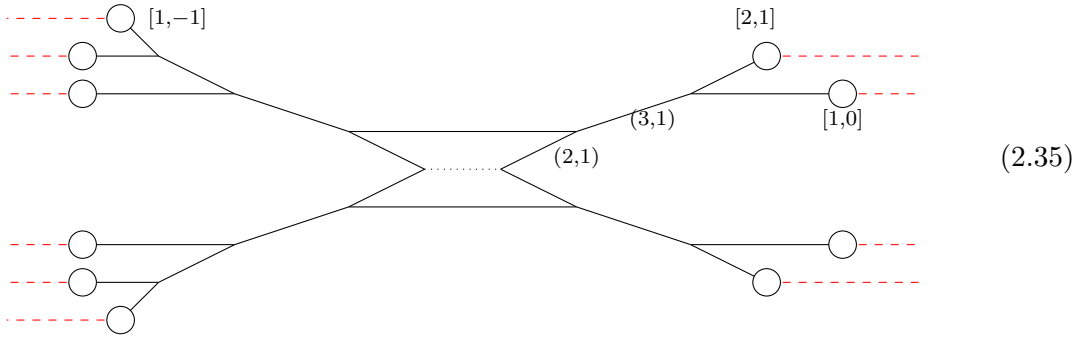
\begin{align}
\raisebox{-.5\height}{
    \begin{tikzpicture}
    \OPlus{0,0}{1,0}
    \Dbrane{0,0}{-1,0.5}
    \Dbrane{1,0}{2,0.5}
     \draw (2,0.15) node {$\scriptstyle{(2,1)}$};
    \Dbrane{-1,0.5}{2,0.5}
    \Dbrane{-1,0.5}{-2.5,1}
    \Dbrane{2,0.5}{3.5,1}
    \draw (3,0.65) node {$\scriptstyle{(3,1)}$};
    \Dbrane{-2.5,1}{-4.5,1}
    \Dbrane{3.5,1}{5.5,1}
     \draw (5.5,0.65) node {$\scriptstyle{[1,0]}$};
    \Dbrane{-2.5,1}{-3.5,1.5}
    \Dbrane{3.5,1}{4.5,1.5}
    \draw (4.35,2) node {$\scriptstyle{[2,1]}$};
    \Dbrane{-3.5,1.5}{-4.5,1.5}
    \Dbrane{-3.5,1.5}{-4,2}
    \draw (-3.25,2) node {$\scriptstyle{[1,-1]}$};
    \MonoCut{-4.5,1.5}{-5.5,1.5}
    \MonoCut{-4.5,1}{-5.5,1}
    \MonoCut{5.5,1}{6.5,1}
    \MonoCut{4.5,1.5}{6.5,1.5}
    \MonoCut{-4,2}{-5.5,2}
    \SevenB{-4,2}
    \SevenB{4.5,1.5}
    \SevenB{-4.5,1.5}
    \SevenB{-4.5,1}
    \SevenB{5.5,1}
    \begin{scope}[yscale=-1,xscale=1]
    \Dbrane{0,0}{-1,0.5}
    \Dbrane{1,0}{2,0.5}
    \Dbrane{-1,0.5}{2,0.5}
    \Dbrane{-1,0.5}{-2.5,1}
    \Dbrane{2,0.5}{3.5,1}
    \Dbrane{-2.5,1}{-4.5,1}
    \Dbrane{3.5,1}{5.5,1}
    \Dbrane{-2.5,1}{-3.5,1.5}
    \Dbrane{3.5,1}{4.5,1.5}
    \Dbrane{-3.5,1.5}{-4.5,1.5}
    \Dbrane{-3.5,1.5}{-4,2}
    \MonoCut{-4.5,1.5}{-5.5,1.5}
    \MonoCut{-4.5,1}{-5.5,1}
    \MonoCut{5.5,1}{6.5,1}
    \MonoCut{4.5,1.5}{6.5,1.5}
    \MonoCut{-4,2}{-5.5,2}
    \SevenB{-4,2}
    \SevenB{4.5,1.5}
    \SevenB{-4.5,1.5}
    \SevenB{-4.5,1}
    \SevenB{5.5,1}
    \end{scope}
    \end{tikzpicture}
}
\label{eq:web_Sp1_3_Coulomb}
\end{align}
where the separation into $2$ flavours on the left and $1$ flavour on the right 
is simply a convenient choice. In order to transfer the 5-brane web 
\eqref{eq:web_Sp1_3_Coulomb} into the Higgs phase, one moves the system to 
the origin of the $5$d Coulomb branch which is reached when the flavour $[1,0]$ 
7-branes and the gauge $(1,0)$ 5-brane are on the \Of\ plane. 
After merging the half $[1,0]$ 7-branes with their mirrors on the orientifold 
and subsequent splitting along the \Of\ plane, the Higgs branch phase for 
\eqref{eq:web_Sp1_3_Coulomb} is given by
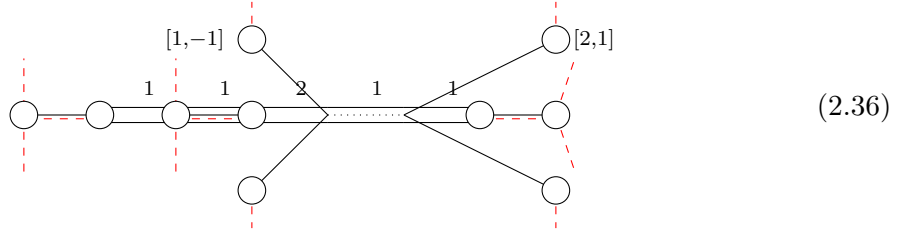
\begin{align}
\raisebox{-.5\height}{
    \begin{tikzpicture}
    \DfiveOPlus{1}{0}
    \DfiveOMinus{2}{-1}
    \DfiveOMinus{1}{1}
    \DfiveOMinusTilde{1}{-2}
    \MonoCut{-2,-0.05}{-1,-0.05}
    \MonoCut{2,-0.05}{3,-0.05}
    \DfiveOMinus{1}{-3}
    \OMinusTilde{-3,0}{-4,0}
    \MonoCut{-3,-0.05}{-4,-0.05}
    \OMinusTilde{2,0}{3,0}
    \Dbrane{0,0}{-1,1}
    \Dbrane{0,0}{-1,-1}
    \draw (-1.75,1) node {$\scriptstyle{[1,-1]}$};
    \Dbrane{1,0}{3,1}
    \Dbrane{1,0}{3,-1}
    \draw (3.5,1) node {$\scriptstyle{[2,1]}$};
    \MonoCut{-2,-0.75}{-2,0.75}
    \MonoCut{-4,-0.75}{-4,0.75}
    \MonoCut{3,0}{3+0.25,0.75}
    \MonoCut{3,0}{3+0.25,-0.75}
    \MonoCut{-1,1}{-1,1.5}
    \MonoCut{-1,-1}{-1,-1.5}
    \MonoCut{3,1}{3,1.5}
    \MonoCut{3,-1}{3,-1.5}
    \SevenB{-1,1}
    \SevenB{3,1}
    \SevenB{-1,-1}
    \SevenB{3,-1}
    \SevenB{-1,0}
    \SevenB{-2,0}
    \SevenB{-3,0}
    \SevenB{-4,0}
    \SevenB{2,0}
    \SevenB{3,0}
    \end{tikzpicture}
}
\label{eq:web_Sp1_3_Higgs}
\end{align}
Inspecting \eqref{eq:web_Sp1_3_Higgs} shows that the $(1,-1)$ 5-brane, in 
contrast to the $(2,1)$ 5-brane, is not a supersymmetric configuration by itself 
and needs to be connected to a $[1,0]$ 7-brane on the left-hand-side via another 
5-brane.
\paragraph{Finite coupling.}
Configuration \eqref{eq:web_Sp1_3_Higgs} can equivalently be described as 
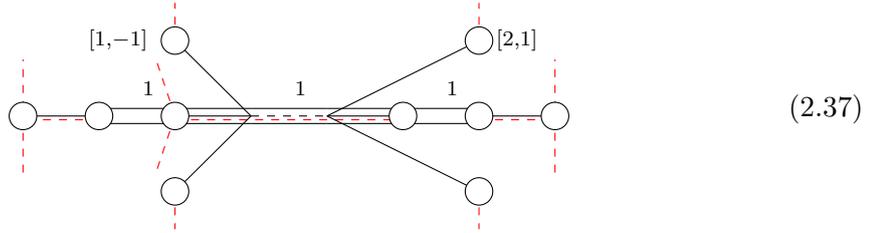
\begin{align}
\raisebox{-.5\height}{
    \begin{tikzpicture}
    \DfiveOMinusTilde{}{0}
    \DfiveOPlusTilde{1}{-1}
    \DfiveOMinus{1}{1}
    \DfiveOMinusTilde{}{-2}
    \MonoCut{-2,-0.05}{1,-0.05}
    \MonoCut{2,-0.05}{3,-0.05}
    \DfiveOMinus{1}{-3}
    \OMinusTilde{-3,0}{-4,0}
    \MonoCut{-3,-0.05}{-4,-0.05}
    \OMinusTilde{2,0}{3,0}
    \Dbrane{-1,0}{-2,1}
    \Dbrane{-1,0}{-2,-1}
    \draw (-2.75,1) node {$\scriptstyle{[1,-1]}$};
    \Dbrane{0,0}{2,1}
    \Dbrane{0,0}{2,-1}
    \draw (2.5,1) node {$\scriptstyle{[2,1]}$};
    \MonoCut{-2,0}{-2-0.25,0.75}
    \MonoCut{-2,0}{-2-0.25,-0.75}
    \MonoCut{-4,-0.75}{-4,0.75}
    \MonoCut{3,-0.75}{3,0.75}
    \MonoCut{-2,1}{-2,1.5}
    \MonoCut{-2,-1}{-2,-1.5}
    \MonoCut{2,1}{2,1.5}
    \MonoCut{2,-1}{2,-1.5}
    \SevenB{-2,1}
    \SevenB{2,1}
    \SevenB{-2,-1}
    \SevenB{2,-1}
    \SevenB{1,0}
    \SevenB{-2,0}
    \SevenB{-3,0}
    \SevenB{-4,0}
    \SevenB{2,0}
    \SevenB{3,0}
    \end{tikzpicture}
}
\label{eq:web_Sp1_3_finite}
\end{align}
by transitioning the $(p,q)$ 5-brane outside the orientifold through the 7-branes and accounting for brane creation and annihilation. As above, the $(1,-1)$ 5-branes between \Ofpt\ and \Ofmt\ are consistent configurations by themselves, but these are not Higgs branch degrees for freedom. The $(2,1)$ 5-branes behave essentially like the $(1,-1)$ branes, because the half monodromy cut turns the $(2,1)$ close to the orientifold into a $(1,1)$ brane.
In contrast, the collection of $(1,0)$ 5-branes suspended between different 7-branes are all independent subwebs, representing non-trivial Higgs branch degrees of freedom. From \eqref{eq:web_Sp1_3_finite}, the magnetic quiver is read off to be
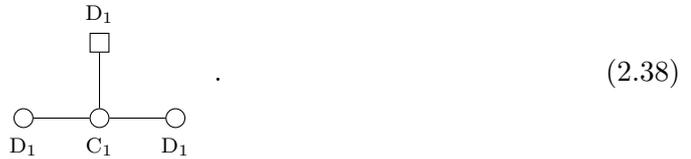
\begin{align}
 \raisebox{-.5\height}{
 	\begin{tikzpicture}
 	\tikzset{node distance = 1cm};
	\tikzstyle{gauge} = [circle, draw,inner sep=2.5pt];
	\tikzstyle{flavour} = [regular polygon,regular polygon sides=4,inner 
sep=2.5pt, draw];
	\node (g1) [gauge,label=below:{\dd{1}}] {};
	\node (g2) [gauge,right of=g1,label=below:{\cc{1}}] {};
	\node (g3) [gauge,right of=g2,label=below:{\dd{1}}] {};
	\node (f2) [flavour,above of=g2,label=above:{\dd{1}}] {};
	\draw (g1)--(g2) (g2)--(g3) (g2)--(f2);
	\end{tikzpicture}
	} \,.
	\label{eq:magQuiv_Sp1_3_finite}
\end{align}
Straightforwardly computing the Coulomb branch dimension and global symmetry, see Appendix \ref{app:Coulomb_branch}, yields
\begin{align}
\dim_{\HH} \Coulomb \eqref{eq:magQuiv_Sp1_3_finite} = 3 \;, \qquad 
G =\sorm(6)
\end{align}
which matches the Higgs branch dimension as well as the non-abelian part of the global symmetry correctly.
\paragraph{Infinite coupling.}
Proceeding to infinite coupling is realised by setting the distance between \NS\ branes to zero. In other words, the $(1,1)$ and $(1,-1)$ 5-branes become coincident on the \Of\ plane and the brane web is given by 
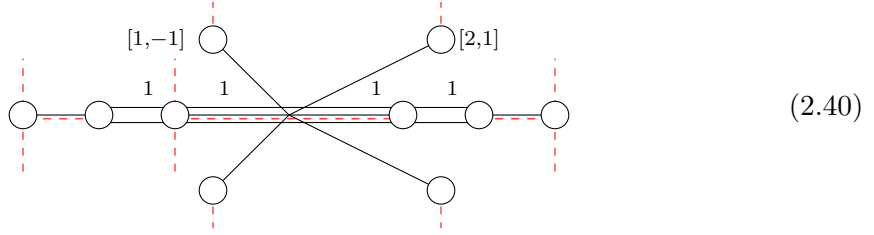
\begin{align}
\raisebox{-.5\height}{
    \begin{tikzpicture}
    \DfiveOMinusTilde{1}{0}
    \DfiveOMinusTilde{}{-1}
    \DfiveOMinus{1}{1}
    \DfiveOMinusTilde{1}{-2}
    \MonoCut{-2,-0.05}{1,-0.05}
    \MonoCut{2,-0.05}{3,-0.05}
    \DfiveOMinus{1}{-3}
    \OMinusTilde{-3,0}{-4,0}
    \MonoCut{-3,-0.05}{-4,-0.05}
    \OMinusTilde{2,0}{3,0}
    \Dbrane{-0.5,0}{-1.5,1}
    \Dbrane{-0.5,0}{-1.5,-1}
    \draw (-2.25,1) node {$\scriptstyle{[1,-1]}$};
    \Dbrane{-0.5,0}{1.5,1}
    \Dbrane{-0.5,0}{1.5,-1}
    \draw (2,1) node {$\scriptstyle{[2,1]}$};
    \MonoCut{-2,-0.75}{-2,0.75}
    \MonoCut{-4,-0.75}{-4,0.75}
    \MonoCut{3,-0.75}{3,0.75}
    \MonoCut{-1.5,1}{-1.5,1.5}
    \MonoCut{-1.5,-1}{-1.5,-1.5}
    \MonoCut{1.5,1}{1.5,1.5}
    \MonoCut{1.5,-1}{1.5,-1.5}
    \SevenB{-1.5,1}
    \SevenB{1.5,1}
    \SevenB{-1.5,-1}
    \SevenB{1.5,-1}
    \SevenB{1,0}
    \SevenB{-2,0}
    \SevenB{-3,0}
    \SevenB{-4,0}
    \SevenB{2,0}
    \SevenB{3,0}
    \end{tikzpicture}
}
\label{eq:web_Sp1_3_infinite}
\end{align}
Crucially, this opens up a new Higgs branch direction, because the combination 
of the $(2,1)$ and $(1,-1)$ 5-brane is a consistent subweb and free to move 
along the $x^{7,8,9}$ direction. As this subweb passes through the \Of\ plane, 
but is mapped to itself, the associated magnetic gauge multiplet is of $\uo$ 
type. Moreover, the intersection number with the central \Dfive\ brane is one, 
so there is one magnetic bi-fundamental hypermultiplet between the $\uo$ and 
the $\calg_1$ node. 
As above, one needs to examine interplay of the $(1,-1)$ and $(2,1)$ 5-branes closely. The intersection number of the 5-branes evaluates to $3$, but becomes corrected by common 7-branes such that the generalised intersection number is $2$.
Therefore, novel matter multiplets arise form the non-trivial generalised intersection number. To see what these might be, it is instructive to imagine the $(1,-1)$ and $(2,1)$ 5-brane without the \Of\ plane. Then each 5-brane would lead to a $\uo$ magnetic vector multiplet and the intersection number dictates the number of copies of bifundamental matter that connects these magnetic gauge nodes. Including the orientifold then leads to an identification of these magnetic vector multiplets and it is suggestive that the two bifundamental magnetic hypermultiplets turn into \emph{one magnetic hypermultiplet of charge} $2$ for the magnetic $\uo$ gauge node.

Since the remainder of the web has not changed compared to the finite coupling case \eqref{eq:web_Sp1_3_finite}, the magnetic quiver for infinite coupling is read off to be
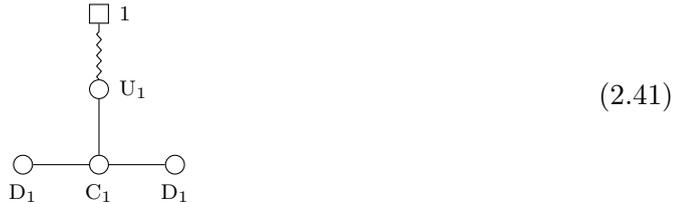
\begin{align}
 \raisebox{-.5\height}{
 	\begin{tikzpicture}
 	\tikzset{node distance = 1cm};
	\tikzstyle{gauge} = [circle, draw,inner sep=2.5pt];
	\tikzstyle{flavour} = [regular polygon,regular polygon sides=4,inner 
sep=2.5pt, draw];
\tikzstyle{fun} = [regular polygon,regular polygon sides=3,inner 
sep=2pt, draw];
	\node (g1) [gauge,label=below:{\dd{1}}] {};
	\node (g2) [gauge,right of=g1,label=below:{\cc{1}}] {};
	\node (g3) [gauge,right of=g2,label=below:{\dd{1}}] {};
	\node (g4) [gauge,above of=g2,label=right:{\uu{1}}] {};
	\node (g5) [flavour,above of=g4,label=right:{$\scriptstyle{1}$}] {};
	\draw (g1)--(g2) (g2)--(g3) (g2)--(g4);
	\draw [line join=round,decorate, decoration={zigzag, segment length=4,amplitude=.9,post=lineto,post length=2pt}]  (g4) -- (g5);
	\end{tikzpicture}
	} 
	\label{eq:magQuiv_Sp1_3_infinite}
\end{align}
where the wiggly line denotes a hypermultiplet of charge $2$.
The Coulomb branch dimension becomes
\begin{align}
    \dim_{\HH} \Coulomb\eqref{eq:magQuiv_Sp1_3_infinite} = 4\,.
\end{align}
Again, the choice of magnetic gauge groups is expanded on in 
\cite{Bourget:2020xdz}; 
the subsequent Hilbert series analysis suggest that the Coulomb branch satisfies 
$\Coulomb\eqref{eq:magQuiv_Sp1_3_infinite}=\clorbit_{D_4}^{\min }$.

One may wonder why the identification in \eqref{eq:magQuiv_Sp1_3_infinite} is 
one hypermultiplet of charge two and not two hypermultiplets of charge one. The 
brane system for both looks the same. The answer to this is coming form the 
computation of the Higgs branch of \eqref{eq:magQuiv_Sp1_3_infinite} which 
should have dimension 1 and not 2. 
%
\subsubsection{\texorpdfstring{$E_3$}{E3}: Sp(1) with 2 flavours}
Next, considering $\sprm(1)$ with $N_f=2$ flavours, the relevant 5-brane web is given by
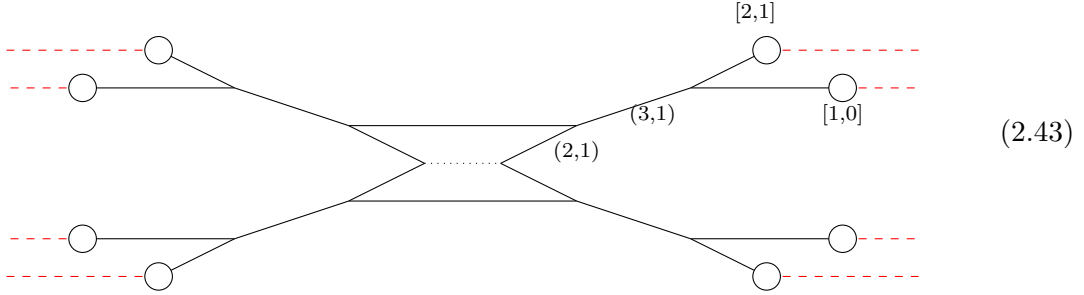
\begin{align}
\raisebox{-.5\height}{
    \begin{tikzpicture}
    \OPlus{0,0}{1,0}
    \Dbrane{0,0}{-1,0.5}
    \Dbrane{1,0}{2,0.5}
     \draw (2,0.15) node {$\scriptstyle{(2,1)}$};
    \Dbrane{-1,0.5}{2,0.5}
    \Dbrane{-1,0.5}{-2.5,1}
    \Dbrane{2,0.5}{3.5,1}
    \draw (3,0.65) node {$\scriptstyle{(3,1)}$};
    \Dbrane{-2.5,1}{-4.5,1}
    \Dbrane{3.5,1}{5.5,1}
     \draw (5.5,0.65) node {$\scriptstyle{[1,0]}$};
    \Dbrane{-2.5,1}{-3.5,1.5}
    \Dbrane{3.5,1}{4.5,1.5}
    \draw (4.35,2) node {$\scriptstyle{[2,1]}$};
    \MonoCut{-4.5,1}{-5.5,1}
    \MonoCut{5.5,1}{6.5,1}
    \MonoCut{4.5,1.5}{6.5,1.5}
    \MonoCut{-3.5,1.5}{-5.5,1.5}
    \SevenB{-3.5,1.5}
    \SevenB{4.5,1.5}
    \SevenB{-4.5,1}
    \SevenB{5.5,1}
    \begin{scope}[yscale=-1,xscale=1]
    \Dbrane{0,0}{-1,0.5}
    \Dbrane{1,0}{2,0.5}
    \Dbrane{-1,0.5}{2,0.5}
    \Dbrane{-1,0.5}{-2.5,1}
    \Dbrane{2,0.5}{3.5,1}
    \Dbrane{-2.5,1}{-4.5,1}
    \Dbrane{3.5,1}{5.5,1}
    \Dbrane{-2.5,1}{-3.5,1.5}
    \Dbrane{3.5,1}{4.5,1.5}
    \MonoCut{-4.5,1}{-5.5,1}
    \MonoCut{5.5,1}{6.5,1}
    \MonoCut{4.5,1.5}{6.5,1.5}
    \MonoCut{-3.5,1.5}{-5.5,1.5}
    \SevenB{-3.5,1.5}
    \SevenB{4.5,1.5}
    \SevenB{-4.5,1}
    \SevenB{5.5,1}
    \end{scope}
    \end{tikzpicture}
}
\label{eq:web_Sp1_2_Coulomb}
\end{align}
where the two flavours have been chosen symmetrically. Transitioning into the Higgs branch phase, the gauge and flavour \Dfive s align with the \Of\ plane and the flavour half 7-branes are split along the orientifold. Then, the brane web becomes
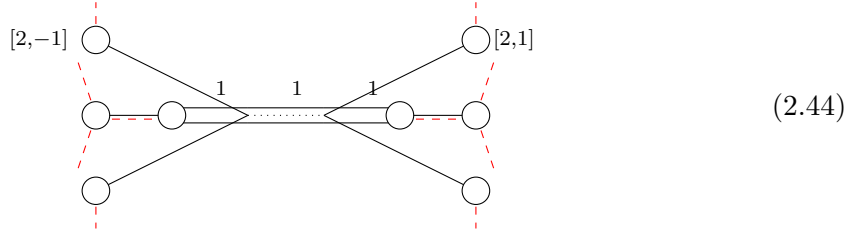
\begin{align}
\raisebox{-.5\height}{
    \begin{tikzpicture}
    \DfiveOPlus{1}{0}
    \DfiveOMinus{1}{-1}
    \DfiveOMinus{1}{1}
    \OMinusTilde{-2,0}{-1,0}
    \MonoCut{-2,-0.05}{-1,-0.05}
    \MonoCut{2,-0.05}{3,-0.05}
    \OMinusTilde{2,0}{3,0}
    \Dbrane{0,0}{-2,1}
    \Dbrane{0,0}{-2,-1}
    \draw (-2.75,1) node {$\scriptstyle{[2,-1]}$};
    \Dbrane{1,0}{3,1}
    \Dbrane{1,0}{3,-1}
    \draw (3.5,1) node {$\scriptstyle{[2,1]}$};
    \MonoCut{-2,0}{-2-0.25,0.75}
    \MonoCut{-2,0}{-2-0.25,-0.75}
    \MonoCut{3,0}{3+0.25,0.75}
    \MonoCut{3,0}{3+0.25,-0.75}
    \MonoCut{-2,1}{-2,1.5}
    \MonoCut{-2,-1}{-2,-1.5}
    \MonoCut{3,1}{3,1.5}
    \MonoCut{3,-1}{3,-1.5}
    \SevenB{-2,1}
    \SevenB{3,1}
    \SevenB{-2,-1}
    \SevenB{3,-1}
    \SevenB{-1,0}
    \SevenB{-2,0}
    \SevenB{2,0}
    \SevenB{3,0}
    \end{tikzpicture}
}
\label{eq:web_Sp1_2_Higgs}
\end{align}
and one observes that the $(2,\pm1)$ 5-brane are already free configurations.
\paragraph{Finite coupling.}
To read off the magnetic quiver at finite coupling, the $(2,\pm1)$ 5-branes in \eqref{eq:web_Sp1_2_Higgs} can stay where they are, as one can read a magnetic \Ofm\ plane with two magnetic flavours. Thus, the magnetic quiver associated to \eqref{eq:web_Sp1_2_Higgs} is found to be
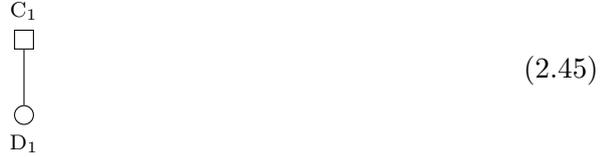
\begin{align}
 \raisebox{-.5\height}{
 	\begin{tikzpicture}
 	\tikzset{node distance = 1cm};
	\tikzstyle{gauge} = [circle, draw,inner sep=2.5pt];
	\tikzstyle{flavour} = [regular polygon,regular polygon sides=4,inner 
sep=2.5pt, draw];
	\node (g1) [gauge,label=below:{\dd{1}}] {};
	\node (f1) [flavour,above of=g1,label=above:{\cc{1}}] {};
	\draw (g1)--(f1);
	\end{tikzpicture}
	} 
	\label{eq:magQuiv_Sp1_2_finite}
\end{align}
which has a Coulomb branch of quaternionic dimension one. In fact, it is known  
that
\begin{align}
    \Coulomb \eqref{eq:magQuiv_Sp1_2_finite}  = 
\clorbit_{\dalg_2}^{\min } \,.
\end{align}
As known from \cite{Ferlito:2016grh}, the finite coupling Higgs branch is a 
union of two (isomorphic) cones. Correspondingly, one would expect two maximal 
inequivalent subdivision of the brane web \eqref{eq:web_Sp1_2_Higgs}, but 
the current observations show only one.
\paragraph{Infinite coupling.}
To reach the infinite coupling phase, the $(2,\pm 1)$ 5-branes in \eqref{eq:web_Sp1_2_Higgs} are made coincident along the orientifold such that the web becomes
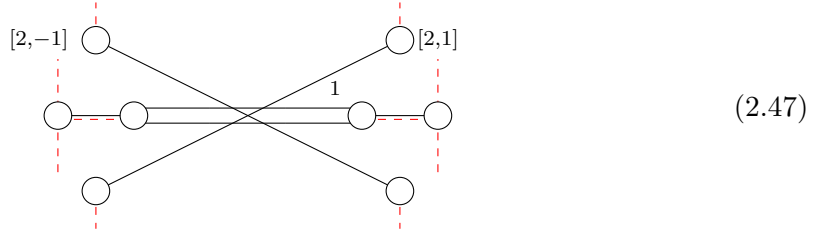
\begin{align}
\raisebox{-.5\height}{
    \begin{tikzpicture}
    \DfiveOMinus{}{-1}
    \DfiveOMinus{}{0}
    \DfiveOMinus{1}{1}
    \OMinusTilde{-2,0}{-1,0}
    \MonoCut{-2,-0.05}{-1,-0.05}
    \OMinusTilde{2,0}{3,0}
    \MonoCut{2,-0.05}{3,-0.05}
    \Dbrane{-1.5,1}{2.5,-1}
    \Dbrane{-1.5,-1}{2.5,1}
    \draw (-2.25,1) node {$\scriptstyle{[2,-1]}$};
    \draw (3,1) node {$\scriptstyle{[2,1]}$};
    \MonoCut{-2,-0.75}{-2,0.75}
    \MonoCut{3,-0.75}{3,0.75}
    \MonoCut{-1.5,1}{-1.5,1.5}
    \MonoCut{-1.5,-1}{-1.5,-1.5}
    \MonoCut{2.5,1}{2.5,1.5}
    \MonoCut{2.5,-1}{2.5,-1.5}
    \SevenB{-1.5,1}
    \SevenB{2.5,1}
    \SevenB{-1.5,-1}
    \SevenB{2.5,-1}
    \SevenB{-1,0}
    \SevenB{-2,0}
    \SevenB{2,0}
    \SevenB{3,0}
    \end{tikzpicture}
}
\label{eq:web_Sp1_2_infinite}
\end{align}
and one can proceed to read off the Higgs branch degrees of freedom. The subweb 
consisting of a \Dfive\ on top of an \Ofm\ plane provides a $\dalg_1$ magnetic 
node. The subweb formed by the $(2,\pm1)$ branes yields a $\uo$ node, that is 
connected by a bifundamental to the $\dalg_1$ node because the intersection 
number is $1$. In addition, the self-intersection of the $(2,\pm1)$ branes needs 
be considered. The intersection number of the $(2,\pm1)$ 5-branes is four, but 
there are also common 7-branes to be take into account. In total, the 
generalised intersection number is $2$, which then suggest that the $\uo$ 
magnetic gauge node needs to be supplemented by \emph{one magnetic 
hypermultiplet of charge} $2$. Thus, the magnetic quiver is given by
\begin{align}
 \raisebox{-.5\height}{
 	\begin{tikzpicture}
 	\tikzset{node distance = 1cm};
	\tikzstyle{gauge} = [circle, draw,inner sep=2.5pt];
	\tikzstyle{flavour} = [regular polygon,regular polygon sides=4,inner 
sep=2.5pt, draw];
\tikzstyle{fun} = [regular polygon,regular polygon sides=3,inner 
sep=2pt, draw];
	\node (g1) [gauge,label=below:{\dd{1}}] {};
	\node (g2) [gauge,above of=g1,label=right:{\uu{1}}] {};
	\node (f2) [flavour,above of=g2,label=right:{$\scriptstyle{1}$}] {};
	\draw (g1)--(g2);
	\draw [line join=round,decorate, decoration={zigzag, segment length=4,amplitude=.9,post=lineto,post length=2pt}]  (g2) -- (f2);
	\end{tikzpicture}
	} 
	\label{eq:magQuiv_Sp1_2_infinite}
\end{align}
Notice that only a $U_1\subset C_1$ subalgebra of the $C_1$ flavour symmetry in \eqref{eq:magQuiv_Sp1_2_finite} is gauged. The moduli space becomes
\begin{align}
    \Coulomb \eqref{eq:magQuiv_Sp1_2_infinite}  = 
\clorbit_{A_2}^{\min}
\end{align}
which is also denoted as $a_2$. 

It should be noted that the expectation is that there should be two maximal subdivisions \cite[Sec.\ 2.1]{Cabrera:2018jxt} in order to reflect the existence of two cones $e_3\cong a_2\cup a_1$ \cite{Seiberg:1996bd,Morrison:1996xf}. However, the 5-brane web \eqref{eq:web_Sp1_2_infinite} does not show any signs for a second subdivision.
\subsubsection{\texorpdfstring{$E_2$}{E2}: Sp(1) with 1 flavour}
Reducing the number of flavours to $N_f=1$, the 5-brane web is given by
\begin{align}
\raisebox{-.5\height}{
    \begin{tikzpicture}
    \OPlus{0,0}{1,0}
    \Dbrane{0,0}{-1,0.5}
    \Dbrane{1,0}{2,0.5}
     \draw (2,0.15) node {$\scriptstyle{(2,1)}$};
    \Dbrane{-1,0.5}{2,0.5}
    \Dbrane{-1,0.5}{-2.5,1}
    \Dbrane{2,0.5}{3.5,1}
    \draw (3.5,1.5) node {$\scriptstyle{[3,1]}$};
    \Dbrane{-2.5,1}{-4.5,1}
    \Dbrane{-2.5,1}{-3.5,1.5}
    \draw (-3.65,2) node {$\scriptstyle{[2,-1]}$};
    \MonoCut{-4.5,1}{-5.5,1}
    \MonoCut{3.5,1}{4.5,1}
    \MonoCut{-3.5,1.5}{-5.5,1.5}
    \SevenB{-3.5,1.5}
    \SevenB{3.5,1}
    \SevenB{-4.5,1}
    \begin{scope}[yscale=-1,xscale=1]
    \Dbrane{0,0}{-1,0.5}
    \Dbrane{1,0}{2,0.5}
    \Dbrane{-1,0.5}{2,0.5}
    \Dbrane{-1,0.5}{-2.5,1}
    \Dbrane{2,0.5}{3.5,1}
    \Dbrane{-2.5,1}{-4.5,1}
    \Dbrane{-2.5,1}{-3.5,1.5}
    \MonoCut{-4.5,1}{-5.5,1}
    \MonoCut{3.5,1}{4.5,1}
    \MonoCut{-3.5,1.5}{-5.5,1.5}
    \SevenB{-3.5,1.5}
    \SevenB{3.5,1}
    \SevenB{-4.5,1}
    \end{scope}
    \end{tikzpicture}
}
\label{eq:web_Sp1_1_Coulomb}
\end{align}
and one transitions to the Higgs branch phase by the same steps as above. Thus, one arrives at
\begin{align}
\raisebox{-.5\height}{
    \begin{tikzpicture}
    \DfiveOPlus{1}{0}
    \DfiveOMinus{1}{-1}
    \OMinusTilde{-2,0}{-1,0}
    \MonoCut{-2,-0.05}{-1,-0.05}
    \Dbrane{0,0}{-2,1}
    \Dbrane{0,0}{-2,-1}
    \draw (-2.75,1) node {$\scriptstyle{[2,-1]}$};
    \Dbrane{1,0}{4,1}
    \Dbrane{1,0}{4,-1}
    \draw (4.5,1) node {$\scriptstyle{[3,1]}$};
    \MonoCut{-2,0}{-2-0.25,0.75}
    \MonoCut{-2,0}{-2-0.25,-0.75}
    \MonoCut{-2,1}{-2,1.5}
    \MonoCut{-2,-1}{-2,-1.5}
    \MonoCut{4,1}{4,1.5}
    \MonoCut{4,-1}{4,-1.5}
    \SevenB{-2,1}
    \SevenB{4,1}
    \SevenB{-2,-1}
    \SevenB{4,-1}
    \SevenB{-1,0}
    \SevenB{-2,0}
\end{tikzpicture}
}
\label{eq:web_Sp1_1_Higgs}
\end{align}
where the $(2,-1)$ brane is a free configuration. However, the $(3,1)$ 5-brane is connected via a \Dfive\ brane to the 7-branes on the left-hand-side.
\paragraph{Finite coupling.}
Eliminating the non-dynamical branes via brane annihilation leads to a clean Higgs branch phase at finite coupling:
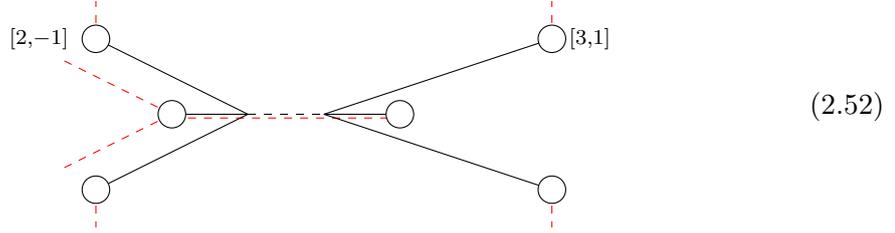
\begin{align}
\raisebox{-.5\height}{
    \begin{tikzpicture}
    \OMinusTilde{-1,0}{0,0}
    \OPlusTilde{0,0}{1,0}
    \OMinusTilde{1,0}{2,0}
    \MonoCut{-1,-0.05}{2,-0.05}
    \Dbrane{0,0}{-2,1}
    \Dbrane{0,0}{-2,-1}
    \draw (-2.75,1) node {$\scriptstyle{[2,-1]}$};
    \Dbrane{1,0}{4,1}
    \Dbrane{1,0}{4,-1}
    \draw (4.5,1) node {$\scriptstyle{[3,1]}$};
    \MonoCut{-1,0}{-1-1.5,0.75}
    \MonoCut{-1,0}{-1-1.5,-0.75}
    \MonoCut{-2,1}{-2,1.5}
    \MonoCut{-2,-1}{-2,-1.5}
    \MonoCut{4,1}{4,1.5}
    \MonoCut{4,-1}{4,-1.5}
    \SevenB{-2,1}
    \SevenB{4,1}
    \SevenB{-2,-1}
    \SevenB{4,-1}
    \SevenB{-1,0}
    \SevenB{2,0}
\end{tikzpicture}
}
\label{eq:web_Sp1_1_finite}
\end{align}
and one realises that there are no Higgs branch directions. Hence, the magnetic quiver is empty.
\paragraph{Infinite coupling.}
From \eqref{eq:web_Sp1_1_finite} one takes the gauge coupling to infinity by 
making the $(2,-1)$ and $(3,1)$ 5-brane coincident on the orientifold. The 
5-brane web is
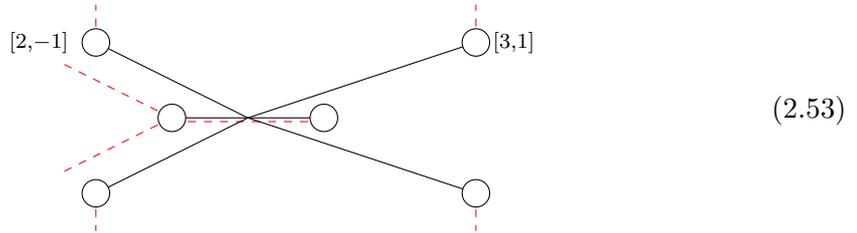
\begin{align}
\raisebox{-.5\height}{
    \begin{tikzpicture}
    \OMinusTilde{-1,0}{1,0}
    \MonoCut{-1,-0.05}{1,-0.05}
    \Dbrane{0,0}{-2,1}
    \Dbrane{0,0}{-2,-1}
    \draw (-2.75,1) node {$\scriptstyle{[2,-1]}$};
    \Dbrane{0,0}{3,1}
    \Dbrane{0,0}{3,-1}
    \draw (3.5,1) node {$\scriptstyle{[3,1]}$};
    \MonoCut{-1,0}{-1-1.5,0.75}
    \MonoCut{-1,0}{-1-1.5,-0.75}
    \MonoCut{-2,1}{-2,1.5}
    \MonoCut{-2,-1}{-2,-1.5}
    \MonoCut{3,1}{3,1.5}
    \MonoCut{3,-1}{3,-1.5}
    \SevenB{-2,1}
    \SevenB{3,1}
    \SevenB{-2,-1}
    \SevenB{3,-1}
    \SevenB{-1,0}
    \SevenB{1,0}
\end{tikzpicture}
}
\label{eq:web_Sp1_1_infinite}
\end{align}
and one recognises a new Higgs branch direction. The subweb consisting of the $(2,-1)$ and $(3,1)$ brane yield a $\uo$ magnetic gauge group, and the self-intersection of $4$ contributes \emph{two magnetic hypermultiplets of charge} $2$. Therefore, the magnetic quiver becomes
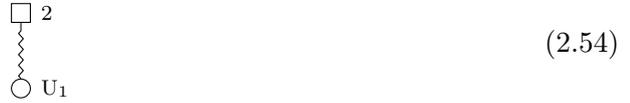
\begin{align}
 \raisebox{-.5\height}{
 	\begin{tikzpicture}
 	\tikzset{node distance = 1cm};
	\tikzstyle{gauge} = [circle, draw,inner sep=2.5pt];
	\tikzstyle{flavour} = [regular polygon,regular polygon sides=4,inner 
sep=2.5pt, draw];
\tikzstyle{fun} = [regular polygon,regular polygon sides=3,inner 
sep=2pt, draw];
	\node (g1) [gauge,label=right:{\uu{1}}] {};
	\node (f1) [flavour,above of=g1,label=right:{$\scriptstyle{2}$}] {};
	\draw [line join=round,decorate, decoration={zigzag, segment length=4,amplitude=.9,post=lineto,post length=2pt}]  (g1) -- (f1);
	\end{tikzpicture}
	} 
	\label{eq:magQuiv_Sp1_1_infinite}
\end{align}
such that the Coulomb branch is
\begin{align}
\Coulomb  \eqref{eq:magQuiv_Sp1_1_infinite}  = \C^2 \slash \Z_2\,,
\end{align}
which equals the $A_1$ part of the $e_2 \cong A_1\cup\mathbb{Z}_2$ moduli space. The $\mathbb{Z}_2$ factor is due to a nilpotent operator in the infinite coupling Higgs branch chiral ring \cite{Cremonesi:2015lsa} and hence not detectable with current magnetic quiver methods.
\subsubsection{\texorpdfstring{$E_1$}{E1}: pure Sp(1) theory}
Now, consider pure $\sprm(1)$ gauge theory with brane web given by
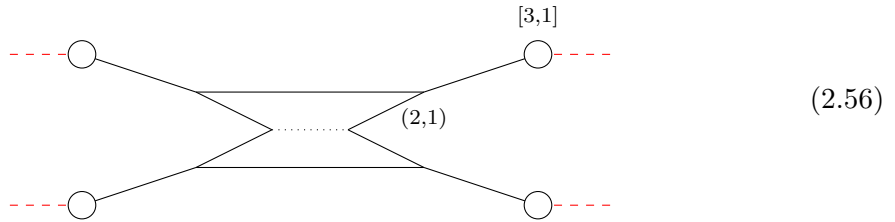
\begin{align}
\raisebox{-.5\height}{
    \begin{tikzpicture}
    \OPlus{0,0}{1,0}
    \Dbrane{0,0}{-1,0.5}
    \Dbrane{1,0}{2,0.5}
     \draw (2,0.15) node {$\scriptstyle{(2,1)}$};
    \Dbrane{-1,0.5}{2,0.5}
    \Dbrane{-1,0.5}{-2.5,1}
    \Dbrane{2,0.5}{3.5,1}
    \draw (3.5,1.5) node {$\scriptstyle{[3,1]}$};
    \MonoCut{3.5,1}{4.5,1}
    \MonoCut{-2.5,1}{-3.5,1}
    \SevenB{-2.5,1}
    \SevenB{3.5,1}
    \begin{scope}[yscale=-1,xscale=1]
    \Dbrane{0,0}{-1,0.5}
    \Dbrane{1,0}{2,0.5}
    \Dbrane{-1,0.5}{2,0.5}
    \Dbrane{-1,0.5}{-2.5,1}
    \Dbrane{2,0.5}{3.5,1}
    \MonoCut{3.5,1}{4.5,1}
    \MonoCut{-2.5,1}{-3.5,1}
    \SevenB{-2.5,1}
    \SevenB{3.5,1}
    \end{scope}
    \end{tikzpicture}
}
\label{eq:web_Sp1_0_Coulomb}
\end{align}
with no flavours.
\paragraph{Finite coupling.}
The Higgs branch phase is realised by aligning the \Dfive\ with the orientifold. Hence, the web becomes
\begin{align}
\raisebox{-.5\height}{
    \begin{tikzpicture}
    \DfiveOPlus{1}{0}
    \Dbrane{-3,1}{0,0}
    \Dbrane{-3,-1}{0,0}
    \Dbrane{1,0}{4,1}
    \Dbrane{1,0}{4,-1}
    \draw (-3.75,1) node {$\scriptstyle{[3,-1]}$};
    \draw (4.5,1) node {$\scriptstyle{[3,1]}$};
    \MonoCut{-3,1}{-3,1.5}
    \MonoCut{-3,-1}{-3,-1.5}
    \MonoCut{4,1}{4,1.5}
    \MonoCut{4,-1}{4,-1.5}
    \SevenB{-3,1}
    \SevenB{4,1}
    \SevenB{-3,-1}
    \SevenB{4,-1}
    \end{tikzpicture}
}
\label{eq:web_Sp1_0_finite}
\end{align}
and shows that there are no Higgs branch directions at finite coupling. Consequently, the magnetic quiver is empty.
\paragraph{Infinite coupling.}
Transitioning \eqref{eq:web_Sp1_0_finite} to infinite coupling, by making the $(3,\pm1)$ branes coincident, the brane web reads
\begin{align}
\raisebox{-.5\height}{
    \begin{tikzpicture}
    \Dbrane{-2.5,1}{3.5,-1}
    \Dbrane{-2.5,-1}{3.5,1}
    \draw (-3.25,1) node {$\scriptstyle{[3,-1]}$};
    \draw (4,1) node {$\scriptstyle{[3,1]}$};
    \MonoCut{-2.5,1}{-2.5,1.5}
    \MonoCut{-2.5,-1}{-2.5,-1.5}
    \MonoCut{3.5,1}{3.5,1.5}
    \MonoCut{3.5,-1}{3.5,-1.5}
    \SevenB{-2.5,1}
    \SevenB{3.5,1}
    \SevenB{-2.5,-1}
    \SevenB{3.5,-1}
    \end{tikzpicture}
}
\label{eq:web_Sp1_0_infinite}
\end{align}
and one recognises one additional Higgs branch direction. The magnetic gauge group is a $\uo$, and the generalised self-intersection of $4$ between the $(3\pm1)$ 5-branes suggest that there are \emph{two magnetic hypermultiplets of charge} $2$ . The associated magnetic quiver becomes
\begin{align}
 \raisebox{-.5\height}{
 	\begin{tikzpicture}
 	\tikzset{node distance = 1cm};
	\tikzstyle{gauge} = [circle, draw,inner sep=2.5pt];
	\tikzstyle{flavour} = [regular polygon,regular polygon sides=4,inner 
sep=2.5pt, draw];
\tikzstyle{fun} = [regular polygon,regular polygon sides=3,inner 
sep=2pt, draw];
	\node (g1) [gauge,label=right:{\uu{1}}] {};
	\node (f1) [flavour,above of=g1,label=right:{$\scriptstyle{2}$}] {};
	\draw [line join=round,decorate, decoration={zigzag, segment length=4,amplitude=.9,post=lineto,post length=2pt}]  (g1) -- (f1);
	\end{tikzpicture}
	} 
	\label{eq:magQuiv_Sp1_0_infinite}
\end{align}
and the moduli space is
\begin{align}
    \Coulomb \eqref{eq:magQuiv_Sp1_0_infinite} = \C^2 \slash \Z_2 
\,,
\end{align}
i.e.\ the $e_1 \cong A_1$ space; in agreement with 
\cite{Cremonesi:2015lsa,Cabrera:2018jxt}.

\paragraph{The $\widetilde{E}_1$ theory.} In \cite{Hayashi:2017btw} an \Of\ 
construction for the $\widetilde{E}_1$ theory, whose Higgs branch is known to be 
a single point \cite{Aharony:1997ju}, was proposed; however, this is not 
elaborated here and the reader is referred to the 
discussion in \cite{Hayashi:2017btw}.
%
%

\subsubsection{\texorpdfstring{$E_0$}{E0} theory}
Finally, to conclude this subsection, it is natural to introduce an $E_0$ theory. It does not admit a gauge theory description, as the number of flavours would need to be negative. However, as shown in Tables \ref{tab:Efamilies}-\ref{tab:CompareFiniteHWG}, higher rank theories allow to define $E_n$ families for arbitrary $- \infty < n \leq 8$. 

The family of interest here is $E_{4-2l}$, specifically for $l=2$. Brane systems, magnetic quivers and Hasse diagrams are derived in full generality in the coming sections for rank $k$ satisfying $k \geq l$; for $k=1$ one can attempt an analytic continuation. As shown in Table \ref{tab:CompareFiniteHasse2}, the dimension of the Higgs branch at infinite coupling is $0$, consistent with the expectation that the $E_0$ theory has a Higgs branch which is a single point \cite{Aharony:1997ju}. Correspondingly the Hasse diagram is reduced to a single trivial leaf, which is a point. 

\subsection{Rules for magnetic quivers}
\label{sec:rules}
After establishing the validity of the magnetic quiver proposal by using magnetic orientifolds of Table \ref{tab:magnetic_orientifold}, one has to formulate the proposal for a generic 5-brane web in the presence of \Of\ planes. 
Building on Section \ref{sec:SU2}, the magnetic quivers for a given 5-brane web in the Higgs branch phase are derived as follows:

Consider a 5-brane web with \Of\ planes in which each external half $(p,q)$ 5-brane is terminated on a half $[p,q]$ 7-brane.
\begin{myConj}[Magnetic Quiver]
\label{conj:rules}
The Higgs branch phase is realised when all the flavour and all gauge \Dfive\ branes are coincident with the \Of\ plane, such that the resulting physical $[1,0]$ 7-branes are split along the orientifold. Then, for each inequivalent maximal subdivision into subwebs suspended between 7-branes, which is compatible with supersymmetry, the associated magnetic quiver can be derived by the following set of rules: 
\begin{itemize}
    \item \textbf{Gauge nodes:} A stack of $k$ identical subwebs which are free to move in the $x^{7,8,9}$ direction contribute a magnetic gauge multiplet. 
    \begin{compactenum}[(a)]
        \item A stack of \Dfive s, on top of an \Of\ plane, the magnetic gauge algebra is the one naturally associate to the magnetic orientifold.
        \item A stack of \NS s,  on top of an \Of\ plane, the magnetic gauge algebra is determined by, firstly, transitioning to the magnetic orientifold of Table \ref{tab:magnetic_orientifold} and, secondly, using the corresponding flavour algebra of Table \ref{tab:orientifold} as magnetic gauge algebra. 
        \item A stack of any other kind of subwebs contributes a $\urm(k)$ 
magnetic gauge multiplet. In addition, Observation \ref{obs:matter} may apply.
    \end{compactenum}
    \item \textbf{Flavour nodes:} The 5-branes that are allowed by 
supersymmetry, but are not free to move along the $x^{7,8,9}$ direction, only 
contribute flavour nodes to the magnetic quiver. The type of flavour is dictated 
by the magnetic orientifold. Most notably, a $(1,\pm1)$ 5-brane between a 
$\Ofpt$ and $\Ofmt$ plane contributes a $\balg_0$ magnetic flavour, 
denoting a single half-hypermultiplet.
    \item \textbf{Matter:} Two non-identical subwebs are linked by a number of bi-fundamental magnetic hypermultiplets, determined by the generalised intersection number.
\end{itemize}
\end{myConj}
In addition, there are infinite coupling configurations that involve $(p,q)$ 5-branes which are neither \Dfive\ nor \NS\ branes. The corresponding contributions to magnetic quivers are observed to be as follows:
\begin{myObs}[Charged matter]
\label{obs:matter}
For a pair of $(r,\pm1)$ 5-branes on a \Ofm\ plane, the stable intersection 
between of the two mirror half 5-branes is $2r$, while the common 7-branes 
contribute a $-2$ such that the generalised intersection number is 
$2r-2$.
The $\uo$ node is supplemented by $\tfrac{2r-2}{2}$ charge $2$ 
hypermultiplets and the magnetic quiver reads
\begin{subequations}
\begin{align}
    \raisebox{-.5\height}{
    \begin{tikzpicture}
    \DfiveOMinus{}{0}
    \DfiveOMinus{}{-1}
    \DfiveOMinus{$k$}{1}
    \Dbrane{0.5,0}{2.5,1}
    \draw (3,1) node {$\scriptstyle{[r,1]}$};
    \Dbrane{0.5,0}{2.5,-1}
    \Dbrane{0.5,0}{-1.5,1}
    \draw (-2.5,1) node {$\scriptstyle{[r,-1]}$};
    \Dbrane{0.5,0}{-1.5,-1}
    \MonoCut{2.5,1}{2.5,1.5}
    \MonoCut{2.5,-1}{2.5,-1.5}
    \MonoCut{-1.5,1}{-1.5,1.5}
    \MonoCut{-1.5,-1}{-1.5,-1.5}
    \SevenB{2.5,1}
    \SevenB{-1.5,1}
    \SevenB{2.5,-1}
    \SevenB{-1.5,-1}
    \SevenB{-1,0}
    \SevenB{2,0}
    \end{tikzpicture}
}
\qquad \longrightarrow\qquad 
 \raisebox{-.5\height}{
 	\begin{tikzpicture}
 	\tikzset{node distance = 1cm};
	\tikzstyle{gauge} = [circle, draw,inner sep=2.5pt];
	\tikzstyle{flavour} = [regular polygon,regular polygon sides=4,inner 
sep=2.5pt, draw];
\tikzstyle{fun} = [regular polygon,regular polygon sides=3,inner 
sep=2pt, draw];
	\node (g1) [gauge,label=right:{\dd{k}}] {};
	\node (g2) [gauge,above of=g1,label=right:{\uu{1}}] {};%
	\node (g3) [flavour,above of=g2,label=right:{$\scriptstyle{r{-}1}$}] {};
	\draw (g1)--(g2);
	\draw [line join=round,decorate, decoration={zigzag, segment length=4,
    amplitude=.9,post=lineto,post length=2pt}]  (g2) -- (g3);
	\end{tikzpicture}
	} 
\end{align}
Analogously, for a pair of  $(r,-1)$ and $(r+1,1)$ 5-branes on a \Ofmt\ plane, 
the generalised intersection number of the two mirror half 5-branes is $2r-1$. 
The \Ofmt\ plane is suspected to additionally modify the intersection number by 
$+1$ due to the stuck half \Dfive. 
The $\uo$ node has $r$ charge $2$ magnetic hypermultiplets such that the 
contribution to the magnetic quiver then becomes
\begin{align}
    \raisebox{-.5\height}{
    \begin{tikzpicture}
    \DfiveOMinusTilde{$k$}{1}
    \DfiveOMinusTilde{}{0}
    \DfiveOMinusTilde{}{-1}
    \DfiveOMinusTilde{}{-2}
    \Dbrane{-0.5,0}{2.5,1}
    \draw (3.5,1) node {$\scriptstyle{[r{+}1,1]}$};
    \Dbrane{-0.5,0}{2.5,-1}
    \Dbrane{-0.5,0}{-2.5,1}
    \draw (-3.5,1) node {$\scriptstyle{[r,-1]}$};
    \Dbrane{-0.5,0}{-2.5,-1}
    \MonoCut{-2,-0.05}{2,-0.05}
    \MonoCut{2.5,1}{2.5,1.5}
    \MonoCut{2.5,-1}{2.5,-1.5}
    \MonoCut{-2.5,1}{-2.5,1.5}
    \MonoCut{-2.5,-1}{-2.5,-1.5}
    \MonoCut{-2,0}{-2-0.75,0.75}
    \MonoCut{-2,0}{-2-0.75,-0.75}
    \SevenB{2.5,1}
    \SevenB{-2.5,1}
    \SevenB{2.5,-1}
    \SevenB{-2.5,-1}
    \SevenB{2,0}
    \SevenB{-2,0}
    \end{tikzpicture}
}
\qquad \longrightarrow\qquad 
 \raisebox{-.5\height}{
 	\begin{tikzpicture}
 	\tikzset{node distance = 1cm};
	\tikzstyle{gauge} = [circle, draw,inner sep=2.5pt];
	\tikzstyle{flavour} = [regular polygon,regular polygon sides=4,inner 
sep=2.5pt, draw];
    \tikzstyle{fun} = [regular polygon,regular polygon sides=3,inner 
sep=2pt, draw];
	\node (g1) [gauge,label=right:{\cc{k}}] {};
	\node (g2) [gauge,above of=g1,label=right:{\uu{1}}] {};%
	\node (g3) [flavour,above of=g2,label=right:{$\scriptstyle{r}$}] {};
	\draw (g1)--(g2);
	\draw [line join=round,decorate, decoration={zigzag, segment length=4,
    amplitude=.9,post=lineto,post length=2pt}]  (g2) -- (g3);
	\end{tikzpicture}
	} 
\end{align}
\end{subequations}
\end{myObs}

As a comment, contrary to 5-brane web without orientifolds 
\cite{Cabrera:2018jxt}, the examples considered in this paper do not show any 
signs of multiple inequivalent subdivisions for a given 5-brane web. However, 
the possibility is covered by Conjecture \ref{conj:rules}.
%
%
\section{Single Sp(k) gauge group}
\label{sec:Spk}
Having developed the magnetic quiver proposal for 5-brane webs with \Of\ 
planes, this section focuses on brane constructions for $\sprm(k)$ gauge 
theories with fundamental hypermultiplets.
To begin with, known facts from field theory and 5-brane webs are collected and 
contrasted.
\paragraph{Field theory.}
For a 5d $\Ncal =1$ $\sprm(k)$ gauge theory with $N_f$ fundamental flavours, a non-trivial interacting fixed point has been argued to exist for $N_f \leq 2k+4$ in \cite[Sec.\ 4]{Intriligator:1997pq}. 

Suppose $N_f \geq 2k$, considering the finite coupling Higgs branch yields
\begin{align}
    \dim_{\HH}\Higgs = \frac{1}{2} \cdot 2N_f \cdot 2k -\dim \sprm(k) = k (2N_f - 2k -1) \,.
    \label{eq:Higgs_dim_Sp_classical}
\end{align}
For $N_f \geq 2k $, complete Higgsing is possible and, according to 
\cite{Ferlito:2016grh}, the expectation for the finite coupling Higgs branch is 
as follows:
\begin{equation}
\begin{aligned}
    \Higgs &= \left\{ M \in \mathrm{Mat}_{2N_f\times 2N_f} \big| M+M^T =0 \, , 
\; M^2 =0   \, , \; \mathrm{rank}(M)\leq 2k  \right\}  \\
    &=\clorbit_{\dalg_{N_f}}^{[2^{2k},1^{2N_f-4k}]}
\end{aligned}
\end{equation}
\begin{compactitem}
\item $N_f > 2k$: The Higgs branch is a single cone which is a nilpotent orbit 
closure of $\sormL(2N_f)$.
\item $N_f =2k$: The Higgs branch is a union of two cones.
\end{compactitem}
For $N_f <2k$, only partial Higgsing is possible and the claim is as follows:
\begin{compactitem}
\item $N_f$ even: partial Higgsing to
\begin{subequations}
\label{eq:part_Higgs_Nf_even}
\begin{align}
 \text{$\sprm(k)$ with $N_f$ flavours }   \longrightarrow \text{pure $\sprm\left(k-\tfrac{N_f}{2} \right)$}   
\end{align}
such that the Higgs branch dimension becomes
\begin{align}
    \dim_{\HH} \Higgs &= \frac{1}{2} 2N_f \cdot 2k  - \left(\dim \sprm(k)- \dim \sprm\left(k-\tfrac{N_f}{2}\right) \right) \notag \\ 
    &= \frac{1}{2} N_f (N_f-1)\,.
\end{align}
Again, the Higgs branch is a union of two cones
\begin{align}
 \Higgs = \clorbit_{\dalg_{N_f}}^{[2^{N_f}]} \,.
 \end{align}
 \end{subequations}
\item $N_f$ odd: partial Higgsing to
\begin{subequations}
\label{eq:part_Higgs_Nf_odd}
\begin{align}
 \text{$\sprm(k)$ with $N_f$ flavours}  \longrightarrow  
 \text{ $\sprm\left(k-\tfrac{N_f-1}{2} \right)$ with  $1$ flavour} 
 \end{align}
and the Higgs branch dimension is computed to be
\begin{align}
\dim_{\HH} \Higgs &= \frac{1}{2} 
    \left(  2N_f \cdot 2k - 2\cdot \left(k-\tfrac{N_f-1}{2}\right) \right) 
    - \left(\dim \sprm(k)- \dim\sprm\left(k-\tfrac{N_f-1}{2}\right) \right)  \notag\\
    &= \frac{1}{2} N_f (N_f-1)\,.
\end{align}
In fact, the Higgs branch equals
\begin{align}
 \Higgs = \clorbit_{\dalg_{N_f}}^{[2^{N_f-1},1^2]} \,.
 \end{align}
\end{subequations}
\end{compactitem}
\paragraph{5-brane web.}
The requirement that the external 5-branes do not intersect, which would introduce additional massless degrees of freedom, leads to the constraint \cite[Sec.\ 3.4]{Brunner:1997gk} that $N_f \leq 2k+4$. It has been argued in \cite[Sec.\ 2]{Bergman:2015dpa} that the case $N_f =2k+5$ also gives rise to a consistent 5d theory, which for $k=1$ yields the familiar $E_8$ theory.

The change in Higgs branch dimension when approaching the fixed point has been studied in \cite{Zafrir:2015ftn} and can be summarised as follows:
\begin{compactitem}
\item $0\leq N_f\leq 2k$: The number of \Dfive s in between the branes with 
non-zero NS charge is larger than on the outside. Then exactly one 
extra dimension is gained by detaching the 5-branes from the \Of\ plane after 
collapsing the gauge 5-branes.
\item $2k+1\leq N_f\leq 2k+3$: The number of \Dfive s in between the branes 
with non-zero NS charge is smaller compared to the outside. Then at infinite 
coupling, there exists one extra dimension due to the $(p,q)$ 5-branes with 
$q>0$. Moreover, there exist further additional Higgs branch directions 
originating from the \Dfive s that can reconnect once the gauge branes 
collapsed.
\item $N_f=2k+4$: There are new Higgs branch directions due to \Dfive s that 
reconnect once the gauge branes collapsed. 
Moreover, one extra dimension is gained by detaching the 5-branes from the \Of\ 
plane after collapsing the gauge 5-branes; the external \NS\ branes are parallel 
and an additional direction becomes accessible by breaking one of them on a 
7-brane.
\item $N_f=2k+5$: The external 5-branes intersect and the resolution thereof 
yields several additional directions. One extra dimension is gained by 
detaching the 5-branes from the \Of\ plane after collapsing the gauge 5-branes; 
while several other directions open up due to \Dfive\ branes that can reconnect 
once the gauge branes collapsed.
\end{compactitem}
In addition, \cite{Zafrir:2015uaa,Bergman:2015dpa} investigated the potential enhancement of the classical $\sorm(2N_f) \times \uo_I$ global symmetry and the claim is that
\begin{compactitem}
\item $N_f<2k+4$: no enhancement.
\item $N_f =2k+4$: the $\uo_I$ factor enhances to a $\su$.
\item $N_f=2k+5$: the symmetry enhances as $\sorm(4k+10 ) \times \uo_I \to 
\sorm(4k+12)$.
\end{compactitem}

\subsection{Sp(k) with fundamental flavours and \texorpdfstring{$\Ofp $ }{O5}}
\label{sec:Sp_OPlus}
The construction of the 5-brane web for a $5$d $\Ncal=1$ $\sprm(k)$ gauge theory 
with $N_f$ fundamental flavours follows \cite{Zafrir:2015ftn}. In detail, the 
brane web for $N_R$ flavour branes on the right-hand-side and $N_L$ flavour 
brane on the left-hand-side starts from the Coulomb branch phase with identical 
masses given by
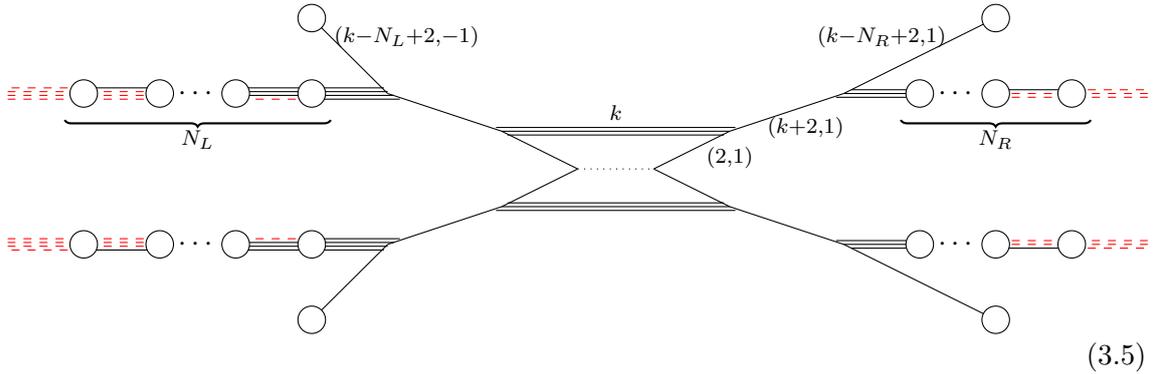
\begin{align}
 \raisebox{-.5\height}{
    \begin{tikzpicture}
    \OPlus{0,0}{1,0}
    \Dbrane{0,0}{-1,0.5}
    \Dbrane{1,0}{2,0.5}
    \draw (2,0.15) node {$\scriptstyle{(2,1)}$};
    \Dbrane{-0.925,0.45}{1.924,0.45}
    \Dbrane{-1,0.5}{2,0.5}
    \Dbrane{-1.075,0.55}{2.075,0.55}
     \draw (0.5,0.75) node {$\scriptstyle{k}$};
    \Dbrane{-1,0.5}{-2.5,1}
    \Dbrane{2,0.5}{3.5,1}
    \draw (3,0.55) node {$\scriptstyle{(k{+}2,1)}$};
    \Dbrane{-2.55,1.075}{-3.5,1.075}
    \Dbrane{-2.5,1.025}{-3.5,1.025}
    \Dbrane{-2.42,0.975}{-3.5,0.975}
    \Dbrane{-2.34,0.925}{-3.5,0.925}
    \Dbrane{-3.5,0.975}{-4.5,0.975}
    \Dbrane{-3.5,1.025}{-4.5,1.025}
    \Dbrane{-3.5,1.075}{-4.5,1.075}
    \Dbrane{-5.5,1.075}{-6.5,1.075}
    \draw (-5,1) node {$\cdots$};
    \Dbrane{3.5,1}{4.5,1}
    \Dbrane{3.6,1.05}{4.5,1.05}
    \Dbrane{3.4,0.95}{4.5,0.95}
    \Dbrane{5.5,1.05}{6.5,1.05}
    \draw (5,1) node {$\cdots$};
    \Dbrane{3.5,1}{5.5,2}
    \draw (4,1.75) node {$\scriptstyle{(k{-}N_R{+}2,1)}$};
    \Dbrane{-2.5,1}{-3.5,2}
    \draw (-2.25,1.75) node {$\scriptstyle{(k{-}N_L{+}2,-1)}$};
    \MonoCut{-3.5,0.925}{-4.5,0.925}
    \MonoCut{-5.5,0.925}{-7.5,0.925}
    \MonoCut{-5.5,0.975}{-7.5,0.975}
    \MonoCut{-5.5,1.025}{-7.5,1.025}
    \MonoCut{-6.5,1.075}{-7.5,1.075}
    \MonoCut{6.5,1.05}{7.5,1.05}
    \MonoCut{5.5,0.95}{7.5,0.95}
    \MonoCut{5.5,1.0}{7.5,1.0}
    \SevenB{-3.5,2}
    \SevenB{5.5,2}
    \SevenB{-3.5,1}
    \SevenB{-4.5,1}
    \SevenB{-5.5,1}
    \SevenB{-6.5,1}
    \SevenB{4.5,1}
    \SevenB{5.5,1}
    \SevenB{6.5,1}
    \draw[decoration={brace,mirror,raise=10pt},decorate,thick]
  (-6.75,1) -- node[below=10pt] {$\scriptstyle{N_L}$ } (-3.25,1);
  \draw[decoration={brace,mirror,raise=10pt},decorate,thick]
  (4.25,1) -- node[below=10pt] {$\scriptstyle{N_R}$ } (6.75,1);
    \begin{scope}[yscale=-1,xscale=1]
    \Dbrane{0,0}{-1,0.5}
    \Dbrane{1,0}{2,0.5}
    \Dbrane{-0.925,0.45}{1.924,0.45}
    \Dbrane{-1,0.5}{2,0.5}
    \Dbrane{-1.075,0.55}{2.075,0.55}
    \Dbrane{-1,0.5}{-2.5,1}
    \Dbrane{2,0.5}{3.5,1}
    \Dbrane{-2.55,1.075}{-3.5,1.075}
    \Dbrane{-2.5,1.025}{-3.5,1.025}
    \Dbrane{-2.42,0.975}{-3.5,0.975}
    \Dbrane{-2.34,0.925}{-3.5,0.925}
    \Dbrane{-3.5,0.975}{-4.5,0.975}
    \Dbrane{-3.5,1.025}{-4.5,1.025}
    \Dbrane{-3.5,1.075}{-4.5,1.075}
    \Dbrane{-5.5,1.075}{-6.5,1.075}
    \draw (-5,1) node {$\cdots$};
    \Dbrane{3.5,1}{4.5,1}
    \Dbrane{3.6,1.05}{4.5,1.05}
    \Dbrane{3.4,0.95}{4.5,0.95}
    \Dbrane{5.5,1.05}{6.5,1.05}
    \draw (5,1) node {$\cdots$};
    \Dbrane{3.5,1}{5.5,2}
    \Dbrane{-2.5,1}{-3.5,2}
    \MonoCut{-3.5,0.925}{-4.5,0.925}
    \MonoCut{-5.5,0.925}{-7.5,0.925}
    \MonoCut{-5.5,0.975}{-7.5,0.975}
    \MonoCut{-5.5,1.025}{-7.5,1.025}
    \MonoCut{-6.5,1.075}{-7.5,1.075}
    \MonoCut{6.5,1.05}{7.5,1.05}
    \MonoCut{5.5,0.95}{7.5,0.95}
    \MonoCut{5.5,1.0}{7.5,1.0}
    \SevenB{-3.5,2}
    \SevenB{5.5,2}
    \SevenB{-3.5,1}
    \SevenB{-4.5,1}
    \SevenB{-5.5,1}
    \SevenB{-6.5,1}
    \SevenB{4.5,1}
    \SevenB{5.5,1}
    \SevenB{6.5,1}
    \end{scope}
    \end{tikzpicture}
}
    \end{align}
such that the transition to the Higgs branch phase is again achieved in several 
steps: firstly, the flavour half $[1,0]$ 7-branes as well as the gauge $(1,0)$ 
5-branes move towards the \Of\ plane such that the half 7-branes merge with 
their mirror images. Secondly, the newly formed physical 7-branes can be split 
along the orientifold such that the 5-brane web becomes
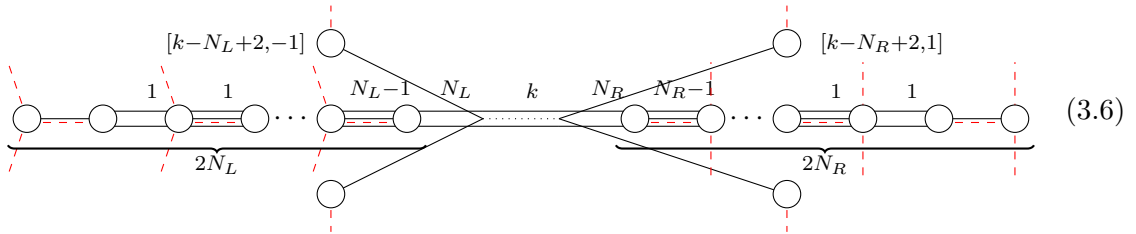
\begin{align}
 \raisebox{-.5\height}{
    \begin{tikzpicture}
    \DfiveOPlus{$k$}{0}
    \DfiveOMinus{$N_L$}{-1}
    \DfiveOMinus{$N_R$}{1}
    \DfiveOMinusTilde{$N_L{-}1$}{-2}
    \MonoCut{-2,-0.05}{-1,-0.05}
    \DfiveOMinusTilde{$N_R{-}1$}{2}
    \MonoCut{2,-0.05}{3,-0.05}
    \DfiveOMinusTilde{1}{-4}
    \MonoCut{-4,-0.05}{-3,-0.05}
    \DfiveOMinusTilde{1}{4}
    \MonoCut{4,-0.05}{5,-0.05}
    \DfiveOMinus{1}{-5}
    \DfiveOMinus{1}{5}
    \OMinusTilde{6,0}{7,0}
    \MonoCut{6,-0.05}{7,-0.05}
    \OMinusTilde{-6,0}{-5,0}
    \MonoCut{-6,-0.05}{-5,-0.05}
    \draw (-2.5,0) node {$\cdots$};
    \draw (3.5,0) node {$\cdots$};
    \Dbrane{1,0}{4,1}
    \draw (5.25,1) node {$\scriptstyle{[k{-}N_R{+}2,1]}$};
    \Dbrane{1,0}{4,-1}
    \Dbrane{0,0}{-2,1}
    \draw (-3.25,1) node {$\scriptstyle{[k{-}N_L{+}2,-1]}$};
    \Dbrane{0,0}{-2,-1}
    \MonoCut{-2,0}{-2-0.25,0.75}
    \MonoCut{-2,0}{-2-0.25,-0.75}
    \MonoCut{-4,0}{-4-0.25,0.75}
    \MonoCut{-4,0}{-4-0.25,-0.75}
    \MonoCut{-6,0}{-6-0.25,0.75}
    \MonoCut{-6,0}{-6-0.25,-0.75}
    \MonoCut{3,-0.75}{3,0.75}
    \MonoCut{5,-0.75}{5,0.75}
    \MonoCut{7,-0.75}{7,0.75}
    \MonoCut{4,1}{4,1.5}
    \MonoCut{4,-1}{4,-1.5}
    \MonoCut{-2,1}{-2,1.5}
    \MonoCut{-2,-1}{-2,-1.5}
    \SevenB{4,1}
    \SevenB{-2,1}
    \SevenB{4,-1}
    \SevenB{-2,-1}
    \SevenB{-1,0}
    \SevenB{-2,0}
    \SevenB{-3,0}
    \SevenB{-4,0}
    \SevenB{-5,0}
    \SevenB{-6,0}
    \SevenB{2,0}
    \SevenB{3,0}
    \SevenB{4,0}
    \SevenB{5,0}
    \SevenB{6,0}
    \SevenB{7,0}
    \draw[decoration={brace,mirror,raise=10pt},decorate,thick]
  (-6.25,0) -- node[below=10pt] {$\scriptstyle{2N_L}$ } (-0.75,0);
  \draw[decoration={brace,mirror,raise=10pt},decorate,thick]
  (1.75,0) -- node[below=10pt] {$\scriptstyle{2N_R}$ } (7.25,0);
    \end{tikzpicture}
}
\label{eq:web_Spk_Nf_Higgs_general}
\end{align}
using the \Of\ plane notation of Table \ref{tab:orientifold}.
Without loss of generality, it is sufficient to consider two cases 
\begin{compactenum}[(i)]
\item $N_f$ even: $N_L = N_R = \tfrac{N_f}{2}$,
\item $N_f$ odd: $N_L =N_R +1 = \tfrac{N_f+1}{2}$.
\end{compactenum}
This is because any other configuration can be transitioned into one of the two cases by moving sufficiently many flavour branes from one side to the other, due to brane annihilation and creation.
\subsubsection{Finite coupling}
\paragraph{$\mathbf{N_f}$ even.}
For $N_L=N_R$ the brane configuration \eqref{eq:web_Spk_Nf_Higgs_general} simplifies to 
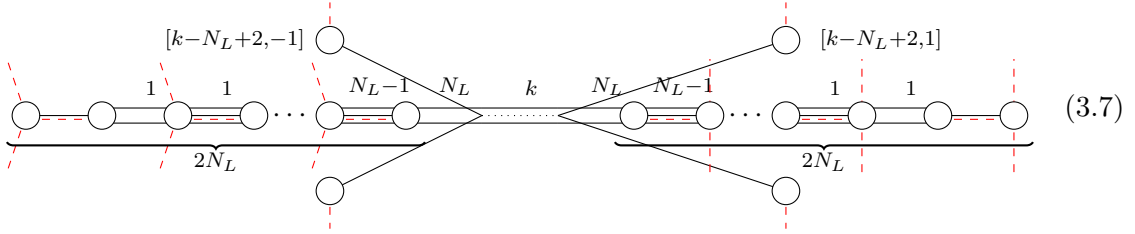
\begin{align}
 \raisebox{-.5\height}{
    \begin{tikzpicture}
    \DfiveOPlus{$k$}{0}
    \DfiveOMinus{$N_L$}{-1}
    \DfiveOMinus{$N_L$}{1}
    \DfiveOMinusTilde{$N_L{-}1$}{-2}
    \MonoCut{-2,-0.05}{-1,-0.05}
    \DfiveOMinusTilde{$N_L{-}1$}{2}
    \MonoCut{2,-0.05}{3,-0.05}
    \DfiveOMinusTilde{1}{-4}
    \MonoCut{-4,-0.05}{-3,-0.05}
    \DfiveOMinusTilde{1}{4}
    \MonoCut{4,-0.05}{5,-0.05}
    \DfiveOMinus{1}{-5}
    \DfiveOMinus{1}{5}
    \OMinusTilde{6,0}{7,0}
    \MonoCut{6,-0.05}{7,-0.05}
    \OMinusTilde{-6,0}{-5,0}
    \MonoCut{-6,-0.05}{-5,-0.05}
    \draw (-2.5,0) node {$\cdots$};
    \draw (3.5,0) node {$\cdots$};
    \Dbrane{1,0}{4,1}
    \draw (5.25,1) node {$\scriptstyle{[k{-}N_L{+}2,1]}$};
    \Dbrane{1,0}{4,-1}
    \Dbrane{0,0}{-2,1}
    \draw (-3.25,1) node {$\scriptstyle{[k{-}N_L{+}2,-1]}$};
    \Dbrane{0,0}{-2,-1}
    \MonoCut{-2,0}{-2-0.25,0.75}
    \MonoCut{-2,0}{-2-0.25,-0.75}
    \MonoCut{-4,0}{-4-0.25,0.75}
    \MonoCut{-4,0}{-4-0.25,-0.75}
    \MonoCut{-6,0}{-6-0.25,0.75}
    \MonoCut{-6,0}{-6-0.25,-0.75}
    \MonoCut{3,-0.75}{3,0.75}
    \MonoCut{5,-0.75}{5,0.75}
    \MonoCut{7,-0.75}{7,0.75}
    \MonoCut{4,1}{4,1.5}
    \MonoCut{4,-1}{4,-1.5}
    \MonoCut{-2,1}{-2,1.5}
    \MonoCut{-2,-1}{-2,-1.5}
    \SevenB{4,1}
    \SevenB{-2,1}
    \SevenB{4,-1}
    \SevenB{-2,-1}
    \SevenB{-1,0}
    \SevenB{-2,0}
    \SevenB{-3,0}
    \SevenB{-4,0}
    \SevenB{-5,0}
    \SevenB{-6,0}
    \SevenB{2,0}
    \SevenB{3,0}
    \SevenB{4,0}
    \SevenB{5,0}
    \SevenB{6,0}
    \SevenB{7,0}
    \draw[decoration={brace,mirror,raise=10pt},decorate,thick]
  (-6.25,0) -- node[below=10pt] {$\scriptstyle{2N_L}$ } (-0.75,0);
  \draw[decoration={brace,mirror,raise=10pt},decorate,thick]
  (1.75,0) -- node[below=10pt] {$\scriptstyle{2N_L}$ } (7.25,0);
    \end{tikzpicture}
}
\label{eq:web_Spk_Nf_Higgs_even_finite}
\end{align}
If $N_L \leq k$ or, equivalently, $N_f \leq 2k $ then only $N_L$ 5-branes in the central segment of $k$ physical $(1,0)$ 5-branes are dynamical. In particular, $k-N_L$ of the central 5-branes remain suspended between \NS\ branes such that there is a residual $\sprm\left(k-\tfrac{N_f}{2}\right)$ gauge group left. For the 5-branes that contribute to the Higgs branch, the associated magnetic quiver reads 
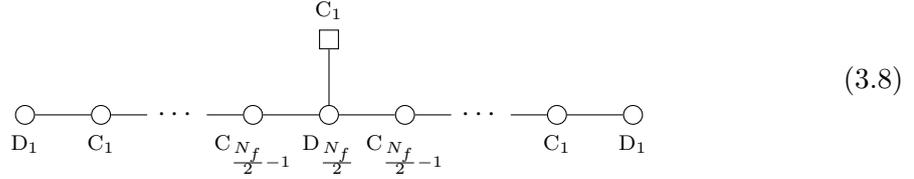
\begin{align}
 \raisebox{-.5\height}{
 	\begin{tikzpicture}
 	\tikzset{node distance = 1cm};
	\tikzstyle{gauge} = [circle, draw,inner sep=2.5pt];
	\tikzstyle{flavour} = [regular polygon,regular polygon sides=4,inner 
sep=2.5pt, draw];
	\node (g1) [gauge,label=below:{\dd{1}}] {};
	\node (g2) [gauge,right of=g1,label=below:{\cc{1}}] {};
	\node (g3) [right of=g2] {$\cdots$};
	\node (g4) [gauge,right of=g3,label=below:{\cc{\frac{N_f}{2}{-}1}}] {};
	\node (g5) [gauge,right of=g4,label=below:{\dd{\frac{N_f}{2}}}] {};
	\node (g6) [gauge,right of=g5,label=below:{\cc{\frac{N_f}{2}{-}1}}] {};
	\node (g7) [right of=g6] {$\cdots$};
	\node (g8) [gauge,right of=g7,label=below:{\cc{1}}] {};
	\node (g9) [gauge,right of=g8,label=below:{\dd{1}}] {};
	\node (f1) [flavour,above of=g5,label=above:{\cc{1}}] {};
	\draw (g1)--(g2) (g2)--(g3) (g3)--(g4) (g4)--(g5) (g5)--(g6) (g6)--(g7) (g7)--(g8) (g8)--(g9) (g5)--(f1);
	\end{tikzpicture}
	} 
	\label{eq:magQuiv_Spk_Nf_even_small}
\end{align}
and one computes
\begin{align}
    \dim_{\HH} \Coulomb \eqref{eq:magQuiv_Spk_Nf_even_small} = \frac{1}{2} N_f (N_f-1) 
    \; ,\qquad
    G = \sorm(2N_f)
    \label{eq:results_Spk_Nf_even_small}
\end{align}
because all gauge nodes are balanced, cf.\ Appendix \ref{app:Coulomb_branch}. The nilpotent orbit $\mathcal{O}^{[2^{N_f}]}_{\dalg_{N_f}}=\mathcal{O}^{[2^{N_f}]I}_{\dalg_{N_f}}\cup\mathcal{O}^{[2^{N_f}]II}_{D_{N_f}}$ of $\sormL(2N_f)$ is a union of two nilpotent orbits \cite{collingwood1993nilpotent}, the Coulomb branch of \eqref{eq:magQuiv_Spk_Nf_even_small} is the closure of one of these orbits. In particular, the properties \eqref{eq:results_Spk_Nf_even_small} agree with the expectation \eqref{eq:part_Higgs_Nf_even} and the brane web \eqref{eq:web_Spk_Nf_Higgs_even_finite} displays the residual gauge theory explicitly. As in the case of $\sprm(1)$ only one cone can be identified from the web, however a direct computation of the hyper-K\"ahler quotient shows, that there are indeed two cones. See Section \ref{sec:prob2cones} for more on the two cone problem.

If $N_L >k$ or, equivalently, $N_f \geq 2k+2$ then all of the $k$ central branes 
are dynamical Higgs branch degrees of freedom. The $(k-N_L+1,\pm1)$ 5-branes are 
connected to 7-branes via $N_L-k$ branes, which can be annihilated when moving 
the $(k-N_L+1,\pm1)$ 5-branes through sufficiently many 7-branes, as discussed 
above. Hence, the magnetic quiver becomes
\begin{align}
 \raisebox{-.5\height}{
 	\begin{tikzpicture}
 	\tikzset{node distance = 1cm};
	\tikzstyle{gauge} = [circle, draw,inner sep=2.5pt];
	\tikzstyle{flavour} = [regular polygon,regular polygon sides=4,inner 
sep=2.5pt, draw];
	\node (g1) [gauge,label=below:{\dd{1}}] {};
	\node (g2) [gauge,right of=g1,label=below:{\cc{1}}] {};
	\node (g3) [right of=g2] {$\cdots$};
	\node (g4) [gauge,right of=g3,label=below:{\dd{k}}] {};
	\node (g5) [gauge,right of=g4,label=below:{\cc{k}}] {};
	\node (g6) [gauge,right of=g5,label=below:{\bb{k}}] {};
	\node (g7) [right of=g6] {$\cdots$};
	\node (g8) [gauge,right of=g7,label=below:{\bb{k}}] {};
	\node (g9) [gauge,right of=g8,label=below:{\cc{k}}] {};
	\node (g10) [gauge,right of=g9,label=below:{\dd{k}}] {};
	\node (g11) [right of=g10] {$\cdots$};
	\node (g12) [gauge,right of=g11,label=below:{\cc{1}}] {};
	\node (g13) [gauge,right of=g12,label=below:{\dd{1}}] {};
	\node (f1) [flavour,above of=g5,label=above:{\bb{0}}] {};
	\node (f2) [flavour,above of=g9,label=above:{\bb{0}}] {};
	\draw (g1)--(g2) (g2)--(g3) (g3)--(g4) (g4)--(g5) (g5)--(g6) (g6)--(g7) (g7)--(g8) (g8)--(g9) (g9)--(g10) (g10)--(g11) (g11)--(g12) (g12)--(g13) (g5)--(f1) (g9)--(f2);
	\draw[decoration={brace,mirror,raise=10pt},decorate,thick]
  (3.9,-0.25) -- node[below=10pt] {$\scriptstyle{2N_f-4k-1}$ } (8.1,-0.25);
	\end{tikzpicture}
	} 
		\label{eq:magQuiv_Spk_Nf_even_large}
\end{align}
and one verifies
\begin{subequations}
\label{eq:results_Spk_Nf_even_large}
\begin{align}
    \dim_{\HH} \Coulomb \eqref{eq:magQuiv_Spk_Nf_even_large} = k(2N_f-2k-1) 
    \;, \qquad
    G = \sorm(2N_f)
\end{align}
because all gauge nodes are balanced, cf.\ Appendix \ref{app:Coulomb_branch}. Moreover, the Coulomb branch geometry is known to be
\begin{align}
    \Coulomb\eqref{eq:magQuiv_Spk_Nf_even_large} = \clorbit_{\dalg_{N_f}}^{[2^{2k},1^{2N_f-4k}]}
\end{align}
\end{subequations}
i.e.\ a nilpotent orbit closure for $\sormL(2N_f)$. Consequently, the properties \eqref{eq:results_Spk_Nf_even_large} correctly reproduce the finite coupling Higgs branch.
\paragraph{$\mathbf{N_f}$ odd.}
For $N_L=N_R+1$ the brane configuration \eqref{eq:web_Spk_Nf_Higgs_general} specialises to
\begin{align}
 \raisebox{-.5\height}{
    \begin{tikzpicture}
    \DfiveOPlus{$k$}{0}
    \DfiveOMinus{$N_L$}{-1}
    \DfiveOMinus{$N_L{-}1$}{1}
    \DfiveOMinusTilde{$N_L{-}1$}{-2}
    \MonoCut{-2,-0.05}{-1,-0.05}
    \DfiveOMinusTilde{$N_L{-}2$}{2}
    \MonoCut{2,-0.05}{3,-0.05}
    \DfiveOMinusTilde{1}{-4}
    \MonoCut{-4,-0.05}{-3,-0.05}
    \DfiveOMinusTilde{1}{4}
    \MonoCut{4,-0.05}{5,-0.05}
    \DfiveOMinus{1}{-5}
    \DfiveOMinus{1}{5}
    \OMinusTilde{6,0}{7,0}
    \MonoCut{6,-0.05}{7,-0.05}
    \OMinusTilde{-6,0}{-5,0}
    \MonoCut{-6,-0.05}{-5,-0.05}
    \draw (-2.5,0) node {$\cdots$};
    \draw (3.5,0) node {$\cdots$};
    \Dbrane{1,0}{4,1}
    \draw (5.25,1) node {$\scriptstyle{[k{-}N_L{+}2{+}1,1]}$};
    \Dbrane{1,0}{4,-1}
    \Dbrane{0,0}{-2,1}
    \draw (-3.25,1) node {$\scriptstyle{[k{-}N_L{+}2,-1]}$};
    \Dbrane{0,0}{-2,-1}
    \MonoCut{-2,0}{-2-0.25,0.75}
    \MonoCut{-2,0}{-2-0.25,-0.75}
    \MonoCut{-4,0}{-4-0.25,0.75}
    \MonoCut{-4,0}{-4-0.25,-0.75}
    \MonoCut{-6,0}{-6-0.25,0.75}
    \MonoCut{-6,0}{-6-0.25,-0.75}
    \MonoCut{3,0}{3+1.25,0.5}
    \MonoCut{3,0}{3+1.25,-0.5}
    \MonoCut{5,0}{5+1.25,0.5}
    \MonoCut{5,0}{5+1.25,-0.5}
    \MonoCut{7,0}{7+1.25,0.5}
    \MonoCut{7,0}{7+1.25,-0.5}
    \MonoCut{4,1}{4,1.5}
    \MonoCut{4,-1}{4,-1.5}
    \MonoCut{-2,1}{-2,1.5}
    \MonoCut{-2,-1}{-2,-1.5}
    \SevenB{4,1}
    \SevenB{-2,1}
    \SevenB{4,-1}
    \SevenB{-2,-1}
    \SevenB{-1,0}
    \SevenB{-2,0}
    \SevenB{-3,0}
    \SevenB{-4,0}
    \SevenB{-5,0}
    \SevenB{-6,0}
    \SevenB{2,0}
    \SevenB{3,0}
    \SevenB{4,0}
    \SevenB{5,0}
    \SevenB{6,0}
    \SevenB{7,0}
    \draw[decoration={brace,mirror,raise=10pt},decorate,thick]
  (-6.25,0) -- node[below=10pt] {$\scriptstyle{2N_L}$ } (-0.75,0);
  \draw[decoration={brace,mirror,raise=10pt},decorate,thick]
  (1.75,0) -- node[below=10pt] {$\quad\scriptstyle{2N_L-2}$ } (7.25,0);
    \end{tikzpicture}
}
\label{eq:web_Spk_Nf_Higgs_odd_finite}
\end{align}
Regardless of the value of $N_f$, after moving both, the $(k{-}N_L{+}2{-}1,1)$ as well as the $(k{-}N_L{+}2,-1)$ 5-brane through the first 7-brane on the left-hand side, the brane configuration becomes
\begin{align}
 \raisebox{-.5\height}{
    \begin{tikzpicture}
    \DfiveOMinusTilde{$N_L{-}1$}{0}
    \DfiveOPlusTilde{$k$}{-1}
    \DfiveOMinus{$N_L{-}1$}{1}
    \DfiveOMinusTilde{$N_L{-}1$}{-2}
    \MonoCut{-2,-0.05}{1,-0.05}
    \DfiveOMinusTilde{$N_L{-}2$}{2}
    \MonoCut{2,-0.05}{3,-0.05}
    \DfiveOMinusTilde{1}{-4}
    \MonoCut{-4,-0.05}{-3,-0.05}
    \DfiveOMinusTilde{1}{4}
    \MonoCut{4,-0.05}{5,-0.05}
    \DfiveOMinus{1}{-5}
    \DfiveOMinus{1}{5}
    \OMinusTilde{6,0}{7,0}
    \MonoCut{6,-0.05}{7,-0.05}
    \OMinusTilde{-6,0}{-5,0}
    \MonoCut{-6,-0.05}{-5,-0.05}
    \draw (-2.5,0) node {$\cdots$};
    \draw (3.5,0) node {$\cdots$};
    \Dbrane{0,0}{3,1}
    \draw (4.25,1) node {$\scriptstyle{[k{-}N_L{+}3,1]}$};
    \Dbrane{0,0}{3,-1}
    \Dbrane{-1,0}{-3,1}
    \draw (-4.25,1) node {$\scriptstyle{[k{-}N_L{+}2,-1]}$};
    \Dbrane{-1,0}{-3,-1}
    \MonoCut{-2,0}{-2-1.25,0.5}
    \MonoCut{-2,0}{-2-1.25,-0.5}
    \MonoCut{-4,0}{-4-1.25,0.5}
    \MonoCut{-4,0}{-4-1.25,-0.5}
    \MonoCut{-6,0}{-6-1.25,0.5}
    \MonoCut{-6,0}{-6-1.25,-0.5}
    \MonoCut{3,0}{3+0.25,0.75}
    \MonoCut{3,0}{3+0.25,-0.75}
    \MonoCut{5,0}{5+0.25,0.75}
    \MonoCut{5,0}{5+0.25,-0.75}
    \MonoCut{7,0}{7+0.25,0.75}
    \MonoCut{7,0}{7+0.25,-0.75}
    \MonoCut{3,1}{3,1.5}
    \MonoCut{3,-1}{3,-1.5}
    \MonoCut{-3,1}{-3,1.5}
    \MonoCut{-3,-1}{-3,-1.5}
    \SevenB{3,1}
    \SevenB{-3,1}
    \SevenB{3,-1}
    \SevenB{-3,-1}
    \SevenB{1,0}
    \SevenB{-2,0}
    \SevenB{-3,0}
    \SevenB{-4,0}
    \SevenB{-5,0}
    \SevenB{-6,0}
    \SevenB{2,0}
    \SevenB{3,0}
    \SevenB{4,0}
    \SevenB{5,0}
    \SevenB{6,0}
    \SevenB{7,0}
    \draw[decoration={brace,mirror,raise=10pt},decorate,thick]
  (-6.25,0) -- node[below=10pt] {$\scriptstyle{2N_L-1}$ } (-1.75,0);
  \draw[decoration={brace,mirror,raise=10pt},decorate,thick]
  (0.75,0) -- node[below=10pt] {$\scriptstyle{2N_L-1}$ } (7.25,0);
    \end{tikzpicture}
}
\end{align}
If $N_L-1 \leq k$ or, equivalently, $N_f \leq 2k+1$ only $\frac{N_f-1}{2}$ $k$ of the $k$ central 5-branes are dynamical Higgs branch degrees of freedom. Hence, $k-\tfrac{N_f-1}{2}$ 5-branes remain suspended between \NS\ branes and indicate a residual $\sprm\left( k-\tfrac{N_f-1}{2}\right)$ gauge group. In addition, the half 7-branes closest to the \NS\ branes contribute a single flavour to the residual gauge group. From the remaining 5-branes that contribute to the Higgs branch one reads off the magnetic quiver as follows
\begin{align}
 \raisebox{-.5\height}{
 	\begin{tikzpicture}
 	\tikzset{node distance = 1cm};
	\tikzstyle{gauge} = [circle, draw,inner sep=2.5pt];
	\tikzstyle{flavour} = [regular polygon,regular polygon sides=4,inner 
sep=2.5pt, draw];
	\node (g1) [gauge,label=below:{\dd{1}}] {};
	\node (g2) [gauge,right of=g1,label=below:{\cc{1}}] {};
	\node (g3) [right of=g2] {$\cdots$};
	\node (g4) [gauge,right of=g3,label=below:{\dd{\frac{N_f{-1}}{2}}}] {};
	\node (g5) [gauge,right of=g4,label=below:{\cc{\frac{N_f{-}1}{2}}}] {};
	\node (g6) [gauge,right of=g5,label=below:{\dd{\frac{N_f{-}1}{2}}}] {};
	\node (g7) [right of=g6] {$\cdots$};
	\node (g8) [gauge,right of=g7,label=below:{\cc{1}}] {};
	\node (g9) [gauge,right of=g8,label=below:{\dd{1}}] {};
	\node (f1) [flavour,above of=g5,label=above:{\dd{1}}] {};
	\draw (g1)--(g2) (g2)--(g3) (g3)--(g4) (g4)--(g5) (g5)--(g6) (g6)--(g7) (g7)--(g8) (g8)--(g9) (g5)--(f1);
	\end{tikzpicture}
	} 
	\label{eq:magQuiv_Spk_Nf_odd_small}
\end{align}
and one can compute
\begin{align}
\label{eq:results_Spk_Nf_odd_small}
    \dim_{\HH} \Coulomb \eqref{eq:magQuiv_Spk_Nf_odd_small} = \frac{1}{2} N_f(N_f-1) 
    \;, \qquad
    G = \sorm(2N_f)
\end{align}
because all gauge nodes are balanced, cf.\ Appendix \ref{app:Coulomb_branch}. The Coulomb branch of \eqref{eq:magQuiv_Spk_Nf_odd_small} is the closure of the nilpotent orbit $[2^{N_f-1},1^2]$ of $\mathfrak{so}(2N_f)$. As a consequence, the properties \eqref{eq:results_Spk_Nf_odd_small} agree with the expectation \eqref{eq:part_Higgs_Nf_odd} and the brane web displays the residual gauge theory explicitly.

On the other hand, if $N_L-1 > k$ or, equivalently, $N_f > 2k+1$ then all of the $k$ central 5-branes are Higgs branch degrees of freedom.  As dictated by the S-rule, the $(k{-}N_L{+}2,-1)$ and  $(k{-}N_L{+}3,1)$ 5-branes are connected to 7-branes via sufficiently many $(1,0)$ 5-branes, which can be annihilated by moving the non-flavour 5-branes through 7-branes. As a result, one finds the following magnetic quiver
\begin{align}
 \raisebox{-.5\height}{
 	\begin{tikzpicture}
 	\tikzset{node distance = 1cm};
	\tikzstyle{gauge} = [circle, draw,inner sep=2.5pt];
	\tikzstyle{flavour} = [regular polygon,regular polygon sides=4,inner 
sep=2.5pt, draw];
	\node (g1) [gauge,label=below:{\dd{1}}] {};
	\node (g2) [gauge,right of=g1,label=below:{\cc{1}}] {};
	\node (g3) [right of=g2] {$\cdots$};
	\node (g4) [gauge,right of=g3,label=below:{\dd{k}}] {};
	\node (g5) [gauge,right of=g4,label=below:{\cc{k}}] {};
	\node (g6) [gauge,right of=g5,label=below:{\bb{k}}] {};
	\node (g7) [right of=g6] {$\cdots$};
	\node (g8) [gauge,right of=g7,label=below:{\bb{k}}] {};
	\node (g9) [gauge,right of=g8,label=below:{\cc{k}}] {};
	\node (g10) [gauge,right of=g9,label=below:{\dd{k}}] {};
	\node (g11) [right of=g10] {$\cdots$};
	\node (g12) [gauge,right of=g11,label=below:{\cc{1}}] {};
	\node (g13) [gauge,right of=g12,label=below:{\dd{1}}] {};
	\node (f1) [flavour,above of=g5,label=above:{\bb{0}}] {};
	\node (f2) [flavour,above of=g9,label=above:{\bb{0}}] {};
	\draw (g1)--(g2) (g2)--(g3) (g3)--(g4) (g4)--(g5) (g5)--(g6) (g6)--(g7) (g7)--(g8) (g8)--(g9) (g9)--(g10) (g10)--(g11) (g11)--(g12) (g12)--(g13) (g5)--(f1) (g9)--(f2);
	\draw[decoration={brace,mirror,raise=10pt},decorate,thick]
  (3.9,-0.25) -- node[below=10pt] {$\scriptstyle{2N_f-4k-1}$ } (8.1,-0.25);
	\end{tikzpicture}
	} 
	\label{eq:magQuiv_Spk_Nf_odd_large}
\end{align}
which is the same as for $N_f$ even and $N_f\geq 2k+1$. A computation shows
\begin{subequations}
\label{eq:results_Spk_Nf_odd_large}
\begin{align}
    \dim_{\HH} \Coulomb\eqref{eq:magQuiv_Spk_Nf_odd_large}=k(2N_f-2k-1)
    \;,\qquad
    G = \sorm(2N_f)
\end{align}
because all gauge nodes are balanced, cf.\ Appendix \ref{app:Coulomb_branch}. In fact
\begin{align}
    \Coulomb\eqref{eq:magQuiv_Spk_Nf_odd_large} = \clorbit_{\dalg_{N_f}}^{[2^{2k},1^{2N_f-4k}]} \;,
\end{align}
\end{subequations}
which is a nilpotent orbit closure of $\sormL(2N_f)$. Therefore, the properties \eqref{eq:results_Spk_Nf_odd_large} match the finite coupling Higgs branch.
%
%
\subsubsection{Infinite coupling}
\paragraph{$\mathbf{N_f}$ even}
For $N_L=N_R$ the 5-brane web at infinite coupling has to be considered for two distinct cases. Firstly, if $N_L<k+2$ or, equivalently, $ N_f < 2k+4$, the web becomes
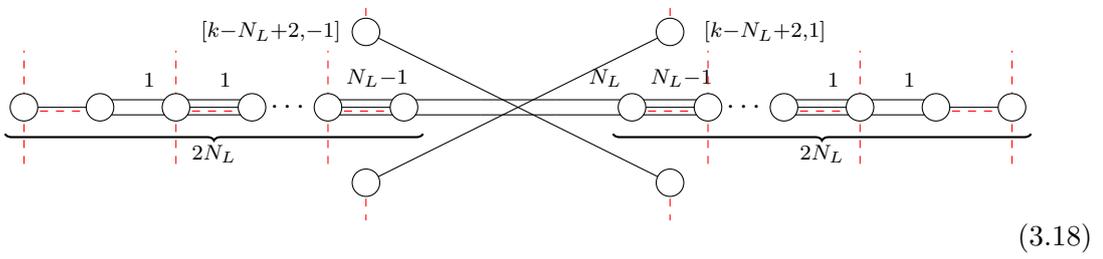
\begin{align}
 \raisebox{-.5\height}{
    \begin{tikzpicture}
    \DfiveOMinus{}{0}
    \DfiveOMinus{}{-1}
    \DfiveOMinus{$N_L$}{1}
    \DfiveOMinusTilde{$N_L{-}1$}{-2}
    \MonoCut{-2,-0.05}{-1,-0.05}
    \DfiveOMinusTilde{$N_L{-}1$}{2}
    \MonoCut{2,-0.05}{3,-0.05}
    \DfiveOMinusTilde{1}{-4}
    \MonoCut{-4,-0.05}{-3,-0.05}
    \DfiveOMinusTilde{1}{4}
    \MonoCut{4,-0.05}{5,-0.05}
    \DfiveOMinus{1}{-5}
    \DfiveOMinus{1}{5}
    \OMinusTilde{6,0}{7,0}
    \MonoCut{6,-0.05}{7,-0.05}
    \OMinusTilde{-6,0}{-5,0}
    \MonoCut{-6,-0.05}{-5,-0.05}
    \draw (-2.5,0) node {$\cdots$};
    \draw (3.5,0) node {$\cdots$};
    \Dbrane{0.5,0}{2.5,1}
    \draw (3.75,1) node {$\scriptstyle{[k{-}N_L{+}2,1]}$};
    \Dbrane{0.5,0}{2.5,-1}
    \Dbrane{0.5,0}{-1.5,1}
    \draw (-2.75,1) node {$\scriptstyle{[k{-}N_L{+}2,-1]}$};
    \Dbrane{0.5,0}{-1.5,-1}
    \MonoCut{-2,-0.75}{-2,0.75}
    \MonoCut{-4,-0.75}{-4,0.75}
    \MonoCut{-6,-0.75}{-6,0.75}
    \MonoCut{3,-0.75}{3,0.75}
    \MonoCut{5,-0.75}{5,0.75}
    \MonoCut{7,-0.75}{7,0.75}
    \MonoCut{2.5,1}{2.5,1.5}
    \MonoCut{2.5,-1}{2.5,-1.5}
    \MonoCut{-1.5,1}{-1.5,1.5}
    \MonoCut{-1.5,-1}{-1.5,-1.5}
    \SevenB{2.5,1}
    \SevenB{-1.5,1}
    \SevenB{2.5,-1}
    \SevenB{-1.5,-1}
    \SevenB{-1,0}
    \SevenB{-2,0}
    \SevenB{-3,0}
    \SevenB{-4,0}
    \SevenB{-5,0}
    \SevenB{-6,0}
    \SevenB{2,0}
    \SevenB{3,0}
    \SevenB{4,0}
    \SevenB{5,0}
    \SevenB{6,0}
    \SevenB{7,0}
    \draw[decoration={brace,mirror,raise=10pt},decorate,thick]
  (-6.25,0) -- node[below=10pt] {$\scriptstyle{2N_L}$ } (-0.75,0);
  \draw[decoration={brace,mirror,raise=10pt},decorate,thick]
  (1.75,0) -- node[below=10pt] {$\scriptstyle{2N_L}$ } (7.25,0);
    \end{tikzpicture}
}
\label{eq:web_Spk_Nf_Higgs_even_infinite_1}
\end{align}
and the magnetic quiver is read off to be
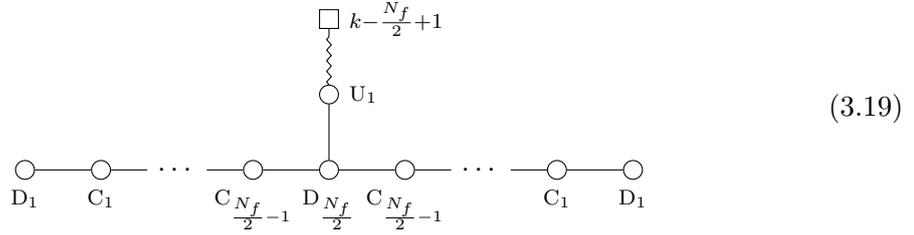
\begin{align}
 \raisebox{-.5\height}{
 	\begin{tikzpicture}
 	\tikzset{node distance = 1cm};
	\tikzstyle{gauge} = [circle, draw,inner sep=2.5pt];
	\tikzstyle{flavour} = [regular polygon,regular polygon sides=4,inner 
sep=2.5pt, draw];
	\node (g1) [gauge,label=below:{\dd{1}}] {};
	\node (g2) [gauge,right of=g1,label=below:{\cc{1}}] {};
	\node (g3) [right of=g2] {$\cdots$};
	\node (g4) [gauge,right of=g3,label=below:{\cc{\frac{N_f}{2}{-}1}}] {};
	\node (g5) [gauge,right of=g4,label=below:{\dd{\frac{N_f}{2}}}] {};
	\node (g6) [gauge,right of=g5,label=below:{\cc{\frac{N_f}{2}{-}1}}] {};
	\node (g7) [right of=g6] {$\cdots$};
	\node (g8) [gauge,right of=g7,label=below:{\cc{1}}] {};
	\node (g9) [gauge,right of=g8,label=below:{\dd{1}}] {};
	\node (g10) [gauge,above of=g5,label=right:{\uu{1}}] {};
	\node (f10) [flavour,above of=g10,label=right:{$\scriptstyle{k{-}\frac{N_f}{2}{+}1}$}] {};
	\draw (g1)--(g2) (g2)--(g3) (g3)--(g4) (g4)--(g5) (g5)--(g6) (g6)--(g7) (g7)--(g8) (g8)--(g9) (g5)--(g10);
	\draw [line join=round,decorate, decoration={zigzag, segment length=4,amplitude=.9,post=lineto,post length=2pt}]  (g10) -- (f10);
	\end{tikzpicture}
	} 
	\label{eq:magQuiv_Spk_infinite_Nf_even_small}
\end{align}
and one verifies
\begin{align}
    \dim_{\HH} \Coulomb\eqref{eq:magQuiv_Spk_infinite_Nf_even_small} = \frac{N_f(N_f-1)}{2} +1 \,.
\end{align}
The charge two magnetic hypermultiplets are expected from the self-intersection of the $(k-N_L+2,\pm1)$ 5-branes, as detailed in Observation \ref{obs:matter}.
Moreover, one notices that the family of magnetic quivers \eqref{eq:magQuiv_Spk_infinite_Nf_even_small} contains the $E_5$ case \eqref{eq:magQuiv_Sp1_4_infinite}, the $E_3$ case \eqref{eq:magQuiv_Sp1_2_infinite} and the $E_1$ case \eqref{eq:magQuiv_Sp1_0_infinite}.

Secondly, if $N_L=k+2$ or, equivalently, $ N_f = 2k+4$, the web at infinite coupling reads
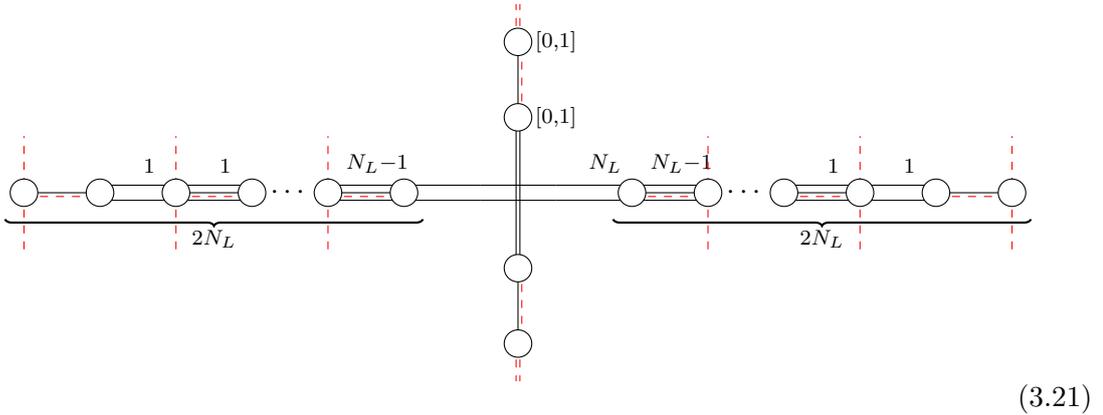
\begin{align}
 \raisebox{-.5\height}{
    \begin{tikzpicture}
    \DfiveOMinus{}{0}
    \DfiveOMinus{}{-1}
    \DfiveOMinus{$N_L$}{1}
    \DfiveOMinusTilde{$N_L{-}1$}{-2}
    \MonoCut{-2,-0.05}{-1,-0.05}
    \DfiveOMinusTilde{$N_L{-}1$}{2}
    \MonoCut{2,-0.05}{3,-0.05}
    \DfiveOMinusTilde{1}{-4}
    \MonoCut{-4,-0.05}{-3,-0.05}
    \DfiveOMinusTilde{1}{4}
    \MonoCut{4,-0.05}{5,-0.05}
    \DfiveOMinus{1}{-5}
    \DfiveOMinus{1}{5}
    \OMinusTilde{6,0}{7,0}
    \MonoCut{6,-0.05}{7,-0.05}
    \OMinusTilde{-6,0}{-5,0}
    \MonoCut{-6,-0.05}{-5,-0.05}
    \draw (-2.5,0) node {$\cdots$};
    \draw (3.5,0) node {$\cdots$};
    \Dbrane{0.525,0}{0.525,1}
    \Dbrane{0.475,0}{0.475,1}
    \draw (1,1) node {$\scriptstyle{[0,1]}$};
    \Dbrane{0.525,0}{0.525,-1}
    \Dbrane{0.475,0}{0.475,-1}
    \Dbrane{0.5,1}{0.5,2}
    \draw (1,2) node {$\scriptstyle{[0,1]}$};
    \Dbrane{0.5,-1}{0.5,-2}
    \MonoCut{-2,-0.75}{-2,0.75}
    \MonoCut{-4,-0.75}{-4,0.75}
    \MonoCut{-6,-0.75}{-6,0.75}
    \MonoCut{3,-0.75}{3,0.75}
    \MonoCut{5,-0.75}{5,0.75}
    \MonoCut{7,-0.75}{7,0.75}
    \MonoCut{0.55,1}{0.55,2}
    \MonoCut{0.55,-1}{0.55,-2}
    \MonoCut{0.525,2}{0.525,2.5}
    \MonoCut{0.475,2}{0.475,2.5}
    \MonoCut{0.525,-2}{0.525,-2.5}
    \MonoCut{0.475,-2}{0.475,-2.5}
    \SevenB{0.5,1}
    \SevenB{0.5,2}
    \SevenB{0.5,-1}
    \SevenB{0.5,-2}
    \SevenB{-1,0}
    \SevenB{-2,0}
    \SevenB{-3,0}
    \SevenB{-4,0}
    \SevenB{-5,0}
    \SevenB{-6,0}
    \SevenB{2,0}
    \SevenB{3,0}
    \SevenB{4,0}
    \SevenB{5,0}
    \SevenB{6,0}
    \SevenB{7,0}
    \draw[decoration={brace,mirror,raise=10pt},decorate,thick]
  (-6.25,0) -- node[below=10pt] {$\scriptstyle{2N_L}$ } (-0.75,0);
  \draw[decoration={brace,mirror,raise=10pt},decorate,thick]
  (1.75,0) -- node[below=10pt] {$\scriptstyle{2N_L}$ } (7.25,0);
    \end{tikzpicture}
}
\label{eq:web_Spk_Nf_Higgs_even_infinite_2}
\end{align}
such that the magnetic quiver is 
\begin{align}
 \raisebox{-.5\height}{
 	\begin{tikzpicture}
 	\tikzset{node distance = 1cm};
	\tikzstyle{gauge} = [circle, draw,inner sep=2.5pt];
	\tikzstyle{flavour} = [regular polygon,regular polygon sides=4,inner 
sep=2.5pt, draw];
	\node (g1) [gauge,label=below:{\dd{1}}] {};
	\node (g2) [gauge,right of=g1,label=below:{\cc{1}}] {};
	\node (g3) [right of=g2] {$\cdots$};
	\node (g4) [gauge,right of=g3,label=below:{\cc{\frac{N_f}{2}{-}1}}] {};
	\node (g5) [gauge,right of=g4,label=below:{\dd{\frac{N_f}{2}}}] {};
	\node (g6) [gauge,right of=g5,label=below:{\cc{\frac{N_f}{2}{-}1}}] {};
	\node (g7) [right of=g6] {$\cdots$};
	\node (g8) [gauge,right of=g7,label=below:{\cc{1}}] {};
	\node (g9) [gauge,right of=g8,label=below:{\dd{1}}] {};
	\node (g10) [gauge,above of=g5,label=right:{\cc{1}}] {};
	\node (g11) [gauge,above of=g10,label=right:{\uu{1}}] {};
	\draw (g1)--(g2) (g2)--(g3) (g3)--(g4) (g4)--(g5) (g5)--(g6) (g6)--(g7) (g7)--(g8) (g8)--(g9) (g5)--(g10) (g10)--(g11);
	\end{tikzpicture}
	} 
	\label{eq:magQuiv_Spk_infinite_Nf=2k+4}
\end{align}
and the dimension is 
\begin{align}
    \dim_{\HH} \Coulomb \eqref{eq:magQuiv_Spk_infinite_Nf=2k+4} = \frac{N_f (N_f -1)}{2} +2 \,.
\end{align}
One observes that this family contains the $E_7$ case \eqref{eq:magQuiv_Sp1_6_infinite}.
\paragraph{$\mathbf{N_f}$ odd.}
For $N_L=N_R+1$ there exists two distinct cases for the brane web at infinite coupling. Firstly, for $N_L <k+3$ or, equivalently, $N_f<2k+5$ the finite coupling 5-brane web \eqref{eq:web_Spk_Nf_Higgs_odd_finite} can be used to make the $(k-N_L+2,-1)$ and $(k-N_L+3,1)$ 5-brane become coincident on the \Of\ plane. Doing so results in
\begin{align}
 \raisebox{-.5\height}{
    \begin{tikzpicture}
    \DfiveOMinusTilde{}{0}
    \DfiveOMinusTilde{}{-1}
    \DfiveOMinus{$N_L{-}1$}{1}
    \DfiveOMinusTilde{$N_L{-}1$}{-2}
    \MonoCut{-2,-0.05}{1,-0.05}
    \DfiveOMinusTilde{$N_L{-}2$}{2}
    \MonoCut{2,-0.05}{3,-0.05}
    \DfiveOMinusTilde{1}{-4}
    \MonoCut{-4,-0.05}{-3,-0.05}
    \DfiveOMinusTilde{1}{4}
    \MonoCut{4,-0.05}{5,-0.05}
    \DfiveOMinus{1}{-5}
    \DfiveOMinus{1}{5}
    \OMinusTilde{6,0}{7,0}
    \MonoCut{6,-0.05}{7,-0.05}
    \OMinusTilde{-6,0}{-5,0}
    \MonoCut{-6,-0.05}{-5,-0.05}
    \draw (-2.5,0) node {$\cdots$};
    \draw (3.5,0) node {$\cdots$};
    \Dbrane{-0.5,0}{2.5,1}
    \draw (4,1) node {$\scriptstyle{[k{-}N_L{+}3,1]}$};
    \Dbrane{-0.5,0}{2.5,-1}
    \Dbrane{-0.5,0}{-2.5,1}
    \draw (-4,1) node {$\scriptstyle{[k{-}N_L{+}2,-1]}$};
    \Dbrane{-0.5,0}{-2.5,-1}
    \MonoCut{-2,-0.75}{-2,0.75}
    \MonoCut{-4,-0.75}{-4,0.75}
    \MonoCut{-6,-0.75}{-6,0.75}
    \MonoCut{3,-0.75}{3,0.75}
    \MonoCut{5,-0.75}{5,0.75}
    \MonoCut{7,-0.75}{7,0.75}
    \MonoCut{2.5,1}{2.5,1.5}
    \MonoCut{2.5,-1}{2.5,-1.5}
    \MonoCut{-2.5,1}{-2.5,1.5}
    \MonoCut{-2.5,-1}{-2.5,-1.5}
    \SevenB{2.5,1}
    \SevenB{-2.5,1}
    \SevenB{2.5,-1}
    \SevenB{-2.5,-1}
    \SevenB{1,0}
    \SevenB{-2,0}
    \SevenB{-3,0}
    \SevenB{-4,0}
    \SevenB{-5,0}
    \SevenB{-6,0}
    \SevenB{2,0}
    \SevenB{3,0}
    \SevenB{4,0}
    \SevenB{5,0}
    \SevenB{6,0}
    \SevenB{7,0}
    \draw[decoration={brace,mirror,raise=10pt},decorate,thick]
  (-6.25,0) -- node[below=10pt] {$\scriptstyle{2N_L-1}$ } (-1.75,0);
  \draw[decoration={brace,mirror,raise=10pt},decorate,thick]
  (0.75,0) -- node[below=10pt] {$\scriptstyle{2N_L-1}$ } (7.25,0);
    \end{tikzpicture}
}
\label{eq:web_Spk_Nf_Higgs_odd_infinite}
\end{align}
such that the magnetic quiver can be read off 
\begin{align}
 \raisebox{-.5\height}{
 	\begin{tikzpicture}
 	\tikzset{node distance = 1cm};
	\tikzstyle{gauge} = [circle, draw,inner sep=2.5pt];
	\tikzstyle{flavour} = [regular polygon,regular polygon sides=4,inner 
sep=2.5pt, draw];
	\node (g1) [gauge,label=below:{\dd{1}}] {};
	\node (g2) [gauge,right of=g1,label=below:{\cc{1}}] {};
	\node (g3) [right of=g2] {$\cdots$};
	\node (g4) [gauge,right of=g3,label=below:{\dd{\frac{N_f{-}1}{2}}}] {};
	\node (g5) [gauge,right of=g4,label=below:{\cc{\frac{N_f{-}1}{2}}}] {};
	\node (g6) [gauge,right of=g5,label=below:{\dd{\frac{N_f{-}1}{2}}}] {};
	\node (g7) [right of=g6] {$\cdots$};
	\node (g8) [gauge,right of=g7,label=below:{\cc{1}}] {};
	\node (g9) [gauge,right of=g8,label=below:{\dd{1}}] {};
	\node (g10) [gauge,above of=g5,label=right:{\uu{1}}] {};
	\node (f10) [flavour,above of=g10,label=right:{$\scriptstyle{k-\frac{N_f+1}{2}+2}$}] {};
	\draw (g1)--(g2) (g2)--(g3) (g3)--(g4) (g4)--(g5) (g5)--(g6) (g6)--(g7) (g7)--(g8) (g8)--(g9) (g5)--(g10);
	\draw [line join=round,decorate, decoration={zigzag, segment length=4,amplitude=.9,post=lineto,post length=2pt}]  (g10) -- (f10);
	\end{tikzpicture}
	} 
	\label{eq:magQuiv_Spk_infinite_Nf_odd_small}
\end{align}
and the Coulomb branch dimension is
\begin{align}
    \dim_{\HH} \Coulomb \eqref{eq:magQuiv_Spk_infinite_Nf_odd_small} = \frac{N_f(N_f-1)}{2} +1 \,.
\end{align}
The charge two magnetic hypermultiplet is an expected consequence from the intersection of the $(k-N_L+2,-1)$ and $(k-N_L+3,1)$ 5-brane, as suggested in Observation \ref{obs:matter}.
One notices that the family \eqref{eq:magQuiv_Spk_infinite_Nf_odd_small} of 
quivers contains the $E_6$ case \eqref{eq:magQuiv_Sp1_5_infinite}, the $E_4$ 
case \eqref{eq:magQuiv_Sp1_3_infinite} and the $E_2$ case 
\eqref{eq:magQuiv_Sp1_1_infinite}. 

Secondly, for $N_L=k+3$ or, equivalently, $N_f =2k+5$ the finite coupling brane web can be written as
\begin{align}
 \raisebox{-.5\height}{
    \begin{tikzpicture}
    \DfiveOPlus{$k$}{0}
    \DfiveOMinus{$N_L$}{-1}
    \DfiveOMinus{$N_L{-}1$}{1}
    \DfiveOMinusTilde{$N_L{-}1$}{-2}
    \MonoCut{-2,-0.05}{-1,-0.05}
    \DfiveOMinusTilde{$N_L{-}2$}{2}
    \MonoCut{2,-0.05}{3,-0.05}
    \DfiveOMinusTilde{1}{-4}
    \MonoCut{-4,-0.05}{-3,-0.05}
    \DfiveOMinusTilde{1}{4}
    \MonoCut{4,-0.05}{5,-0.05}
    \DfiveOMinus{1}{-5}
    \DfiveOMinus{1}{5}
    \OMinusTilde{6,0}{7,0}
    \MonoCut{6,-0.05}{7,-0.05}
    \OMinusTilde{-6,0}{-5,0}
    \MonoCut{-6,-0.05}{-5,-0.05}
    \draw (-2.5,0) node {$\cdots$};
    \draw (3.5,0) node {$\cdots$};
    \Dbrane{1,0}{1,2}
    \Dbrane{1.05,1}{1.05,2}
    \draw (0.5,2) node {$\scriptstyle{[0,1]}$};
    \Dbrane{1,0}{1,-2}
    \Dbrane{1.05,-1}{1.05,-2}
    \Dbrane{0,0}{1,1}    
    \Dbrane{1,1}{1.5,1}
    \draw (2,1) node {$\scriptstyle{[1,0]}$};
    \Dbrane{0,0}{1,-1}    
    \Dbrane{1,-1}{1.5,-1}
    \MonoCut{-2,-0.75}{-2,0.75}
    \MonoCut{-4,-0.75}{-4,0.75}
    \MonoCut{-6,-0.75}{-6,0.75}
    \MonoCut{3,-0.75}{3,0.75}
    \MonoCut{5,-0.75}{5,0.75}
    \MonoCut{7,-0.75}{7,0.75}
    \MonoCut{1,2}{1,2.5}
    \MonoCut{1.5,1}{1.5,2.5}
    \MonoCut{1,-2}{1,-2.5}
    \MonoCut{1.5,-1}{1.5,-2.5}
    \SevenB{1,2}
    \SevenB{1.5,1}
    \SevenB{1,-2}
    \SevenB{1.5,-1}
    \SevenB{-1,0}
    \SevenB{-2,0}
    \SevenB{-3,0}
    \SevenB{-4,0}
    \SevenB{-5,0}
    \SevenB{-6,0}
    \SevenB{2,0}
    \SevenB{3,0}
    \SevenB{4,0}
    \SevenB{5,0}
    \SevenB{6,0}
    \SevenB{7,0}
    \draw[decoration={brace,mirror,raise=10pt},decorate,thick]
  (-6.25,0) -- node[below=10pt] {$\scriptstyle{2N_L}$ } (-0.75,0);
  \draw[decoration={brace,mirror,raise=10pt},decorate,thick]
  (1.75,0) -- node[below=10pt] {$\scriptstyle{2N_L-2}$ } (7.25,0);
    \end{tikzpicture}
}
\label{eq:web_Spk_Nf=2k+5_var1}
\end{align}
such that the infinite coupling configuration is reached once the $[1,0]$ 7-brane moves onto the \Of\ plane, and the physical 7-brane splits along the orientifold. In detail, the brane web becomes
\begin{align}
 \raisebox{-.5\height}{
    \begin{tikzpicture}
    \DfiveOMinus{}{0}
    \DfiveOMinus{$N_L$}{-1}
    \DfiveOMinus{}{1}
    \DfiveOMinusTilde{$N_L{-}1$}{-2}
    \MonoCut{-2,-0.05}{-1,-0.05}
    \DfiveOMinusTilde{$N_L{-}1$}{2}
    \MonoCut{2,-0.05}{3,-0.05}
    \DfiveOMinusTilde{1}{-4}
    \MonoCut{-4,-0.05}{-3,-0.05}
    \DfiveOMinusTilde{1}{4}
    \MonoCut{4,-0.05}{5,-0.05}
    \DfiveOMinus{1}{-5}
    \DfiveOMinus{1}{5}
    \OMinusTilde{6,0}{7,0}
    \MonoCut{6,-0.05}{7,-0.05}
    \OMinusTilde{-6,0}{-5,0}
    \MonoCut{-6,-0.05}{-5,-0.05}
    \draw (-2.5,0) node {$\cdots$};
    \draw (3.5,0) node {$\cdots$};
    \Dbrane{0.45,-1}{0.45,1}
    \draw (1,1) node {$\scriptstyle{[0,1]}$};
    \Dbrane{0.55,-1}{0.55,1}
    \MonoCut{-2,-0.75}{-2,0.75}
    \MonoCut{-4,-0.75}{-4,0.75}
    \MonoCut{-6,-0.75}{-6,0.75}
    \MonoCut{3,-0.75}{3,0.75}
    \MonoCut{5,-0.75}{5,0.75}
    \MonoCut{7,-0.75}{7,0.75}
    \MonoCut{0.5,1}{0.5,1.5}
    \MonoCut{0.5,-1}{0.5,-1.5}
    \SevenB{0.5,1}
    \SevenB{0.5,-1}
    \SevenB{-1,0}
    \SevenB{-2,0}
    \SevenB{-3,0}
    \SevenB{-4,0}
    \SevenB{-5,0}
    \SevenB{-6,0}
    \SevenB{2,0}
    \SevenB{3,0}
    \SevenB{4,0}
    \SevenB{5,0}
    \SevenB{6,0}
    \SevenB{7,0}
    \draw[decoration={brace,mirror,raise=10pt},decorate,thick]
  (-6.25,0) -- node[below=10pt] {$\scriptstyle{2N_L}$ } (-0.75,0);
  \draw[decoration={brace,mirror,raise=10pt},decorate,thick]
  (1.75,0) -- node[below=10pt] {$\scriptstyle{2N_L}$ } (7.25,0);
    \end{tikzpicture}
}
\label{eq:web_Spk_Nf=2k+5_var2}
\end{align}
and the magnetic quiver is straightforwardly read off to be
\begin{align}
 \raisebox{-.5\height}{
 	\begin{tikzpicture}
 	\tikzset{node distance = 1cm};
	\tikzstyle{gauge} = [circle, draw,inner sep=2.5pt];
	\tikzstyle{flavour} = [regular polygon,regular polygon sides=4,inner 
sep=2.5pt, draw];
	\node (g1) [gauge,label=below:{\dd{1}}] {};
	\node (g2) [gauge,right of=g1,label=below:{\cc{1}}] {};
	\node (g3) [right of=g2] {$\cdots$};
	\node (g4) [gauge,right of=g3,label=below:{\cc{\frac{N_f{-1}}{2}}}] {};
	\node (g5) [gauge,right of=g4,label=below:{\dd{\frac{N_f{+}1}{2}}}] {};
	\node (g6) [gauge,right of=g5,label=below:{\cc{\frac{N_f{-}1}{2}}}] {};
	\node (g7) [right of=g6] {$\cdots$};
	\node (g8) [gauge,right of=g7,label=below:{\cc{1}}] {};
	\node (g9) [gauge,right of=g8,label=below:{\dd{1}}] {};
	\node (g10) [gauge,above of=g5,label=above:{\cc{1}}] {};
	\draw (g1)--(g2) (g2)--(g3) (g3)--(g4) (g4)--(g5) (g5)--(g6) (g6)--(g7) (g7)--(g8) (g8)--(g9) (g5)--(g10);
	\end{tikzpicture}
	} 
	\label{eq:magQuiv_Spk_infinite_Nf_2k=5}
\end{align}
and 
\begin{align}
    \dim_{\HH} \Coulomb \eqref{eq:magQuiv_Spk_infinite_Nf_2k=5} = \frac{N_f(N_f+1)}{2} +1 \,.
\end{align}
One notices that this family contains the $E_8$ case \eqref{eq:magQuiv_Sp1_7_infinite}.
%
%
\subsection{Sp(k) with fundamental flavours and \texorpdfstring{$\Ofpt $ }{O5}}
\label{sec:Sp_OPlustilde}
Since an \Ofpt\ plane does also give rise to symplectic low-energy gauge 
theories, it is tempting to generalise the analysis to this case. 
From the brane-web perspective, a convenient starting point is the 5-brane web 
with an 
\Ofp\ plane for a $\sprm(k)$ theory with $N_f=N_L +N_R+1$ fundamental flavours. 
Without loss of generality, the additional flavour is chosen to be on the 
right-hand-side, such that 
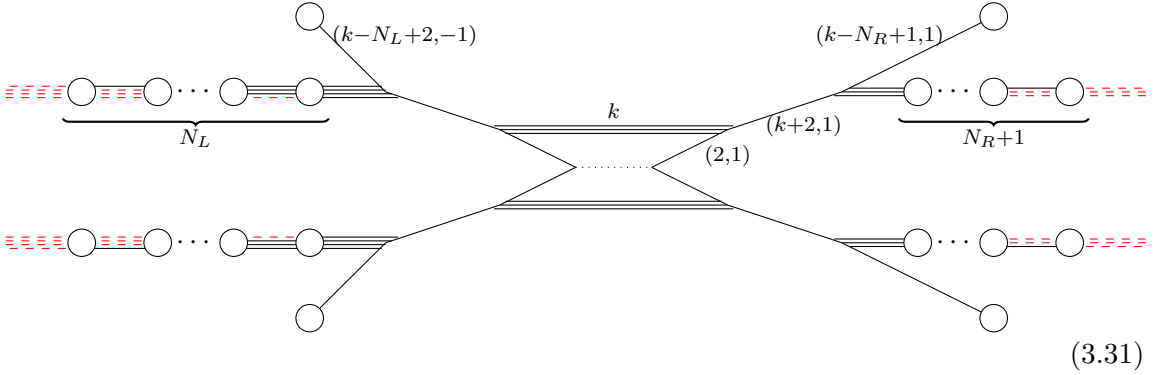
\begin{align}
 \raisebox{-.5\height}{
    \begin{tikzpicture}
    \OPlus{0,0}{1,0}
    \Dbrane{0,0}{-1,0.5}
    \Dbrane{1,0}{2,0.5}
    \draw (2,0.15) node {$\scriptstyle{(2,1)}$};
    \Dbrane{-0.925,0.45}{1.924,0.45}
    \Dbrane{-1,0.5}{2,0.5}
    \Dbrane{-1.075,0.55}{2.075,0.55}
     \draw (0.5,0.75) node {$\scriptstyle{k}$};
    \Dbrane{-1,0.5}{-2.5,1}
    \Dbrane{2,0.5}{3.5,1}
    \draw (3,0.55) node {$\scriptstyle{(k{+}2,1)}$};
    \Dbrane{-2.55,1.075}{-3.5,1.075}
    \Dbrane{-2.5,1.025}{-3.5,1.025}
    \Dbrane{-2.42,0.975}{-3.5,0.975}
    \Dbrane{-2.34,0.925}{-3.5,0.925}
    \Dbrane{-3.5,0.975}{-4.5,0.975}
    \Dbrane{-3.5,1.025}{-4.5,1.025}
    \Dbrane{-3.5,1.075}{-4.5,1.075}
    \Dbrane{-5.5,1.075}{-6.5,1.075}
    \draw (-5,1) node {$\cdots$};
    \Dbrane{3.5,1}{4.5,1}
    \Dbrane{3.6,1.05}{4.5,1.05}
    \Dbrane{3.4,0.95}{4.5,0.95}
    \Dbrane{5.5,1.05}{6.5,1.05}
    \draw (5,1) node {$\cdots$};
    \Dbrane{3.5,1}{5.5,2}
    \draw (4,1.75) node {$\scriptstyle{(k{-}N_R{+}1,1)}$};
    \Dbrane{-2.5,1}{-3.5,2}
    \draw (-2.25,1.75) node {$\scriptstyle{(k{-}N_L{+}2,-1)}$};
    \MonoCut{-3.5,0.925}{-4.5,0.925}
    \MonoCut{-5.5,0.925}{-7.5,0.925}
    \MonoCut{-5.5,0.975}{-7.5,0.975}
    \MonoCut{-5.5,1.025}{-7.5,1.025}
    \MonoCut{-6.5,1.075}{-7.5,1.075}
    \MonoCut{6.5,1.05}{7.5,1.05}
    \MonoCut{5.5,0.95}{7.5,0.95}
    \MonoCut{5.5,1.0}{7.5,1.0}
    \SevenB{-3.5,2}
    \SevenB{5.5,2}
    \SevenB{-3.5,1}
    \SevenB{-4.5,1}
    \SevenB{-5.5,1}
    \SevenB{-6.5,1}
    \SevenB{4.5,1}
    \SevenB{5.5,1}
    \SevenB{6.5,1}
    \draw[decoration={brace,mirror,raise=10pt},decorate,thick]
  (-6.75,1) -- node[below=10pt] {$\scriptstyle{N_L}$ } (-3.25,1);
  \draw[decoration={brace,mirror,raise=10pt},decorate,thick]
  (4.25,1) -- node[below=10pt] {$\scriptstyle{N_R+1}$ } (6.75,1);
    \begin{scope}[yscale=-1,xscale=1]
    \Dbrane{0,0}{-1,0.5}
    \Dbrane{1,0}{2,0.5}
    \Dbrane{-0.925,0.45}{1.924,0.45}
    \Dbrane{-1,0.5}{2,0.5}
    \Dbrane{-1.075,0.55}{2.075,0.55}
    \Dbrane{-1,0.5}{-2.5,1}
    \Dbrane{2,0.5}{3.5,1}
    \Dbrane{-2.55,1.075}{-3.5,1.075}
    \Dbrane{-2.5,1.025}{-3.5,1.025}
    \Dbrane{-2.42,0.975}{-3.5,0.975}
    \Dbrane{-2.34,0.925}{-3.5,0.925}
    \Dbrane{-3.5,0.975}{-4.5,0.975}
    \Dbrane{-3.5,1.025}{-4.5,1.025}
    \Dbrane{-3.5,1.075}{-4.5,1.075}
    \Dbrane{-5.5,1.075}{-6.5,1.075}
    \draw (-5,1) node {$\cdots$};
    \Dbrane{3.5,1}{4.5,1}
    \Dbrane{3.6,1.05}{4.5,1.05}
    \Dbrane{3.4,0.95}{4.5,0.95}
    \Dbrane{5.5,1.05}{6.5,1.05}
    \draw (5,1) node {$\cdots$};
    \Dbrane{3.5,1}{5.5,2}
    \Dbrane{-2.5,1}{-3.5,2}
    \MonoCut{-3.5,0.925}{-4.5,0.925}
    \MonoCut{-5.5,0.925}{-7.5,0.925}
    \MonoCut{-5.5,0.975}{-7.5,0.975}
    \MonoCut{-5.5,1.025}{-7.5,1.025}
    \MonoCut{-6.5,1.075}{-7.5,1.075}
    \MonoCut{6.5,1.05}{7.5,1.05}
    \MonoCut{5.5,0.95}{7.5,0.95}
    \MonoCut{5.5,1.0}{7.5,1.0}
    \SevenB{-3.5,2}
    \SevenB{5.5,2}
    \SevenB{-3.5,1}
    \SevenB{-4.5,1}
    \SevenB{-5.5,1}
    \SevenB{-6.5,1}
    \SevenB{4.5,1}
    \SevenB{5.5,1}
    \SevenB{6.5,1}
    \end{scope}
    \end{tikzpicture}
}
    \end{align}
    and then one can move one of the flavours on the right-hand-side onto the orientifold.
The resulting physical 7-brane, from the merging of the half 7-brane and its mirror, can be split along the orientifold.
Then, moving only one half 7-brane through the 5-branes to the left-hand-side results in
    \begin{align}
 \raisebox{-.5\height}
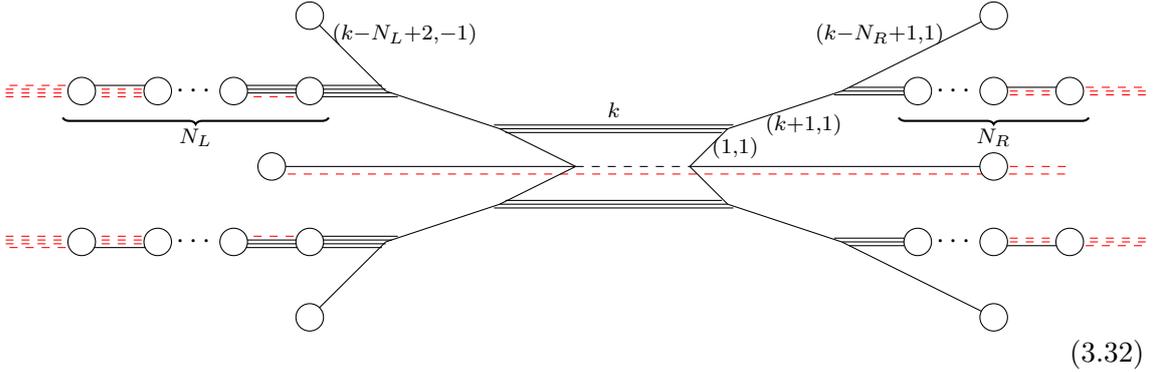
{
    \begin{tikzpicture}
    \OMinusTilde{-4,0}{0,0}
    \OMinusTilde{1.5,0}{5.5,0}
    \MonoCut{-4,-0.1}{6.5,-0.1}
    \MonoCut{5.5,0}{6.5,0}
    \SevenB{5.5,0}
    \SevenB{-4,0}
    \OPlusTilde{0,0}{1.5,0}
    \Dbrane{0,0}{-1,0.5}
    \Dbrane{1.5,0}{2,0.5}
    \draw (2.1,0.25) node {$\scriptstyle{(1,1)}$};
    \Dbrane{-0.925,0.45}{1.924,0.45}
    \Dbrane{-1,0.5}{2,0.5}
    \Dbrane{-1.075,0.55}{2.075,0.55}
     \draw (0.5,0.75) node {$\scriptstyle{k}$};
    \Dbrane{-1,0.5}{-2.5,1}
    \Dbrane{2,0.5}{3.5,1}
    \draw (3,0.55) node {$\scriptstyle{(k{+}1,1)}$};
    \Dbrane{-2.55,1.075}{-3.5,1.075}
    \Dbrane{-2.5,1.025}{-3.5,1.025}
    \Dbrane{-2.42,0.975}{-3.5,0.975}
    \Dbrane{-2.34,0.925}{-3.5,0.925}
    \Dbrane{-3.5,0.975}{-4.5,0.975}
    \Dbrane{-3.5,1.025}{-4.5,1.025}
    \Dbrane{-3.5,1.075}{-4.5,1.075}
    \Dbrane{-5.5,1.075}{-6.5,1.075}
    \draw (-5,1) node {$\cdots$};
    \Dbrane{3.5,1}{4.5,1}
    \Dbrane{3.6,1.05}{4.5,1.05}
    \Dbrane{3.4,0.95}{4.5,0.95}
    \Dbrane{5.5,1.05}{6.5,1.05}
    \draw (5,1) node {$\cdots$};
    \Dbrane{3.5,1}{5.5,2}
    \draw (4,1.75) node {$\scriptstyle{(k{-}N_R{+}1,1)}$};
    \Dbrane{-2.5,1}{-3.5,2}
    \draw (-2.25,1.75) node {$\scriptstyle{(k{-}N_L{+}2,-1)}$};
    \MonoCut{-3.5,0.925}{-4.5,0.925}
    \MonoCut{-5.5,0.925}{-7.5,0.925}
    \MonoCut{-5.5,0.975}{-7.5,0.975}
    \MonoCut{-5.5,1.025}{-7.5,1.025}
    \MonoCut{-6.5,1.075}{-7.5,1.075}
    \MonoCut{6.5,1.05}{7.5,1.05}
    \MonoCut{5.5,0.95}{7.5,0.95}
    \MonoCut{5.5,1.0}{7.5,1.0}
    \SevenB{-3.5,2}
    \SevenB{5.5,2}
    \SevenB{-3.5,1}
    \SevenB{-4.5,1}
    \SevenB{-5.5,1}
    \SevenB{-6.5,1}
    \SevenB{4.5,1}
    \SevenB{5.5,1}
    \SevenB{6.5,1}
    \draw[decoration={brace,mirror,raise=10pt},decorate,thick]
  (-6.75,1) -- node[below=10pt] {$\scriptstyle{N_L}$ } (-3.25,1);
  \draw[decoration={brace,mirror,raise=10pt},decorate,thick]
  (4.25,1) -- node[below=10pt] {$\scriptstyle{N_R}$ } (6.75,1);
    \begin{scope}[yscale=-1,xscale=1]
    \Dbrane{0,0}{-1,0.5}
    \Dbrane{1.5,0}{2,0.5}
    \Dbrane{-0.925,0.45}{1.924,0.45}
    \Dbrane{-1,0.5}{2,0.5}
    \Dbrane{-1.075,0.55}{2.075,0.55}
    \Dbrane{-1,0.5}{-2.5,1}
    \Dbrane{2,0.5}{3.5,1}
    \Dbrane{-2.55,1.075}{-3.5,1.075}
    \Dbrane{-2.5,1.025}{-3.5,1.025}
    \Dbrane{-2.42,0.975}{-3.5,0.975}
    \Dbrane{-2.34,0.925}{-3.5,0.925}
    \Dbrane{-3.5,0.975}{-4.5,0.975}
    \Dbrane{-3.5,1.025}{-4.5,1.025}
    \Dbrane{-3.5,1.075}{-4.5,1.075}
    \Dbrane{-5.5,1.075}{-6.5,1.075}
    \draw (-5,1) node {$\cdots$};
    \Dbrane{3.5,1}{4.5,1}
    \Dbrane{3.6,1.05}{4.5,1.05}
    \Dbrane{3.4,0.95}{4.5,0.95}
    \Dbrane{5.5,1.05}{6.5,1.05}
    \draw (5,1) node {$\cdots$};
    \Dbrane{3.5,1}{5.5,2}
    \Dbrane{-2.5,1}{-3.5,2}
    \MonoCut{-3.5,0.925}{-4.5,0.925}
    \MonoCut{-5.5,0.925}{-7.5,0.925}
    \MonoCut{-5.5,0.975}{-7.5,0.975}
    \MonoCut{-5.5,1.025}{-7.5,1.025}
    \MonoCut{-6.5,1.075}{-7.5,1.075}
    \MonoCut{6.5,1.05}{7.5,1.05}
    \MonoCut{5.5,0.95}{7.5,0.95}
    \MonoCut{5.5,1.0}{7.5,1.0}
    \SevenB{-3.5,2}
    \SevenB{5.5,2}
    \SevenB{-3.5,1}
    \SevenB{-4.5,1}
    \SevenB{-5.5,1}
    \SevenB{-6.5,1}
    \SevenB{4.5,1}
    \SevenB{5.5,1}
    \SevenB{6.5,1}
    \end{scope}
    \end{tikzpicture}
}
    \end{align}
and one can send the two half 7-branes which are on the orientifold towards $\pm \infty$. Via these steps, one has achieve a 5-brane construction for a symplectic gauge group using an \Ofpt\ plane.
Consequently, the brane web proposal for $\sprm(k)$ with $\tilde{N}_f = N_L +N_R 
= N_f -1$ from an \Ofp\ plane is equivalent to that of $\sprm(k)$ with $N_f$ 
flavours from an \Ofpt\ plane. Hence, there does not appear to be a new theory, 
which agrees with the findings of \cite{Zafrir:2015ftn} and the $3$d $\mathcal{N}=4$ analysis in \cite{Feng:2000eq}.
%
%
\section{Comparison with \texorpdfstring{$O7^-$}{O7-} construction}
\label{sec:O7_plane}
\begin{table}[t]
    \centering
    \begin{tabular}{c|cccccccccc}
\toprule
Type IIB & $x^0$ & $x^1$ & $x^2$ & $x^3$ & $x^4$ & $x^5$ & $x^6$ & $x^7$ & $x^8$ & $x^9$\\
\midrule
\NS & $\times$ & $\times$ & $\times$ & $\times$ & $\times$ & $\times$ & & & &  \\
\Dfive & $\times$ & $\times$ & $\times$ & $\times$ & $\times$ & & $\times$ & & & \\
$(p,q)$ 5-brane & $\times$ & $\times$ & $\times$ & $\times$ & $\times$ & \multicolumn{2}{c}{angle $\alpha$} & & & \\
O$7^-$ & $\times$ & $\times$ & $\times$ & $\times$ & $\times$ & & & $\times$ & $\times$ & $\times$ \\
$[p,q]$ 7-brane & $\times$ & $\times$ & $\times$ & $\times$ & $\times$ & & & $\times$ & $\times$ & $\times$ \\
\bottomrule
\end{tabular}
    \caption{Type IIB 5-brane web set-up with O$7^-$ plane: $\times$ indicates the space-time directions spanned by the various branes and the orientifold plane. A $(p,q)$ 5-brane is a line of slope $\tan(\alpha)=q \tau_2/(p+q\tau_1)$ in the $x^{5,6}$ plane where the axiodilaton is $\tau = \tau_1 + i \tau_2$. In this paper all the brane webs are drawn for the value $\tau = i$, so that $\tan(\alpha)=q/p$.}
    \label{tab:O7space}
\end{table}
While the main focus of this work are constructions of $5$d $\mathcal{N}=1$ gauge theories and their UV fixed points using brane webs with \Of\ planes, it is instructive to compare the results to constructions with an O$7$ plane \cite{Bergman:2015dpa}. The space-time occupation of the various branes in this setup are summarised in Table \ref{tab:O7space}. 
In this section, the brane webs with an O$7^-$ plane realising an $\sprm(k)$ gauge theory with $N_f\leq2k+5$ fundamental hypermultiplets are presented. The starting point is the massive Coulomb phase, which for $k$ being the rank of the gauge group and $N_f=N_L+N_R$ can be depicted as follows:
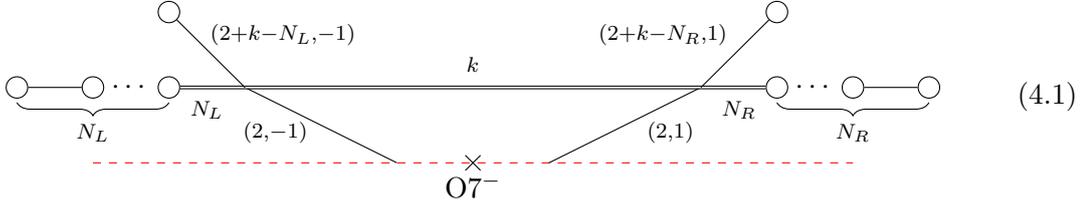
\begin{equation}
 \raisebox{-.5\height}{
    \begin{tikzpicture}
        \coordinate (1) at (-1,0);
        \coordinate (2) at (1,0);
        \coordinate (3) at (-3,1);
        \coordinate (4) at (3,1);
        \draw (-0.1,-0.1)--(0.1,0.1) (0.1,-0.1)--(-0.1,0.1);
        \node at (-2.6,0.4) {$\scriptstyle{(2,-1)}$};
        \node at (2.6,0.4) {$\scriptstyle{(2,1)}$};
        \draw[dashed,red] (-5,0)--(5,0);
        \draw (1)--(3) (2)--(4);
        \draw[double] (3)--(4);
        \node[gauge] (5) at (-4,1) {};
        \node[gauge] (6) at (-5,1) {};
        \node[gauge] (7) at (-6,1) {};
        \node[gauge] (9) at (4,1) {};
        \node[gauge] (10) at (5,1) {};
        \node[gauge] (11) at (6,1) {};
        \draw (7)--(6) (10)--(11);
        \node at (-4.5,1) {$\cdots$};
        \node at (4.5,1) {$\cdots$};
        \draw[double] (5)--(3) (9)--(4);
        \node[gauge] (8) at (-4,2) {};
        \draw (8)--(3);
        \node[gauge] (12) at (4,2) {};
        \draw (12)--(4);
        \node at (-2.5,1.7) {$\scriptstyle{(2+k-N_L,-1)}$};
        \node at (2.5,1.7) {$\scriptstyle{(2+k-N_R,1)}$};
        \node at (0,1.3) {$\scriptstyle{k}$};
        \draw [decorate,decoration={brace,amplitude=5pt},xshift=0pt,yshift=-0.2cm]
        (-4,1)--(-6,1) node [black,midway,yshift=-0.4cm] {$\scriptstyle{N_L}$};
        \draw [decorate,decoration={brace,amplitude=5pt},xshift=0pt,yshift=-0.2cm]
        (6,1)--(4,1) node [black,midway,yshift=-0.4cm] {$\scriptstyle{N_R}$};
        \node at (0,-0.3) {$\mathrm{O}7^-$};
        \node at (-3.5,0.7) {$\scriptstyle{N_L}$};
        \node at (3.5,0.7) {$\scriptstyle{N_R}$};
    \end{tikzpicture}
    }
\end{equation}
and masses of the hypermultiplets are chosen to be equal for convenience. 
For the purposes of this section, the decomposition of the number of flavours 
into $N_f=N_L+N_R$ is a convenient choice and has no impact on the results.
As shown in \cite{Sen:1996vd}, an O$7^-$ orientifold is quantum mechanically 
resolved into a pair of mutually non-local $[p,q]$ 7-branes, such that the 
combined monodromy equals that of the O$7^-$. One common choice is given by a  
$[1,1]$ 7-brane together with a $[1,-1]$ 7-brane, which is denoted as blue and 
red respectively. Thus, then quantum mechanically corrected 5-brane web becomes
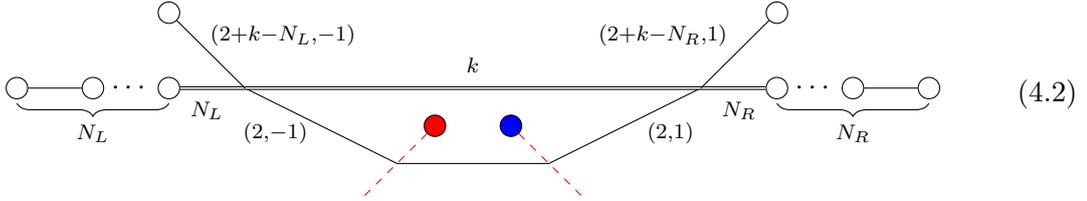
\begin{equation}
 \raisebox{-.5\height}{
    \begin{tikzpicture}
        \coordinate (1) at (-1,0);
        \coordinate (2) at (1,0);
        \coordinate (3) at (-3,1);
        \coordinate (4) at (3,1);
        \node at (-2.6,0.4) {$\scriptstyle{(2,-1)}$};
        \node at (2.6,0.4) {$\scriptstyle{(2,1)}$};
        \draw (3)--(1)--(2)--(4);
        \draw[double] (3)--(4);
        \node[gauge] (5) at (-4,1) {};
        \node[gauge] (6) at (-5,1) {};
        \node[gauge] (7) at (-6,1) {};
        \node[gauge] (9) at (4,1) {};
        \node[gauge] (10) at (5,1) {};
        \node[gauge] (11) at (6,1) {};
        \draw (7)--(6) (10)--(11);
        \node at (-4.5,1) {$\cdots$};
        \node at (4.5,1) {$\cdots$};
        \draw[double] (5)--(3) (9)--(4);
        \node[gauge] (8) at (-4,2) {};
        \draw (8)--(3);
        \node[gauge] (12) at (4,2) {};
        \draw (12)--(4);
        \node at (-2.5,1.7) {$\scriptstyle{(2+k-N_L,-1)}$};
        \node at (2.5,1.7) {$\scriptstyle{(2+k-N_R,1)}$};
        \node at (0,1.3) {$\scriptstyle{k}$};
        \draw [decorate,decoration={brace,amplitude=5pt},xshift=0pt,yshift=-0.2cm]
        (-4,1)--(-6,1) node [black,midway,yshift=-0.4cm] {$\scriptstyle{N_L}$};
        \draw [decorate,decoration={brace,amplitude=5pt},xshift=0pt,yshift=-0.2cm]
        (6,1)--(4,1) node [black,midway,yshift=-0.4cm] {$\scriptstyle{N_R}$};
        \node[gauger] (13) at (-0.5,0.5) {};
        \node[gaugeb] (14) at (0.5,0.5) {};
        \draw[dashed, red] (13)--(-1.5,-0.5) (14)--(1.5,-0.5);
        \node at (-3.5,0.7) {$\scriptstyle{N_L}$};
        \node at (3.5,0.7) {$\scriptstyle{N_R}$};
    \end{tikzpicture}
    }
\end{equation}
which can be brought into a more convenient form by moving the monodromy cuts. Accounting for the effects of the monodromy \eqref{eq:monodromy}, one finds
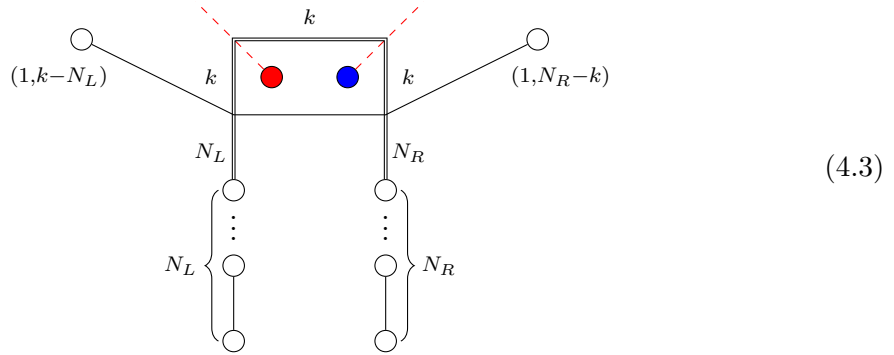
\begin{equation}
 \raisebox{-.5\height}{
    \begin{tikzpicture}
        \node[gauger] (1) at (-0.5,0.5) {};
        \node[gaugeb] (2) at (0.5,0.5) {};
        \draw[red,dashed] (1)--(-1.5,1.5) (2)--(1.5,1.5);
        \coordinate (3) at (-1,1);
        \coordinate (4) at (1,1);
        \coordinate (5) at (-1,0);
        \coordinate (6) at (1,0);
        \node[gauge] (7) at (-3,1) {};
        \node[gauge] (8) at (3,1) {};
        \node[gauge] (9) at (-1,-1) {};
        \node[gauge] (10) at (-1,-2) {};
        \node[gauge] (11) at (-1,-3) {};
        \node[gauge] (12) at (1,-1) {};
        \node[gauge] (13) at (1,-2) {};
        \node[gauge] (14) at (1,-3) {};
        \draw[double] (9)--(5)--(3)--(4)--(6)--(12);
        \draw (7)--(5)--(6)--(8) (10)--(11) (13)--(14);
        \node at (-1,-1.40) {$\vdots$};
        \node at (1,-1.40) {$\vdots$};
        \node at (-3.3,0.5) {$\scriptstyle{(1,k-N_L)}$};
        \node at (3.3,0.5) {$\scriptstyle{(1,N_R-k)}$};
        \node at (-1.3,0.5) {$\scriptstyle{k}$};
        \node at (0,1.3) {$\scriptstyle{k}$};
        \node at (1.3,0.5) {$\scriptstyle{k}$};
        \node at (-1.3,-0.5) {$\scriptstyle{N_L}$};
        \node at (1.3,-0.5) {$\scriptstyle{N_R}$};
        \draw [decorate,decoration={brace,amplitude=5pt},yshift=0pt,xshift=-0.2cm]
        (-1,-3)--(-1,-1) node [black,midway,xshift=-0.5cm] {$\scriptstyle{N_L}$};
        \draw [decorate,decoration={brace,amplitude=5pt},yshift=0pt,xshift=0.2cm]
        (1,-1)--(1,-3) node [black,midway,xshift=0.5cm] {$\scriptstyle{N_R}$};
    \end{tikzpicture}
    }
\end{equation}
and the $[1,\pm1]$ 7-branes inside the web can be moved through the 5-branes by accounting for brane creation, as in \eqref{eq:brane_creation}. Therefore, the 5-brane web with all 7-branes outside is given by
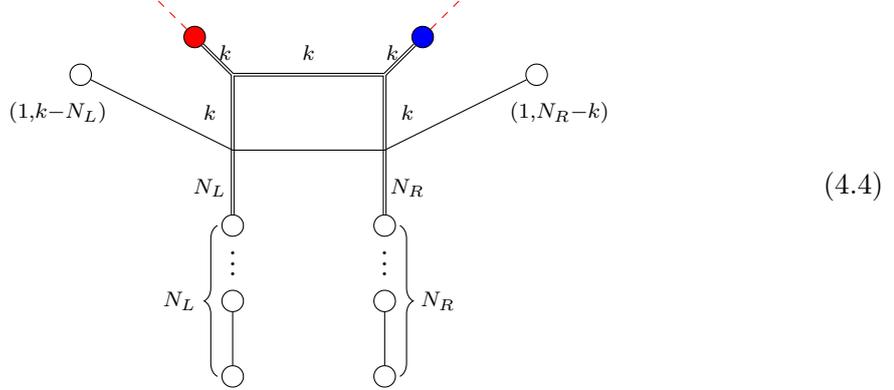
\begin{equation}
 \raisebox{-.5\height}{
    \begin{tikzpicture}
        \node[gauger] (1) at (-1.5,1.5) {};
        \node[gaugeb] (2) at (1.5,1.5) {};
        \draw[red,dashed] (1)--(-2,2) (2)--(2,2);
        \coordinate (5) at (-1,0);
        \coordinate (6) at (1,0);
        \node[gauge] (7) at (-3,1) {};
        \node[gauge] (8) at (3,1) {};
        \node[gauge] (9) at (-1,-1) {};
        \node[gauge] (10) at (-1,-2) {};
        \node[gauge] (11) at (-1,-3) {};
        \node[gauge] (12) at (1,-1) {};
        \node[gauge] (13) at (1,-2) {};
        \node[gauge] (14) at (1,-3) {};
        \draw[double] (9)--(5)--(-1,1)--(1,1)--(6)--(12) (-1,1)--(1) (1,1)--(2);
        \draw (7)--(5)--(6)--(8) (10)--(11) (13)--(14);
        \node at (-1,-1.4) {$\vdots$};
        \node at (1,-1.4) {$\vdots$};
        \node at (-3.3,0.5) {$\scriptstyle{(1,k-N_L)}$};
        \node at (3.3,0.5) {$\scriptstyle{(1,N_R-k)}$};
        \node at (-1.3,0.5) {$\scriptstyle{k}$};
        \node at (0,1.3) {$\scriptstyle{k}$};
        \node at (1.3,0.5) {$\scriptstyle{k}$};
        \node at (-1.1,1.3) {$\scriptstyle{k}$};
        \node at (1.1,1.3) {$\scriptstyle{k}$};
        \node at (-1.3,-0.5) {$\scriptstyle{N_L}$};
        \node at (1.3,-0.5) {$\scriptstyle{N_R}$};
        \draw [decorate,decoration={brace,amplitude=5pt},yshift=0pt,xshift=-0.2cm]
        (-1,-3)--(-1,-1) node [black,midway,xshift=-0.5cm] {$\scriptstyle{N_L}$};
        \draw [decorate,decoration={brace,amplitude=5pt},yshift=0pt,xshift=0.2cm]
        (1,-1)--(1,-3) node [black,midway,xshift=0.5cm] {$\scriptstyle{N_R}$};
    \end{tikzpicture}
    }
\end{equation}
such that one can finally transition to the Higgs branch phase. For this, all masses are set to zero and one ends up with
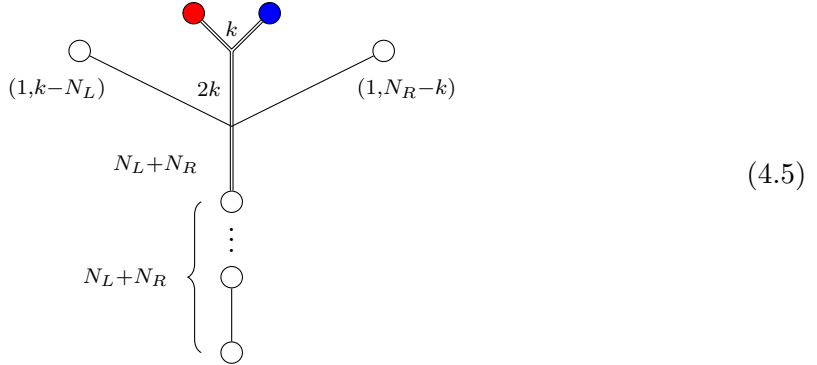
\begin{equation}
 \raisebox{-.5\height}{
    \begin{tikzpicture}
        \node[gauger] (1) at (-0.5,1.5) {};
        \node[gaugeb] (2) at (0.5,1.5) {};
        \node[gauge] (7) at (-2,1) {};
        \node[gauge] (8) at (2,1) {};
        \node[gauge] (9) at (0,-1) {};
        \node[gauge] (10) at (0,-2) {};
        \node[gauge] (11) at (0,-3) {};
        \coordinate (3) at (0,0);
        \draw[double] (1)--(0,1)--(2) (0,1)--(3) (3)--(9);
        \draw (10)--(11);
        \draw (7)--(3)--(8);
        \node at (0,-1.4) {$\vdots$};
        \node at (-2.3,0.5) {$\scriptstyle{(1,k-N_L)}$};
        \node at (2.3,0.5) {$\scriptstyle{(1,N_R-k)}$};
        \node at (-0.3,0.5) {$\scriptstyle{2k}$};
        \node at (0,1.3) {$\scriptstyle{k}$};
        \node at (-1,-0.5) {$\scriptstyle{N_L+N_R}$};
        \draw [decorate,decoration={brace,amplitude=5pt},yshift=0pt,xshift=-0.4cm]
        (0,-3)--(0,-1) node [black,midway,xshift=-1cm] {$\scriptstyle{N_L+N_R}$};
    \end{tikzpicture}
    }
    \label{eq:web_Sp_O7_finite}
\end{equation}
which is a 5-brane web ending on 7-branes without \Of\ planes. Therefore, the results of \cite{Cabrera:2018jxt} apply. In other words, starting from \eqref{eq:web_Sp_O7_finite}, the associated magnetic quivers with only unitary magnetic gauge nodes are read off by \cite[Conj.\ 1]{Cabrera:2018jxt}.
\subsection{Finite coupling}
In the following, unitary magnetic quivers are computed for $\sprm(k)$ theories at finite coupling.
\paragraph{$\mathbf{N_f\leq 2k+1}$.}
For small numbers of flavours the magnetic quiver depends on $N_f$ being even or 
odd. For even number of flavours one finds the following two maximal 
subdivisions:
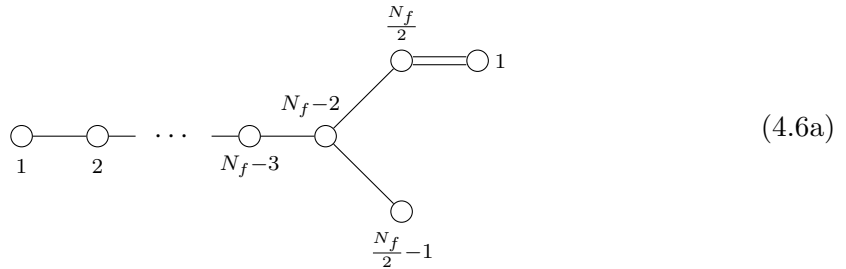
\begin{subequations}
\label{eq:magQuiv_Sp_O7_finite_Nf_small_even}
    \begin{align}
     \raisebox{-.5\height}{
    \begin{tikzpicture}
        \node[gauge] (1) at (-1,0) {};
        \node[gauge] (2) at (0,0) {};
        \node[gauge] (3) at (2,0) {};
        \node[gauge] (4) at (3,0) {};
        \node[gauge] (5) at (4,1) {};
        \node[gauge] (6) at (4,-1) {};
        \node[gauge] (7) at (5,1) {};
        \draw (1)--(2)--(0.5,0) (1.5,0)--(3)--(4)--(5) (6)--(4);
        \draw[transform canvas={yshift=-1.5pt}] (5)--(7);
        \draw[transform canvas={yshift=1.5pt}] (5)--(7);
        \node at (-1,-0.4) {$\scriptstyle{1}$};
        \node at (0,-0.4) {$\scriptstyle{2}$};
        \node at (2,-0.4) {$\scriptstyle{N_f {-} 3}$};
        \node at (2.8,0.4) {$\scriptstyle{N_f {-} 2}$};
        \node at (4,-1.5) {$\scriptstyle{\frac{N_f}{2} {-} 1}$};
        \node at (4,1.5) {$\scriptstyle{\frac{N_f}{2}}$};
        \node at (5.3,1) {$\scriptstyle{1}$};
        \node at (1,0) {$\cdots$};
    \end{tikzpicture}
    }
    \label{eq:magQuiv_Sp_O7_finite_Nf_small_even_I}
    \\
 \raisebox{-.5\height}
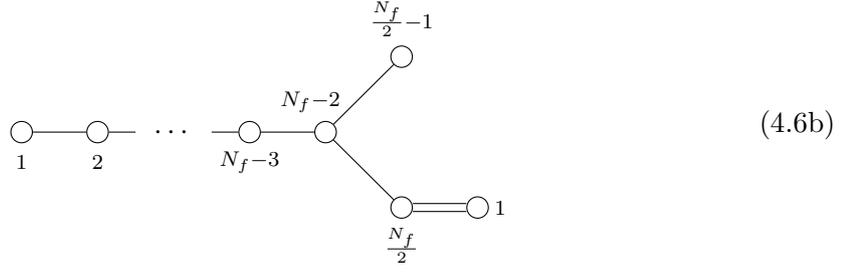
{
    \begin{tikzpicture}
        \node[gauge] (1) at (-1,0) {};
        \node[gauge] (2) at (0,0) {};
        \node[gauge] (3) at (2,0) {};
        \node[gauge] (4) at (3,0) {};
        \node[gauge] (5) at (4,1) {};
        \node[gauge] (6) at (4,-1) {};
        \node[gauge] (7) at (5,-1) {};
        \draw (1)--(2)--(0.5,0) (1.5,0)--(3)--(4)--(5) (6)--(4);
        \draw[transform canvas={yshift=-1.5pt}] (6)--(7);
        \draw[transform canvas={yshift=1.5pt}] (6)--(7);
        \node at (-1,-0.4) {$\scriptstyle{1}$};
        \node at (0,-0.4) {$\scriptstyle{2}$};
        \node at (2,-0.4) {$\scriptstyle{N_f {-} 3}$};
        \node at (2.8,0.4) {$\scriptstyle{N_f {-} 2}$};
        \node at (4,-1.5) {$\scriptstyle{\frac{N_f}{2}}$};
        \node at (4,1.5) {$\scriptstyle{\frac{N_f}{2} {-} 1}$};
        \node at (5.3,-1) {$\scriptstyle{1}$};
        \node at (1,0) {$\cdots$};
    \end{tikzpicture}
    }
    \label{eq:magQuiv_Sp_O7_finite_Nf_small_even_II}
    \end{align}
and the intersection of the cones associated to 
\eqref{eq:magQuiv_Sp_O7_finite_Nf_small_even_I} and 
\eqref{eq:magQuiv_Sp_O7_finite_Nf_small_even_II} in the Higgs branch is 
described by
    \begin{align}
    \raisebox{-.5\height}
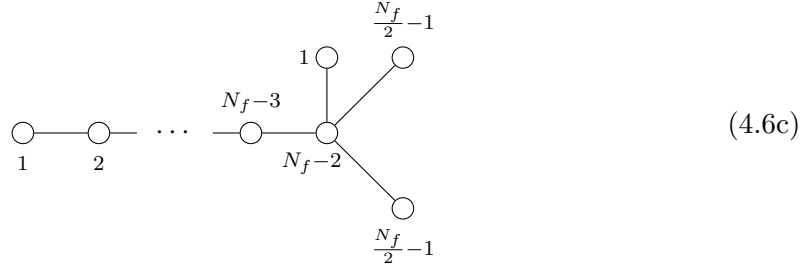
{
    \begin{tikzpicture}
        \node[gauge] (1) at (-1,0) {};
        \node[gauge] (2) at (0,0) {};
        \node[gauge] (3) at (2,0) {};
        \node[gauge] (4) at (3,0) {};
        \node[gauge] (5) at (4,1) {};
        \node[gauge] (6) at (4,-1) {};
        \node[gauge] (7) at (3,1) {};
        \draw (1)--(2)--(0.5,0) (1.5,0)--(3)--(4)--(5) (6)--(4)--(7);
        \node at (-1,-0.4) {$\scriptstyle{1}$};
        \node at (0,-0.4) {$\scriptstyle{2}$};
        \node at (2,0.4) {$\scriptstyle{N_f {-} 3}$};
        \node at (2.8,-0.4) {$\scriptstyle{N_f {-} 2}$};
        \node at (4,-1.5) {$\scriptstyle{\frac{N_f}{2} {-} 1}$};
        \node at (4,1.5) {$\scriptstyle{\frac{N_f}{2} {-} 1}$};
        \node at (2.7,1) {$\scriptstyle{1}$};
        \node at (1,0) {$\cdots$};
    \end{tikzpicture}
    }
       \label{eq:magQuiv_Sp_O7_finite_Nf_small_even_int}
\end{align}
\end{subequations}
In contrast, for odd number of flavours there is only one maximal subdivision and the magnetic quiver becomes
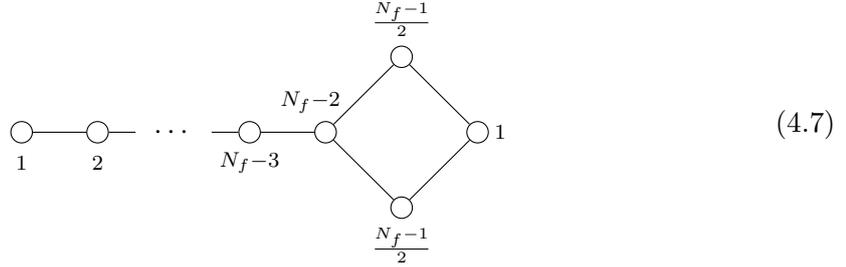
\begin{equation}
 \raisebox{-.5\height}{
    \begin{tikzpicture}
        \node[gauge] (1) at (-1,0) {};
        \node[gauge] (2) at (0,0) {};
        \node[gauge] (3) at (2,0) {};
        \node[gauge] (4) at (3,0) {};
        \node[gauge] (5) at (4,1) {};
        \node[gauge] (6) at (4,-1) {};
        \node[gauge] (7) at (5,0) {};
        \draw (1)--(2)--(0.5,0) (1.5,0)--(3)--(4)--(5)--(7)--(6)--(4);
        \node at (-1,-0.4) {$\scriptstyle{1}$};
        \node at (0,-0.4) {$\scriptstyle{2}$};
        \node at (2,-0.4) {$\scriptstyle{N_f {-} 3}$};
        \node at (2.8,0.4) {$\scriptstyle{N_f {-} 2}$};
        \node at (4,-1.5) {$\scriptstyle{\frac{N_f-1}{2}}$};
        \node at (4,1.5) {$\scriptstyle{\frac{N_f-1}{2}}$};
        \node at (5.3,0) {$\scriptstyle{1}$};
        \node at (1,0) {$\cdots$};
    \end{tikzpicture}
    }
    \label{eq:magQuiv_Sp_O7_finite_Nf_small_odd}
\end{equation}
In both cases, one straightforwardly computes
\begin{align}
\dim_{\HH}\Coulomb \eqref{eq:magQuiv_Sp_O7_finite_Nf_small_even} = \dim_{\HH}\Coulomb \eqref{eq:magQuiv_Sp_O7_finite_Nf_small_odd} = \frac{N_f (N_f-1)}{2} 
\quad 
\text{and}
\quad 
G = \sorm(2N_f)
\end{align}
due to the subset of balanced nodes, see \ref{app:Coulomb_branch}.
In fact, the moduli spaces are known \cite{Ferlito:2016grh} to satisfy
\begin{subequations}
\begin{align}
 \Coulomb \eqref{eq:magQuiv_Sp_O7_finite_Nf_small_even_I} \cup  
 \Coulomb\eqref{eq:magQuiv_Sp_O7_finite_Nf_small_even_II} &= 
\clorbit_{\dalg}^{[2^{N_f}]} 
\;, \qquad 
 \Coulomb \eqref{eq:magQuiv_Sp_O7_finite_Nf_small_even_I} \cap  
 \Coulomb\eqref{eq:magQuiv_Sp_O7_finite_Nf_small_even_II} = 
 \Coulomb \eqref{eq:magQuiv_Sp_O7_finite_Nf_small_even_int} =
 \clorbit_{\dalg}^{[2^{N_f-2},1^4]}
\,, 
\\
\Coulomb \eqref{eq:magQuiv_Sp_O7_finite_Nf_small_odd} &=  
\clorbit_{\dalg}^{[2^{N_f-1},1^2]} \,.
\end{align}
\end{subequations}
As a remark, incomplete Higgsing has not been detailed in 
\cite{Ferlito:2016grh}; however, the magnetic quivers still have the same 
structure and thus the results apply.
In particular, these results agree with the Higgs branch expectations 
\eqref{eq:part_Higgs_Nf_even}-\eqref{eq:part_Higgs_Nf_odd} as well as the 
magnetic quivers results \eqref{eq:results_Spk_Nf_even_small} and 
\eqref{eq:results_Spk_Nf_odd_small} derived from \Of\ planes.
\paragraph{$\mathbf{N_f> 2k+1}$.}
For large number of flavours, the magnetic quiver shows uniform behaviour and is read off to be
\begin{equation}
 \raisebox{-.5\height}{
    \begin{tikzpicture}
        \node[gauge] (1) at (-1,0) {};
        \node[gauge] (2) at (0,0) {};
        \node[gauge] (3) at (2,0) {};
        \node[gauge] (4) at (3,0) {};
        \node[gauge] (5) at (5,0) {};
        \node[gauge] (6) at (6,1) {};
        \node[gauge] (7) at (6,-1) {};
        \node[gauge] (8) at (3,1) {};
        \draw (1)--(2)--(0.5,0) (1.5,0)--(3)--(4)--(3.5,0) (4.5,0)--(5)--(6) (5)--(7) (4)--(8);
        \node at (1,0) {$\cdots$};
        \node at (4,0) {$\cdots$};
        \node at (-1,-0.4) {$\scriptstyle{1}$};
        \node at (0,-0.4) {$\scriptstyle{2}$};
        \node at (2,-0.4) {$\scriptstyle{2k{-}1}$};
        \node at (3,-0.4) {$\scriptstyle{2k}$};
        \node at (5,-0.4) {$\scriptstyle{2k}$};
        \node at (6.5,1) {$\scriptstyle{k}$};
        \node at (6.5,-1) {$\scriptstyle{k}$};
        \node at (2.6,1) {$\scriptstyle{1}$};
        \draw [decorate,decoration={brace,amplitude=5pt},yshift=-0.8cm,xshift=0cm]
        (5,0)--(3,0) node [black,midway,yshift=-0.5cm] {$\scriptstyle{N_f-2k-1}$};
    \end{tikzpicture}
    }
    \label{eq:magQuiv_Sp_O7_finite_Nf_large}
\end{equation}
and one confirms that
\begin{align}
    \dim_{\HH} \Coulomb\eqref{eq:magQuiv_Sp_O7_finite_Nf_large} =     k \left(2 N_f - 2 k -1 \right)
    \quad \text{and} \quad 
    G = \sorm(2N_f)
\end{align}
which agrees with the Higgs branch expectation \eqref{eq:Higgs_dim_Sp_classical} as well as the magnetic quiver result \eqref{eq:results_Spk_Nf_even_large} and \eqref{eq:results_Spk_Nf_odd_large} derived from \Of\ planes.
In addition, one notes that \eqref{eq:magQuiv_Sp_O7_finite_Nf_large} agrees with 
the $3$d $\Ncal=4$ mirror theory for $\sprm(k)$ gauge theory with $N_f \geq 2k$ 
flavours derived in \cite{Hanany:1999sj} \cite[Fig.\ 4]{Ferlito:2016grh}. The moduli space 
\begin{align}
 \Coulomb \eqref{eq:magQuiv_Sp_O7_finite_Nf_large} = 
\clorbit_{\dalg}^{[2^{2k},1^{2N_f-4k}]}  
\end{align}
is known to be a nilpotent orbit closure.
\subsection{Infinite coupling}
In the following unitary magnetic quivers are computed for $\sprm(k)$ theories at infinite coupling.
\paragraph{$\mathbf{N_f\leq2k}$.}
Suppose the number of flavours satisfies $N_f\leq2k$, then the infinite coupling regime is reached via global deformation of the brane web
\begin{equation}
 \raisebox{-.5\height}{
    \begin{tikzpicture}
        \node[gauger] (1) at (-1.5,1.5) {};
        \node[gaugeb] (2) at (1.5,1.5) {};
        \node[gauge] (7) at (-2,-1) {};
        \node[gauge] (8) at (2,-1) {};
        \node[gauge] (9) at (0,-1) {};
        \node[gauge] (10) at (0,-2) {};
        \node[gauge] (11) at (0,-3) {};
        \coordinate (3) at (0,0);
        \draw[double] (1)--(3)--(2) (3)--(9);
        \draw (10)--(11);
        \draw (7)--(3)--(8);
        \node at (0,-1.4) {$\vdots$};
        \node at (-2.3,-0.5) {$\scriptstyle{(1,k-N_L)}$};
        \node at (2.3,-0.5) {$\scriptstyle{(1,N_R-k)}$};
        \node at (-1,1.3) {$\scriptstyle{k}$};
        \node at (1,1.3) {$\scriptstyle{k}$};
        \node at (-0.6,-0.75) {$\scriptstyle{N_L{+}N_R}$};
        \draw [decorate,decoration={brace,amplitude=5pt},yshift=0pt,xshift=-0.4cm]
        (0,-3)--(0,-1) node [black,midway,xshift=-1cm] {$\scriptstyle{N_L+N_R}$};
    \end{tikzpicture}
    }
\end{equation}
and magnetic quivers at infinite coupling can be read off by the usual rules. However, the result depends on $N_f$ being even or odd, similar to the finite coupling case. 
To begin with, suppose $N_f$ is even, then one finds two maximal subdivision with the following magnetic quivers:
    \begin{subequations}
     \label{eq:magQuiv_Sp_O7_infinite_Nf_small_even}
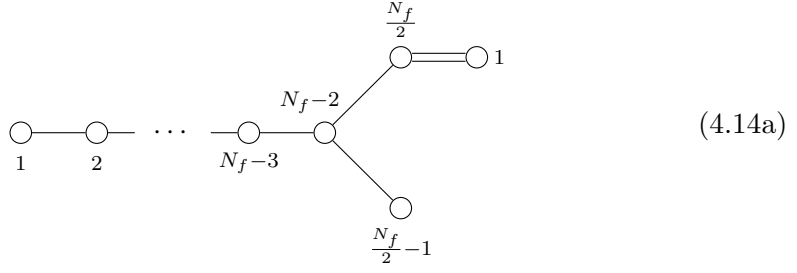
    \begin{align}
     \raisebox{-.5\height}{
    \begin{tikzpicture}
        \node[gauge] (1) at (-1,0) {};
        \node[gauge] (2) at (0,0) {};
        \node[gauge] (3) at (2,0) {};
        \node[gauge] (4) at (3,0) {};
        \node[gauge] (5) at (4,1) {};
        \node[gauge] (6) at (4,-1) {};
        \node[gauge] (7) at (5,1) {};
        \draw (1)--(2)--(0.5,0) (1.5,0)--(3)--(4)--(5) (6)--(4);
        \draw[transform canvas={yshift=-1.5pt}] (5)--(7);
        \draw[transform canvas={yshift=1.5pt}] (5)--(7);
        \node at (-1,-0.4) {$\scriptstyle{1}$};
        \node at (0,-0.4) {$\scriptstyle{2}$};
        \node at (2,-0.4) {$\scriptstyle{N_f {-} 3}$};
        \node at (2.8,0.4) {$\scriptstyle{N_f {-} 2}$};
        \node at (4,-1.5) {$\scriptstyle{\frac{N_f}{2} {-} 1}$};
        \node at (4,1.5) {$\scriptstyle{\frac{N_f}{2}}$};
        \node at (5.3,1) {$\scriptstyle{1}$};
        \node at (1,0) {$\cdots$};
    \end{tikzpicture}
    }
    \label{eq:magQuiv_Sp_O7_infinite_Nf_small_even_1}
    \\
 \raisebox{-.5\height}
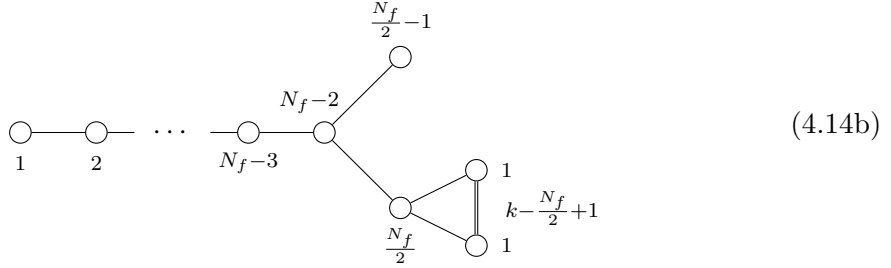
{
    \begin{tikzpicture}
        \node[gauge] (1) at (-1,0) {};
        \node[gauge] (2) at (0,0) {};
        \node[gauge] (3) at (2,0) {};
        \node[gauge] (4) at (3,0) {};
        \node[gauge] (5) at (4,1) {};
        \node[gauge] (6) at (4,-1) {};
        \node[gauge] (8) at (5,-0.5) {};
        \node[gauge] (7) at (5,-1.5) {};
        \draw (1)--(2)--(0.5,0) (1.5,0)--(3)--(4)--(5) (6)--(4) (6)--(7) (6)--(8);
        \draw[double] (7)--(8);
        \node at (6,-1) {$\scriptstyle{ k {-} \frac{N_f}{2} {+} 1}$};
        \node at (-1,-0.4) {$\scriptstyle{1}$};
        \node at (0,-0.4) {$\scriptstyle{2}$};
        \node at (2,-0.4) {$\scriptstyle{N_f{-}3}$};
        \node at (2.8,0.4) {$\scriptstyle{N_f{-}2}$};
        \node at (4,-1.5) {$\scriptstyle{\frac{N_f}{2}}$};
        \node at (4,1.5) {$\scriptstyle{\frac{N_f}{2}{-}1}$};
        \node at (1,0) {$\cdots$};
        \node at (5.4,-0.5) {$\scriptstyle{1}$};
        \node at (5.4,-1.5) {$\scriptstyle{1}$};
    \end{tikzpicture}
    }
        \label{eq:magQuiv_Sp_O7_infinite_Nf_small_even_2}
\end{align}
\end{subequations}
and the moduli space becomes the union of two cones. Note that \eqref{eq:magQuiv_Sp_O7_infinite_Nf_small_even_1} coincides with the finite coupling results, while for \eqref{eq:magQuiv_Sp_O7_infinite_Nf_small_even_2} one computes the following 
\begin{align}
    \dim_{\HH}\Coulomb \eqref{eq:magQuiv_Sp_O7_infinite_Nf_small_even_2} = \frac{1}{2}N_f(N_f-1)+1
    \quad \text{} \quad 
     G= \sorm(2N_f) \times \uo \,.
\end{align}
For odd number of flavours $N_f$, there exists only one maximal subdivision and the magnetic quivers is read off to be
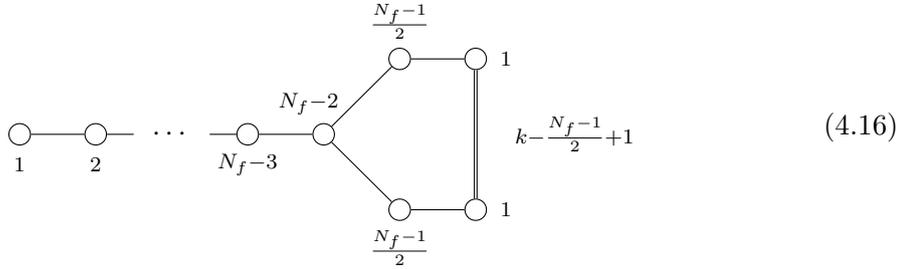
\begin{equation}
 \raisebox{-.5\height}{
    \begin{tikzpicture}
        \node[gauge] (1) at (-1,0) {};
        \node[gauge] (2) at (0,0) {};
        \node[gauge] (3) at (2,0) {};
        \node[gauge] (4) at (3,0) {};
        \node[gauge] (5) at (4,1) {};
        \node[gauge] (6) at (4,-1) {};
        \node[gauge] (7) at (5,1) {};
        \node[gauge] (8) at (5,-1) {};
        \draw (1)--(2)--(0.5,0) (1.5,0)--(3)--(4)--(5)--(7) (8)--(6)--(4);
        \draw[double] (7)--(8);
        \node at (6.3,0) {$\scriptstyle{k-\frac{N_f-1}{2}+1}$};
        \node at (-1,-0.4) {$\scriptstyle{1}$};
        \node at (0,-0.4) {$\scriptstyle{2}$};
        \node at (2,-0.4) {$\scriptstyle{N_f{-}3}$};
        \node at (2.8,0.4) {$\scriptstyle{N_f{-}2}$};
        \node at (4,-1.5) {$\scriptstyle{\frac{N_f-1}{2}}$};
        \node at (4,1.5) {$\scriptstyle{\frac{N_f-1}{2}}$};
        \node at (1,0) {$\cdots$};
        \node at (5.4,1) {$\scriptstyle{1}$};
        \node at (5.4,-1) {$\scriptstyle{1}$};
    \end{tikzpicture}
    }
     \label{eq:magQuiv_Sp_O7_infinite_Nf_small_odd}
\end{equation}
and, as above, one concludes that 
\begin{align}
   \dim_{\HH}\Coulomb \eqref{eq:magQuiv_Sp_O7_infinite_Nf_small_odd} = \frac{1}{2}N_f(N_f-1)+1
    \quad \text{and} \quad 
     G_J= \sorm(2N_f) \times \uo \,.
\end{align}
Both cases \eqref{eq:magQuiv_Sp_O7_infinite_Nf_small_even}-\eqref{eq:magQuiv_Sp_O7_infinite_Nf_small_odd} correctly reproduce the number of additional Higgs branch directions at infinite coupling \cite{Bergman:2015dpa}.
Comparing to the results \eqref{eq:magQuiv_Spk_infinite_Nf_even_small} and \eqref{eq:magQuiv_Spk_infinite_Nf_odd_small} from the \Of\ plane construction, the O$7^-$ results do indicate the existence of multiple cones, which are not currently visible in the \Of\ analysis.
\paragraph{$\mathbf{N_f=2k+1}$.} 
Increasing the number of flavours to $N_f=2k+1$, the 5-brane web has to be evaluated more carefully. In detail, it becomes
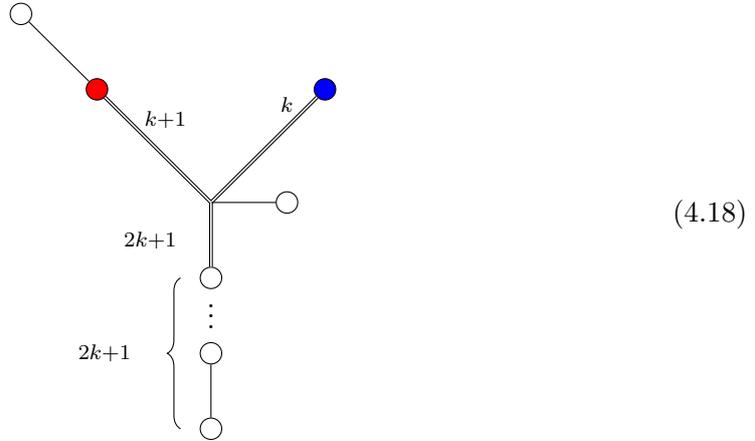
\begin{equation}
 \raisebox{-.5\height}{
    \begin{tikzpicture}
        \node[gauger] (1) at (-1.5,1.5) {};
        \node[gaugeb] (2) at (1.5,1.5) {};
        \node[gauge] (7) at (-2.5,2.5) {};
        \node[gauge] (8) at (1,0) {};
        \node[gauge] (9) at (0,-1) {};
        \node[gauge] (10) at (0,-2) {};
        \node[gauge] (11) at (0,-3) {};
        \coordinate (3) at (0,0);
        \draw[double] (1)--(3)--(2) (3)--(9);
        \draw (10)--(11);
        \draw (7)--(1) (3)--(8);
        \node at (0,-1.4) {$\vdots$};
        \node at (-0.6,1.1) {$\scriptstyle{k+1}$};
        \node at (1,1.3) {$\scriptstyle{k}$};
        \node at (-0.8,-0.5) {$\scriptstyle{2k+1}$};
        \draw [decorate,decoration={brace,amplitude=5pt},yshift=0pt,xshift=-0.4cm]
        (0,-3)--(0,-1) node [black,midway,xshift=-1cm] {$\scriptstyle{2k+1}$};
    \end{tikzpicture}
    }
\end{equation}
and the associated magnetic quiver is
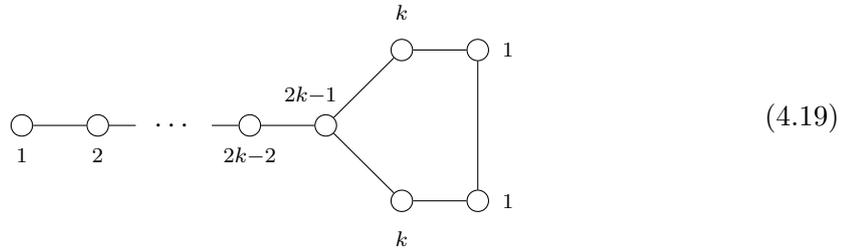
\begin{equation}
 \raisebox{-.5\height}{
    \begin{tikzpicture}
        \node[gauge] (1) at (-1,0) {};
        \node[gauge] (2) at (0,0) {};
        \node[gauge] (3) at (2,0) {};
        \node[gauge] (4) at (3,0) {};
        \node[gauge] (5) at (4,1) {};
        \node[gauge] (6) at (4,-1) {};
        \node[gauge] (7) at (5,1) {};
        \node[gauge] (8) at (5,-1) {};
        \draw (1)--(2)--(0.5,0) (1.5,0)--(3)--(4)--(5)--(7)--(8)--(6)--(4);
        \node at (-1,-0.4) {$\scriptstyle{1}$};
        \node at (0,-0.4) {$\scriptstyle{2}$};
        \node at (2,-0.4) {$\scriptstyle{2k {-} 2}$};
        \node at (2.8,0.4) {$\scriptstyle{2k {-} 1}$};
        \node at (4,-1.5) {$\scriptstyle{k}$};
        \node at (4,1.5) {$\scriptstyle{k}$};
        \node at (1,0) {$\cdots$};
        \node at (5.4,1) {$\scriptstyle{1}$};
        \node at (5.4,-1) {$\scriptstyle{1}$};
    \end{tikzpicture}
    }
       \label{eq:magQuiv_Sp_O7_infinite_Nf=2k+1}
\end{equation}
The Coulomb branch dimension is readily computed 
\begin{align}
    \dim_{\HH}\Coulomb\eqref{eq:magQuiv_Sp_O7_infinite_Nf=2k+1} = \frac{1}{2}N_f(N_f-1)+1
    \quad \text{and} \quad 
    G = \sorm(2N_f) \times \uo \,.
\end{align}
The addition in Higgs branch dimension agrees with the expectation \cite{Bergman:2015dpa} as well as the dimension of the magnetic quiver \eqref{eq:magQuiv_Spk_infinite_Nf_odd_small} from the \Of\ plane construction. Furthermore, \eqref{eq:magQuiv_Sp_O7_infinite_Nf=2k+1} reduces for $k=1$ to the affine $a_4$ quiver whose Coulomb branch equals $\clorbit_{A_4}^{\min}$.
\paragraph{$\mathbf{N_f=2k+2}$.} 
Next, for $N_f=2k+2$ fundamental flavours the 5-brane web is given by
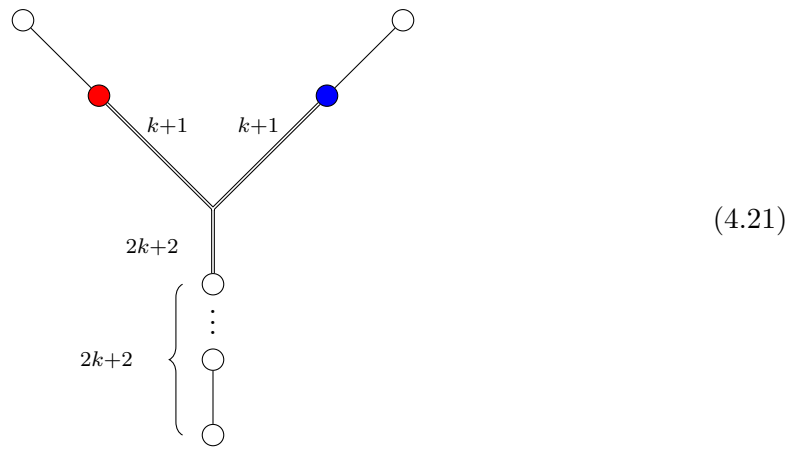
\begin{equation}
 \raisebox{-.5\height}{
    \begin{tikzpicture}
        \node[gauger] (1) at (-1.5,1.5) {};
        \node[gaugeb] (2) at (1.5,1.5) {};
        \node[gauge] (7) at (-2.5,2.5) {};
        \node[gauge] (8) at (2.5,2.5) {};
        \node[gauge] (9) at (0,-1) {};
        \node[gauge] (10) at (0,-2) {};
        \node[gauge] (11) at (0,-3) {};
        \coordinate (3) at (0,0);
        \draw[double] (1)--(3)--(2) (3)--(9);
        \draw (10)--(11);
        \draw (7)--(1) (2)--(8);
        \node at (0,-1.4) {$\vdots$};
        \node at (-0.6,1.1) {$\scriptstyle{k+1}$};
        \node at (0.6,1.1) {$\scriptstyle{k+1}$};
        \node at (-0.8,-0.5) {$\scriptstyle{2k+2}$};
        \draw [decorate,decoration={brace,amplitude=5pt},yshift=0pt,xshift=-0.4cm]
        (0,-3)--(0,-1) node [black,midway,xshift=-1cm] {$\scriptstyle{2k+2}$};
    \end{tikzpicture}
    }
\end{equation}
and the corresponding magnetic quiver is determined to be
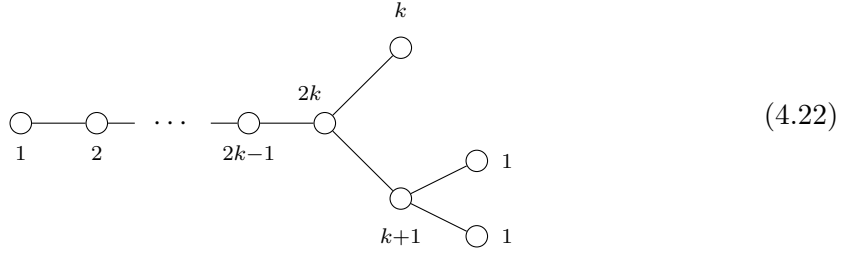
\begin{equation}
 \raisebox{-.5\height}{
    \begin{tikzpicture}
        \node[gauge] (1) at (-1,0) {};
        \node[gauge] (2) at (0,0) {};
        \node[gauge] (3) at (2,0) {};
        \node[gauge] (4) at (3,0) {};
        \node[gauge] (5) at (4,1) {};
        \node[gauge] (6) at (4,-1) {};
        \node[gauge] (8) at (5,-0.5) {};
        \node[gauge] (7) at (5,-1.5) {};
        \draw (1)--(2)--(0.5,0) (1.5,0)--(3)--(4)--(5) (6)--(4) (6)--(7) (6)--(8);
        \node at (-1,-0.4) {$\scriptstyle{1}$};
        \node at (0,-0.4) {$\scriptstyle{2}$};
        \node at (2,-0.4) {$\scriptstyle{2 k {-} 1}$};
        \node at (2.8,0.4) {$\scriptstyle{2 k}$};
        \node at (4,-1.5) {$\scriptstyle{k {+} 1}$};
        \node at (4,1.5) {$\scriptstyle{k}$};
        \node at (1,0) {$\cdots$};
        \node at (5.4,-0.5) {$\scriptstyle{1}$};
        \node at (5.4,-1.5) {$\scriptstyle{1}$};
    \end{tikzpicture}
    }
     \label{eq:magQuiv_Sp_O7_infinite_Nf=2k+2}
\end{equation}
The associated Coulomb branch is of dimension
\begin{align}
    \dim_{\HH} \Coulomb \eqref{eq:magQuiv_Sp_O7_infinite_Nf=2k+2} = 
    \frac{1}{2}N_f (N_f-1)+1 
    \quad \text{and} \quad 
    G = \sorm(2N_f) \times \uo
    \,.
\end{align}
Note that the case $k=1$ reduces to the affine $d_5$ quiver, i.e.\ the quiver whose Coulomb branch equals $\clorbit_{\dalg_{5}}^{\min }$.
\paragraph{$\mathbf{N_f=2k+3}$.} 
Increasing the number of flavours further $N_f=2k+3$, the brane web is given by
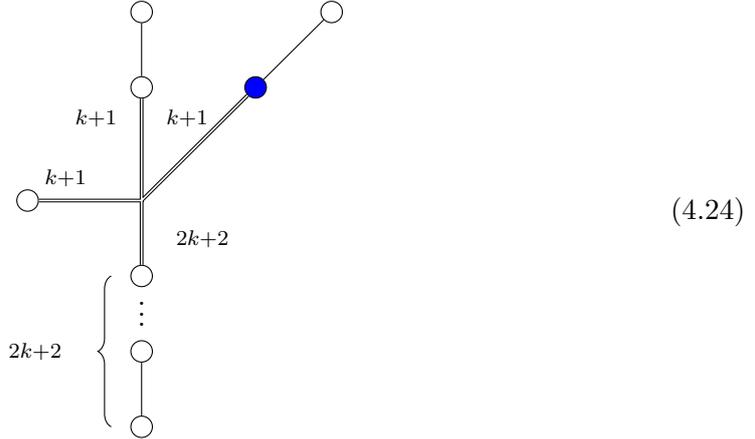
\begin{equation}
 \raisebox{-.5\height}{
    \begin{tikzpicture}
        \node[gauge] (1) at (-1.5,0) {};
        \node[gaugeb] (2) at (1.5,1.5) {};
        \node[gauge] (4) at (0,1.5) {};
        \node[gauge] (5) at (0,2.5) {};
        \node[gauge] (8) at (2.5,2.5) {};
        \node[gauge] (9) at (0,-1) {};
        \node[gauge] (10) at (0,-2) {};
        \node[gauge] (11) at (0,-3) {};
        \coordinate (3) at (0,0);
        \draw[double] (1)--(3)--(2) (3)--(9) (3)--(4);
        \draw (10)--(11) (4)--(5);
        \draw (2)--(8);
        \node at (0,-1.4) {$\vdots$};
        \node at (-1,0.3) {$\scriptstyle{k+1}$};
        \node at (-0.6,1.1) {$\scriptstyle{k+1}$};
        \node at (0.6,1.1) {$\scriptstyle{k+1}$};
        \node at (0.8,-0.5) {$\scriptstyle{2k+2}$};
        \draw [decorate,decoration={brace,amplitude=5pt},yshift=0pt,xshift=-0.4cm]
        (0,-3)--(0,-1) node [black,midway,xshift=-1cm] {$\scriptstyle{2k+2}$};
    \end{tikzpicture}
    }
\end{equation}
such that the magnetic quiver reads
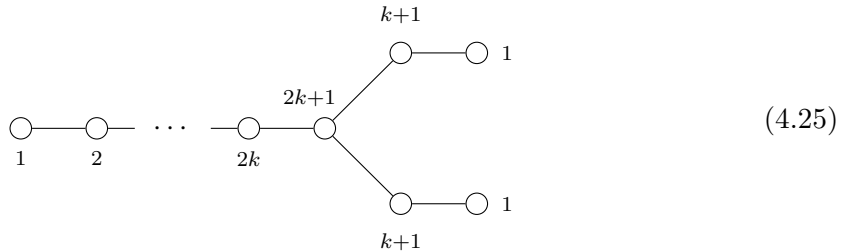
\begin{equation}
 \raisebox{-.5\height}{
    \begin{tikzpicture}
        \node[gauge] (1) at (-1,0) {};
        \node[gauge] (2) at (0,0) {};
        \node[gauge] (3) at (2,0) {};
        \node[gauge] (4) at (3,0) {};
        \node[gauge] (5) at (4,1) {};
        \node[gauge] (6) at (4,-1) {};
        \node[gauge] (7) at (5,1) {};
        \node[gauge] (8) at (5,-1) {};
        \draw (1)--(2)--(0.5,0) (1.5,0)--(3)--(4)--(5)--(7) (8)--(6)--(4);
        \node at (-1,-0.4) {$\scriptstyle{1}$};
        \node at (0,-0.4) {$\scriptstyle{2}$};
        \node at (2,-0.4) {$\scriptstyle{2k}$};
        \node at (2.8,0.4) {$\scriptstyle{2k {+} 1}$};
        \node at (4,-1.5) {$\scriptstyle{k {+} 1}$};
        \node at (4,1.5) {$\scriptstyle{k {+} 1}$};
        \node at (1,0) {$\cdots$};
        \node at (5.4,1) {$\scriptstyle{1}$};
        \node at (5.4,-1) {$\scriptstyle{1}$};
    \end{tikzpicture}
    }
     \label{eq:magQuiv_Sp_O7_infinite_Nf=2k+3}
\end{equation}
One readily confirms that 
\begin{align}
    \dim_{\HH}\Coulomb\eqref{eq:magQuiv_Sp_O7_infinite_Nf=2k+3} = \frac{1}{2} N_f(N_f-1) +1
    \quad \text{and} \quad 
    G = \sorm(2N_f) \times \uo \,,
\end{align}
which indicates one additional Higgs branch direction at infinite coupling, in agreement with \cite{Bergman:2015dpa}.
Moreover, note that the $k=1$ case reduces \eqref{eq:magQuiv_Sp_O7_infinite_Nf=2k+3} to the affine $e_6$ quiver, i.e.\ the quiver with Coulomb branch $\clorbit_{E_6}^{\min}$.
\paragraph{$\mathbf{N_f=2k+4}$.}
The brane web for the field theory limit $N_f=2k+4$ is given by
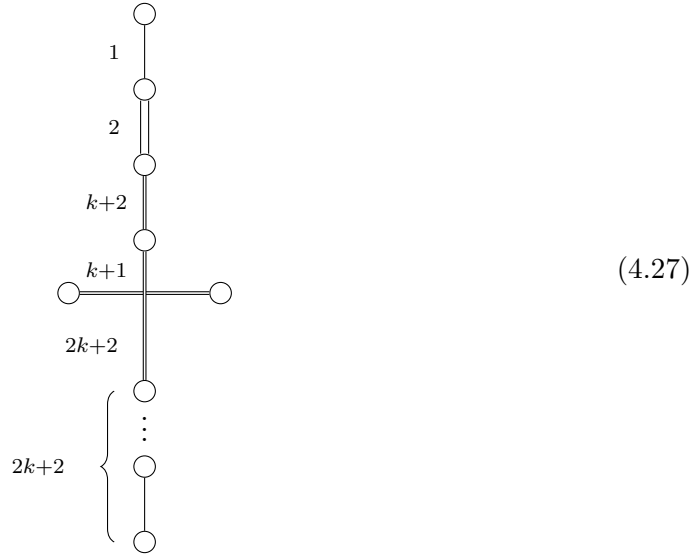
\begin{equation}
 \raisebox{-.5\height}{
    \begin{tikzpicture}
        \node[gauge] (1) at (0,4) {};
        \node[gauge] (2) at (0,3) {};
        \node[gauge] (3) at (0,2) {};
        \node[gauge] (4) at (0,1) {};
        \node[gauge] (5) at (-1,0.3) {};
        \node[gauge] (6) at (1,0.3) {};
        \node[gauge] (9) at (0,-1) {};
        \node[gauge] (10) at (0,-2) {};
        \node[gauge] (11) at (0,-3) {};
        \draw[double] (6)--(5);
        \draw[double] (9)--(4)--(3);
        \draw[transform canvas={xshift=-1.5pt}] (2)--(3);
        \draw[transform canvas={xshift=1.5pt}] (2)--(3);
        \draw (1)--(2) (10)--(11);
        \node at (-0.7,-0.4) {$\scriptstyle{2k+2}$};
        \draw [decorate,decoration={brace,amplitude=5pt},yshift=0pt,xshift=-0.4cm]
        (0,-3)--(0,-1) node [black,midway,xshift=-1cm] {$\scriptstyle{2k+2}$};
        \node at (-0.5,1.5) {$\scriptstyle{k+2}$};
        \node at (-0.4,2.5) {$\scriptstyle{2}$};
        \node at (-0.4,3.5) {$\scriptstyle{1}$};
        \node at (-0.5,0.6) {$\scriptstyle{k {+} 1}$};
        \node at (0,-1.4) {$\vdots$};
    \end{tikzpicture}
    }
\end{equation}
and the associated magnetic quiver is 
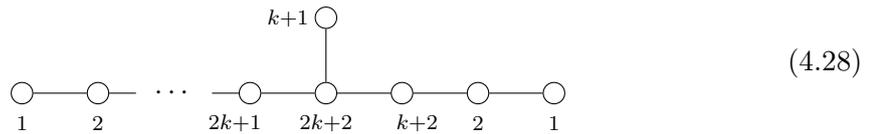
\begin{equation}
 \raisebox{-.5\height}{
    \begin{tikzpicture}
        \node[gauge] (1) at (-1,0) {};
        \node[gauge] (2) at (0,0) {};
        \node[gauge] (3) at (2,0) {};
        \node[gauge] (4) at (3,0) {};
        \node[gauge] (5) at (4,0) {};
        \node[gauge] (6) at (5,0) {};
        \node[gauge] (7) at (6,0) {};
        \node[gauge] (8) at (3,1) {};
        \draw (1)--(2)--(0.5,0) (1.5,0)--(3)--(4)--(5)--(6)--(7) (8)--(4);
        \node at (1,0) {$\cdots$};
        \node at (-1,-0.4) {$\scriptstyle{1}$};
        \node at (0,-0.4) {$\scriptstyle{2}$};
        \node at (1.8,-0.4) {$\scriptstyle{2k {+} 1}$};
        \node at (3,-0.4) {$\scriptstyle{2k {+} 2}$};
        \node at (4.2,-0.4) {$\scriptstyle{k {+} 2}$};
        \node at (5,-0.4) {$\scriptstyle{2}$};
        \node at (6,-0.4) {$\scriptstyle{1}$};
        \node at (2.5,1) {$\scriptstyle{k {+} 1}$};
    \end{tikzpicture}
    }
     \label{eq:magQuiv_Sp_O7_infinite_Nf=2k+4}
\end{equation}
One readily computes
\begin{align}
    \dim_{\HH} \Coulomb \eqref{eq:magQuiv_Sp_O7_infinite_Nf=2k+4} = 
    \frac{1}{2}N_f(N_f-1)+2
    \quad \text{and} \quad 
    G = \sorm(2N_f) \times \surm(2)
\end{align}
which indicates two additional Higgs branch directions in comparison to the finite coupling case.
This agrees with the expectation of \cite{Bergman:2015dpa} as well as the magnetic quiver \eqref{eq:magQuiv_Spk_infinite_Nf=2k+4} from \Of\ plane construction. In particular, note that for $k=1$ the quiver \eqref{eq:magQuiv_Sp_O7_infinite_Nf=2k+4} reduces to the affine $e_7$ quiver, meaning that its Coulomb branch is $\clorbit_{E_7}^{\min}$.
\paragraph{$\mathbf{N_f=2k+5}$.} 
Lastly, considering the maximal number of flavours, the brane web is given by
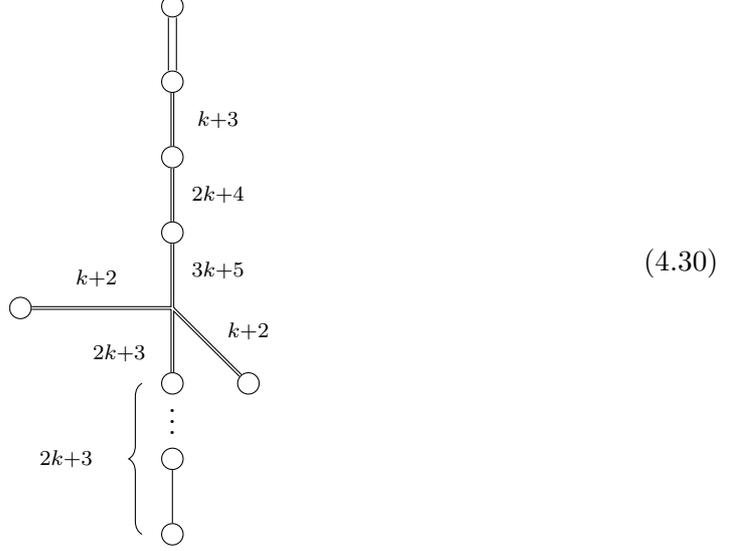
\begin{equation}
 \raisebox{-.5\height}{
    \begin{tikzpicture}
        \node[gauge] (1) at (0,4) {};
        \node[gauge] (2) at (0,3) {};
        \node[gauge] (3) at (0,2) {};
        \node[gauge] (4) at (0,1) {};
        \node[gauge] (5) at (-2,0) {};
        \node[gauge] (6) at (1,-1) {};
        \coordinate (12) at (0,0);
        \node[gauge] (9) at (0,-1) {};
        \node[gauge] (10) at (0,-2) {};
        \node[gauge] (11) at (0,-3) {};
        \draw[double] (12)--(9) (5)--(12)--(6) (12)--(4)--(3)--(2);
        \draw[transform canvas={xshift=-1.5pt}] (1)--(2);
        \draw[transform canvas={xshift=1.5pt}] (1)--(2);
        \draw (10)--(11);
        \node at (-0.7,-0.6) {$\scriptstyle{2k+3}$};
        \draw [decorate,decoration={brace,amplitude=5pt},yshift=0pt,xshift=-0.4cm]
        (0,-3)--(0,-1) node [black,midway,xshift=-1cm] {$\scriptstyle{2k+3}$};
        \node at (0,-1.4) {$\vdots$};
        \node at (-1,0.4) {$\scriptstyle{k+2}$};
        \node at (1,-0.3) {$\scriptstyle{k+2}$};
        \node at (0.6,0.5) {$\scriptstyle{3k+5}$};
        \node at (0.6,1.5) {$\scriptstyle{2k+4}$};
        \node at (0.6,2.5) {$\scriptstyle{k+3}$};
    \end{tikzpicture}
    }
\end{equation}
and one derives the magnetic quiver to be
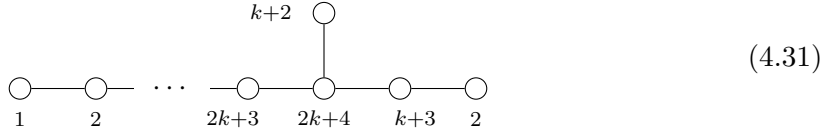
\begin{equation}
 \raisebox{-.5\height}{
    \begin{tikzpicture}
        \node[gauge] (1) at (-1,0) {};
        \node[gauge] (2) at (0,0) {};
        \node[gauge] (3) at (2,0) {};
        \node[gauge] (4) at (3,0) {};
        \node[gauge] (5) at (4,0) {};
        \node[gauge] (6) at (5,0) {};
        \node[gauge] (8) at (3,1) {};
        \draw (1)--(2)--(0.5,0) (1.5,0)--(3)--(4)--(5)--(6) (8)--(4);
        \node at (1,0) {$\cdots$};
        \node at (-1,-0.4) {$\scriptstyle{1}$};
        \node at (0,-0.4) {$\scriptstyle{2}$};
        \node at (1.8,-0.4) {$\scriptstyle{2k {+} 3}$};
        \node at (3,-0.4) {$\scriptstyle{2k {+} 4}$};
        \node at (4.2,-0.4) {$\scriptstyle{k {+} 3}$};
        \node at (5,-0.4) {$\scriptstyle{2}$};
        \node at (2.3,1) {$\scriptstyle{k {+} 2}$};
    \end{tikzpicture}
    }
      \label{eq:magQuiv_Sp_O7_infinite_Nf=2k+5}
\end{equation}
Computing the moduli space properties reveals
\begin{align}
    \dim_{\HH}\Coulomb \eqref{eq:magQuiv_Sp_O7_infinite_Nf=2k+5} =\frac{1}{2} N_f (N_f+1)+1 
    \quad \text{and} \quad
    G = \sorm(2N_f+2)\,,
\end{align}
which matches the dimension of \eqref{eq:magQuiv_Spk_infinite_Nf_2k=5} from the \Of\ construction. Moreover, for $k=1$ one observes that \eqref{eq:magQuiv_Sp_O7_infinite_Nf=2k+5} reduces to the affine $e_8$ quiver such that its Coulomb branch equals $\clorbit_{E_8}^{\min}$.
\subsection{Duality with 5d SQCD}
The unitary magnetic quiver descriptions for the 5d $\sprm(k)$ theories derived above admit an interesting property. Recalling the infinite coupling Higgs branch descriptions for 5d SQCD in \cite[Sec.\ 5.4]{Cabrera:2018jxt} one finds that the following theories have the same infinite coupling Higgs branch descriptions in terms of magnetic quivers
\begin{compactitem}
\item $E_8$ family \eqref{eq:magQuiv_Sp_O7_infinite_Nf=2k+5}: 5d $\sprm(k)$ with $N_f=2k+5$ flavours $\leftrightarrow$ 5d $\surm(k+1)$ with $N_f$ flavours and CS-level $\kappa=\pm\tfrac{1}{2}$, \cite[Tab.\ 20]{Cabrera:2018jxt}
\item $E_7$ family \eqref{eq:magQuiv_Sp_O7_infinite_Nf=2k+4}: 5d $\sprm(k)$ with $N_f=2k+4$ flavours $\leftrightarrow$ 5d $\surm(k+1)$ with $N_f$ flavours and CS-level $\kappa=\pm1$, \cite[Tab.\ 19]{Cabrera:2018jxt}  
\item $E_6$ family \eqref{eq:magQuiv_Sp_O7_infinite_Nf=2k+3}: 5d $\sprm(k)$ with $N_f=2k+3$ flavours $\leftrightarrow$ 5d $\surm(k+1)$ with $N_f$ flavours and CS-level $\kappa=\pm\tfrac{3}{2}$, \cite[Tab.\ 18]{Cabrera:2018jxt}  
\item $E_5$ family \eqref{eq:magQuiv_Sp_O7_infinite_Nf=2k+2}: 5d $\sprm(k)$ with $N_f=2k+2$ flavours $\leftrightarrow$ 5d $\surm(k+1)$ with $N_f$ flavours and CS-level $\kappa=\pm2$, \cite[Tab.\ 17]{Cabrera:2018jxt}  
\item $E_4$ family \eqref{eq:magQuiv_Sp_O7_infinite_Nf=2k+1}: 5d $\sprm(k)$ with $N_f=2k+1$ flavours $\leftrightarrow$ 5d $\surm(k+1)$ with $N_f$ flavours and CS-level $\kappa=\pm\tfrac{5}{2}$, \cite[Tab.\ 16 (V)]{Cabrera:2018jxt}  
\item Case \eqref{eq:magQuiv_Sp_O7_infinite_Nf_small_even}: 5d $\sprm(k)$ with even number of flavours $N_f \leq 2k$  $\leftrightarrow$ 5d $\surm(k+1)$ with $N_f$ flavours and CS-level $\kappa= \pm (k+3-\tfrac{N_f}{2})$, \cite[Tab.\ 16 (IV) \& (V)]{Cabrera:2018jxt} 
\item Case \eqref{eq:magQuiv_Sp_O7_infinite_Nf_small_odd}: 5d $\sprm(k)$ with odd number of flavours $N_f \leq 2k$  $\leftrightarrow$ 5d $\surm(k+1)$ with $N_f$ flavours and CS-level $\kappa= \pm (k+3-\tfrac{N_f}{2})$, \cite[Tab.\ 16 (V)]{Cabrera:2018jxt}
\end{compactitem}
As summarised in Table \ref{tab:Efamilies}, the magnetic quivers hint on the $5$d $\Ncal=1$ duality 
\begin{align}
    \text{$\sprm(k)$, $N_f$ flavours} 
    \quad \longleftrightarrow \quad 
    \text{$\surm(k+1)_{\pm (k+3 -
 \frac{N_f}{2})}$ , $N_f$ flavours}
 \label{eq:Sp_SU_duality}
\end{align}
In fact, the duality \eqref{eq:Sp_SU_duality} has been conjectured in \cite{Gaiotto:2015una} and observed from 5-brane webs in \cite{Hayashi:2015zka} or from geometric studies in \cite[Eq.\ (4.2)]{Jefferson:2018irk} and \cite[Eq.\ (2.225)]{Bhardwaj:2020gyu}.

\subsection{Two realisations}
Another immediate observations follows from the construction of 5d $\sprm(k)$ theories via \Of\ planes or O$7$ planes. In other words, for each 5d $\sprm(k)$ theory with $N_f$ flavours, there exists
\begin{compactenum}[(i)]
\item a unitary-orthosymplectic (or orthosymplectic) magnetic quiver from the \Ofm\ plane construction of Section \ref{sec:Spk}, and 
\item a unitary magnetic quiver from the O$7^-$ construction of Section \ref{sec:O7_plane}.
\end{compactenum}
Table \ref{tab:Efamilies} contrasts the quivers in each case.
For consistency, one needs to verify that the moduli spaces associated to both 
types of magnetic quiver are the same, or at least isomorphic. In a companion 
paper \cite{Bourget:2020xdz}, the Hilbert series analysis is presented and the 
results confirm the suspected agreement. 

\subsection{Problem of two cones}
\label{sec:prob2cones}
From the construction of unitary magnetic quivers, it is clear that the 
(classical as well as infinite coupling) Higgs branch of $\sprm(k)$ with an even 
number of flavours $N_f\leq2k$, is a union of two cones. However, in the \Of\ 
construction of Section \ref{sec:Spk} only one cone is visible. Note that a 
similar problem exists for constructions of 3 dimensional theories in Type IIB 
systems of \NS, \Dthree, \Dfive\ with an \Ot\ plane versus an \Of\ plane 
construction. One crucial difference between the unitary and orthosymplectic 
quivers for D-type global symmetry is the visibility of the spinor nodes in the 
unitary construction. To be specific, focus on the finite coupling case. The 
classical Higgs branch is the closure of a \emph{very even} height two 
$\orm(2N_f)$ nilpotent orbit closure. The $\orm(2N_f)$ very even orbit is 
known to be a union of two $\sorm(2N_f)$ orbits 
\cite{collingwood1993nilpotent}, and the Higgs branch is a union of two cones. 
For a unitary quiver, one is able to make a clear distinction between the two 
cones and also detail the intersection of the cones. The quivers representing 
the two cones are related by a non-trivial $\Z_2$-action, and the individual 
cones are isomorphic. The non-trivial intersection is given by a quiver 
invariant under the $\Z_2$-action. The relevant quivers are as follows:
\begin{align}
        \raisebox{-.5\height}{
    \begin{tikzpicture}
        \node[gauge] (1) at (-1,0) {};
        \node[gauge] (2) at (0,0) {};
        \node[gauge] (3) at (2,0) {};
        \node[gauge] (4) at (3,0) {};
        \node[gauge] (5) at (4,1) {};
        \node[gauge] (6) at (4,-1) {};
        \node[gauge] (7) at (5,1) {};
        \draw (1)--(2)--(0.5,0) (1.5,0)--(3)--(4)--(5) (6)--(4);
        \draw[transform canvas={yshift=-1.5pt}] (5)--(7);
        \draw[transform canvas={yshift=1.5pt}] (5)--(7);
        \node at (-1,-0.4) {$\scriptstyle{1}$};
        \node at (0,-0.4) {$\scriptstyle{2}$};
        \node at (2,-0.4) {$\scriptstyle{N_f {-} 3}$};
        \node at (2.8,0.4) {$\scriptstyle{N_f {-} 2}$};
        \node at (4,-1.5) {$\scriptstyle{\frac{N_f}{2} {-} 1}$};
        \node at (4,1.5) {$\scriptstyle{\frac{N_f}{2}}$};
        \node at (5,1.5) {$\scriptstyle{1}$};
        \node at (1,0) {$\cdots$};
    \end{tikzpicture}
    }
    \quad \xleftrightarrow{\quad \Z_2 \quad} \quad 
    \raisebox{-.5\height}{
    \begin{tikzpicture}
        \node[gauge] (1) at (-1,0) {};
        \node[gauge] (2) at (0,0) {};
        \node[gauge] (3) at (2,0) {};
        \node[gauge] (4) at (3,0) {};
        \node[gauge] (5) at (4,1) {};
        \node[gauge] (6) at (4,-1) {};
        \node[gauge] (7) at (5,-1) {};
        \draw (1)--(2)--(0.5,0) (1.5,0)--(3)--(4)--(5) (6)--(4);
        \draw[transform canvas={yshift=-1.5pt}] (6)--(7);
        \draw[transform canvas={yshift=1.5pt}] (6)--(7);
        \node at (-1,-0.4) {$\scriptstyle{1}$};
        \node at (0,-0.4) {$\scriptstyle{2}$};
        \node at (2,-0.4) {$\scriptstyle{N_f {-} 3}$};
        \node at (2.8,0.4) {$\scriptstyle{N_f {-} 2}$};
        \node at (4,-1.5) {$\scriptstyle{\frac{N_f}{2}}$};
        \node at (4,1.5) {$\scriptstyle{\frac{N_f}{2} {-} 1}$};
        \node at (5,-1.5) {$\scriptstyle{1}$};
        \node at (1,0) {$\cdots$};
    \end{tikzpicture}
    }
   \notag  \\
    \raisebox{-.5\height}{
    \begin{tikzpicture}
        \node[gauge] (1) at (-1,0) {};
        \node[gauge] (2) at (0,0) {};
        \node[gauge] (3) at (2,0) {};
        \node[gauge] (4) at (3,0) {};
        \node[gauge] (5) at (4,1) {};
        \node[gauge] (6) at (4,-1) {};
        \node[gauge] (7) at (3,1) {};
        \draw (1)--(2)--(0.5,0) (1.5,0)--(3)--(4)--(5) (6)--(4)--(7);
        \node at (-1,-0.4) {$\scriptstyle{1}$};
        \node at (0,-0.4) {$\scriptstyle{2}$};
        \node at (2,0.4) {$\scriptstyle{N_f {-} 3}$};
        \node at (2.8,-0.4) {$\scriptstyle{N_f {-} 2}$};
        \node at (4,-1.5) {$\scriptstyle{\frac{N_f}{2} {-} 1}$};
        \node at (4,1.5) {$\scriptstyle{\frac{N_f}{2} {-} 1}$};
        \node at (2.7,1) {$\scriptstyle{1}$};
        \node at (1,0) {$\cdots$};
        \node at (10,0) {};
        \node at (-3,0) {Intersection:};
    \end{tikzpicture}
    }
\end{align}
In contrast, the construction of the two cones as Coulomb branches of orthosymplectic quivers is slightly different. Here, the two cones cannot be made distinct, the relevant quivers are as follows:
\begin{align}
        \raisebox{-.5\height}{
    \begin{tikzpicture}
 	\tikzset{node distance = 1cm};
	\tikzstyle{gauge} = [circle, draw,inner sep=2.5pt];
	\tikzstyle{flavour} = [regular polygon,regular polygon sides=4,inner 
sep=2.5pt, draw];
	\node (g1) [gauge,label=below:{\dd{1}}] {};
 	\node (g3) [right of=g1] {$\cdots$};%
	\node (g4) [gauge,right of=g3,label=below:{\cc{\frac{N_f}{2}{-}1}}] {};
	\node (g5) [gauge,right of=g4,label=below:{\dd{\frac{N_f}{2}}}] {};
	\node (g6) [gauge,right of=g5,label=below:{\cc{\frac{N_f}{2}{-}1}}] {};
	\node (g7) [right of=g6] {$\cdots$};
	\node (g9) [gauge,right of=g7,label=below:{\dd{1}}] {};%
	\node (f1) [flavour,above of=g5,label=above:{\cc{1}}] {};
	\draw (g1)--(g3) (g3)--(g4)--(g5) (g5)--(g6)--(g7) (g7)--(g9) (g5)--(f1);%
	\end{tikzpicture}
    }
    \quad \xleftrightarrow{\quad = \quad} \quad 
    \raisebox{-.5\height}{
    \begin{tikzpicture}
 	\tikzset{node distance = 1cm};
	\tikzstyle{gauge} = [circle, draw,inner sep=2.5pt];
	\tikzstyle{flavour} = [regular polygon,regular polygon sides=4,inner 
sep=2.5pt, draw];
	\node (g1) [gauge,label=below:{\dd{1}}] {};
 	\node (g3) [right of=g1] {$\cdots$};%
	\node (g4) [gauge,right of=g3,label=below:{\cc{\frac{N_f}{2}{-}1}}] {};
	\node (g5) [gauge,right of=g4,label=below:{\dd{\frac{N_f}{2}}}] {};
	\node (g6) [gauge,right of=g5,label=below:{\cc{\frac{N_f}{2}{-}1}}] {};
	\node (g7) [right of=g6] {$\cdots$};
	\node (g9) [gauge,right of=g7,label=below:{\dd{1}}] {};%
	\node (f1) [flavour,above of=g5,label=above:{\cc{1}}] {};
	\draw (g1)--(g3) (g3)--(g4)--(g5) (g5)--(g6)--(g7) (g7)--(g9) (g5)--(f1);%
	\end{tikzpicture}
    }
    \notag \\
    \raisebox{-.5\height}{
    \begin{tikzpicture}
 	\tikzset{node distance = 1cm};
	\tikzstyle{gauge} = [circle, draw,inner sep=2.5pt];
	\tikzstyle{flavour} = [regular polygon,regular polygon sides=4,inner 
sep=2.5pt, draw];
	\node (g1) at (-1,0) [gauge,label=below:{\dd{1}}] {};
	\node (g3) [right of=g1] {$\cdots$};%
	\node (g5) [gauge,right of=g3,label=below:{\cc{\frac{N_f}{2}{-}1}}] {};%
	\node (g8) [gauge,right of=g5,label=below:{\bb{\frac{N_f}{2}{-}1}}] {};
	\node (g9) [gauge,right of=g8,label=below:{\cc{\frac{N_f}{2}{-}1}}] {};
	\node (g11) [right of=g9] {$\cdots$};%
	\node (g13) [gauge,right of=g11,label=below:{\dd{1}}] {};%
	\node (f1) [flavour,above of=g5,label=above:{\bb{0}}] {};
	\node (f2) [flavour,above of=g9,label=above:{\bb{0}}] {};
	\draw (g1)--(g3) (g3)--(g5) (g5)--(g8) (g8)--(g9) (g9)--(g11) (g11)--(g13) (g5)--(f1) (g9)--(f2);
        \node at (10,0) {};
        \node at (-3,0) {Intersection:};
	\end{tikzpicture}
    }
\end{align}
So the inability to distinguish two cones may come from the inability to distinguish the magnetic quivers in the first place. This changes at infinite coupling, where only one of the cones grows with respect to the finite coupling case, and one can in principle make a distinction between the quivers representing the two cones. Still there is no way to identify which of the cones was enhanced and only the magnetic quiver representing the enhanced cone can be identified.
%
%
\section{Hasse diagrams and quiver subtraction}
\label{sec:Hasse}
This section is dedicated to the study of Hasse diagrams of the theories discussed in this note. The finite coupling Hasse diagrams are known \cite{Kraft1982}, so only the infinite coupling Hasse diagrams are computed in this section. The results for both finite and infinite coupling are juxtaposed in Tables \ref{tab:CompareFiniteHasse} and \ref{tab:CompareFiniteHasse2}.

Following the spirit of \cite{Bourget:2019aer} one can derive the Higgs branch \emph{Hasse diagrams} of the theories considered by using the brane web construction of Section \ref{sec:Spk}. In principle, one could also use the algorithm of quiver subtraction, as introduced in \cite{Cabrera:2017njm,Cabrera:2018ann} and further developed for unitary quivers in \cite{Bourget:2019aer} and for \emph{framed} (flavoured) orthosymplectic quivers in \cite{Hanany:2019tji}. However, the orthosymplectic quivers appearing in Table \ref{tab:Efamilies} are not framed, and may involve unitary gauge groups, and as a consequence the currently known algorithm needs to be extended. After one transition, framed quivers appear and the quiver subtraction algorithm of \cite{Cabrera:2017njm,Cabrera:2018ann,Hanany:2019tji} yields the rest of the Hasse diagram. 

The results of Section \ref{sec:SU2} and \ref{sec:Spk} show that 
unitary-orthosymplectic magnetic quivers naturally arise and create the need for 
an adaptation of quiver subtraction to this class of theories.
In this section, the quiver subtraction algorithm is derived from brane set-ups, 
which then allows one to derive the Hasse diagram.
\subsection{\texorpdfstring{$E_8$}{E8} family} 
Consider the brane web \eqref{eq:web_Spk_Nf=2k+5_var2}, which describes the Higgs branch of $\sprm(k)$ with $N_f=2k+5$ flavours at infinite coupling.
Opening up a Coulomb branch modulus one has to be careful to take the S-rule as well as charge conservation into account. In order to respect the S-rule from the point of view of the $[0,1]$ $7$-brane, the two $(0,1)$ $5$-branes have to split into a $(2,1)$ and $(2,-1)$ $5$-brane. Moreover, charge conservation on the orientifold implies that there have to be 4 full $(1,0)$ $5$-branes on the left-hand-side and right-hand-side. The S-rule further determines that every half $(1,0)$ $5$-brane has to end on its own half $[1,0]$ $7$-brane. The central part of the resulting web is
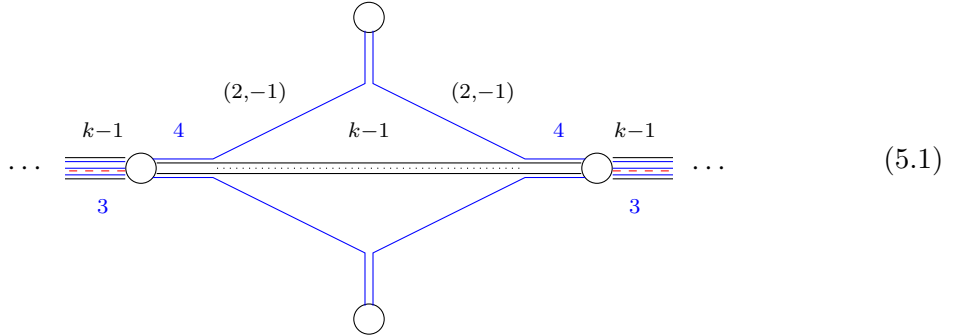
\begin{equation}
 \raisebox{-.5\height}{
    \begin{tikzpicture}
    \node[sev] (1) at (0,2) {};
    \node[sev] (2) at (0,-2) {};
    \node[sev] (3) at (-3,0) {};
    \node[sev] (4) at (3,0) {};
    \draw[blue,transform canvas={yshift=3.5pt,xshift=-1.5pt}] (0,1.7)--(0,1)--(-2,0)--(3);
    \draw[blue,transform canvas={yshift=3.5pt,xshift=1.5pt}] (0,1.7)--(0,1)--(2,0)--(4);
    \draw[blue,transform canvas={yshift=-3.5pt,xshift=-1.5pt}] (0,-1.7)--(0,-1)--(-2,0)--(3);
    \draw[blue,transform canvas={yshift=-3.5pt,xshift=1.5pt}] (0,-1.7)--(0,-1)--(2,0)--(4);
    \draw[dotted] (-2,0)--(2,0);
    \draw[transform canvas={yshift=2pt}] (3)--(4);
    \draw[transform canvas={yshift=-2pt}] (3)--(4);
    \node at (4.5,0) {$\cdots$};
    \node at (-4.5,0) {$\cdots$};
    \node at (1.5,1) {$\scriptstyle{(2,-1)}$};
    \node at (-1.5,1) {$\scriptstyle{(2,-1)}$};
    \node at (2.5,0.5) {\color{blue}$\scriptstyle{4}$};
    \node at (-2.5,0.5) {\color{blue}$\scriptstyle{4}$};
    \node at (0,0.5) {$\scriptstyle{k-1}$};
    \draw[dashed, red, transform canvas={yshift=-1pt}] (3)--(-4,0) (4)--(4,0);
    \draw[blue] (3)--(-4,0) (4)--(4,0);
    \draw[blue, transform canvas={yshift=2.5pt}] (3)--(-4,0) (4)--(4,0);
    \draw[blue, transform canvas={yshift=-2.5pt}] (3)--(-4,0) (4)--(4,0);
    \draw[transform canvas={yshift=4pt}] (3)--(-4,0) (4)--(4,0);
    \draw[transform canvas={yshift=-4pt}] (3)--(-4,0) (4)--(4,0);
    \node at (-3.5,0.5) {$\scriptstyle{k-1}$};
    \node at (3.5,0.5) {$\scriptstyle{k-1}$};
    \node at (-3.5,-0.5) {\color{blue}$\scriptstyle{3}$};
    \node at (3.5,-0.5) {\color{blue}$\scriptstyle{3}$};
    \end{tikzpicture}
    }
    \label{eq:web_E8family_1st_transition}
\end{equation}
such that the blue 5-branes subweb can no longer split and is stuck to the orientifold. Thus, this subweb contributes as a flavour to both, the magnetic gauge group on its left and right. In more detail, the magnetic quiver corresponding to \eqref{eq:web_E8family_1st_transition} is read off to be 
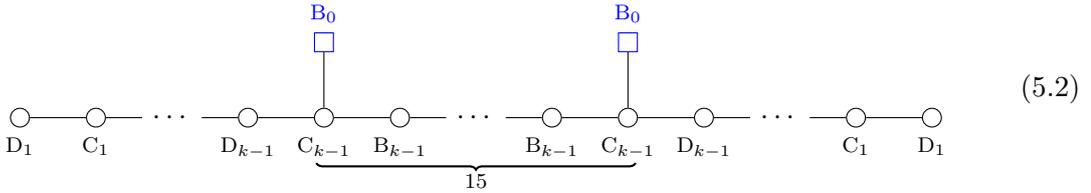
\begin{align}
 \raisebox{-.5\height}{
 	\begin{tikzpicture}
 	\tikzset{node distance = 1cm};
	\tikzstyle{gauge} = [circle, draw,inner sep=2.5pt];
	\tikzstyle{flavour} = [regular polygon,regular polygon sides=4,inner 
sep=2.5pt, draw];
	\node (g1) [gauge,label=below:{\dd{1}}] {};
	\node (g2) [gauge,right of=g1,label=below:{\cc{1}}] {};
	\node (g3) [right of=g2] {$\cdots$};
	\node (g4) [gauge,right of=g3,label=below:{\dd{k-1}}] {};
	\node (g5) [gauge,right of=g4,label=below:{\cc{k-1}}] {};
	\node (g6) [gauge,right of=g5,label=below:{\bb{k-1}}] {};
	\node (g7) [right of=g6] {$\cdots$};
	\node (g8) [gauge,right of=g7,label=below:{\bb{k-1}}] {};
	\node (g9) [gauge,right of=g8,label=below:{\cc{k-1}}] {};
	\node (g10) [gauge,right of=g9,label=below:{\dd{k-1}}] {};
	\node (g11) [right of=g10] {$\cdots$};
	\node (g12) [gauge,right of=g11,label=below:{\cc{1}}] {};
	\node (g13) [gauge,right of=g12,label=below:{\dd{1}}] {};
	{\color{blue}\node (f1) [flavour,above of=g5,label=above:{\bb{0}}] {};
	\node (f2) [flavour,above of=g9,label=above:{\bb{0}}] {};}
	\draw (g1)--(g2) (g2)--(g3) (g3)--(g4) (g4)--(g5) (g5)--(g6) (g6)--(g7) (g7)--(g8) (g8)--(g9) (g9)--(g10) (g10)--(g11) (g11)--(g12) (g12)--(g13) (g5)--(f1) (g9)--(f2);
	\draw[decoration={brace,mirror,raise=10pt},decorate,thick]
  (3.9,-0.25) -- node[below=10pt] {$\scriptstyle{15}$ } (8.1,-0.25);
	\end{tikzpicture}
	}
\end{align}
The blue piece can now be considered on its own and can be recognised as the $E_8$ brane system \eqref{eq:web_Sp1_7_infinite}, with magnetic quiver given in \eqref{eq:magQuiv_Sp1_7_infinite}. The procedure on the brane web above corresponds to a quiver subtraction of unframed quivers resulting in a quiver with framing, which was discussed for this $e_8$ transition in \cite{Hanany:2018uhm}.

Having reduced the problem to a familiar setting, one can now employ the rules for quiver subtraction of framed linear orthosymplectic quivers \cite{Cabrera:2017njm,Hanany:2019tji}. As a result, one obtains the following Hasse diagram:
\begin{equation}
 \raisebox{-.5\height}{
\begin{tikzpicture}
    \node[hasse] (1) at (0,0) {};
    \node[hasse] (2) at (0,-1) {};
    \node[hasse] (3) at (0,-2) {};
    \node[hasse] (4) at (0,-3) {};
    \node[hasse] (5) at (0,-4) {};
    \node[hasse] (6) at (0,-5) {};
    \draw (1)--(2)--(3)--(4) (5)--(6);
    \node at (0.4,-0.5) {$e_8$};
    \node at (0.4,-1.5) {$d_{10}$};
    \node at (0.4,-2.5) {$d_{12}$};
    \node at (0.6,-4.5) {$d_{2k+6}$};
    \node at (0,-3.5) {$\vdots$};
\end{tikzpicture}
}
\end{equation}
which matches the Hasse diagram \cite[Tab.\ 30]{Bourget:2019aer} obtained from the unitary magnetic quiver \eqref{eq:magQuiv_Sp_O7_infinite_Nf=2k+5}. For $k=1$, the Hasse diagram reduces to a single $e_8$ transition. 
\subsection{\texorpdfstring{$E_7$}{E7} family} 
Next, consider the brane web \eqref{eq:web_Spk_Nf_Higgs_even_infinite_2} for the Higgs branch of $\sprm(k)$ with $N_f=2k+4$ flavours at infinite coupling.
In this case there are two ways to open up a single Coulomb branch modulus. On the one hand, one can repeat the same transition as for the $E_8$ family, given that $k>1$:
\begin{equation}
 \raisebox{-.5\height}{
    \begin{tikzpicture}
    \node[sev] (1) at (0,2) {};
    \node[sev] (2) at (0,-2) {};
    \node[sev] (3) at (-3,0) {};
    \node[sev] (4) at (3,0) {};
    \node[sev] (10) at (0,3) {};
    \node[sev] (11) at (0,-3) {};
    \draw (10)--(1) (11)--(2);
    \draw[blue,transform canvas={yshift=3.5pt,xshift=-1.5pt}] (0,1.7)--(0,1)--(-2,0)--(3);
    \draw[blue,transform canvas={yshift=3.5pt,xshift=1.5pt}] (0,1.7)--(0,1)--(2,0)--(4);
    \draw[blue,transform canvas={yshift=-3.5pt,xshift=-1.5pt}] (0,-1.7)--(0,-1)--(-2,0)--(3);
    \draw[blue,transform canvas={yshift=-3.5pt,xshift=1.5pt}] (0,-1.7)--(0,-1)--(2,0)--(4);
    \draw[dotted] (-2,0)--(2,0);
    \draw[transform canvas={yshift=2pt}] (3)--(4);
    \draw[transform canvas={yshift=-2pt}] (3)--(4);
    \node at (4.5,0) {$\cdots$};
    \node at (-4.5,0) {$\cdots$};
    \node at (1.5,1) {$\scriptstyle{(2,-1)}$};
    \node at (-1.5,1) {$\scriptstyle{(2,1)}$};
    \node at (2.5,0.5) {\color{blue}$\scriptstyle{4}$};
    \node at (-2.5,0.5) {\color{blue}$\scriptstyle{4}$};
    \node at (0,0.5) {$\scriptstyle{k-2}$};
    \draw[dashed, red, transform canvas={yshift=-1pt}] (3)--(-4,0) (4)--(4,0);
    \draw[blue] (3)--(-4,0) (4)--(4,0);
    \draw[blue, transform canvas={yshift=2.5pt}] (3)--(-4,0) (4)--(4,0);
    \draw[blue, transform canvas={yshift=-2.5pt}] (3)--(-4,0) (4)--(4,0);
    \draw[transform canvas={yshift=4pt}] (3)--(-4,0) (4)--(4,0);
    \draw[transform canvas={yshift=-4pt}] (3)--(-4,0) (4)--(4,0);
    \node at (-3.5,0.5) {$\scriptstyle{k-2}$};
    \node at (3.5,0.5) {$\scriptstyle{k-2}$};
    \node at (-3.5,-0.5) {\color{blue}$\scriptstyle{3}$};
    \node at (3.5,-0.5) {\color{blue}$\scriptstyle{3}$};
    \end{tikzpicture}
    }
    \label{eq:web_E7family_1st_transition_1}
\end{equation}
Consequently, the Hasse diagram has a $e_8$ transition at the top. The structure of the brane web \eqref{eq:web_E7family_1st_transition_1} implies that the Higgs branch is described by a magnetic quiver which is a disjoint union of two quivers
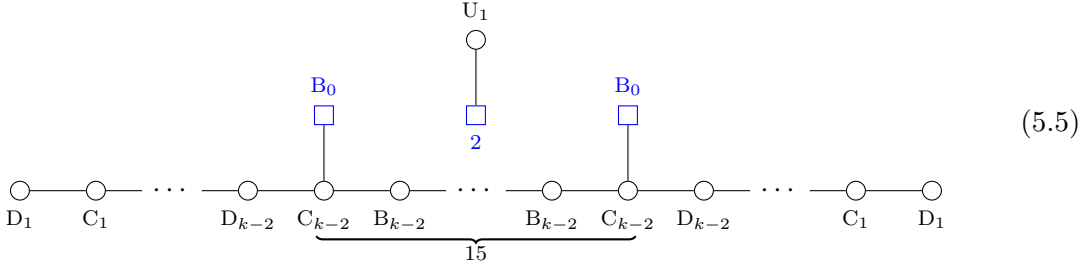
\begin{align}
 \raisebox{-.5\height}{
 	\begin{tikzpicture}
 	\tikzset{node distance = 1cm};
	\tikzstyle{gauge} = [circle, draw,inner sep=2.5pt];
	\tikzstyle{flavour} = [regular polygon,regular polygon sides=4,inner 
sep=2.5pt, draw];
	\node (g1) [gauge,label=below:{\dd{1}}] {};
	\node (g2) [gauge,right of=g1,label=below:{\cc{1}}] {};
	\node (g3) [right of=g2] {$\cdots$};
	\node (g4) [gauge,right of=g3,label=below:{\dd{k-2}}] {};
	\node (g5) [gauge,right of=g4,label=below:{\cc{k-2}}] {};
	\node (g6) [gauge,right of=g5,label=below:{\bb{k-2}}] {};
	\node (g7) [right of=g6] {$\cdots$};
	\node (g8) [gauge,right of=g7,label=below:{\bb{k-2}}] {};
	\node (g9) [gauge,right of=g8,label=below:{\cc{k-2}}] {};
	\node (g10) [gauge,right of=g9,label=below:{\dd{k-2}}] {};
	\node (g11) [right of=g10] {$\cdots$};
	\node (g12) [gauge,right of=g11,label=below:{\cc{1}}] {};
	\node (g13) [gauge,right of=g12,label=below:{\dd{1}}] {};
	{\color{blue}\node (f1) [flavour,above of=g5,label=above:{\bb{0}}] {};
	\node (f2) [flavour,above of=g9,label=above:{\bb{0}}] {};}
	\draw (g1)--(g2) (g2)--(g3) (g3)--(g4) (g4)--(g5) (g5)--(g6) (g6)--(g7) (g7)--(g8) (g8)--(g9) (g9)--(g10) (g10)--(g11) (g11)--(g12) (g12)--(g13) (g5)--(f1) (g9)--(f2);
	\draw[decoration={brace,mirror,raise=10pt},decorate,thick]
  (3.9,-0.25) -- node[below=10pt] {$\scriptstyle{15}$ } (8.1,-0.25); 
    \node[gauge,label=above:{\uu{1}}] (1) at (6,2) {};
    {\color{blue}\node[flavour,label=below:{$\scriptstyle{2}$}] (2) at (6,1) {};}
    \draw (1)--(2);
	\end{tikzpicture}
	}
	\label{eq:magQuiv_E7family_1st_transition_1}
\end{align}
which is a framed orthosymplectic quiver together with a framed unitary quiver.

On the other hand, for $k \geq 1$ one can open up a different Coulomb branch modulus in the brane web \eqref{eq:web_Spk_Nf_Higgs_even_infinite_2} via 
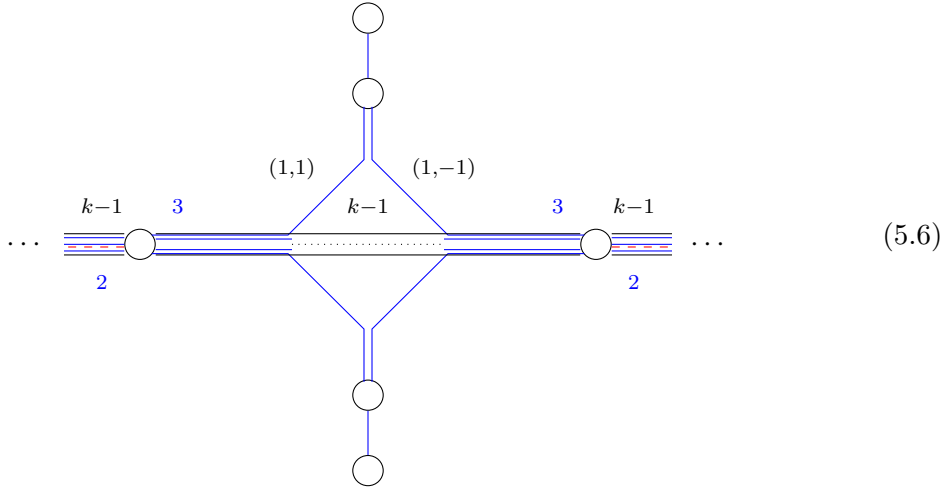
\begin{equation}
 \raisebox{-.5\height}{
    \begin{tikzpicture}
    \node[sev] (1) at (0,3) {};
    \node[sev] (2) at (0,2) {};
    \node[sev] (3) at (0,-2) {};
    \node[sev] (4) at (0,-3) {};
    \node[sev] (5) at (-3,0) {};
    \node[sev] (6) at (3,0) {};
    \draw[blue] (1)--(2) (3)--(4);
    \draw[blue,transform canvas={yshift=3.5pt,xshift=-1.5pt}] (0,1.7)--(0,1)--(-1,0)--(5);
    \draw[blue,transform canvas={yshift=3.5pt,xshift=1.5pt}] (0,1.7)--(0,1)--(1,0)--(6);
    \draw[blue,transform canvas={yshift=-3.5pt,xshift=-1.5pt}] (0,-1.7)--(0,-1)--(-1,0)--(5);
    \draw[blue,transform canvas={yshift=-3.5pt,xshift=1.5pt}] (0,-1.7)--(0,-1)--(1,0)--(6);
    \draw[dotted] (-1,0)--(1,0);
    \draw[blue, transform canvas={yshift=2pt}] (5)--(-1,0) (1,0)--(6);
    \draw[blue, transform canvas={yshift=-2pt}] (5)--(-1,0) (1,0)--(6);
    \draw[transform canvas={yshift=4pt}] (5)--(6);
    \draw[transform canvas={yshift=-4pt}] (5)--(6);
    \node at (1,1) {$\scriptstyle{(1,-1)}$};
    \node at (-1,1) {$\scriptstyle{(1,1)}$};
    \node at (2.5,0.5) {\color{blue}$\scriptstyle{3}$};
    \node at (-2.5,0.5) {\color{blue}$\scriptstyle{3}$};
    \node at (0,0.5) {$\scriptstyle{k-1}$};
    \draw[dashed, red, transform canvas={yshift=-1pt}] (5)--(-4,0) (6)--(4,0);
    \draw[blue] (5)--(-4,0) (6)--(4,0);
    \draw[blue, transform canvas={yshift=2.5pt}] (5)--(-4,0) (6)--(4,0);
    \draw[blue, transform canvas={yshift=-2.5pt}] (5)--(-4,0) (6)--(4,0);
    \draw[transform canvas={yshift=4pt}] (5)--(-4,0) (6)--(4,0);
    \draw[transform canvas={yshift=-4pt}] (5)--(-4,0) (6)--(4,0);
    \node at (-3.5,0.5) {$\scriptstyle{k-1}$};
    \node at (3.5,0.5) {$\scriptstyle{k-1}$};
    \node at (-3.5,-0.5) {\color{blue}$\scriptstyle{2}$};
    \node at (3.5,-0.5) {\color{blue}$\scriptstyle{2}$};
    \node at (4.5,0) {$\cdots$};
    \node at (-4.5,0) {$\cdots$};
    \end{tikzpicture}
    }
    \label{eq:web_E7family_1st_transition_2}
\end{equation}
such that the blue brane subweb corresponds to the $E_7$ brane configuration \eqref{eq:web_Sp1_6_infinite} whose magnetic quiver is \eqref{eq:magQuiv_Sp1_6_infinite}. For the Higgs branch described by the brane web \eqref{eq:web_E7family_1st_transition_2} one reads off the following magnetic quiver:
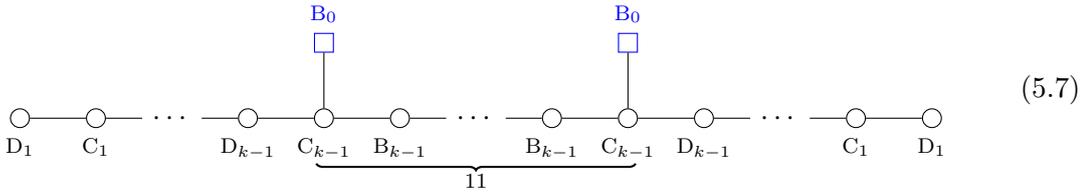
\begin{align}
 \raisebox{-.5\height}{
 	\begin{tikzpicture}
 	\tikzset{node distance = 1cm};
	\tikzstyle{gauge} = [circle, draw,inner sep=2.5pt];
	\tikzstyle{flavour} = [regular polygon,regular polygon sides=4,inner 
sep=2.5pt, draw];
	\node (g1) [gauge,label=below:{\dd{1}}] {};
	\node (g2) [gauge,right of=g1,label=below:{\cc{1}}] {};
	\node (g3) [right of=g2] {$\cdots$};
	\node (g4) [gauge,right of=g3,label=below:{\dd{k-1}}] {};
	\node (g5) [gauge,right of=g4,label=below:{\cc{k-1}}] {};
	\node (g6) [gauge,right of=g5,label=below:{\bb{k-1}}] {};
	\node (g7) [right of=g6] {$\cdots$};
	\node (g8) [gauge,right of=g7,label=below:{\bb{k-1}}] {};
	\node (g9) [gauge,right of=g8,label=below:{\cc{k-1}}] {};
	\node (g10) [gauge,right of=g9,label=below:{\dd{k-1}}] {};
	\node (g11) [right of=g10] {$\cdots$};
	\node (g12) [gauge,right of=g11,label=below:{\cc{1}}] {};
	\node (g13) [gauge,right of=g12,label=below:{\dd{1}}] {};
	{\color{blue}\node (f1) [flavour,above of=g5,label=above:{\bb{0}}] {};
	\node (f2) [flavour,above of=g9,label=above:{\bb{0}}] {};}
	\draw (g1)--(g2) (g2)--(g3) (g3)--(g4) (g4)--(g5) (g5)--(g6) (g6)--(g7) (g7)--(g8) (g8)--(g9) (g9)--(g10) (g10)--(g11) (g11)--(g12) (g12)--(g13) (g5)--(f1) (g9)--(f2);
	\draw[decoration={brace,mirror,raise=10pt},decorate,thick]
  (3.9,-0.25) -- node[below=10pt] {$\scriptstyle{11}$ } (8.1,-0.25); 
	\end{tikzpicture}
	}
	\label{eq:magQuiv_E7family_1st_transition_2}
\end{align}
which is again a framed orthosymplectic quiver.

As a consequence, both ways to open up the minimal amount of Coulomb branch directions result in Higgs branches that are describable by framed orthosymplectic (or unitary) magnetic quivers \eqref{eq:magQuiv_E7family_1st_transition_1} and \eqref{eq:magQuiv_E7family_1st_transition_2}. Therefore, the known algorithms for quiver subtraction then allow to derive the full Hasse diagram
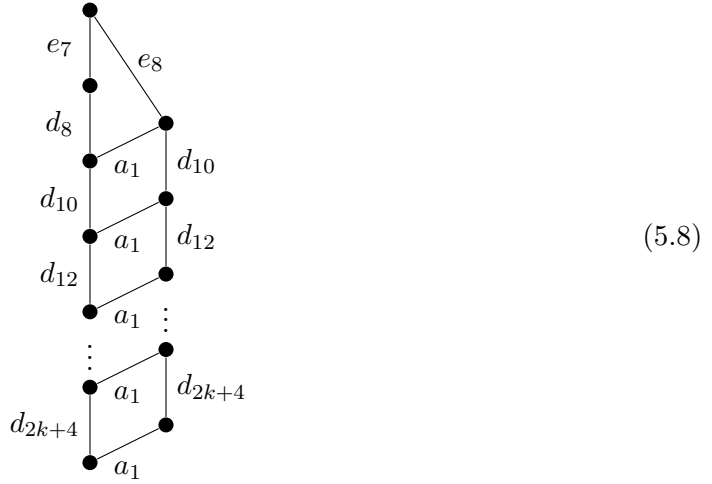
\begin{equation}
 \raisebox{-.5\height}{
\begin{tikzpicture}
    \node[hasse] (1) at (-1,0.5) {};
    \node[hasse] (2) at (0,-1) {};
    \node[hasse] (3) at (0,-2) {};
    \node[hasse] (4) at (0,-3) {};
    \node[hasse] (5) at (0,-4) {};
    \node[hasse] (6) at (0,-5) {};
    \draw (1)--(2)--(3)--(4) (5)--(6);
    \node at (-0.2,-0.2) {$e_8$};
    \node at (0.4,-1.5) {$d_{10}$};
    \node at (0.4,-2.5) {$d_{12}$};
    \node at (0.6,-4.5) {$d_{2k+4}$};
    \node at (0,-3.5) {$\vdots$};
    \node[hasse] (11) at (-1,-0.5) {};
    \node[hasse] (12) at (-1,-1.5) {};
    \node[hasse] (13) at (-1,-2.5) {};
    \node[hasse] (14) at (-1,-3.5) {};
    \node[hasse] (15) at (-1,-4.5) {};
    \node[hasse] (16) at (-1,-5.5) {};
    \draw (1)--(11)--(12)--(13)--(14) (15)--(16) (2)--(12) (3)--(13) (4)--(14) (5)--(15) (6)--(16);
    \node at (-1,-4) {$\vdots$};
    \node at (-1.4,0) {$e_7$};
    \node at (-1.4,-1) {$d_8$};
    \node at (-1.4,-2) {$d_{10}$};
    \node at (-1.4,-3) {$d_{12}$};
    \node at (-1.6,-5) {$d_{2k+4}$};
    \node at (-0.5,-1.6) {$a_1$};
    \node at (-0.5,-2.6) {$a_1$};
    \node at (-0.5,-3.6) {$a_1$};
    \node at (-0.5,-4.6) {$a_1$};
    \node at (-0.5,-5.6) {$a_1$};
\end{tikzpicture}
}
\end{equation}
which agrees with the Hasse diagram \cite[Tab.\ 29]{Bourget:2019aer} computed from the unitary magnetic quiver \eqref{eq:magQuiv_Sp_O7_infinite_Nf=2k+4}. For $k=1$, the Hasse diagram reduces to a single $e_7$ transition. 
\subsection{\texorpdfstring{$E_6$}{E6} family}
Moving on to the brane web \eqref{eq:web_Spk_Nf_Higgs_odd_infinite} for the Higgs branch of $\sprm(k)$ with $N_f=2k+3$ flavours at infinite coupling, one recognises that there is exactly one option to open up a minimal Coulomb branch direction. In the brane web, the central part becomes
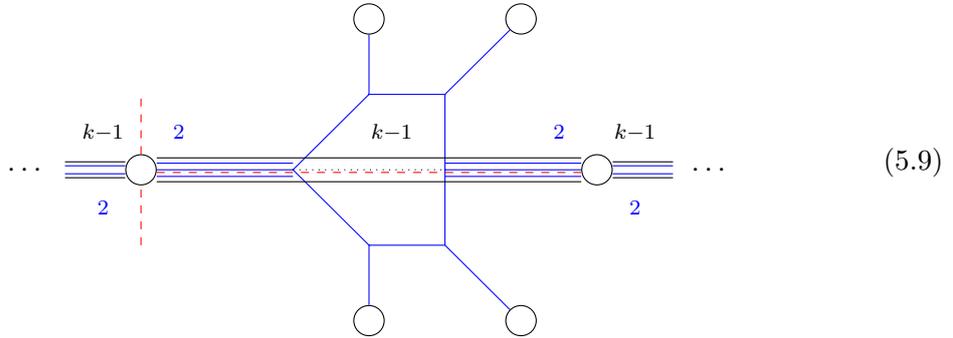
\begin{equation}
 \raisebox{-.5\height}{
    \begin{tikzpicture}
    \node[sev] (1) at (0,2) {};
    \node[sev] (2) at (2,2) {};
    \node[sev] (3) at (0,-2) {};
    \node[sev] (4) at (2,-2) {};
    \node[sev] (5) at (-3,0) {};
    \node[sev] (6) at (3,0) {};
    \draw[blue] (1)--(0,1)--(1,1)--(2) (0,1)--(-1,0)--(0,-1)--(3) (0,-1)--(1,-1)--(4) (1,-1)--(1,1) (5)--(-1,0) (1,0)--(6);
    \draw[dotted] (-1,0)--(1,0);
    \draw[dashed, red, transform canvas={yshift=-1pt}] (5)--(6);
    \draw[blue, transform canvas={yshift=2.5pt}] (5)--(-1,0) (1,0)--(6);
    \draw[blue, transform canvas={yshift=-2.5pt}] (5)--(-1,0) (1,0)--(6);
    \draw[transform canvas={yshift=4.5pt}] (5)--(6);
    \draw[transform canvas={yshift=-4.5pt}] (5)--(6);
    \node at (2.5,0.5) {\color{blue}$\scriptstyle{2}$};
    \node at (-2.5,0.5) {\color{blue}$\scriptstyle{2}$};
    \node at (0.3,0.5) {$\scriptstyle{k-1}$};
    \draw[blue, transform canvas={yshift=1.5pt}] (5)--(-4,0) (6)--(4,0);
    \draw[blue, transform canvas={yshift=-1.5pt}] (5)--(-4,0) (6)--(4,0);
    \draw[transform canvas={yshift=3pt}] (5)--(-4,0) (6)--(4,0);
    \draw[transform canvas={yshift=-3pt}] (5)--(-4,0) (6)--(4,0);
    \node at (-3.5,0.5) {$\scriptstyle{k-1}$};
    \node at (3.5,0.5) {$\scriptstyle{k-1}$};
    \node at (-3.5,-0.5) {\color{blue}$\scriptstyle{2}$};
    \node at (3.5,-0.5) {\color{blue}$\scriptstyle{2}$};
    \node at (4.5,0) {$\cdots$};
    \node at (-4.5,0) {$\cdots$};
    \draw[dashed, red] (-3,-1)--(5)--(-3,1);
    \end{tikzpicture}
    }
        \label{eq:web_E6family_1st_transition}
\end{equation}
and the blue brane subweb is identified as the $E_6$ configuration 
\eqref{eq:web_Sp1_5_infinite}, whose magnetic quiver is 
\eqref{eq:magQuiv_Sp1_5_infinite}. For the remaining Higgs branch degrees of 
freedom in \eqref{eq:web_E6family_1st_transition} one derives the following 
magnetic quiver
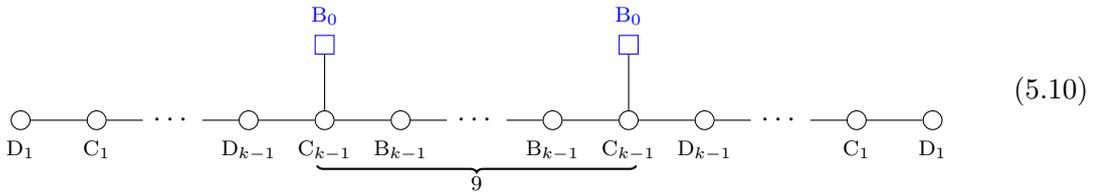
\begin{align}
 \raisebox{-.5\height}{
 	\begin{tikzpicture}
 	\tikzset{node distance = 1cm};
	\tikzstyle{gauge} = [circle, draw,inner sep=2.5pt];
	\tikzstyle{flavour} = [regular polygon,regular polygon sides=4,inner 
sep=2.5pt, draw];
	\node (g1) [gauge,label=below:{\dd{1}}] {};
	\node (g2) [gauge,right of=g1,label=below:{\cc{1}}] {};
	\node (g3) [right of=g2] {$\cdots$};
	\node (g4) [gauge,right of=g3,label=below:{\dd{k-1}}] {};
	\node (g5) [gauge,right of=g4,label=below:{\cc{k-1}}] {};
	\node (g6) [gauge,right of=g5,label=below:{\bb{k-1}}] {};
	\node (g7) [right of=g6] {$\cdots$};
	\node (g8) [gauge,right of=g7,label=below:{\bb{k-1}}] {};
	\node (g9) [gauge,right of=g8,label=below:{\cc{k-1}}] {};
	\node (g10) [gauge,right of=g9,label=below:{\dd{k-1}}] {};
	\node (g11) [right of=g10] {$\cdots$};
	\node (g12) [gauge,right of=g11,label=below:{\cc{1}}] {};
	\node (g13) [gauge,right of=g12,label=below:{\dd{1}}] {};
	{\color{blue}\node (f1) [flavour,above of=g5,label=above:{\bb{0}}] {};
	\node (f2) [flavour,above of=g9,label=above:{\bb{0}}] {};}
	\draw (g1)--(g2) (g2)--(g3) (g3)--(g4) (g4)--(g5) (g5)--(g6) (g6)--(g7) (g7)--(g8) (g8)--(g9) (g9)--(g10) (g10)--(g11) (g11)--(g12) (g12)--(g13) (g5)--(f1) (g9)--(f2);
	\draw[decoration={brace,mirror,raise=10pt},decorate,thick]
  (3.9,-0.25) -- node[below=10pt] {$\scriptstyle{9}$ } (8.1,-0.25); 
	\end{tikzpicture}
	}
	\label{eq:magQuiv_E6family_1st_transition}
\end{align}
which is a framed orthosymplectic quiver.

Again, as the Higgs branch after the $e_6$ transition is described by a framed 
quiver \eqref{eq:magQuiv_E6family_1st_transition}, the quiver subtraction 
algorithm allows one to complete the Higgs branch Hasse diagram. In 
detail, one obtains
\begin{equation}
 \raisebox{-.5\height}{
\begin{tikzpicture}
    \node[hasse] (1) at (0,0) {};
    \node[hasse] (2) at (0,-1) {};
    \node[hasse] (3) at (0,-2) {};
    \node[hasse] (4) at (0,-3) {};
    \node[hasse] (5) at (0,-4) {};
    \node[hasse] (6) at (0,-5) {};
    \draw (1)--(2)--(3)--(4) (5)--(6);
    \node at (0.4,-0.5) {$e_6$};
    \node at (0.4,-1.5) {$d_{7}$};
    \node at (0.4,-2.5) {$d_{9}$};
    \node at (0.6,-4.5) {$d_{2k+3}$};
    \node at (0,-3.5) {$\vdots$};
\end{tikzpicture}
    }
\end{equation}
and one can verify the agreement with the Hasse diagram \cite[Tab.\ 28]{Bourget:2019aer} obtained from the unitary magnetic quiver \eqref{eq:magQuiv_Sp_O7_infinite_Nf=2k+3}. For $k=1$, the Hasse diagram reduces to a single $e_6$ transition. 
\subsection{\texorpdfstring{$E_5$}{E5} family}
Consider the brane web \eqref{eq:web_Spk_Nf_Higgs_even_infinite_1} for the Higgs branch of $\sprm(k)$ with $N_f=2k+2$ flavours at infinite coupling. In order to open up the minimal number of Coulomb branch moduli, one finds that there is only one possibility
\begin{equation}
 \raisebox{-.5\height}{
    \begin{tikzpicture}
    \node[sev] (1) at (-2,2) {};
    \node[sev] (2) at (2,2) {};
    \node[sev] (3) at (-2,-2) {};
    \node[sev] (4) at (2,-2) {};
    \node[sev] (5) at (-3,0) {};
    \node[sev] (6) at (3,0) {};
    \draw[blue] (1)--(-1,1)--(1,1)--(2) (-1,1)--(-1,-1)--(3) (-1,-1)--(1,-1)--(4) (1,-1)--(1,1);
    \draw[dotted] (-1,0)--(1,0);
    \draw[blue, transform canvas={yshift=2pt}] (5)--(-1,0) (1,0)--(6);
    \draw[blue, transform canvas={yshift=-2pt}] (5)--(-1,0) (1,0)--(6);
    \draw[transform canvas={yshift=4pt}] (5)--(6);
    \draw[transform canvas={yshift=-4pt}] (5)--(6);
    \node at (2.5,0.5) {\color{blue}$\scriptstyle{2}$};
    \node at (-2.5,0.5) {\color{blue}$\scriptstyle{2}$};
    \node at (0,0.5) {$\scriptstyle{k-1}$};
    \draw[dashed, red, transform canvas={yshift=-1pt}] (5)--(-4,0) (6)--(4,0);
    \draw[blue] (5)--(-4,0) (6)--(4,0);
    \draw[blue, transform canvas={yshift=2.5pt}] (5)--(-4,0) (6)--(4,0);
    \draw[blue, transform canvas={yshift=-2.5pt}] (5)--(-4,0) (6)--(4,0);
    \draw[transform canvas={yshift=4pt}] (5)--(-4,0) (6)--(4,0);
    \draw[transform canvas={yshift=-4pt}] (5)--(-4,0) (6)--(4,0);
    \node at (-3.5,0.5) {$\scriptstyle{k-1}$};
    \node at (3.5,0.5) {$\scriptstyle{k-1}$};
    \node at (-3.5,-0.5) {\color{blue}$\scriptstyle{1}$};
    \node at (3.5,-0.5) {\color{blue}$\scriptstyle{1}$};
    \node at (4.5,0) {$\cdots$};
    \node at (-4.5,0) {$\cdots$};
    \end{tikzpicture}
    }
     \label{eq:web_E5family_1st_transition}
\end{equation}
where the blue brane subweb corresponds to the $E_5$ configuration \eqref{eq:web_Sp1_4_infinite}, with magnetic quiver \eqref{eq:magQuiv_Sp1_4_infinite}. For the remaining Higgs branch directions in \eqref{eq:web_E5family_1st_transition} one finds the magnetic quiver
\begin{align}
 \raisebox{-.5\height}{
 	\begin{tikzpicture}
 	\tikzset{node distance = 1cm};
	\tikzstyle{gauge} = [circle, draw,inner sep=2.5pt];
	\tikzstyle{flavour} = [regular polygon,regular polygon sides=4,inner 
sep=2.5pt, draw];
	\node (g1) [gauge,label=below:{\dd{1}}] {};
	\node (g2) [gauge,right of=g1,label=below:{\cc{1}}] {};
	\node (g3) [right of=g2] {$\cdots$};
	\node (g4) [gauge,right of=g3,label=below:{\dd{k-1}}] {};
	\node (g5) [gauge,right of=g4,label=below:{\cc{k-1}}] {};
	\node (g6) [gauge,right of=g5,label=below:{\bb{k-1}}] {};
	\node (g7) [right of=g6] {$\cdots$};
	\node (g8) [gauge,right of=g7,label=below:{\bb{k-1}}] {};
	\node (g9) [gauge,right of=g8,label=below:{\cc{k-1}}] {};
	\node (g10) [gauge,right of=g9,label=below:{\dd{k-1}}] {};
	\node (g11) [right of=g10] {$\cdots$};
	\node (g12) [gauge,right of=g11,label=below:{\cc{1}}] {};
	\node (g13) [gauge,right of=g12,label=below:{\dd{1}}] {};
	{\color{blue}\node (f1) [flavour,above of=g5,label=above:{\bb{0}}] {};
	\node (f2) [flavour,above of=g9,label=above:{\bb{0}}] {};}
	\draw (g1)--(g2) (g2)--(g3) (g3)--(g4) (g4)--(g5) (g5)--(g6) (g6)--(g7) (g7)--(g8) (g8)--(g9) (g9)--(g10) (g10)--(g11) (g11)--(g12) (g12)--(g13) (g5)--(f1) (g9)--(f2);
	\draw[decoration={brace,mirror,raise=10pt},decorate,thick]
  (3.9,-0.25) -- node[below=10pt] {$\scriptstyle{7}$ } (8.1,-0.25); 
	\end{tikzpicture}
	}
	 \label{eq:magQuiv_E5family_1st_transition}
\end{align}
which is again a framed orthosymplectic quiver.

Hence, the full Higgs branch Hasse diagram can then be derived from the framed starting point \eqref{eq:magQuiv_E5family_1st_transition} via quiver subtraction. The Hasse diagram reads 
\begin{equation}
 \raisebox{-.5\height}{
\begin{tikzpicture}
    \node[hasse] (1) at (0,0) {};
    \node[hasse] (2) at (0,-1) {};
    \node[hasse] (3) at (0,-2) {};
    \node[hasse] (4) at (0,-3) {};
    \node[hasse] (5) at (0,-4) {};
    \node[hasse] (6) at (0,-5) {};
    \draw (1)--(2)--(3)--(4) (5)--(6);
    \node at (0.8,-0.5) {$e_5=d_5$};
    \node at (0.4,-1.5) {$d_{6}$};
    \node at (0.4,-2.5) {$d_{8}$};
    \node at (0.6,-4.5) {$d_{2k+2}$};
    \node at (0,-3.5) {$\vdots$};
\end{tikzpicture}
}
\end{equation}
which coincides with the Hasse diagram \cite[Tab.\ 27]{Bourget:2019aer} derived from the unitary magnetic quiver \eqref{eq:magQuiv_Sp_O7_infinite_Nf=2k+2}. For $k=1$, the Hasse diagram reduces to a single $e_5=d_5$ transition. 
\subsection{\texorpdfstring{$E_4$}{E4} family}
Next, moving on to the brane web \eqref{eq:web_Spk_Nf_Higgs_odd_infinite} for the Higgs branch of $\sprm(k)$ with $N_f=2k+1$ flavours at infinite coupling. The minimal choice of opening up Coulomb branch directions is 
\begin{equation}
\raisebox{-.5\height}{
    \begin{tikzpicture}
    \node[sev] (1) at (-2,2) {};
    \node[sev] (2) at (4,2) {};
    \node[sev] (3) at (-2,-2) {};
    \node[sev] (4) at (4,-2) {};
    \node[sev] (5) at (-3,0) {};
    \node[sev] (6) at (3,0) {};
    \draw[blue] (1)--(-1,1)--(2,1)--(2) (-1,1)--(-1,-1)--(3) (-1,-1)--(2,-1)--(4) (2,-1)--(1,0)--(2,1) (5)--(-1,0) (1,0)--(6);
    \draw[dotted] (-1,0)--(1,0);
    \draw[dashed, red, transform canvas={yshift=-1pt}] (5)--(6);
    \draw[blue, transform canvas={yshift=2.5pt}] (5)--(-1,0) (1,0)--(6);
    \draw[blue, transform canvas={yshift=-2.5pt}] (5)--(-1,0) (1,0)--(6);
    \draw[transform canvas={yshift=4.5pt}] (5)--(6);
    \draw[transform canvas={yshift=-4.5pt}] (5)--(6);
    \node at (2.5,0.5) {\color{blue}$\scriptstyle{1}$};
    \node at (-2.5,0.5) {\color{blue}$\scriptstyle{1}$};
    \node at (0.3,0.5) {$\scriptstyle{k{-}1}$};
    \draw[blue, transform canvas={yshift=1.5pt}] (5)--(-4,0) (6)--(4,0);
    \draw[blue, transform canvas={yshift=-1.5pt}] (5)--(-4,0) (6)--(4,0);
    \draw[transform canvas={yshift=3pt}] (5)--(-4,0) (6)--(4,0);
    \draw[transform canvas={yshift=-3pt}] (5)--(-4,0) (6)--(4,0);
    \node at (-3.5,0.5) {$\scriptstyle{k{-}1}$};
    \node at (3.5,0.5) {$\scriptstyle{k{-}1}$};
    \node at (-3.5,-0.5) {\color{blue}$\scriptstyle{1}$};
    \node at (3.5,-0.5) {\color{blue}$\scriptstyle{1}$};
    \node at (4.5,0) {$\cdots$};
    \node at (-4.5,0) {$\cdots$};
    \draw[dashed, red] (-3,-1)--(5)--(-3,1);
    \node at (2.5,1.5) {$\scriptstyle{(2,1)}$};
    \end{tikzpicture}
    }
    \label{eq:web_E4family_1st_transition}
\end{equation}
where the blue brane subweb corresponds to the $E_4$ configuration 
\eqref{eq:web_Sp1_3_infinite}, with magnetic quiver 
\eqref{eq:magQuiv_Sp1_3_infinite}. The magnetic quiver for 
\eqref{eq:web_E4family_1st_transition} reads
\begin{align}
 \raisebox{-.5\height}{
 	\begin{tikzpicture}
 	\tikzset{node distance = 1cm};
	\tikzstyle{gauge} = [circle, draw,inner sep=2.5pt];
	\tikzstyle{flavour} = [regular polygon,regular polygon sides=4,inner 
sep=2.5pt, draw];
	\node (g1) [gauge,label=below:{\dd{1}}] {};
	\node (g2) [gauge,right of=g1,label=below:{\cc{1}}] {};
	\node (g3) [right of=g2] {$\cdots$};
	\node (g4) [gauge,right of=g3,label=below:{\dd{k-1}}] {};
	\node (g5) [gauge,right of=g4,label=below:{\cc{k-1}}] {};
	\node (g6) [gauge,right of=g5,label=below:{\bb{k-1}}] {};
	\node (g7) [gauge,right of=g6,label=below:{\cc{k-1}}] {};
	\node (g8) [gauge,right of=g7,label=below:{\bb{k-1}}] {};
	\node (g9) [gauge,right of=g8,label=below:{\cc{k-1}}] {};
	\node (g10) [gauge,right of=g9,label=below:{\dd{k-1}}] {};
	\node (g11) [right of=g10] {$\cdots$};
	\node (g12) [gauge,right of=g11,label=below:{\cc{1}}] {};
	\node (g13) [gauge,right of=g12,label=below:{\dd{1}}] {};
	{\color{blue}\node (f1) [flavour,above of=g5,label=above:{\bb{0}}] {};
	\node (f2) [flavour,above of=g9,label=above:{\bb{0}}] {};}
	\draw (g1)--(g2) (g2)--(g3) (g3)--(g4) (g4)--(g5) (g5)--(g6) (g6)--(g7) (g7)--(g8) (g8)--(g9) (g9)--(g10) (g10)--(g11) (g11)--(g12) (g12)--(g13) (g5)--(f1) (g9)--(f2);
	\end{tikzpicture}
	}
	 \label{eq:magQuiv_E4family_1st_transition}
\end{align}
The remaining Hasse diagram can be derived by the same techniques as discussed above and one finds
\begin{equation}
\raisebox{-.5\height}{
\begin{tikzpicture}
    \node[hasse] (1) at (0,0) {};
    \node[hasse] (2) at (0,-1) {};
    \node[hasse] (3) at (0,-2) {};
    \node[hasse] (4) at (0,-3) {};
    \node[hasse] (5) at (0,-4) {};
    \node[hasse] (6) at (0,-5) {};
    \draw (1)--(2)--(3)--(4) (5)--(6);
    \node at (0.8,-0.5) {$e_4=a_4$};
    \node at (0.4,-1.5) {$d_{5}$};
    \node at (0.4,-2.5) {$d_{7}$};
    \node at (0.6,-4.5) {$d_{2k+1}$};
    \node at (0,-3.5) {$\vdots$};
\end{tikzpicture}
}
\end{equation}
which agrees with the Hasse diagram \cite[Tab.\ 26]{Bourget:2019aer} obtained from the unitary magnetic quiver \eqref{eq:magQuiv_Sp_O7_infinite_Nf=2k+1}.  For $k=1$, the Hasse diagram reduces to a single $e_4=a_4$ transition. 
\subsection{\texorpdfstring{$E_3$}{E3} family}
In the case of the generalised $E_3$ family, the brane web \eqref{eq:web_Spk_Nf_Higgs_even_infinite_1} for the Higgs branch of $\sprm(k)$ with $N_f=2k$ flavours at infinite coupling does only give rise to one cone from the expected union of two cones.
It is then not surprising that the web only allows one to identify 
the $a_2$ part of $e_3=a_2\cup a_1$ transition. In more detail, opening up the 
minimal number of Coulomb branch directions looks as follows:
\begin{equation}
\raisebox{-.5\height}{
    \begin{tikzpicture}
    \node[sev] (1) at (-4,2) {};
    \node[sev] (2) at (4,2) {};
    \node[sev] (3) at (-4,-2) {};
    \node[sev] (4) at (4,-2) {};
    \node[sev] (5) at (-3,0) {};
    \node[sev] (6) at (3,0) {};
    \draw[blue] (1)--(-2,1)--(2,1)--(2) (-2,1)--(-1,0)--(-2,-1)--(3) (-2,-1)--(2,-1)--(4) (2,-1)--(1,0)--(2,1);
    \draw[dotted] (-1,0)--(1,0);
    \draw[blue, transform canvas={yshift=2pt}] (5)--(-1,0) (1,0)--(6);
    \draw[blue, transform canvas={yshift=-2pt}] (5)--(-1,0) (1,0)--(6);
    \draw[transform canvas={yshift=4pt}] (5)--(6);
    \draw[transform canvas={yshift=-4pt}] (5)--(6);
    \node at (2.5,0.5) {\color{blue}$\scriptstyle{1}$};
    \node at (-2.5,0.5) {\color{blue}$\scriptstyle{1}$};
    \node at (0,0.5) {$\scriptstyle{k{-}1}$};
    \draw[dashed, red, transform canvas={yshift=-1pt}] (5)--(-4,0) (6)--(4,0);
    \draw[blue] (5)--(-4,0) (6)--(4,0);
    \draw[transform canvas={yshift=4pt}] (5)--(-4,0) (6)--(4,0);
    \draw[transform canvas={yshift=-4pt}] (5)--(-4,0) (6)--(4,0);
    \node at (-3.5,0.5) {$\scriptstyle{k{-}1}$};
    \node at (3.5,0.5) {$\scriptstyle{k{-}1}$};
    \node at (-3.5,-0.5) {\color{blue}$\scriptstyle{0}$};
    \node at (3.5,-0.5) {\color{blue}$\scriptstyle{0}$};
    \node at (4.5,0) {$\cdots$};
    \node at (-4.5,0) {$\cdots$};
    \node at (2.5,1.5) {$\scriptstyle{(2,1)}$};
    \node at (-2.4,1.5) {$\scriptstyle{(2,-1)}$};
    \end{tikzpicture}
    }
\label{eq:web_E3family_1st_transition}
\end{equation}
where the blue brane subweb corresponds to the $E_3$ configuration \eqref{eq:web_Sp1_2_infinite}, with magnetic quiver \eqref{eq:magQuiv_Sp1_2_infinite}. The magnetic quiver for \ref{eq:web_E3family_1st_transition} reads
\begin{align}
 \raisebox{-.5\height}{
 	\begin{tikzpicture}
 	\tikzset{node distance = 1cm};
	\tikzstyle{gauge} = [circle, draw,inner sep=2.5pt];
	\tikzstyle{flavour} = [regular polygon,regular polygon sides=4,inner 
sep=2.5pt, draw];
	\node (g1) [gauge,label=below:{\dd{1}}] {};
	\node (g2) [gauge,right of=g1,label=below:{\cc{1}}] {};
	\node (g3) [right of=g2] {$\cdots$};
	\node (g4) [gauge,right of=g3,label=below:{\dd{k-1}}] {};
	\node (g5) [gauge,right of=g4,label=below:{\cc{k-1}}] {};
	\node (g8) [gauge,right of=g5,label=below:{\bb{k-1}}] {};
	\node (g9) [gauge,right of=g8,label=below:{\cc{k-1}}] {};
	\node (g10) [gauge,right of=g9,label=below:{\dd{k-1}}] {};
	\node (g11) [right of=g10] {$\cdots$};
	\node (g12) [gauge,right of=g11,label=below:{\cc{1}}] {};
	\node (g13) [gauge,right of=g12,label=below:{\dd{1}}] {};
	{\color{blue}\node (f1) [flavour,above of=g5,label=above:{\bb{0}}] {};
	\node (f2) [flavour,above of=g9,label=above:{\bb{0}}] {};}
	\draw (g1)--(g2) (g2)--(g3) (g3)--(g4) (g4)--(g5) (g5)--(g8) (g8)--(g9) (g9)--(g10) (g10)--(g11) (g11)--(g12) (g12)--(g13) (g5)--(f1) (g9)--(f2);
	\end{tikzpicture}
	}
	 \label{eq:magQuiv_E3family_1st_transition}
\end{align}
Taking \eqref{eq:web_E3family_1st_transition} as a starting point, using quiver 
subtraction of framed orthosymplectic quivers, one obtains the black part of the 
following Hasse diagram: 
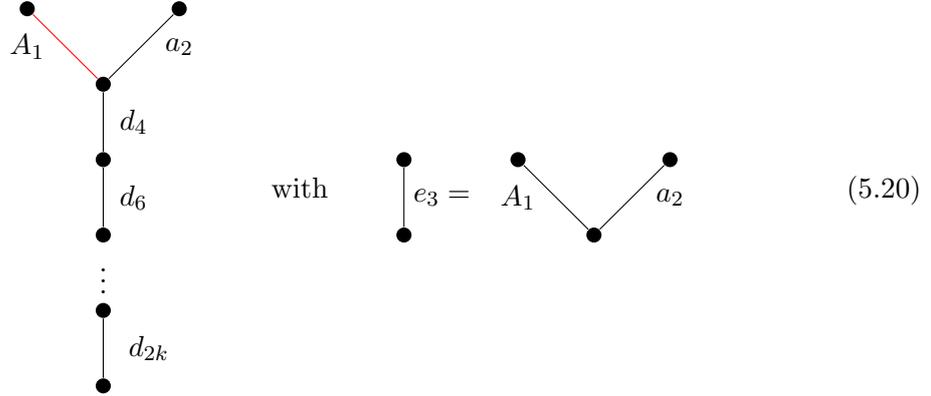
\begin{align}
\raisebox{-.5\height}{
\begin{tikzpicture}
    \node[hasse] (0) at (-1,0) {};
    \node[hasse] (1) at (1,0) {};
    \node[hasse] (2) at (0,-1) {};
    \node[hasse] (3) at (0,-2) {};
    \node[hasse] (4) at (0,-3) {};
    \node[hasse] (5) at (0,-4) {};
    \node[hasse] (6) at (0,-5) {};
    \draw[red] (0)--(2);
    \draw (1)--(2)--(3)--(4) (5)--(6);
    \node at (1,-0.5) {$a_2$};
    \node at (-1,-0.5) {$A_1$};
    \node at (0.4,-1.5) {$d_{4}$};
    \node at (0.4,-2.5) {$d_{6}$};
    \node at (0.6,-4.5) {$d_{2k}$};
    \node at (0,-3.5) {$\vdots$};
\end{tikzpicture}
}
\qquad 
\text{with}
\qquad 
\raisebox{-.5\height}{
    \begin{tikzpicture}
    \node[hasse] (-1) at (-2.5,0) {};
    \node[hasse] (-2) at (-2.5,-1) {};
    \draw (-1)--(-2);
    \node at (-2,-0.5) {$e_3=$};
    \node[hasse] (0) at (-1,0) {};
    \node[hasse] (1) at (1,0) {};
    \node[hasse] (2) at (0,-1) {};
    \draw (0)--(2) (1)--(2);
    \node at (1,-0.5) {$a_2$};
    \node at (-1,-0.5) {$A_1$};
    \end{tikzpicture}
    }
    \label{eq:HasseforE3}
\end{align}
The $A_1$ transition indicated in \textcolor{red}{red} seems beyond the 
reach of the brane web \eqref{eq:web_Spk_Nf_Higgs_even_infinite_1} and it is
currently not possible to identify it in this construction. Comparing with the 
Hasse diagram \cite[Tab.\ 6]{Bourget:2019aer} obtained from the unitary 
magnetic quivers, one finds the black part of \eqref{eq:HasseforE3} from 
\eqref{eq:magQuiv_Sp_O7_infinite_Nf_small_even_2} and the \textcolor{red}{red} 
transition from \eqref{eq:magQuiv_Sp_O7_infinite_Nf_small_even_1}. For $k=1$, the Hasse diagram reduces to a single $e_3$ transition. 
\subsection{\texorpdfstring{$E_2$}{E2} family}
For the Higgs branch of $\sprm(k)$ with $N_f=2k-1$ flavours at infinite coupling, the brane web is given in \eqref{eq:web_Spk_Nf_Higgs_odd_infinite}. The expectation is to find an $a_1$ transition on top, since $e_2=a_1\cup \mathbb{Z}_2$, where $\mathbb{Z}_2$ is an abuse of notation to denote the `fat point' non-reduced algebraic scheme \cite{Cremonesi:2015lsa}. The following Coulomb branch modulus can be opened up:
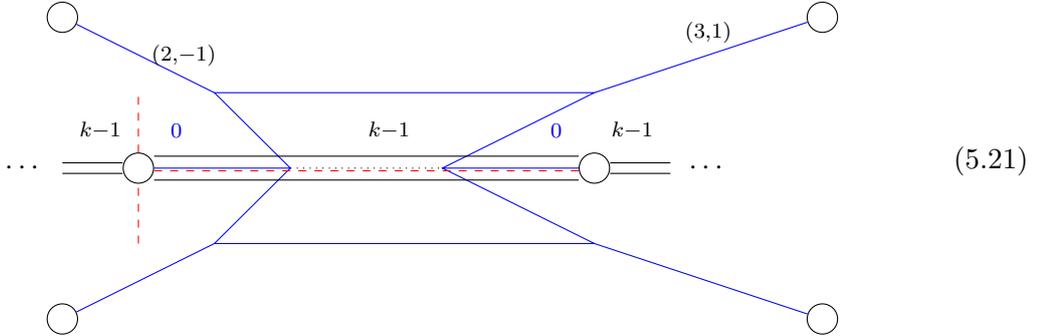
\begin{equation}
\raisebox{-.5\height}{
    \begin{tikzpicture}
    \node[sev] (1) at (-4,2) {};
    \node[sev] (2) at (6,2) {};
    \node[sev] (3) at (-4,-2) {};
    \node[sev] (4) at (6,-2) {};
    \node[sev] (5) at (-3,0) {};
    \node[sev] (6) at (3,0) {};
    \draw[blue] (1)--(-2,1)--(3,1)--(2) (-2,1)--(-1,0)--(-2,-1)--(3) (-2,-1)--(3,-1)--(4) (3,-1)--(1,0)--(3,1) (5)--(-1,0) (1,0)--(6);
    \draw[dotted] (-1,0)--(1,0);
    \draw[dashed, red, transform canvas={yshift=-1pt}] (5)--(6);
    \draw[transform canvas={yshift=4.5pt}] (5)--(6);
    \draw[transform canvas={yshift=-4.5pt}] (5)--(6);
    \node at (2.5,0.5) {\color{blue}$\scriptstyle{0}$};
    \node at (-2.5,0.5) {\color{blue}$\scriptstyle{0}$};
    \node at (0.3,0.5) {$\scriptstyle{k{-}1}$};
    \draw[transform canvas={yshift=2pt}] (5)--(-4,0) (6)--(4,0);
    \draw[transform canvas={yshift=-2pt}] (5)--(-4,0) (6)--(4,0);
    \node at (-3.5,0.5) {$\scriptstyle{k{-}1}$};
    \node at (3.5,0.5) {$\scriptstyle{k{-}1}$};
    \node at (4.5,0) {$\cdots$};
    \node at (-4.5,0) {$\cdots$};
    \draw[dashed, red] (-3,-1)--(5)--(-3,1);
    \node at (4.5,1.8) {$\scriptstyle{(3,1)}$};
    \node at (-2.4,1.5) {$\scriptstyle{(2,-1)}$};
    \end{tikzpicture}
    }
    \label{eq:web_E2family_1st_transition}
\end{equation}
where the blue brane subweb corresponds to the $E_2$ configuration \eqref{eq:web_Sp1_1_infinite}, with magnetic quiver \eqref{eq:magQuiv_Sp1_1_infinite}. The magnetic quiver for \eqref{eq:web_E2family_1st_transition} is
\begin{align}
 \raisebox{-.5\height}{
 	\begin{tikzpicture}
 	\tikzset{node distance = 1cm};
	\tikzstyle{gauge} = [circle, draw,inner sep=2.5pt];
	\tikzstyle{flavour} = [regular polygon,regular polygon sides=4,inner 
sep=2.5pt, draw];
	\node (g1) [gauge,label=below:{\dd{1}}] {};
	\node (g2) [gauge,right of=g1,label=below:{\cc{1}}] {};
	\node (g3) [right of=g2] {$\cdots$};
	\node (g4) [gauge,right of=g3,label=below:{\dd{k{-}1}}] {};
	\node (g5) [gauge,right of=g4,label=below:{\cc{k{-}1}}] {};
	\node (g6) [gauge,right of=g5,label=below:{\dd{k{-}1}}] {};
	\node (g7) [right of=g6] {$\cdots$};
	\node (g8) [gauge,right of=g7,label=below:{\cc{1}}] {};
	\node (g9) [gauge,right of=g8,label=below:{\dd{1}}] {};
	{\color{blue}\node (f1) [flavour,above of=g5,label=above:{\dd{1}}] {};}
	\draw (g1)--(g2) (g2)--(g3) (g3)--(g4) (g4)--(g5) (g5)--(g6) (g6)--(g7) (g7)--(g8) (g8)--(g9) (g5)--(f1);
	\end{tikzpicture}
	} 
	 \label{eq:magQuiv_E2family_1st_transition}
\end{align}
The Hasse diagram is
\begin{equation}
\raisebox{-.5\height}{
\begin{tikzpicture}
    \node[hasse] (1) at (0,0) {};
    \node[hasse] (2) at (0,-1) {};
    \node[hasse] (3) at (0,-2) {};
    \node[hasse] (4) at (0,-3) {};
    \node[hasse] (5) at (0,-4) {};
    \node[hasse] (6) at (0,-5) {};
    \draw (1)--(2)--(3)--(4) (5)--(6);
    \node at (0.4,-0.5) {$a_1$};
    \node at (0.8,-1.5) {$d_{3}=a_3$};
    \node at (0.4,-2.5) {$d_{5}$};
    \node at (0.6,-4.5) {$d_{2k-1}$};
    \node at (0,-3.5) {$\vdots$};
\end{tikzpicture}
}
\end{equation}
which matches the Hasse diagram \cite[Tab.\ 26]{Bourget:2019aer} obtained from the unitary magnetic quiver \ref{eq:magQuiv_Sp_O7_infinite_Nf_small_odd}. For $k=1$, the Hasse diagram reduces to a single $e_2$ transition. 
\subsection{\texorpdfstring{$E_1$}{E1} family}
Next, for $\sprm(k)$ with $N_f=2k-2$ flavours at infinite coupling, the brane web is provided in \eqref{eq:web_Spk_Nf_Higgs_even_infinite_1}. To open up the minimal amount of Coulomb branch directions, one finds
\begin{equation}
\raisebox{-.5\height}{
    \begin{tikzpicture}
    \node[sev] (1) at (-6,2) {};
    \node[sev] (2) at (6,2) {};
    \node[sev] (3) at (-6,-2) {};
    \node[sev] (4) at (6,-2) {};
    \node[sev] (5) at (-3,0) {};
    \node[sev] (6) at (3,0) {};
    \draw[blue] (1)--(-3,1)--(3,1)--(2) (-3,1)--(-1,0)--(-3,-1)--(3) (-3,-1)--(3,-1)--(4) (3,-1)--(1,0)--(3,1);
    \draw[dotted] (-1,0)--(1,0);
    \draw[transform canvas={yshift=4pt}] (5)--(6);
    \draw[transform canvas={yshift=-4pt}] (5)--(6);
    \node at (0,0.5) {$\scriptstyle{k{-}1}$};
    \draw[dashed, red, transform canvas={yshift=-1pt}] (5)--(-4,0) (6)--(4,0);
    \draw (5)--(-4,0) (6)--(4,0);
    \draw[transform canvas={yshift=4pt}] (5)--(-4,0) (6)--(4,0);
    \draw[transform canvas={yshift=-4pt}] (5)--(-4,0) (6)--(4,0);
    \node at (-3.5,0.5) {$\scriptstyle{k{-}2}$};
    \node at (3.5,0.5) {$\scriptstyle{k{-}2}$};
    \node at (4.5,0) {$\cdots$};
    \node at (-4.5,0) {$\cdots$};
    \node at (3.5,1.8) {$\scriptstyle{(3,1)}$};
    \node at (-3.3,1.8) {$\scriptstyle{(3,-1)}$};
    \end{tikzpicture}
    }
    \label{eq:web_E1family_1st_transition}
\end{equation}
where the blue brane subweb corresponds to the $E_1$ configuration \eqref{eq:web_Sp1_0_infinite}, with magnetic quiver \eqref{eq:magQuiv_Sp1_0_infinite}. The magnetic quiver  for \eqref{eq:web_E1family_1st_transition} is
\begin{align}
 \raisebox{-.5\height}{
 	\begin{tikzpicture}
 	\tikzset{node distance = 1cm};
	\tikzstyle{gauge} = [circle, draw,inner sep=2.5pt];
	\tikzstyle{flavour} = [regular polygon,regular polygon sides=4,inner 
sep=2.5pt, draw];
	\node (g1) [gauge,label=below:{\dd{1}}] {};
	\node (g2) [gauge,right of=g1,label=below:{\cc{1}}] {};
	\node (g3) [right of=g2] {$\cdots$};
	\node (g4) [gauge,right of=g3,label=below:{\cc{k{-}2}}] {};
	\node (g5) [gauge,right of=g4,label=below:{\dd{k{-}1}}] {};
	\node (g6) [gauge,right of=g5,label=below:{\cc{k{-}2}}] {};
	\node (g7) [right of=g6] {$\cdots$};
	\node (g8) [gauge,right of=g7,label=below:{\cc{1}}] {};
	\node (g9) [gauge,right of=g8,label=below:{\dd{1}}] {};
	{\color{blue}\node (f1) [flavour,above of=g5,label=above:{\cc{1}}] {};}
	\draw (g1)--(g2) (g2)--(g3) (g3)--(g4) (g4)--(g5) (g5)--(g6) (g6)--(g7) (g7)--(g8) (g8)--(g9) (g5)--(f1);
	\end{tikzpicture}
	} 
	 \label{eq:magQuiv_E1family_1st_transition}
\end{align}
However, the problem lies in identifying the second cone from the brane web with \Of\ plane such that only a part of the Hasse diagram is visible. Nevertheless, the missing transition, denoted in \textcolor{red}{red}, can be seen from the O$7^-$ construction of Section \ref{sec:O7_plane}. Then, the full Hasse diagram is
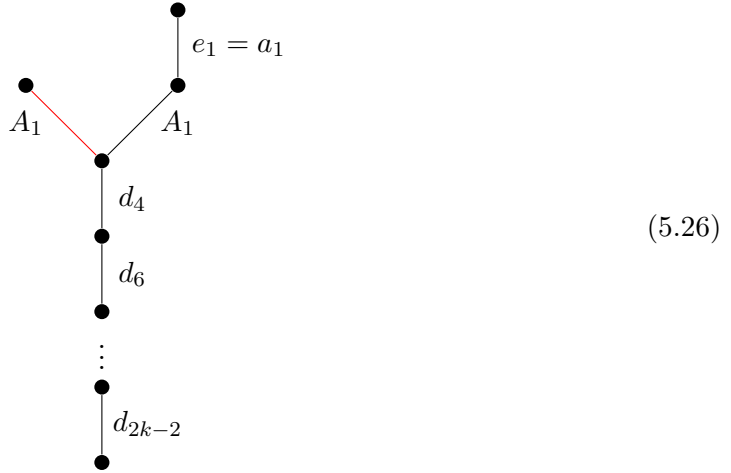
\begin{equation}
\raisebox{-.5\height}{
\begin{tikzpicture}
    \node[hasse] (-1) at (1,1) {};
    \node[hasse] (0) at (-1,0) {};
    \node[hasse] (1) at (1,0) {};
    \node[hasse] (2) at (0,-1) {};
    \node[hasse] (3) at (0,-2) {};
    \node[hasse] (4) at (0,-3) {};
    \node[hasse] (5) at (0,-4) {};
    \node[hasse] (6) at (0,-5) {};
    \draw (-1)--(1)--(2)--(3)--(4) (5)--(6);
    \draw[red] (0)--(2);
    \node at (1.8,0.5) {$e_1=a_1$};
    \node at (1,-0.5) {$A_1$};
    \node at (-1,-0.5) {$A_1$};
    \node at (0.4,-1.5) {$d_{4}$};
    \node at (0.4,-2.5) {$d_{6}$};
    \node at (0.6,-4.5) {$d_{2k-2}$};
    \node at (0,-3.5) {$\vdots$};
\end{tikzpicture}
}
\end{equation}
which matches the Hasse diagram  \cite[Tab.\ 6]{Bourget:2019aer} obtained from the unitary magnetic quivers \eqref{eq:magQuiv_Sp_O7_infinite_Nf_small_even_1} and \eqref{eq:magQuiv_Sp_O7_infinite_Nf_small_even_2}. The \textcolor{red}{red} transition stems from \eqref{eq:magQuiv_Sp_O7_infinite_Nf_small_even_1}.
Note that the two $A_1$ transitions above the $d_4$ can be viewed as a $d_2=A_1\cup A_1$. For $k=2$ the Hasse diagram consists only of $A_1\cup A_1$ with the $e_1$ transition on one cone.
\subsection{\texorpdfstring{$E_{\leq0}$}{El0} families}
For the top transitions in the Higgs branch Hasse diagram of $\sprm(k)$ with $N_f<2k-2$ fundamental flavours at infinite coupling one does a similar transition on the brane web, \eqref{eq:web_Spk_Nf_Higgs_even_infinite_1} or \eqref{eq:web_Spk_Nf_Higgs_odd_infinite},  as for the $E_2$ or $E_1$ family. Thus, one finds a $A_{k-N_f/2}$ transition for $N_f$ even and $A_{k-(N_f-1)/2}$ transition for $N_f$ odd. In the case of $N_f$ even, only one cone is visible, the second cone is denoted in \textcolor{red}{red} as above. In total, one obtains the following Hasse diagram:
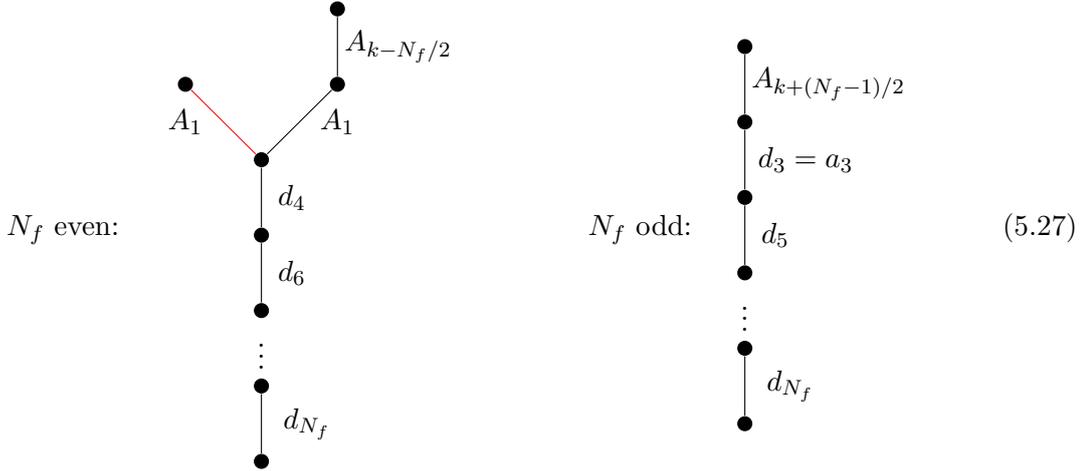
\begin{align}
N_f \text{ even:}\quad
\raisebox{-.5\height}{
\begin{tikzpicture}
    \node[hasse] (-1) at (1,1) {};
    \node[hasse] (0) at (-1,0) {};
    \node[hasse] (1) at (1,0) {};
    \node[hasse] (2) at (0,-1) {};
    \node[hasse] (3) at (0,-2) {};
    \node[hasse] (4) at (0,-3) {};
    \node[hasse] (5) at (0,-4) {};
    \node[hasse] (6) at (0,-5) {};
    \draw (-1)--(1)--(2)--(3)--(4) (5)--(6);
    \draw[red] (0)--(2);
    \node at (1.8,0.5) {$A_{k-N_f/2}$};
    \node at (1,-0.5) {$A_1$};
    \node at (-1,-0.5) {$A_1$};
    \node at (0.4,-1.5) {$d_{4}$};
    \node at (0.4,-2.5) {$d_{6}$};
    \node at (0.6,-4.5) {$d_{N_f}$};
    \node at (0,-3.5) {$\vdots$};
\end{tikzpicture}
}
\qquad \qquad
N_f \text{ odd:}\quad
\raisebox{-.5\height}{
\begin{tikzpicture}
    \node[hasse] (1) at (0,0) {};
    \node[hasse] (2) at (0,-1) {};
    \node[hasse] (3) at (0,-2) {};
    \node[hasse] (4) at (0,-3) {};
    \node[hasse] (5) at (0,-4) {};
    \node[hasse] (6) at (0,-5) {};
    \draw (1)--(2)--(3)--(4) (5)--(6);
    \node at (1.1,-0.5) {$A_{k+(N_f-1)/2}$};
    \node at (0.8,-1.5) {$d_{3}=a_3$};
    \node at (0.4,-2.5) {$d_{5}$};
    \node at (0.6,-4.5) {$d_{N_f}$};
    \node at (0,-3.5) {$\vdots$};
\end{tikzpicture}
}
\end{align}
which matches the Hasse diagram  \cite[Tab.\ 6]{Bourget:2019aer} obtained from 
the unitary magnetic quivers \eqref{eq:magQuiv_Sp_O7_infinite_Nf_small_even_1} 
and \eqref{eq:magQuiv_Sp_O7_infinite_Nf_small_even_2} for $N_f$ even, and 
\eqref{eq:magQuiv_Sp_O7_infinite_Nf_small_odd} for $N_f$ odd.
%
%
%
\section{Linear orthosymplectic quiver theories}
\label{sec:quiver_theories}
Instead of a theory with a single gauge group, one can construct quiver gauge theories with alternating orthogonal and symplectic gauge groups. Consider, for instance, the following 5-brane web
    \begin{align}
 \raisebox{-.5\height}
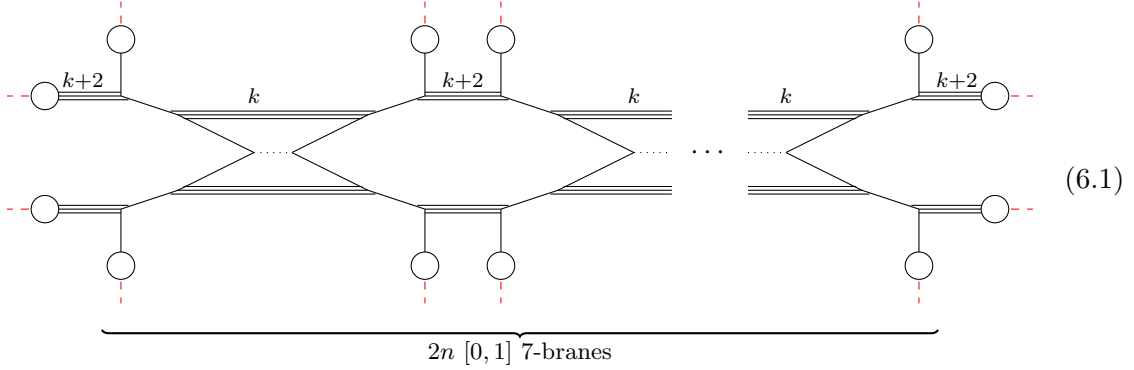
{
    \begin{tikzpicture}
    \OPlus{0.5,0}{1,0}
    \Dbrane{0.5,0}{-0.5,0.5}
    \Dbrane{1,0}{2,0.5}
    \Dbrane{-0.5,0.5}{2,0.5}
    \Dbrane{-0.6,0.55}{2.1,0.55}
    \Dbrane{-0.425,0.45}{1.925,0.45}
    \Dbrane{-0.5,0.5}{-1.25,0.75}
    \Dbrane{2,0.5}{2.75,0.75}
    \Dbrane{-1.25,0.75}{-2.25,0.75}
    \Dbrane{-1.25,0.8}{-2.25,0.8}
    \Dbrane{-1.15,0.7}{-2.25,0.7}
    \Dbrane{-1.25,0.75}{-1.25,1.5}
    \Dbrane{2.75,0.75}{2.75,1.5}
    \Dbrane{2.75,0.75}{3.75,0.75}
    \Dbrane{2.75,0.8}{3.75,0.8}
    \Dbrane{2.65,0.7}{3.85,0.7}
    \Dbrane{3.75,0.75}{4.5,0.5}
    \Dbrane{4.5,0.5}{5.5,0}
    \OPlus{5.5,0}{6,0}
    \Dbrane{4.5,0.5}{6,0.5}
    \Dbrane{4.4,0.55}{6,0.55}
    \Dbrane{4.575,0.45}{6,0.45}
    \Dbrane{3.75,0.75}{3.75,1.5}
    \draw (6.5,0) node {$\cdots$};
    \OPlus{7,0}{7.5,0}
    \Dbrane{7.5,0}{8.5,0.5}
    \Dbrane{8.5,0.5}{9.25,0.75}
    \Dbrane{9.25,0.75}{9.25,1.5}
    \Dbrane{7,0.5}{8.5,0.5}
    \Dbrane{7,0.55}{8.6,0.55}
    \Dbrane{7,0.45}{8.425,0.45}
    \Dbrane{9.25,0.75}{10.25,0.75}
    \Dbrane{9.25,0.8}{10.25,0.8}
    \Dbrane{9.15,0.7}{10.25,0.7}
    \MonoCut{-2.25,0.75}{-2.75,0.75}
    \MonoCut{10.25,0.75}{10.75,0.75}
    \MonoCut{-1.25,1.5}{-1.25,2}
    \MonoCut{2.75,1.5}{2.75,2}
    \MonoCut{3.75,1.5}{3.75,2}
    \MonoCut{9.25,1.5}{9.25,2}
    \SevenB{-2.25,0.75}
    \SevenB{10.25,0.75}
    \SevenB{-1.25,1.5}
    \SevenB{2.75,1.5}
    \SevenB{3.75,1.5}
    \SevenB{9.25,1.5}
    \begin{scope}[yscale=-1,xscale=1]
    \Dbrane{0.5,0}{-0.5,0.5}
    \Dbrane{1,0}{2,0.5}
    \Dbrane{-0.5,0.5}{2,0.5}
    \Dbrane{-0.6,0.55}{2.1,0.55}
    \Dbrane{-0.425,0.45}{1.925,0.45}
    \Dbrane{-0.5,0.5}{-1.25,0.75}
    \Dbrane{2,0.5}{2.75,0.75}
    \Dbrane{-1.25,0.75}{-2.25,0.75}
    \Dbrane{-1.25,0.8}{-2.25,0.8}
    \Dbrane{-1.15,0.7}{-2.25,0.7}
    \Dbrane{-1.25,0.75}{-1.25,1.5}
    \Dbrane{2.75,0.75}{2.75,1.5}
    \Dbrane{2.75,0.75}{3.75,0.75}
    \Dbrane{2.75,0.8}{3.75,0.8}
    \Dbrane{2.65,0.7}{3.85,0.7}
    \Dbrane{3.75,0.75}{4.5,0.5}
    \Dbrane{4.5,0.5}{5.5,0}
    \Dbrane{4.5,0.5}{6,0.5}
    \Dbrane{4.4,0.55}{6,0.55}
    \Dbrane{4.575,0.45}{6,0.45}
    \Dbrane{3.75,0.75}{3.75,1.5}
    \Dbrane{7.5,0}{8.5,0.5}
    \Dbrane{8.5,0.5}{9.25,0.75}
    \Dbrane{9.25,0.75}{9.25,1.5}
    \Dbrane{7,0.5}{8.5,0.5}
    \Dbrane{7,0.55}{8.6,0.55}
    \Dbrane{7,0.45}{8.425,0.45}
    \Dbrane{9.25,0.75}{10.25,0.75}
    \Dbrane{9.25,0.8}{10.25,0.8}
    \Dbrane{9.15,0.7}{10.25,0.7}
    \MonoCut{-2.25,0.75}{-2.75,0.75}
    \MonoCut{10.25,0.75}{10.75,0.75}
    \MonoCut{-1.25,1.5}{-1.25,2}
    \MonoCut{2.75,1.5}{2.75,2}
    \MonoCut{3.75,1.5}{3.75,2}
    \MonoCut{9.25,1.5}{9.25,2}
    \SevenB{-2.25,0.75}
    \SevenB{10.25,0.75}
    \SevenB{-1.25,1.5}
    \SevenB{2.75,1.5}
    \SevenB{3.75,1.5}
    \SevenB{9.25,1.5}
    \end{scope}
    \draw (-1.75,0.95) node {$\scriptstyle{k{+}2}$};
    \draw (0.5,0.75) node {$\scriptstyle{k}$};
    \draw (3.25,0.95) node {$\scriptstyle{k{+}2}$};
    \draw (5.5,0.75) node {$\scriptstyle{k}$};
    \draw (7.5,0.75) node {$\scriptstyle{k}$};
    \draw (9.75,0.95) node {$\scriptstyle{k{+}2}$};
    \draw[decoration={brace,mirror,raise=10pt},decorate,thick]
  (-1.5,-2) -- node[below=10pt] {\footnotesize{$2n$ $[0,1]$ 7-branes} } (9.5,-2);
    \end{tikzpicture}
}
\label{eq:web_SO-Sp_quiver}
    \end{align}
which gives rise to the following 5d $\Ncal=1$ quiver gauge theory
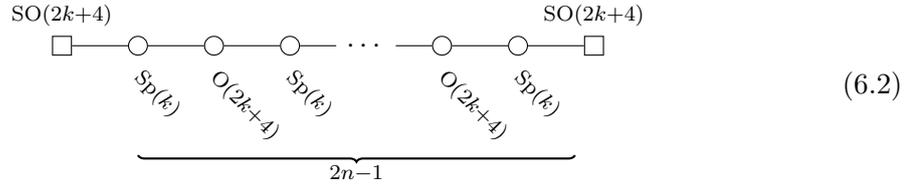
\begin{align}
 \raisebox{-.5\height}{
 	\begin{tikzpicture}
 	\tikzset{node distance = 1cm};
	\tikzstyle{gauge} = [circle, draw,inner sep=2.5pt];
	\tikzstyle{flavour} = [regular polygon,regular polygon sides=4,inner 
sep=2.5pt, draw];
	\node (g1) [gauge,label={[rotate=-45]below right:{$\scriptstyle{\sprm(k)}$}}] {};
	\node (g2) [gauge,right of=g1,label={[rotate=-45]below right:{$\scriptstyle{\orm(2k{+}4)}$}}] {};
	\node (g3) [gauge,right of=g2,label={[rotate=-45]below right:{$\scriptstyle{\sprm(k)}$}}] {};
	\node (g4) [right of=g3] {$\cdots$};
	\node (g5) [gauge,right of=g4,label={[rotate=-45]below right:{$\scriptstyle{\orm(2k{+}4)}$}}] {};
	\node (g6) [gauge,right of=g5,label={[rotate=-45]below right:{$\scriptstyle{\sprm(k)}$}}] {};
	\node (f1) [flavour,left of=g1,label=above:{$\scriptstyle{\sorm(2k{+}4)}$}] {};
	\node (f6) [flavour,right of=g6,label=above:{$\scriptstyle{\sorm(2k{+}4)}$}] {};
	\draw (g1)--(g2) (g2)--(g3) (g3)--(g4) (g4)--(g5) (g5)--(g6)  (f1)--(g1) (f6)--(g6);
	  \draw[decoration={brace,mirror,raise=10pt},decorate,thick]
  (0,-1.1) -- node[below=10pt] {$\scriptstyle{2n-1}$ } (5.75,-1.1);
	\end{tikzpicture}
	} 
	\label{eq:quiver_SO-Sp}
\end{align}
The theory \eqref{eq:quiver_SO-Sp} gives rise to various different Higgs branches. Given that there are $2n-1$ gauge couplings, any of these can either be finite or infinite. In the following, the finite coupling phase, some intermediate, and the fixed point phase are detailed.
\paragraph{Finite coupling.}
Starting from the brane web \eqref{eq:web_SO-Sp_quiver}, the finite coupling Higgs branch phase is accessible as detailed above: aligning flavour and gauge branes on the orientifold, such that the half $[1,0]$ 7-branes merge with their mirror images. The resulting physical $[1,0]$ 7-branes can split along the orientifold and one obtains
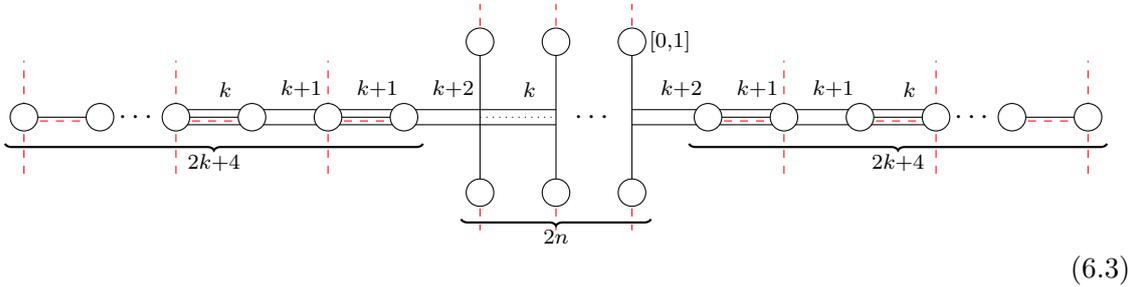
\begin{align}
 \raisebox{-.5\height}{
    \begin{tikzpicture}
    \OMinusTilde{-7,0}{-6,0}
    \MonoCut{-7,-0.05}{-6,-0.05}
    \DfiveOMinusTilde{$k$}{-5}
    \MonoCut{-5,-0.05}{-4,-0.05}
    \DfiveOMinus{$k{+}1$}{-4}
    \DfiveOMinusTilde{$k{+}1$}{-3}
    \MonoCut{-3,-0.05}{-2,-0.05}
    \DfiveOMinus{$k{+}2$}{-2}
    \DfiveOPlus{$k$}{-1}
    \DfiveOMinus{$k{+}2$}{1}
    \DfiveOMinusTilde{$k{+}1$}{2}
    \MonoCut{2,-0.05}{3,-0.05}
    \DfiveOMinus{$k{+}1$}{3}
    \DfiveOMinusTilde{$k$}{4}
    \MonoCut{4,-0.05}{5,-0.05}
    \OMinusTilde{6,0}{7,0}
    \MonoCut{6,-0.05}{7,-0.05}
    \draw (-5.5,0) node {$\cdots$};
    \draw (5.5,0) node {$\cdots$};
    \Dbrane{-1,-1}{-1,1}
    \Dbrane{0,-1}{0,1}
    \Dbrane{1,-1}{1,1}
    \draw (0.5,0) node {$\cdots$};
    \draw (1.5,1) node {$\scriptstyle{[0,1]}$};
    \MonoCut{-7,-0.75}{-7,0.75}
    \MonoCut{-5,-0.75}{-5,0.75}
    \MonoCut{-3,-0.75}{-3,0.75}
    \MonoCut{3,-0.75}{3,0.75}
    \MonoCut{5,-0.75}{5,0.75}
    \MonoCut{7,-0.75}{7,0.75}
    \MonoCut{-1,1}{-1,1.5}
    \MonoCut{-1,-1}{-1,-1.5}
    \MonoCut{0,1}{0,1.5}
    \MonoCut{0,-1}{0,-1.5}
    \MonoCut{1,1}{1,1.5}
    \MonoCut{1,-1}{1,-1.5}
    \SevenB{-1,1}
    \SevenB{-1,-1}
    \SevenB{0,1}
    \SevenB{0,-1}
    \SevenB{1,1}
    \SevenB{1,-1}
    \SevenB{-2,0}
    \SevenB{-3,0}
    \SevenB{-4,0}
    \SevenB{-5,0}
    \SevenB{-6,0}
    \SevenB{-7,0}
    \SevenB{2,0}
    \SevenB{3,0}
    \SevenB{4,0}
    \SevenB{5,0}
    \SevenB{6,0}
    \SevenB{7,0}
    \draw[decoration={brace,mirror,raise=10pt},decorate,thick]
  (-7.25,0) -- node[below=10pt] {$\scriptstyle{2k+4}$ } (-1.75,0);
  \draw[decoration={brace,mirror,raise=10pt},decorate,thick]
  (1.75,0) -- node[below=10pt] {$\scriptstyle{2k+4}$ } (7.25,0);
    \draw[decoration={brace,mirror,raise=10pt},decorate,thick]
  (-1.25,-1) -- node[below=10pt] {$\scriptstyle{2n}$ } (1.25,-1);
    \end{tikzpicture}
}
\label{eq:web_SO-Sp_Higgs}
\end{align}
The classical Higgs branch of \eqref{eq:quiver_SO-Sp} is of dimension
\begin{align}
\label{eq:Higgs_dim_quiver_SO-Sp}
\dim_{\HH} \Higgs \eqref{eq:quiver_SO-Sp} &=n_h - n_v = k(2k+7)= \dim 
\orm(2k+4) -\dim \orm(4) \,, \\
n_h &= \tfrac{1}{2} \cdot 2k\cdot (2k+4) \cdot 2n \;,\notag \\
n_v&= n\cdot \dim \sprm(k)
+ (n-1) \cdot \left( \dim\orm(2k+4) - \dim \orm(4) \right)\notag \,.
\end{align}
Note that there is no complete Higgsing for the orthogonal gauge groups, analogous to the 6d scenario of \cite[Sec.\ 2.4]{Cabrera:2019dob}.

In order to find a suitable configuration to read off the magnetic quiver, one moves the left and right outermost $(0,1)$ 5-branes through three half 7-branes each such that the brane web becomes
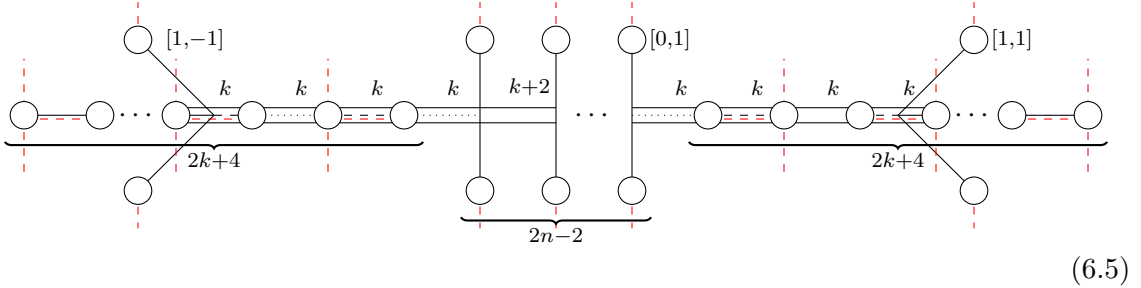
\begin{align}
 \raisebox{-.5\height}{
    \begin{tikzpicture}
    \OMinusTilde{-7,0}{-6,0}
    \MonoCut{-7,-0.05}{-6,-0.05}
    \DfiveOMinus{$k$}{-5}
    \OMinusTilde{-5,0}{-4.5,0}
    \MonoCut{-5,-0.05}{-4,-0.05}
    \OPlusTilde{-4.5,0}{-4,0}
    \DfiveOPlus{$k$}{-4}
    \DfiveOPlusTilde{$k$}{-3}
    \MonoCut{-3,-0.05}{-2,-0.05}
    \DfiveOPlus{$k$}{-2}
    \DfiveOMinus{$k{+}2$}{-1}
    \DfiveOPlus{$k$}{1}
    \DfiveOPlusTilde{$k$}{2}
    \MonoCut{2,-0.05}{3,-0.05}
    \DfiveOMinus{$k$}{3}
    \DfiveOMinus{$k$}{4}
    \OPlusTilde{4,0}{4.5,0}
    \MonoCut{4,-0.05}{5,-0.05}
    \OMinusTilde{4.5,0}{5,0}
    \OMinusTilde{6,0}{7,0}
    \MonoCut{6,-0.05}{7,-0.05}
    \draw (-5.5,0) node {$\cdots$};
    \draw (5.5,0) node {$\cdots$};
    \Dbrane{-4.5,0}{-5.5,1}
    \Dbrane{-4.5,0}{-5.5,-1}
    \draw (-4.75,1) node {$\scriptstyle{[1,-1]}$};
    \Dbrane{4.5,0}{5.5,1}
    \Dbrane{4.5,0}{5.5,-1}
    \draw (6,1) node {$\scriptstyle{[1,1]}$};
    \Dbrane{-1,-1}{-1,1}
    \Dbrane{0,-1}{0,1}
    \Dbrane{1,-1}{1,1}
    \draw (0.5,0) node {$\cdots$};
    \draw (1.5,1) node {$\scriptstyle{[0,1]}$};
    \MonoCut{-7,-0.75}{-7,0.75}
    \MonoCut{-5,-0.75}{-5,0.75}
    \MonoCut{-3,-0.75}{-3,0.75}
    \MonoCut{3,-0.75}{3,0.75}
    \MonoCut{5,-0.75}{5,0.75}
    \MonoCut{7,-0.75}{7,0.75}
    \MonoCut{-1,1}{-1,1.5}
    \MonoCut{-1,-1}{-1,-1.5}
    \MonoCut{0,1}{0,1.5}
    \MonoCut{0,-1}{0,-1.5}
    \MonoCut{1,1}{1,1.5}
    \MonoCut{1,-1}{1,-1.5}
    \MonoCut{-5.5,1}{-5.5,1.5}
    \MonoCut{-5.5,-1}{-5.5,-1.5}
    \MonoCut{5.5,1}{5.5,1.5}
    \MonoCut{5.5,-1}{5.5,-1.5}
    \SevenB{-1,1}
    \SevenB{-1,-1}
    \SevenB{0,1}
    \SevenB{0,-1}
    \SevenB{1,1}
    \SevenB{1,-1}
    \SevenB{-5.5,1}
    \SevenB{-5.5,-1}
    \SevenB{5.5,1}
    \SevenB{5.5,-1}
    \SevenB{-2,0}
    \SevenB{-3,0}
    \SevenB{-4,0}
    \SevenB{-5,0}
    \SevenB{-6,0}
    \SevenB{-7,0}
    \SevenB{2,0}
    \SevenB{3,0}
    \SevenB{4,0}
    \SevenB{5,0}
    \SevenB{6,0}
    \SevenB{7,0}
    \draw[decoration={brace,mirror,raise=10pt},decorate,thick]
  (-7.25,0) -- node[below=10pt] {$\scriptstyle{2k+4}$ } (-1.75,0);
  \draw[decoration={brace,mirror,raise=10pt},decorate,thick]
  (1.75,0) -- node[below=10pt] {$\scriptstyle{2k+4}$ } (7.25,0);
    \draw[decoration={brace,mirror,raise=10pt},decorate,thick]
  (-1.25,-1) -- node[below=10pt] {$\scriptstyle{2n-2}$ } (1.25,-1);
    \end{tikzpicture}
}
\label{eq:web_SO-Sp_quiver_finite}
\end{align}
The central segments deserve some comments: there are $k$ physical \Dfive\ branes that are freely moving subwebs between 7-branes. In addition, there are pairs of $(0,1)$ 5-branes that have $2$ extra \Dfive\ suspended in between; these are interpreted as residual gauge branes. In order words, these are subwebs that remain on the Coulomb branch and, as such, they contribute as flavour to the magnetic quiver
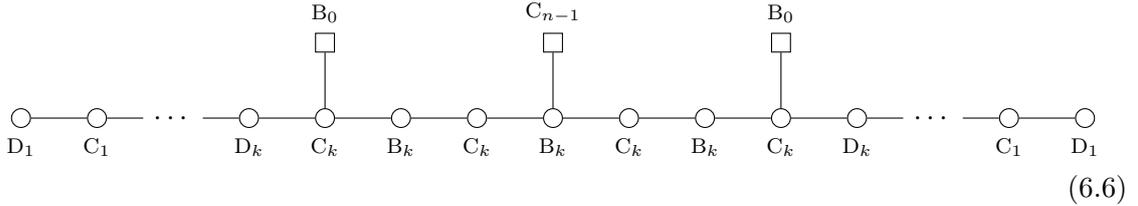
\begin{align}
 \raisebox{-.5\height}{
 	\begin{tikzpicture}
 	\tikzset{node distance = 1cm};
	\tikzstyle{gauge} = [circle, draw,inner sep=2.5pt];
	\tikzstyle{flavour} = [regular polygon,regular polygon sides=4,inner 
sep=2.5pt, draw];
	\node (g1) [gauge,label=below:{\dd{1}}] {};
	\node (g2) [gauge,right of=g1,label=below:{\cc{1}}] {};
	\node (g3) [right of=g2] {$\cdots$};
	\node (g4) [gauge,right of=g3,label=below:{\dd{k}}] {};
	\node (g5) [gauge,right of=g4,label=below:{\cc{k}}] {};
	\node (g6) [gauge,right of=g5,label=below:{\bb{k}}] {};
	\node (g7) [gauge,right of=g6,label=below:{\cc{k}}] {};
	\node (g8) [gauge,right of=g7,label=below:{\bb{k}}] {};
	\node (g9) [gauge,right of=g8,label=below:{\cc{k}}] {};
	\node (g10) [gauge,right of=g9,label=below:{\bb{k}}] {};
	\node (g11) [gauge,right of=g10,label=below:{\cc{k}}] {};
	\node (g12) [gauge,right of=g11,label=below:{\dd{k}}] {};
	\node (g13) [right of=g12] {$\cdots$};
	\node (g14) [gauge,right of=g13,label=below:{\cc{1}}] {};
	\node (g15) [gauge,right of=g14,label=below:{\dd{1}}] {};
	\node (f5) [flavour,above of=g5,label=above:{\bb{0}}] {};
	\node (f11) [flavour,above of=g11,label=above:{\bb{0}}] {};
	\node(f8) [flavour,above of=g8,label=above:{\cc{n-1}}] {};
	\draw (g1)--(g2) (g2)--(g3) (g3)--(g4) (g4)--(g5) (g5)--(g6) (g6)--(g7) (g7)--(g8) (g8)--(g9) (g9)--(g10) (g10)--(g11) (g11)--(g12) (g12)--(g13) (g13)--(g14) (g14)--(g15) (g5)--(f5) (g11)--(f11)  (g8)--(f8);
	\end{tikzpicture}
	} 
	\label{eq:magQuiv_SO-Sp_finite}
\end{align}
for which one can compute
\begin{subequations}
\label{eq:results_SO-Sp_finite}
\begin{align}
    \dim_{\HH} \Coulomb \eqref{eq:magQuiv_SO-Sp_finite} &= \dim\orm (2k+4) -\dim\orm (4) \,, \\
    G &= \sorm(2k+4)\times \sorm(2k+4) \,.
\end{align}
\end{subequations}
The global symmetry is a product because the central $\balg_{k}$ node is not 
balanced, see Appendix \ref{app:Coulomb_branch}. Compared to the classical Higgs 
branch of \eqref{eq:quiver_SO-Sp}, the magnetic quiver 
\eqref{eq:magQuiv_SO-Sp_finite} correctly reproduces dimension 
\eqref{eq:Higgs_dim_quiver_SO-Sp} and the non-abelian global symmetry, but the way the magnetic quiver is read from the brane system is not sensitive to the possibility of multiple cones in the Higgs branch.
Moreover, note that this magnetic quiver is similar to the 6d case \cite[Sec.\ 2.4]{Cabrera:2019dob}.
\paragraph{One infinite gauge coupling.}
Starting from the finite coupling phase, one can now inquire the effects of turning a single gauge coupling to infinity. Inspecting the brane web \eqref{eq:web_SO-Sp_quiver_finite}, one notices there are two potentially different transitions that make two $(0,1)$ 5-brane subwebs become coincident. 
\begin{compactenum}[(i)]
\item Two $(0,1)$ 5-brane subwebs on an \Ofp\ plane, where the $k$ physical \Dfive\ branes are suspended between 7-branes.
\item Two $(0,1)$ 5-brane subwebs on an \Ofm\ plane, where two out of the $k+2$ physical \Dfive\ branes are suspended between the \NS\ branes, forming the residual $O(4)$ gauge group.
\end{compactenum}
To begin with, consider the first case, i.e.\ the brane web is
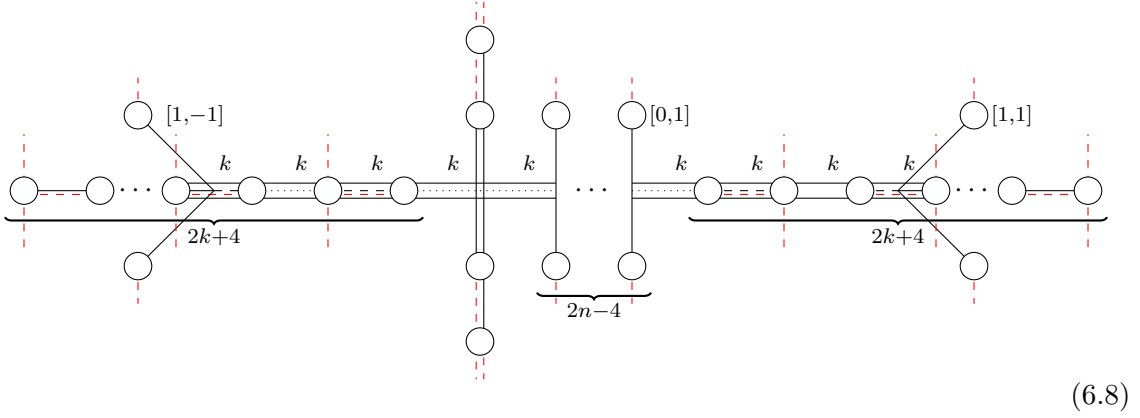
\begin{align}
 \raisebox{-.5\height}{
    \begin{tikzpicture}
    \OMinusTilde{-7,0}{-6,0}
    \MonoCut{-7,-0.05}{-6,-0.05}
    \DfiveOMinus{$k$}{-5}
    \OMinusTilde{-5,0}{-4.5,0}
    \MonoCut{-5,-0.05}{-4,-0.05}
    \OPlusTilde{-4.5,0}{-4,0}
    \DfiveOPlus{$k$}{-4}
    \DfiveOPlusTilde{$k$}{-3}
    \MonoCut{-3,-0.05}{-2,-0.05}
    \DfiveOPlus{$k$}{-2}
    \DfiveOPlus{$k$}{-1}
    \DfiveOPlus{$k$}{1}
    \DfiveOPlusTilde{$k$}{2}
    \MonoCut{2,-0.05}{3,-0.05}
    \DfiveOMinus{$k$}{3}
    \DfiveOMinus{$k$}{4}
    \OPlusTilde{4,0}{4.5,0}
    \MonoCut{4,-0.05}{5,-0.05}
    \OMinusTilde{4.5,0}{5,0}
    \OMinusTilde{6,0}{7,0}
    \MonoCut{6,-0.05}{7,-0.05}
    \draw (-5.5,0) node {$\cdots$};
    \draw (5.5,0) node {$\cdots$};
    \Dbrane{-4.5,0}{-5.5,1}
    \Dbrane{-4.5,0}{-5.5,-1}
    \draw (-4.75,1) node {$\scriptstyle{[1,-1]}$};
    \Dbrane{4.5,0}{5.5,1}
    \Dbrane{4.5,0}{5.5,-1}
    \draw (6,1) node {$\scriptstyle{[1,1]}$};
    \Dbrane{-1.05,-1}{-1.05,1}
    \Dbrane{-0.95,-2}{-0.95,2}
    \Dbrane{0,-1}{0,1}
    \Dbrane{1,-1}{1,1}
    \draw (0.5,0) node {$\cdots$};
    \draw (1.5,1) node {$\scriptstyle{[0,1]}$};
    \MonoCut{-7,-0.75}{-7,0.75}
    \MonoCut{-5,-0.75}{-5,0.75}
    \MonoCut{-3,-0.75}{-3,0.75}
    \MonoCut{3,-0.75}{3,0.75}
    \MonoCut{5,-0.75}{5,0.75}
    \MonoCut{7,-0.75}{7,0.75}
    \MonoCut{-1.05,1}{-1.05,2.5}
    \MonoCut{-1.05,-1}{-1.05,-2.5}
    \MonoCut{-0.95,2}{-0.95,2.5}
    \MonoCut{-0.95,-2}{-0.95,-2.5}
    \MonoCut{0,1}{0,1.5}
    \MonoCut{0,-1}{0,-1.5}
    \MonoCut{1,1}{1,1.5}
    \MonoCut{1,-1}{1,-1.5}
    \MonoCut{-5.5,1}{-5.5,1.5}
    \MonoCut{-5.5,-1}{-5.5,-1.5}
    \MonoCut{5.5,1}{5.5,1.5}
    \MonoCut{5.5,-1}{5.5,-1.5}
    \SevenB{-1,1}
    \SevenB{-1,-1}
    \SevenB{-1,2}
    \SevenB{-1,-2}
    \SevenB{0,1}
    \SevenB{0,-1}
    \SevenB{1,1}
    \SevenB{1,-1}
    \SevenB{-5.5,1}
    \SevenB{-5.5,-1}
    \SevenB{5.5,1}
    \SevenB{5.5,-1}
    \SevenB{-2,0}
    \SevenB{-3,0}
    \SevenB{-4,0}
    \SevenB{-5,0}
    \SevenB{-6,0}
    \SevenB{-7,0}
    \SevenB{2,0}
    \SevenB{3,0}
    \SevenB{4,0}
    \SevenB{5,0}
    \SevenB{6,0}
    \SevenB{7,0}
    \draw[decoration={brace,mirror,raise=10pt},decorate,thick]
  (-7.25,0) -- node[below=10pt] {$\scriptstyle{2k+4}$ } (-1.75,0);
  \draw[decoration={brace,mirror,raise=10pt},decorate,thick]
  (1.75,0) -- node[below=10pt] {$\scriptstyle{2k+4}$ } (7.25,0);
    \draw[decoration={brace,mirror,raise=10pt},decorate,thick]
  (-0.25,-1) -- node[below=10pt] {$\scriptstyle{2n-4}$ } (1.25,-1);
    \end{tikzpicture}
}
\label{eq:web_SO-Sp_quiver_intermediate_2d_var1}
\end{align}
and the change of the magnetic quiver is apparent. The $(0,1)$ 5-branes have entered the Higgs branch and are freely moving subwebs suspended between 7-branes. Thus, \eqref{eq:magQuiv_SO-Sp_finite} changes to
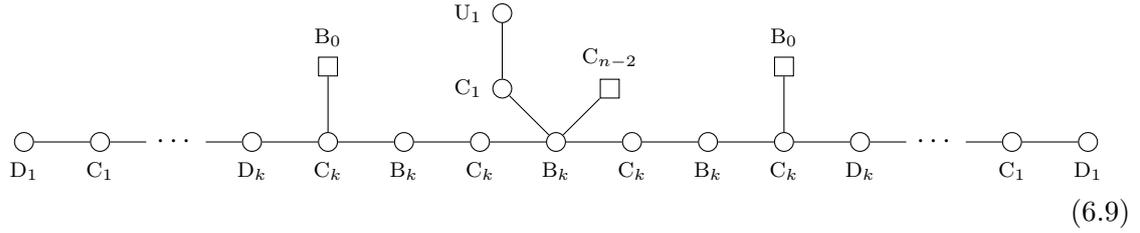
\begin{align}
 \raisebox{-.5\height}{
 	\begin{tikzpicture}
 	\tikzset{node distance = 1cm};
	\tikzstyle{gauge} = [circle, draw,inner sep=2.5pt];
	\tikzstyle{flavour} = [regular polygon,regular polygon sides=4,inner 
sep=2.5pt, draw];
	\node (g1) [gauge,label=below:{\dd{1}}] {};
	\node (g2) [gauge,right of=g1,label=below:{\cc{1}}] {};
	\node (g3) [right of=g2] {$\cdots$};
	\node (g4) [gauge,right of=g3,label=below:{\dd{k}}] {};
	\node (g5) [gauge,right of=g4,label=below:{\cc{k}}] {};
	\node (g6) [gauge,right of=g5,label=below:{\bb{k}}] {};
	\node (g7) [gauge,right of=g6,label=below:{\cc{k}}] {};
	\node (g8) [gauge,right of=g7,label=below:{\bb{k}}] {};
	\node (g9) [gauge,right of=g8,label=below:{\cc{k}}] {};
	\node (g10) [gauge,right of=g9,label=below:{\bb{k}}] {};
	\node (g11) [gauge,right of=g10,label=below:{\cc{k}}] {};
	\node (g12) [gauge,right of=g11,label=below:{\dd{k}}] {};
	\node (g13) [right of=g12] {$\cdots$};
	\node (g14) [gauge,right of=g13,label=below:{\cc{1}}] {};
	\node (g15) [gauge,right of=g14,label=below:{\dd{1}}] {};
	\node (f5) [flavour,above of=g5,label=above:{\bb{0}}] {};
	\node (f11) [flavour,above of=g11,label=above:{\bb{0}}] {};
	\node (f8) [flavour,above right of=g8,label=above:{\cc{n-2}}] {};
	\node (u1) [gauge,above left of=g8,label=left:{\cc{1}}] {};
	\node (u2) [gauge,above of=u1,label=left:{\uu{1}}] {};
	\draw (g1)--(g2) (g2)--(g3) (g3)--(g4) (g4)--(g5) (g5)--(g6) (g6)--(g7) (g7)--(g8) (g8)--(g9) (g9)--(g10) (g10)--(g11) (g11)--(g12) (g12)--(g13) (g13)--(g14) (g14)--(g15) (g5)--(f5) (g11)--(f11)  (g8)--(f8) (g8)--(u1) (u1)--(u2);
	\end{tikzpicture}
	} 
	\label{eq:magQuiv_SO-Sp_intermediate_2d_var1}
\end{align}
and the dimension has increased
\begin{align}
    \dim_{\HH} \Coulomb\eqref{eq:magQuiv_SO-Sp_intermediate_2d_var1} =2+ \dim\orm (2k+4) -\dim\orm(4)\;.
\end{align}
In terms of the electric theory, a residual $\orm(4)$ gauge symmetry vanished.

Considering now the second possibility, the brane web \eqref{eq:web_SO-Sp_quiver_finite} changes to 
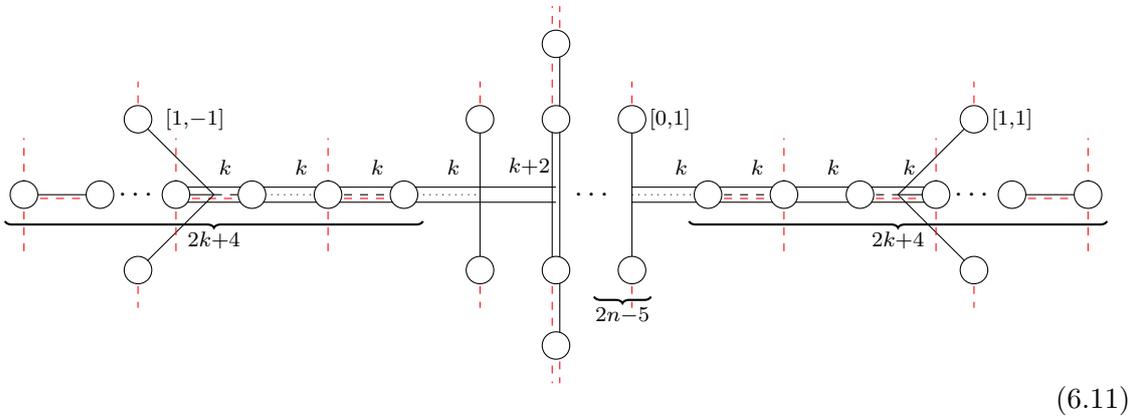
\begin{align}
 \raisebox{-.5\height}{
    \begin{tikzpicture}
    \OMinusTilde{-7,0}{-6,0}
    \MonoCut{-7,-0.05}{-6,-0.05}
    \DfiveOMinus{$k$}{-5}
    \OMinusTilde{-5,0}{-4.5,0}
    \MonoCut{-5,-0.05}{-4,-0.05}
    \OPlusTilde{-4.5,0}{-4,0}
    \DfiveOPlus{$k$}{-4}
    \DfiveOPlusTilde{$k$}{-3}
    \MonoCut{-3,-0.05}{-2,-0.05}
    \DfiveOPlus{$k$}{-2}
    \DfiveOMinus{$k{+}2$}{-1}
    \DfiveOPlus{$k$}{1}
    \DfiveOPlusTilde{$k$}{2}
    \MonoCut{2,-0.05}{3,-0.05}
    \DfiveOMinus{$k$}{3}
    \DfiveOMinus{$k$}{4}
    \OPlusTilde{4,0}{4.5,0}
    \MonoCut{4,-0.05}{5,-0.05}
    \OMinusTilde{4.5,0}{5,0}
    \OMinusTilde{6,0}{7,0}
    \MonoCut{6,-0.05}{7,-0.05}
    \draw (-5.5,0) node {$\cdots$};
    \draw (5.5,0) node {$\cdots$};
    \Dbrane{-4.5,0}{-5.5,1}
    \Dbrane{-4.5,0}{-5.5,-1}
    \draw (-4.75,1) node {$\scriptstyle{[1,-1]}$};
    \Dbrane{4.5,0}{5.5,1}
    \Dbrane{4.5,0}{5.5,-1}
    \draw (6,1) node {$\scriptstyle{[1,1]}$};
    \Dbrane{-0.05,-1}{-0.05,1}
    \Dbrane{0.05,-2}{0.05,2}
    \Dbrane{-1,-1}{-1,1}
    \Dbrane{1,-1}{1,1}
    \draw (0.5,0) node {$\cdots$};
    \draw (1.5,1) node {$\scriptstyle{[0,1]}$};
    \MonoCut{-7,-0.75}{-7,0.75}
    \MonoCut{-5,-0.75}{-5,0.75}
    \MonoCut{-3,-0.75}{-3,0.75}
    \MonoCut{3,-0.75}{3,0.75}
    \MonoCut{5,-0.75}{5,0.75}
    \MonoCut{7,-0.75}{7,0.75}
    \MonoCut{-0.05,1}{-0.05,2.5}
    \MonoCut{-0.05,-1}{-0.05,-2.5}
    \MonoCut{0.05,2}{0.05,2.5}
    \MonoCut{0.05,-2}{0.05,-2.5}
    \MonoCut{-1,1}{-1,1.5}
    \MonoCut{-1,-1}{-1,-1.5}
    \MonoCut{1,1}{1,1.5}
    \MonoCut{1,-1}{1,-1.5}
    \MonoCut{-5.5,1}{-5.5,1.5}
    \MonoCut{-5.5,-1}{-5.5,-1.5}
    \MonoCut{5.5,1}{5.5,1.5}
    \MonoCut{5.5,-1}{5.5,-1.5}
    \SevenB{0,1}
    \SevenB{0,-1}
    \SevenB{0,2}
    \SevenB{0,-2}
    \SevenB{-1,1}
    \SevenB{-1,-1}
    \SevenB{1,1}
    \SevenB{1,-1}
    \SevenB{-5.5,1}
    \SevenB{-5.5,-1}
    \SevenB{5.5,1}
    \SevenB{5.5,-1}
    \SevenB{-2,0}
    \SevenB{-3,0}
    \SevenB{-4,0}
    \SevenB{-5,0}
    \SevenB{-6,0}
    \SevenB{-7,0}
    \SevenB{2,0}
    \SevenB{3,0}
    \SevenB{4,0}
    \SevenB{5,0}
    \SevenB{6,0}
    \SevenB{7,0}
    \draw[decoration={brace,mirror,raise=10pt},decorate,thick]
  (-7.25,0) -- node[below=10pt] {$\scriptstyle{2k+4}$ } (-1.75,0);
  \draw[decoration={brace,mirror,raise=10pt},decorate,thick]
  (1.75,0) -- node[below=10pt] {$\scriptstyle{2k+4}$ } (7.25,0);
    \draw[decoration={brace,mirror,raise=10pt},decorate,thick]
  (0.5,-1) -- node[below=10pt] {$\scriptstyle{2n-5}$ } (1.25,-1);
    \end{tikzpicture}
}
\label{eq:web_SO-Sp_quiver_intermediate_2d_var2}
\end{align}
such that the new Higgs branch degrees of freedom are suspended on an \Of\ interval with a residual $\orm(4)$ gauge symmetry. Analogous to the proposal of \cite[Sec.\ 2.4]{Cabrera:2019dob}, the contribution of this gauge symmetry is argued to be a flavour in the magnetic quiver which reads as 
\begin{align}
 \raisebox{-.5\height}{
 	\begin{tikzpicture}
 	\tikzset{node distance = 1cm};
	\tikzstyle{gauge} = [circle, draw,inner sep=2.5pt];
	\tikzstyle{flavour} = [regular polygon,regular polygon sides=4,inner 
sep=2.5pt, draw];
	\node (g1) [gauge,label=below:{\dd{1}}] {};
	\node (g2) [gauge,right of=g1,label=below:{\cc{1}}] {};
	\node (g3) [right of=g2] {$\cdots$};
	\node (g4) [gauge,right of=g3,label=below:{\dd{k}}] {};
	\node (g5) [gauge,right of=g4,label=below:{\cc{k}}] {};
	\node (g6) [gauge,right of=g5,label=below:{\bb{k}}] {};
	\node (g7) [gauge,right of=g6,label=below:{\cc{k}}] {};
	\node (g8) [gauge,right of=g7,label=below:{\bb{k}}] {};
	\node (g9) [gauge,right of=g8,label=below:{\cc{k}}] {};
	\node (g10) [gauge,right of=g9,label=below:{\bb{k}}] {};
	\node (g11) [gauge,right of=g10,label=below:{\cc{k}}] {};
	\node (g12) [gauge,right of=g11,label=below:{\dd{k}}] {};
	\node (g13) [right of=g12] {$\cdots$};
	\node (g14) [gauge,right of=g13,label=below:{\cc{1}}] {};
	\node (g15) [gauge,right of=g14,label=below:{\dd{1}}] {};
	\node (f5) [flavour,above of=g5,label=above:{\bb{0}}] {};
	\node (f11) [flavour,above of=g11,label=above:{\bb{0}}] {};
	\node (f8) [flavour,above right of=g8,label=above:{\cc{n-2}}] {};
	\node (u1) [gauge,above left of=g8,label=left:{\cc{1}}] {};
	\node (u2) [gauge,above of=u1,label=left:{\uu{1}}] {};
	\node (fu) [flavour,above right of=u1,label=above:{\dd{2}}] {};
	\draw (g1)--(g2) (g2)--(g3) (g3)--(g4) (g4)--(g5) (g5)--(g6) (g6)--(g7) (g7)--(g8) (g8)--(g9) (g9)--(g10) (g10)--(g11) (g11)--(g12) (g12)--(g13) (g13)--(g14) (g14)--(g15) (g5)--(f5) (g11)--(f11)  (g8)--(f8) (g8)--(u1) (u1)--(u2) (u1)--(fu);
	\end{tikzpicture}
	} 
	\label{eq:magQuiv_SO-Sp_intermediate_2d_var2}
\end{align}
and the dimension has increased
\begin{align}
    \dim_{\HH} \Coulomb\eqref{eq:magQuiv_SO-Sp_intermediate_2d_var2} =2+ \dim \orm(2k+4) -\dim\orm(4)\;.
\end{align}
As in the first case \eqref{eq:web_SO-Sp_quiver_intermediate_2d_var1}, the web \eqref{eq:web_SO-Sp_quiver_intermediate_2d_var2} clearly displays the loss of an $\orm(4)$ residual gauge symmetry.
\paragraph{Several infinite couplings.}
Repeatedly taking couplings to infinite, in the form of \eqref{eq:web_SO-Sp_quiver_intermediate_2d_var1} or \eqref{eq:web_SO-Sp_quiver_intermediate_2d_var2}, and assuming that at most two \NS\ branes become coincident, one obtains a generalisation of the notion of \emph{bouquet} for a magnetic quiver \cite{Hanany:2018vph,Hanany:2018cgo,Hanany:2018dvd,Cabrera:2019izd,Cabrera:2019dob}. For instance, allowing $n_1$ transitions of type \eqref{eq:magQuiv_SO-Sp_intermediate_2d_var1} and $n_2$ transitions of type \eqref{eq:magQuiv_SO-Sp_intermediate_2d_var2} such that $n_1+n_2 \leq n-1$ the magnetic quiver reads
\begin{align}
 \raisebox{-.5\height}{
 	\begin{tikzpicture}
 	\tikzset{node distance = 1cm};
	\tikzstyle{gauge} = [circle, draw,inner sep=2.5pt];
	\tikzstyle{flavour} = [regular polygon,regular polygon sides=4,inner 
sep=2.5pt, draw];
	\node (g1) [gauge,label=below:{\dd{1}}] {};
	\node (g2) [gauge,right of=g1,label=below:{\cc{1}}] {};
	\node (g3) [right of=g2] {$\cdots$};
	\node (g4) [gauge,right of=g3,label=below:{\dd{k}}] {};
	\node (g5) [gauge,right of=g4,label=below:{\cc{k}}] {};
	\node (g6) [gauge,right of=g5,label=below:{\bb{k}}] {};
	\node (g7) [gauge,right of=g6,label=below:{\cc{k}}] {};
	\node (g8) [gauge,right of=g7,label=below:{\bb{k}}] {};
	\node (g9) [gauge,right of=g8,label=below:{\cc{k}}] {};
	\node (g10) [gauge,right of=g9,label=below:{\bb{k}}] {};
	\node (g11) [gauge,right of=g10,label=below:{\cc{k}}] {};
	\node (g12) [gauge,right of=g11,label=below:{\dd{k}}] {};
	\node (g13) [right of=g12] {$\cdots$};
	\node (g14) [gauge,right of=g13,label=below:{\cc{1}}] {};
	\node (g15) [gauge,right of=g14,label=below:{\dd{1}}] {};
	\node (f5) [flavour,above of=g5,label=above:{\bb{0}}] {};
	\node (f11) [flavour,above of=g11,label=above:{\bb{0}}] {};
	\node (f8) [flavour,above of=g8,label=above:{\cc{l}}] {};
	\node (b1) [gauge,right of=f5,label=left:{\cc{1}}] {};
	\node (b11) [gauge,above left of=b1,label=above:{\uu{1}}] {};
	\node (b12) [flavour,above of=b1,label=above:{\dd{2}}] {};
	\node (b2) [gauge,right of=b1,label=left:{\cc{1}}] {};
	\node (b21) [gauge,above of=b2,label=above:{\uu{1}}] {};
	\node (b22) [flavour,above right of=b2,label=above:{\dd{2}}] {};
	\draw (5.5,1.5) node {$\cdots$};
	\node (d1) [gauge,left of=f11,label=right:{\cc{1}}] {};
	\node (d11) [gauge,above of=d1,label=above:{\uu{1}}] {};
	\node (d2) [gauge,left of=d1,label=right:{\cc{1}}] {};
	\node (d21) [gauge,above of=d2,label=above:{\uu{1}}] {};
	\draw (8.5,1.5) node {$\cdots$};
	\draw (g1)--(g2) (g2)--(g3) (g3)--(g4) (g4)--(g5) (g5)--(g6) (g6)--(g7) (g7)--(g8) (g8)--(g9) (g9)--(g10) (g10)--(g11) (g11)--(g12) (g12)--(g13) (g13)--(g14) (g14)--(g15) (g5)--(f5) (g11)--(f11)  (g8)--(f8) (g8)--(b1) (b1)--(b11) (b1)--(b12) (g8)--(b2) (b2)--(b21) (b2)--(b22) (g8)--(d1) (d1)--(d11) (g8)--(d2) (d2)--(d21);
	\draw[decoration={brace,mirror,raise=10pt},decorate,thick]
  (7,2.25) -- node[above=10pt] {$\scriptstyle{n_2}$ } (4,2.25);
  \draw[decoration={brace,mirror,raise=10pt},decorate,thick]
  (9,2.25) -- node[above=10pt] {$\scriptstyle{n_1}$ } (8,2.25);
	\end{tikzpicture}
	} 
	\label{eq:magQuiv_SO-Sp_intermediate_bouquet}
\end{align}
with $l=n-1-n_1-n_2 \geq 0$.

So far, the assumption has been to have at most two coinciding \NS\ branes. In order to move towards the fixed point of \eqref{eq:quiver_SO-Sp} one needs to coincide all of the \NS\ simultaneously. As an exemplary case, consider \eqref{eq:web_SO-Sp_quiver_finite} and make the first $2j$ half \NS\ branes on the left-hand-side of the central segment coincident. Then the brane web becomes
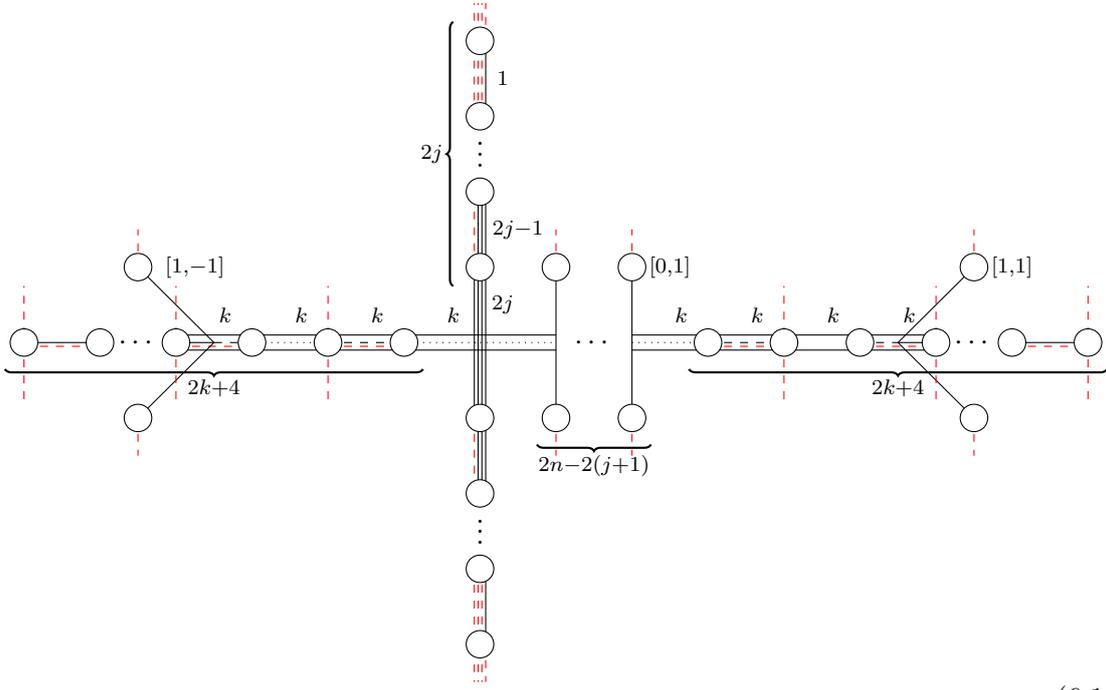
\begin{align}
 \raisebox{-.5\height}{
    \begin{tikzpicture}
    \OMinusTilde{-7,0}{-6,0}
    \MonoCut{-7,-0.05}{-6,-0.05}
    \DfiveOMinus{$k$}{-5}
    \OMinusTilde{-5,0}{-4.5,0}
    \MonoCut{-5,-0.05}{-4,-0.05}
    \OPlusTilde{-4.5,0}{-4,0}
    \DfiveOPlus{$k$}{-4}
    \DfiveOPlusTilde{$k$}{-3}
    \MonoCut{-3,-0.05}{-2,-0.05}
    \DfiveOPlus{$k$}{-2}
    \DfiveOPlus{}{-1}
    \DfiveOPlus{$k$}{1}
    \DfiveOPlusTilde{$k$}{2}
    \MonoCut{2,-0.05}{3,-0.05}
    \DfiveOMinus{$k$}{3}
    \DfiveOMinus{$k$}{4}
    \OPlusTilde{4,0}{4.5,0}
    \MonoCut{4,-0.05}{5,-0.05}
    \OMinusTilde{4.5,0}{5,0}
    \OMinusTilde{6,0}{7,0}
    \MonoCut{6,-0.05}{7,-0.05}
    \draw (-5.5,0) node {$\cdots$};
    \draw (5.5,0) node {$\cdots$};
    \Dbrane{-4.5,0}{-5.5,1}
    \Dbrane{-4.5,0}{-5.5,-1}
    \draw (-4.75,1) node {$\scriptstyle{[1,-1]}$};
    \Dbrane{4.5,0}{5.5,1}
    \Dbrane{4.5,0}{5.5,-1}
    \draw (6,1) node {$\scriptstyle{[1,1]}$};
    \Dbrane{-1.075,-1}{-1.075,1}
    \Dbrane{-1.025,-2}{-1.025,2}
    \Dbrane{-0.975,-2}{-0.975,2}
    \Dbrane{-0.925,-2}{-0.925,2}
    \Dbrane{-0.925,3}{-0.925,4}
    \Dbrane{-0.925,-3}{-0.925,-4}
    \Dbrane{0,-1}{0,1}
    \Dbrane{1,-1}{1,1}
    \draw (0.5,0) node {$\cdots$};
    \draw (1.5,1) node {$\scriptstyle{[0,1]}$};
    \draw (-1,2.6) node {$\vdots$};
    \draw (-1,-2.4) node {$\vdots$};
    \draw (-0.7,0.5) node {$\scriptstyle{2j}$};
    \draw (-0.5,1.5) node {$\scriptstyle{2j{-}1}$};
    \draw (-0.7,3.5) node {$\scriptstyle{1}$};
    \MonoCut{-7,-0.75}{-7,0.75}
    \MonoCut{-5,-0.75}{-5,0.75}
    \MonoCut{-3,-0.75}{-3,0.75}
    \MonoCut{3,-0.75}{3,0.75}
    \MonoCut{5,-0.75}{5,0.75}
    \MonoCut{7,-0.75}{7,0.75}
    \MonoCut{-1.075,1}{-1.075,2}
    \MonoCut{-1.075,-1}{-1.075,-2}
    \MonoCut{-1.075,3}{-1.075,4.5}
    \MonoCut{-1.025,3}{-1.025,4.5}
    \MonoCut{-0.975,3}{-0.975,4.5}
    \MonoCut{-0.925,4}{-0.925,4.5}
    \MonoCut{-1.075,-3}{-1.075,-4.5}
    \MonoCut{-1.025,-3}{-1.025,-4.5}
    \MonoCut{-0.975,-3}{-0.975,-4.5}
    \MonoCut{-0.925,-4}{-0.925,-4.5}
    \MonoCut{0,1}{0,1.5}
    \MonoCut{0,-1}{0,-1.5}
    \MonoCut{1,1}{1,1.5}
    \MonoCut{1,-1}{1,-1.5}
    \MonoCut{-5.5,1}{-5.5,1.5}
    \MonoCut{-5.5,-1}{-5.5,-1.5}
    \MonoCut{5.5,1}{5.5,1.5}
    \MonoCut{5.5,-1}{5.5,-1.5}
    \SevenB{-1,1}
    \SevenB{-1,-1}
    \SevenB{-1,2}
    \SevenB{-1,-2}
    \SevenB{-1,3}
    \SevenB{-1,-3}
    \SevenB{-1,4}
    \SevenB{-1,-4}
    \SevenB{0,1}
    \SevenB{0,-1}
    \SevenB{1,1}
    \SevenB{1,-1}
    \SevenB{-5.5,1}
    \SevenB{-5.5,-1}
    \SevenB{5.5,1}
    \SevenB{5.5,-1}
    \SevenB{-2,0}
    \SevenB{-3,0}
    \SevenB{-4,0}
    \SevenB{-5,0}
    \SevenB{-6,0}
    \SevenB{-7,0}
    \SevenB{2,0}
    \SevenB{3,0}
    \SevenB{4,0}
    \SevenB{5,0}
    \SevenB{6,0}
    \SevenB{7,0}
    \draw[decoration={brace,mirror,raise=10pt},decorate,thick]
  (-7.25,0) -- node[below=10pt] {$\scriptstyle{2k+4}$ } (-1.75,0);
  \draw[decoration={brace,mirror,raise=10pt},decorate,thick]
  (1.75,0) -- node[below=10pt] {$\scriptstyle{2k+4}$ } (7.25,0);
    \draw[decoration={brace,mirror,raise=10pt},decorate,thick]
  (-0.25,-1) -- node[below=10pt] {$\scriptstyle{2n-2(j+1)}$ } (1.25,-1);
  \draw[decoration={brace,mirror,raise=10pt},decorate,thick]
  (-1,4.25) -- node[left=10pt] {$\scriptstyle{2j}$ } (-1,0.75);
    \end{tikzpicture}
}
\label{eq:web_SO-Sp_quiver_intermediate_many_NS}
\end{align}
such that one now has to explain how to read off the magnetic quiver. The \Dfive\ branes suspended between half $[1,0]$ 7-branes along the orientifold are no conceptual challenge, due to the discussion above. In contrast, the half \NS\ branes between half $[0,1]$ 7-branes behave differently. The set of $2j$ identical $(0,1)$ 5-brane subwebs suspended between a half 7-brane and its mirror image is affected by the orientifold projection, because the \NS\ branes cross the \Of\ plane. Hence, this set of identical subwebs contributes a $\calg_j$ magnetic gauge node. In the adjacent segments, there is always a set of $m$  identical \NS\ branes suspended between two half $[0,1]$ 7-branes. Note, however, that these \NS\ branes do not cross the orientifold and, therefore, the associated magnetic gauge node is a $\urm(m)$. Following \cite{Cabrera:2018jxt}, the unitary nodes are connected by magnetic bi-fundamental hypermultiplets. In summary, the magnetic quiver for \eqref{eq:web_SO-Sp_quiver_intermediate_many_NS} reads
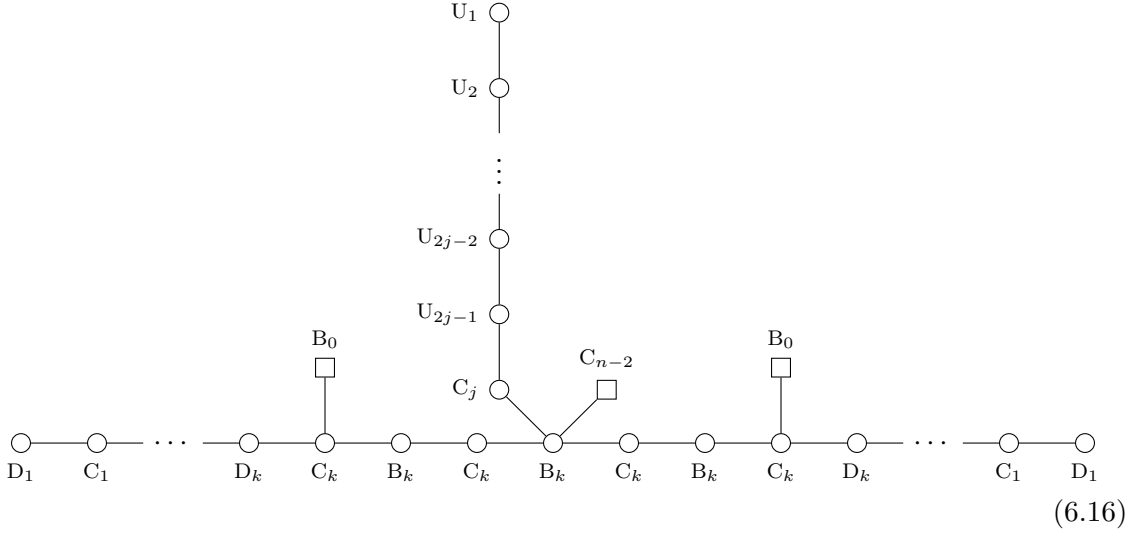
\begin{align}
 \raisebox{-.5\height}{
 	\begin{tikzpicture}
 	\tikzset{node distance = 1cm};
	\tikzstyle{gauge} = [circle, draw,inner sep=2.5pt];
	\tikzstyle{flavour} = [regular polygon,regular polygon sides=4,inner 
sep=2.5pt, draw];
	\node (g1) [gauge,label=below:{\dd{1}}] {};
	\node (g2) [gauge,right of=g1,label=below:{\cc{1}}] {};
	\node (g3) [right of=g2] {$\cdots$};
	\node (g4) [gauge,right of=g3,label=below:{\dd{k}}] {};
	\node (g5) [gauge,right of=g4,label=below:{\cc{k}}] {};
	\node (g6) [gauge,right of=g5,label=below:{\bb{k}}] {};
	\node (g7) [gauge,right of=g6,label=below:{\cc{k}}] {};
	\node (g8) [gauge,right of=g7,label=below:{\bb{k}}] {};
	\node (g9) [gauge,right of=g8,label=below:{\cc{k}}] {};
	\node (g10) [gauge,right of=g9,label=below:{\bb{k}}] {};
	\node (g11) [gauge,right of=g10,label=below:{\cc{k}}] {};
	\node (g12) [gauge,right of=g11,label=below:{\dd{k}}] {};
	\node (g13) [right of=g12] {$\cdots$};
	\node (g14) [gauge,right of=g13,label=below:{\cc{1}}] {};
	\node (g15) [gauge,right of=g14,label=below:{\dd{1}}] {};
	\node (f5) [flavour,above of=g5,label=above:{\bb{0}}] {};
	\node (f11) [flavour,above of=g11,label=above:{\bb{0}}] {};
	\node (f8) [flavour,above right of=g8,label=above:{\cc{n-2}}] {};
	\node (u1) [gauge,above left of=g8,label=left:{\cc{j}}] {};
	\node (u2) [gauge,above of=u1,label=left:{\uu{2j-1}}] {};
	\node (u3) [gauge,above of=u2,label=left:{\uu{2j-2}}] {};
	\node (u4) [above of=u3] {$\vdots$};
	\node (u5) [gauge,above of=u4,label=left:{\uu{2}}] {};
	\node (u6) [gauge,above of=u5,label=left:{\uu{1}}] {};
	\draw (g1)--(g2) (g2)--(g3) (g3)--(g4) (g4)--(g5) (g5)--(g6) (g6)--(g7) (g7)--(g8) (g8)--(g9) (g9)--(g10) (g10)--(g11) (g11)--(g12) (g12)--(g13) (g13)--(g14) (g14)--(g15) (g5)--(f5) (g11)--(f11)  (g8)--(f8) (g8)--(u1) (u1)--(u2) (u2)--(u3) (u3)--(u4) (u4)--(u5) (u5)--(u6);
	\end{tikzpicture}
	} 
	\label{eq:magQuiv_SO-Sp_intermediate_many_NS}
\end{align}
and the dimension is increased as 
\begin{align}
    \dim_{\HH} \Coulomb\eqref{eq:magQuiv_SO-Sp_intermediate_many_NS} = 2j^2+ \dim\orm(2k+4)-\dim\orm(4)\,.
\end{align}
Analogously, if instead one coincides $2j$ half \NS\ branes over a segment with a residual $\orm(4)$ gauge symmetry, the only change compared to \eqref{eq:magQuiv_SO-Sp_intermediate_many_NS} lies in an additional $\dalg_4$ flavour node for the $\calg_j$ gauge node, similar to \eqref{eq:magQuiv_SO-Sp_intermediate_2d_var2}.

\paragraph{Infinite coupling.}
To transfer the finite coupling brane web \eqref{eq:web_SO-Sp_quiver_finite} into the fixed point phase, the various $(0,1)$ 5-branes, i.e.\ \NS\ branes, have to be made coincident. So far, it has been demonstrated how to make at most $2n-2$ half \NS\ brane coincident starting from the brane web \eqref{eq:web_SO-Sp_quiver_finite}. The process proceeds via the two fundamental transitions \eqref{eq:web_SO-Sp_quiver_intermediate_2d_var1}, \eqref{eq:web_SO-Sp_quiver_intermediate_2d_var2} via various intermediate phases \eqref{eq:web_SO-Sp_quiver_intermediate_many_NS}. However, to reach the fixed point phase, the two outermost $(1,\pm1)$ 5-branes need to be transitioned to the central segment again, i.e.\ one starts from \eqref{eq:web_SO-Sp_Higgs}. Next, merging all the half \NS\ branes pairwise, the brane web becomes
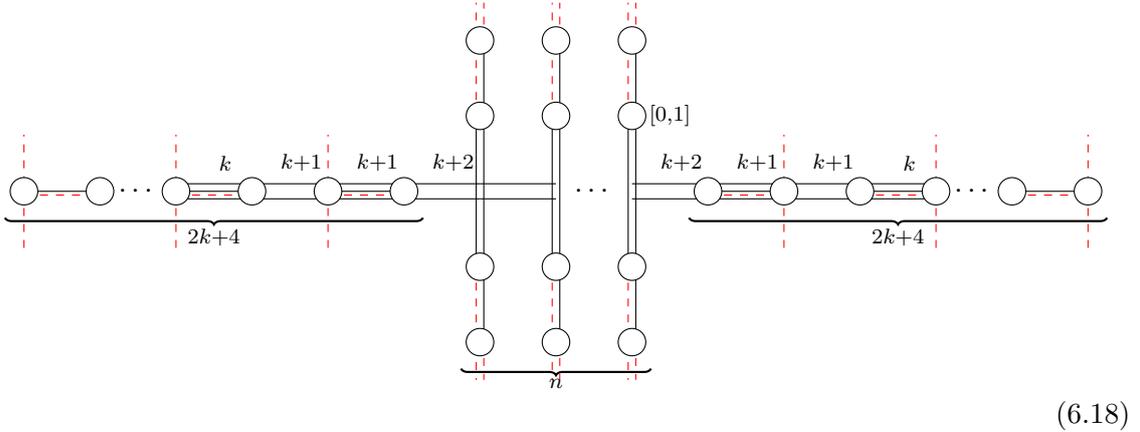
\begin{align}
 \raisebox{-.5\height}{
    \begin{tikzpicture}
    \OMinusTilde{-7,0}{-6,0}
    \MonoCut{-7,-0.05}{-6,-0.05}
    \DfiveOMinusTilde{$k$}{-5}
    \MonoCut{-5,-0.05}{-4,-0.05}
    \DfiveOMinus{$k{+}1$}{-4}
    \DfiveOMinusTilde{$k{+}1$}{-3}
    \MonoCut{-3,-0.05}{-2,-0.05}
    \DfiveOMinus{$k{+}2$}{-2}
    \DfiveOMinus{}{-1}
    \DfiveOMinus{$k{+}2$}{1}
    \DfiveOMinusTilde{$k{+}1$}{2}
    \MonoCut{2,-0.05}{3,-0.05}
    \DfiveOMinus{$k{+}1$}{3}
    \DfiveOMinusTilde{$k$}{4}
    \MonoCut{4,-0.05}{5,-0.05}
    \OMinusTilde{6,0}{7,0}
    \MonoCut{6,-0.05}{7,-0.05}
    \draw (-5.5,0) node {$\cdots$};
    \draw (5.5,0) node {$\cdots$};
    \Dbrane{-1.05,-1}{-1.05,1}
    \Dbrane{-0.95,-2}{-0.95,2}
    \Dbrane{-0.05,-1}{-0.05,1}
    \Dbrane{0.05,-2}{0.05,2}
    \Dbrane{0.95,-1}{0.95,1}
    \Dbrane{1.05,-2}{1.05,2}
    \draw (0.5,0) node {$\cdots$};
    \draw (1.5,1) node {$\scriptstyle{[0,1]}$};
    \MonoCut{-7,-0.75}{-7,0.75}
    \MonoCut{-5,-0.75}{-5,0.75}
    \MonoCut{-3,-0.75}{-3,0.75}
    \MonoCut{3,-0.75}{3,0.75}
    \MonoCut{5,-0.75}{5,0.75}
    \MonoCut{7,-0.75}{7,0.75}
    \MonoCut{-0.95,2}{-0.95,2.5}
    \MonoCut{-0.95,-2}{-0.95,-2.5}
    \MonoCut{0.05,2}{0.05,2.5}
    \MonoCut{0.05,-2}{0.05,-2.5}
    \MonoCut{1.05,2}{1.05,2.5}
    \MonoCut{1.05,-2}{1.05,-2.5}
    \MonoCut{-1.05,1}{-1.05,2.5}
    \MonoCut{-1.05,-1}{-1.05,-2.5}
    \MonoCut{-0.05,1}{-0.05,2.5}
    \MonoCut{-0.05,-1}{-0.05,-2.5}
    \MonoCut{0.95,1}{0.95,2.5}
    \MonoCut{0.95,-1}{0.95,-2.5}
    \SevenB{-1,1}
    \SevenB{-1,-1}
    \SevenB{0,1}
    \SevenB{0,-1}
    \SevenB{1,1}
    \SevenB{1,-1}
    \SevenB{-1,2}
    \SevenB{-1,-2}
    \SevenB{0,2}
    \SevenB{0,-2}
    \SevenB{1,2}
    \SevenB{1,-2}
    \SevenB{-2,0}
    \SevenB{-3,0}
    \SevenB{-4,0}
    \SevenB{-5,0}
    \SevenB{-6,0}
    \SevenB{-7,0}
    \SevenB{2,0}
    \SevenB{3,0}
    \SevenB{4,0}
    \SevenB{5,0}
    \SevenB{6,0}
    \SevenB{7,0}
    \draw[decoration={brace,mirror,raise=10pt},decorate,thick]
  (-7.25,0) -- node[below=10pt] {$\scriptstyle{2k+4}$ } (-1.75,0);
  \draw[decoration={brace,mirror,raise=10pt},decorate,thick]
  (1.75,0) -- node[below=10pt] {$\scriptstyle{2k+4}$ } (7.25,0);
    \draw[decoration={brace,mirror,raise=10pt},decorate,thick]
  (-1.25,-2) -- node[below=10pt] {$\scriptstyle{n}$ } (1.25,-2);
    \end{tikzpicture}
}
\label{eq:web_SO-Sp_almost_infinite}
\end{align}
such that the magnetic quiver is straightforwardly read off 
\begin{align}
 \raisebox{-.5\height}{
 	\begin{tikzpicture}
 	\tikzset{node distance = 1cm};
	\tikzstyle{gauge} = [circle, draw,inner sep=2.5pt];
	\tikzstyle{flavour} = [regular polygon,regular polygon sides=4,inner 
sep=2.5pt, draw];
	\node (g1) [gauge,label=below:{\dd{1}}] {};
	\node (g2) [gauge,right of=g1,label=below:{\cc{1}}] {};
	\node (g3) [right of=g2] {$\cdots$};
	\node (g4) [gauge,right of=g3,label=below:{\dd{k+1}}] {};
	\node (g5) [gauge,right of=g4,label=below:{\cc{k+1}}] {};
	\node (g6) [gauge,right of=g5,label=below:{\dd{k+2}}] {};
	\node (g7) [gauge,right of=g6,label=below:{\cc{k+1}}] {};
	\node (g8) [gauge,right of=g7,label=below:{\dd{k+1}}] {};
	\node (g9) [right of=g8] {$\cdots$};
	\node (g10) [gauge,right of=g9,label=below:{\cc{1}}] {};
	\node (g11) [gauge,right of=g10,label=below:{\dd{1}}] {};
	\node (u1) [gauge,above left of=g6,label=left:{\cc{1}}] {};
	\node (u2) [gauge,above of=u1,label=left:{\uu{1}}] {};
	\node (f1) [above of=g6] {$\cdots$};
	\node (d1) [gauge,above right of=g6,label=right:{\cc{1}}] {};
	\node (d2) [gauge,above of=d1,label=right:{\uu{1}}] {};
	\draw (g1)--(g2) (g2)--(g3) (g3)--(g4) (g4)--(g5) (g5)--(g6) (g6)--(g7) (g7)--(g8) (g8)--(g9) (g9)--(g10) (g10)--(g11) (g6)--(u1) (u1)--(u2) (g6)--(d1) (d1)--(d2);
	\draw[decoration={brace,mirror,raise=10pt},decorate,thick]
  (6.1,1.55) -- node[above=10pt] {$\scriptstyle{n}$ } (3.9,1.55);
	\end{tikzpicture}
	} 
	\label{eq:magQuiv_SO-Sp_almost_infinite}
\end{align}
with a Coulomb branch of dimension
\begin{align}
    \dim_{\HH} \Coulomb \eqref{eq:magQuiv_SO-Sp_almost_infinite}
    = 2n +6 + \dim\orm(2k+4) - \dim\orm(4) \,.
\end{align}
Note that the moduli space dimension of \eqref{eq:magQuiv_SO-Sp_almost_infinite} is increased by two sources: (i) each pair of coincident half \NS\ branes yields $2$ Higgs branch degrees of freedom, and (ii) moving the outermost $(1,\pm1)$ 5-branes from \eqref{eq:web_SO-Sp_quiver_finite} into the central segment leads to another $6$ Higgs branch degrees of freedom due to \Dfive\ branes suspended between 7-branes along the orientifold plane.

Finally, one can make all \NS\ branes coincident. Analogously to \eqref{eq:web_SO-Sp_quiver_intermediate_many_NS}, the $[0,1]$ 7-branes can be vertically displaced because the \NS\ branes can be split between them. As a consequence, the brane web becomes
\begin{align}
 \raisebox{-.5\height}{
    \begin{tikzpicture}
    \OMinusTilde{-7,0}{-6,0}
    \MonoCut{-7,-0.05}{-6,-0.05}
    \DfiveOMinusTilde{$k$}{-5}
    \MonoCut{-5,-0.05}{-4,-0.05}
    \DfiveOMinus{$k{+}1$}{-4}
    \DfiveOMinusTilde{$k{+}1$}{-3}
    \MonoCut{-3,-0.05}{-2,-0.05}
    \DfiveOMinus{$k{+}2$}{-2}
    \DfiveOMinus{}{-1}
    \DfiveOMinus{}{0}
    \DfiveOMinusTilde{$k{+}1$}{1}
    \MonoCut{1,-0.05}{2,-0.05}
    \DfiveOMinus{$k{+}1$}{2}
    \DfiveOMinusTilde{$k$}{3}
    \MonoCut{3,-0.05}{4,-0.05}
    \OMinusTilde{5,0}{6,0}
    \MonoCut{5,-0.05}{6,-0.05}
    \draw (-5.5,0) node {$\cdots$};
    \draw (4.5,0) node {$\cdots$};
    \Dbrane{-0.575,-1}{-0.575,1}
    \Dbrane{-0.525,-2}{-0.525,2}
    \Dbrane{-0.475,-2}{-0.475,2}
    \Dbrane{-0.425,-2}{-0.425,2}
    \MonoCut{-0.575,1}{-0.575,2}
    \MonoCut{-0.575,-1}{-0.575,-2}
    \draw (-0.5,2.6) node {$\vdots$};
    \draw (-0.5,-2.4) node {$\vdots$};
    \MonoCut{-0.575,3}{-0.575,4.5}
    \MonoCut{-0.525,3}{-0.525,4.5}
    \MonoCut{-0.475,3}{-0.475,4.5}
    \Dbrane{-0.425,3}{-0.425,4}
    \MonoCut{-0.425,4}{-0.425,4.5}
    \MonoCut{-0.575,-3}{-0.575,-4.5}
    \MonoCut{-0.525,-3}{-0.525,-4.5}
    \MonoCut{-0.475,-3}{-0.475,-4.5}
    \Dbrane{-0.425,-3}{-0.425,-4}
    \MonoCut{-0.425,-4}{-0.425,-4.5}
    \draw (0.25,0.5) node {$\scriptstyle{2n}$};
    \draw (0.25,1.5) node {$\scriptstyle{2n{-}1}$};
    \draw (0.25,3.5) node {$\scriptstyle{1}$};
    \MonoCut{-7,-0.75}{-7,0.75}
    \MonoCut{-5,-0.75}{-5,0.75}
    \MonoCut{-3,-0.75}{-3,0.75}
    \MonoCut{2,-0.75}{2,0.75}
    \MonoCut{4,-0.75}{4,0.75}
    \MonoCut{6,-0.75}{6,0.75}
    \SevenB{-0.5,1}
    \SevenB{-0.5,-1}
    \SevenB{-0.5,2}
    \SevenB{-0.5,-2}
    \SevenB{-0.5,3}
    \SevenB{-0.5,-3}
    \SevenB{-0.5,4}
    \SevenB{-0.5,-4}
    \SevenB{-2,0}
    \SevenB{-3,0}
    \SevenB{-4,0}
    \SevenB{-5,0}
    \SevenB{-6,0}
    \SevenB{-7,0}
    \SevenB{1,0}
    \SevenB{2,0}
    \SevenB{3,0}
    \SevenB{4,0}
    \SevenB{5,0}
    \SevenB{6,0}
    \draw[decoration={brace,mirror,raise=10pt},decorate,thick]
  (-7.25,0) -- node[below=10pt] {$\scriptstyle{2k+4}$ } (-1.75,0);
  \draw[decoration={brace,mirror,raise=10pt},decorate,thick]
  (0.75,0) -- node[below=10pt] {$\scriptstyle{2k+4}$ } (6.25,0);
    \draw[decoration={brace,mirror,raise=10pt},decorate,thick]
  (-0.75,4.25) -- node[left=10pt] {$\scriptstyle{2n}$ } (-0.75,0.75);
    \end{tikzpicture}
}
\label{eq:web_SO-Sp_quiver_infinite}
\end{align}
The 5-brane webs suspended between 7-branes on the orientifold are no conceptual challenge, due to the discussions above. The $(0,1)$ 5-branes, in contrast, need to be examined in detail. The first $2n$ $(0,1)$ branes that go through the \Of\ plane contribute a $\calg_n$ magnetic gauge node, because the orientifold projection acts on these subwebs. The subsequent $j$ copies of identical $(0,1)$ 5-branes between adjacent half 7-branes are not affected by the orientifold, because they are away from the plane. Hence, these stacks of identical subwebs contribute magnetic $\urm(j)$ gauge nodes. Put differently, away from the orientifold, the rules of \cite{Cabrera:2018jxt} apply without modifications. Putting all the pieces together, the magnetic quiver reads
\begin{align}
 \raisebox{-.5\height}{
 	\begin{tikzpicture}
 	\tikzset{node distance = 1cm};
	\tikzstyle{gauge} = [circle, draw,inner sep=2.5pt];
	\tikzstyle{flavour} = [regular polygon,regular polygon sides=4,inner 
sep=2.5pt, draw];
	\node (g1) [gauge,label=below:{\dd{1}}] {};
	\node (g2) [gauge,right of=g1,label=below:{\cc{1}}] {};
	\node (g3) [right of=g2] {$\cdots$};
	\node (g4) [gauge,right of=g3,label=below:{\dd{k+1}}] {};
	\node (g5) [gauge,right of=g4,label=below:{\cc{k+1}}] {};
	\node (g6) [gauge,right of=g5,label=below:{\dd{k+2}}] {};
	\node (g7) [gauge,right of=g6,label=below:{\cc{k+1}}] {};
	\node (g8) [gauge,right of=g7,label=below:{\dd{k+1}}] {};
	\node (g9) [right of=g8] {$\cdots$};
	\node (g10) [gauge,right of=g9,label=below:{\cc{1}}] {};
	\node (g11) [gauge,right of=g10,label=below:{\dd{1}}] {};
	\node (u1) [gauge,above of=g6,label=right:{\cc{n}}] {};
	\node (u2) [gauge,above of=u1,label=right:{\uu{2n-1}}] {};
	\node (u3) [gauge,above of=u2,label=right:{\uu{2n-2}}] {};
	\node (u4) [above of=u3] {$\vdots$};
	\node (u5) [gauge,above of=u4,label=right:{\uu{2}}] {};
	\node (u6) [gauge,above of=u5,label=right:{\uu{1}}] {};
	\draw (g1)--(g2) (g2)--(g3) (g3)--(g4) (g4)--(g5) (g5)--(g6) (g6)--(g7) (g7)--(g8) (g8)--(g9) (g9)--(g10) (g10)--(g11) (g6)--(u1) (u1)--(u2) (u2)--(u3) (u3)--(u4) (u4)--(u5) (u5)--(u6);
	\end{tikzpicture}
	} 
	\label{eq:magQuiv_SO-Sp_infinite}
\end{align}
and its Coulomb branch dimension is computed to be
\begin{align}
    \dim_{\HH} \Coulomb \eqref{eq:magQuiv_SO-Sp_infinite} = 
    2n^2 +6 + \dim \orm(2k+4) -\dim \orm(4) \,.
\end{align}
Compare to the finite coupling magnetic quiver \eqref{eq:magQuiv_SO-Sp_finite}, the moduli spaces has grown in dimension by $2n^2+6$ quaternionic units. The increase can be traced back to two different origins, similarly to \eqref{eq:magQuiv_SO-Sp_almost_infinite}: (i) the $6$ additional Higgs branch directions originate from \Dfive\ branes suspended between 7-branes along the \Of\ plane; while, (ii) the $2n^2$ new moduli originate from the coincident \NS\ branes.  
\paragraph{Hasse diagram.}
For the quiver theories \eqref{eq:quiver_SO-Sp} one could, in principle, detail the Hasse diagrams as well. Inspecting the infinite coupling magnetic quiver \eqref{eq:magQuiv_SO-Sp_infinite} shows that the challenge lies in the unitary-orthosymplectic quiver with a large number of unitary nodes. From the insights gained in Section \ref{sec:magQuiv_known_cases}, it is apparent that the first minimal transitions that open up a Coulomb branch direction are necessarily $e_7$ or $e_8$ transitions. Thereafter, the number of possible transitions is quite involved and there is no clear gain of attempting to derive the entire Hasse diagram.
%
%
\section{Conclusion}
\label{sec:conclusion}
In this work, the Higgs branches of $5$d $\Ncal=1$ theories with symplectic or orthogonal gauge groups and fundamental matter are investigated at finite and infinite gauge coupling. Based on the 5-brane web realisations in the presence of \Of\ orientifold planes, the key technique for this study is the use of magnetic quivers. More generally, the techniques presented here open a window on the Higgs branch of any 5d SCFT that can be engineered via brane webs involving $(p,q)$ 5-branes and O5 planes, which potentially includes theories with exceptional gauge algebra $G_2$.   

Compared to the magnetic quiver proposal \cite{Cabrera:2018jxt} for brane webs without \Of\ planes, the conceptual challenge is the inclusion of the orientifolds. As detailed in Section \ref{sec:magQuiv_known_cases}, the proposed \emph{magnetic orientifolds} of \cite{Cabrera:2019dob} do consistently reproduce known Higgs branch geometries at finite coupling and some infinite coupling cases. In addition, an unprecedented phenomenon appeared at infinite coupling: due to additional Higgs branch directions the corresponding magnetic quivers become a combination of orthosymplectic and unitary quivers. Therefore, magnetic orientifolds need to be supplemented by unitary magnetic gauge multiplets and corresponding matter fields, in order to address all Higgs branch phases.

In Section \ref{sec:Spk}, the magnetic quiver proposal has been applied to 
$\sprm(k)$ gauge theories with $N_f \leq 2k+5 $ flavours. The finite coupling 
Higgs branches are all described by orthosymplectic magnetic quivers, which 
correctly reproduce moduli space dimensions and the non-abelian part of the 
global symmetry. A further matching between the Higgs branch Hilbert series and 
magnetic quiver Hilbert series is detailed in a companion paper 
\cite{Bourget:2020xdz}. For infinite gauge coupling, the magnetic quivers do 
match 
the moduli space dimensions, in agreement with the additional Higgs branch 
directions at the fixed point found in \cite{Zafrir:2015ftn}. Further Hilbert 
series details will be presented in \cite{Bourget:2020xdz}. Moreover, the 
infinite 
coupling results fall into generalised $E_n$ classes (for $-\infty<n\leq8$), 
summarised in Tables \ref{tab:Efamilies}, \ref{tab:CompareFiniteHasse}, 
\ref{tab:CompareFiniteHasse2} and \ref{tab:CompareFiniteHWG}. It is important to 
note that when the finite coupling Higgs branch consists of two identical cones 
(i.e.\ when $N_f\leq2k$ even), from the \Of\ construction only cone can be 
seen. At infinite coupling only one cone gets enhanced, which is the cone 
obtained from the \Of\ construction.

Besides \Ofp\ planes, one may engineer $\sprm(k)$ theories via O$7^-$ planes as 
detailed in Section \ref{sec:O7_plane}. Since an O$7^-$ plane can be resolved 
into a pair of $[1,\pm1]$ 7-branes, one can analyse these cases with the 
techniques developed in \cite{Cabrera:2018jxt}. Interestingly, this approach 
yields unitary magnetic quiver for the same Higgs branches as discussed in 
Section \ref{sec:Spk}. On the one hand, the unitary magnetic quivers allow to 
immediately verify dimension and global symmetry of the various Higgs branches. 
These agree with known symmetry enhancement results at the fixed point. On the 
other hand, one has to confirm that the unitary-orthosymplectic quiver from the 
\Of\ plane construction and the unitary magnetic quiver from the O$7^-$ 
construction yield the same moduli space. One step further towards a 
verification is subject of a companion paper \cite{Bourget:2020xdz}. Most 
importantly, this provides different realisations for the same Higgs branches, 
as summarised in Table \ref{tab:Efamilies}. 

Finally, the validity of the proposed modification for magnetic quivers from 5-branes on \Of\ planes is further underpinned by deriving the associated Hasse diagrams. As detailed in Section \ref{sec:Hasse} for the generalised $E_n$ families (for $-\infty<n\leq 8$), the Hasse diagrams are derived starting from the 5-brane web with \Of\ at infinite coupling and then opening up a minimal amount of Coulomb branch directions. The transitions found are all exceptional minimal nilpotent orbit closures (or the appropriate substitution for $n \leq 3$), because the transitions are realised by the brane configurations of Section \ref{sec:SU2}. Thereafter, the remaining magnetic quivers are all framed orthosymplectic quivers, such that conventional quiver subtraction suffices to completely determine the Hasse diagram. A non-trivial consistency check is provided by verifying that the Hasse diagrams from the unitary magnetic quivers of Section \ref{sec:O7_plane} agree with those from the unitary-orthosymplectic quivers.

As a next more complicated example, Section \ref{sec:quiver_theories} deals with 
5d $\Ncal=1$ linear orthosymplectic quiver gauge theories. In contrast to the 
single gauge group cases, the quiver theories admit a multitude of Higgs branch 
phases, depending on which subset of gauge couplings has been taking to 
infinity. In particular, there exist two types of infinite coupling transitions 
\eqref{eq:magQuiv_SO-Sp_intermediate_2d_var1} and 
\eqref{eq:magQuiv_SO-Sp_intermediate_2d_var2} that are expected to open up two 
new Higgs branch dimensions. One notes the close analogy to the $D_4$ transition 
of \cite{Cabrera:2019dob}. However, the 5d setting of a generalised bouquet of 
type
\begin{align}
 \raisebox{-.5\height}{
 	\begin{tikzpicture}
 	 \tikzset{node distance = 1cm};
	\tikzstyle{gauge} = [circle, draw,inner sep=2.5pt];
	\tikzstyle{flavour} = [regular polygon,regular polygon sides=4,inner 
sep=2.5pt, draw];
    \node (g1) [gauge,label=below:{\cc{1}}] {};
    \node (g2) [gauge,above of=g1,label=above:{\uu{1}}] {};
    \draw (g1)--(g2);
 	\end{tikzpicture}
}
\qquad \text{or} \qquad 
 \raisebox{-.5\height}{
 	\begin{tikzpicture}
 	 \tikzset{node distance = 1cm};
	\tikzstyle{gauge} = [circle, draw,inner sep=2.5pt];
	\tikzstyle{flavour} = [regular polygon,regular polygon sides=4,inner 
sep=2.5pt, draw];
    \node (g1) [gauge,label=below:{\cc{1}}] {};
    \node (f1) [flavour,above left of=g1,label=above:{\dd{2}}] {};
    \node (g2) [gauge,above right of=g1,label=above:{\uu{1}}] {};
    \draw (g1)--(g2) (g1)--(f1);
 	\end{tikzpicture}
}
\end{align}
is not comparable to the discrete gauge phenomenon 
\cite{Hanany:2018vph,Hanany:2018cgo,Hanany:2018dvd}. If the relevant \NS\ branes 
become coincident, the resulting magnetic quiver is obtained by replacing the 
bouquet by a longer unitary tail.

%
%
\begin{table}[t]
    \centering
    \begin{tabular}{c|c}
    \toprule
\multicolumn{2}{c}{
\raisebox{-.5\height}{
    \begin{tikzpicture}
    \OPlus{-7,0}{-2,0}
    \OPlus{3,0}{7,0}
    \Dbrane{-2,0}{-1,0.5}
    \Dbrane{3,0}{2,0.5}
    \draw (3.5,0.15) node {$\scriptstyle{(2,-1)}$};
    \Dbrane{-0.925,0.55}{1.924,0.55}
    \Dbrane{-1,0.5}{2,0.5}
    \Dbrane{-1.075,0.55}{2.075,0.55}
     \draw (0.5,0.75) node {$\scriptstyle{k}$};
    \Dbrane{-1,0.5}{-2.5,1}
    \Dbrane{2,0.5}{3.5,1}
    \draw (3,0.55) node {$\scriptstyle{(k{-}2,1)}$};
    \Dbrane{-2.55,1.075}{-3.5,1.075}
    \Dbrane{-2.5,1.025}{-3.5,1.025}
    \Dbrane{-2.42,0.975}{-3.5,0.975}
    \Dbrane{-2.34,0.925}{-3.5,0.925}
    \Dbrane{-3.5,0.975}{-4.5,0.975}
    \Dbrane{-3.5,1.025}{-4.5,1.025}
    \Dbrane{-3.5,1.075}{-4.5,1.075}
    \Dbrane{-5.5,1.075}{-6.5,1.075}
    \draw (-5,1) node {$\cdots$};
    \Dbrane{3.5,1}{4.5,1}
    \Dbrane{3.6,1.05}{4.5,1.05}
    \Dbrane{3.4,0.95}{4.5,0.95}
    \Dbrane{5.5,1.05}{6.5,1.05}
    \draw (5,1) node {$\cdots$};
    \Dbrane{3.5,1}{5.5,2}
    \draw (4,1.75) node {$\scriptstyle{(k{-}N_R{-}2,1)}$};
    \Dbrane{-2.5,1}{-3.5,2}
    \draw (-2.25,1.75) node {$\scriptstyle{(k{-}N_L{-}2,-1)}$};
    \MonoCut{-3.5,0.925}{-4.5,0.925}
    \MonoCut{-5.5,0.925}{-7.5,0.925}
    \MonoCut{-5.5,0.975}{-7.5,0.975}
    \MonoCut{-5.5,1.025}{-7.5,1.025}
    \MonoCut{-6.5,1.075}{-7.5,1.075}
    \MonoCut{6.5,1.05}{7.5,1.05}
    \MonoCut{5.5,0.95}{7.5,0.95}
    \MonoCut{5.5,1.0}{7.5,1.0}
    \SevenB{-3.5,2}
    \SevenB{5.5,2}
    \SevenB{-3.5,1}
    \SevenB{-4.5,1}
    \SevenB{-5.5,1}
    \SevenB{-6.5,1}
    \SevenB{4.5,1}
    \SevenB{5.5,1}
    \SevenB{6.5,1}
    \draw[decoration={brace,mirror,raise=10pt},decorate,thick]
  (-6.75,1) -- node[below=10pt] {$\scriptstyle{N_L}$ } (-3.25,1);
  \draw[decoration={brace,mirror,raise=10pt},decorate,thick]
  (4.25,1) -- node[below=10pt] {$\scriptstyle{N_R}$ } (6.75,1);
     \begin{scope}[yscale=-1,xscale=1]
    \Dbrane{-2,0}{-1,0.5}
    \Dbrane{3,0}{2,0.5}
    \Dbrane{-0.925,0.55}{1.924,0.55}
    \Dbrane{-1,0.5}{2,0.5}
    \Dbrane{-1.075,0.55}{2.075,0.55}
    \Dbrane{-1,0.5}{-2.5,1}
    \Dbrane{2,0.5}{3.5,1}
    \Dbrane{-2.55,1.075}{-3.5,1.075}
    \Dbrane{-2.5,1.025}{-3.5,1.025}
    \Dbrane{-2.42,0.975}{-3.5,0.975}
    \Dbrane{-2.34,0.925}{-3.5,0.925}
    \Dbrane{-3.5,0.975}{-4.5,0.975}
    \Dbrane{-3.5,1.025}{-4.5,1.025}
    \Dbrane{-3.5,1.075}{-4.5,1.075}
    \Dbrane{-5.5,1.075}{-6.5,1.075}
    \draw (-5,1) node {$\cdots$};
    \Dbrane{3.5,1}{4.5,1}
    \Dbrane{3.6,1.05}{4.5,1.05}
    \Dbrane{3.4,0.95}{4.5,0.95}
    \Dbrane{5.5,1.05}{6.5,1.05}
    \draw (5,1) node {$\cdots$};
    \Dbrane{3.5,1}{5.5,2}
    \Dbrane{-2.5,1}{-3.5,2}
    \MonoCut{-3.5,0.925}{-4.5,0.925}
    \MonoCut{-5.5,0.925}{-7.5,0.925}
    \MonoCut{-5.5,0.975}{-7.5,0.975}
    \MonoCut{-5.5,1.025}{-7.5,1.025}
    \MonoCut{-6.5,1.075}{-7.5,1.075}
    \MonoCut{6.5,1.05}{7.5,1.05}
    \MonoCut{5.5,0.95}{7.5,0.95}
    \MonoCut{5.5,1.0}{7.5,1.0}
    \SevenB{-3.5,2}
    \SevenB{5.5,2}
    \SevenB{-3.5,1}
    \SevenB{-4.5,1}
    \SevenB{-5.5,1}
    \SevenB{-6.5,1}
    \SevenB{4.5,1}
    \SevenB{5.5,1}
    \SevenB{6.5,1}
    \end{scope}
    \end{tikzpicture}
}}    
    \\ \midrule
      $N_f$ even     & 
         \raisebox{-.5\height}{
 	\begin{tikzpicture}
 	\tikzset{node distance = 1cm};
	\tikzstyle{gauge} = [circle, draw,inner sep=2.5pt];
	\tikzstyle{flavour} = [regular polygon,regular polygon sides=4,inner 
sep=2.5pt, draw];
	\node (g1) [gauge,label=below:{\bb{0}}] {};
	\node (g2) [gauge,right of=g1,label=below:{\cc{1}}] {};
	\node (g3) [gauge,right of=g2,label=below:{\bb{1}}] {};
	\node (g4) [right of=g3] {$\cdots$};
	\node (g5) [gauge,right of=g4,label=below:{\bb{ \frac{N_f}{2} {-1} }}] {};
	\node (g6) [gauge,right of=g5,label=below:{\cc{ \frac{N_f}{2} }}] {};
	\node (g7) [gauge,right of=g6,label=below:{\bb{ \frac{N_f}{2} }}] {};
	\node (g8) [gauge,right of=g7,label=below:{\cc{ \frac{N_f}{2} }}] {};
	\node (g9) [gauge,right of=g8,label=below:{\bb{ \frac{N_f}{2} {-1} }}] {};
	\node (g10) [right of=g9] {$\cdots$};
	\node (g11) [gauge,right of=g10,label=below:{\bb{1}}] {};
	\node (g12) [gauge,right of=g11,label=below:{\cc{1}}] {};
	\node (g13) [gauge,right of=g12,label=below:{\bb{0}}] {};
	\node (f1) [flavour,above of=g7,label=right:{\cc{1}}] {};
	\draw (g1)--(g2) (g2)--(g3) (g3)--(g4) (g4)--(g5) (g5)--(g6) (g6)--(g7) (g7)--(g8) (g8)--(g9) (g9)--(g10) (g10)--(g11) (g11)--(g12) (g12)--(g13) (g7)--(f1); 
	\end{tikzpicture}
	} \\
$N_f$ odd & 
 \raisebox{-.5\height}{
 	\begin{tikzpicture}
 	\tikzset{node distance = 1cm};
	\tikzstyle{gauge} = [circle, draw,inner sep=2.5pt];
	\tikzstyle{flavour} = [regular polygon,regular polygon sides=4,inner 
sep=2.5pt, draw];
	\node (g1) [gauge,label=below:{\bb{0}}] {};
	\node (g2) [gauge,right of=g1,label=below:{\cc{1}}] {};
	\node (g3) [gauge,right of=g2,label=below:{\bb{1}}] {};
	\node (g4) [right of=g3] {$\cdots$};
	\node (g5) [gauge,right of=g4,label=below:{\cc{ \frac{N_f {-1} }{2} }}] {};
	\node (g6) [gauge,right of=g5,label=below:{\bb{ \frac{N_f {-1} }{2} }}] {};
	\node (g7) [gauge,right of=g6,label=below:{\cc{ \frac{N_f {+1} }{2} }}] {};
	\node (g8) [gauge,right of=g7,label=below:{\bb{ \frac{N_f {-1} }{2} }}] {};
	\node (g9) [gauge,right of=g8,label=below:{\cc{ \frac{N_f {-1} }{2} }}] {};
	\node (g10) [right of=g9] {$\cdots$};
	\node (g11) [gauge,right of=g10,label=below:{\bb{1}}] {};
	\node (g12) [gauge,right of=g11,label=below:{\cc{1}}] {};
	\node (g13) [gauge,right of=g12,label=below:{\bb{0}}] {};
	\node (f1) [flavour,above of=g7,label=right:{\dd{1}}] {};
	\draw (g1)--(g2) (g2)--(g3) (g3)--(g4) (g4)--(g5) (g5)--(g6) (g6)--(g7) (g7)--(g8) (g8)--(g9) (g9)--(g10) (g10)--(g11) (g11)--(g12) (g12)--(g13) (g7)--(f1);
	\end{tikzpicture}
	} \\  \bottomrule
    \end{tabular}
    \caption{Brane web for $\orm(2k)$ with $N_f=N_L+N_R$ flavours. On a generic point of the Higgs branch $\orm(2k)$ with $N_f$ flavours is broken to pure $\orm(2k-N_f)$ gauge theory. The magnetic quivers describe then a moduli space of quaternionic dimension $\tfrac{1}{2} N_f(N_f+1)$.}
    \label{tab:O2k_finite}
\end{table}

\begin{table}[t]
    \centering
    \begin{tabular}{c|c}
    \toprule
    \multicolumn{2}{c}{
    \raisebox{-.5\height}{
    \begin{tikzpicture}
    \OPlusTilde{-7,0}{-1.5,0}
    \OPlusTilde{3,0}{7,0}
    \OMinusTilde{-2,0}{3,0}
    \MonoCut{-7,-0.05}{7,-0.05}
    \Dbrane{-1.5,0}{-1,0.5}
    \draw (-1.55,0.35) node {$\scriptstyle{(1,1)}$};
    \Dbrane{3,0}{2,0.5}
    \draw (3.05,0.35) node {$\scriptstyle{(2,{-}1)}$};
    \Dbrane{-0.925,0.55}{1.924,0.55}
    \Dbrane{-1,0.5}{2,0.5}
    \Dbrane{-1.075,0.55}{2.075,0.55}
     \draw (0.5,0.75) node {$\scriptstyle{k}$};
    \Dbrane{-1,0.5}{-2.5,1}
    \draw (-1.5,1) node {$\scriptstyle{(k{-}1,{-}1)}$};
    \Dbrane{2,0.5}{3.5,1}
    \draw (2.5,1) node {$\scriptstyle{(k{-}2,1)}$};
    \Dbrane{-2.55,1.075}{-3.5,1.075}
    \Dbrane{-2.5,1.025}{-3.5,1.025}
    \Dbrane{-2.42,0.975}{-3.5,0.975}
    \Dbrane{-2.34,0.925}{-3.5,0.925}
    \Dbrane{-3.5,0.975}{-4.5,0.975}
    \Dbrane{-3.5,1.025}{-4.5,1.025}
    \Dbrane{-3.5,1.075}{-4.5,1.075}
    \Dbrane{-5.5,1.075}{-6.5,1.075}
    \draw (-5,1) node {$\cdots$};
    \Dbrane{3.5,1}{4.5,1}
    \Dbrane{3.6,1.05}{4.5,1.05}
    \Dbrane{3.4,0.95}{4.5,0.95}
    \Dbrane{5.5,1.05}{6.5,1.05}
    \draw (5,1) node {$\cdots$};
    \Dbrane{3.5,1}{5.5,2}
    \draw (4,1.75) node {$\scriptstyle{(k{-}N_R{-}2,1)}$};
    \Dbrane{-2.5,1}{-3.5,2}
    \draw (-2.25,1.75) node {$\scriptstyle{(k{-}N_L{-}1,-1)}$};
    \MonoCut{-3.5,0.925}{-4.5,0.925}
    \MonoCut{-5.5,0.925}{-7.5,0.925}
    \MonoCut{-5.5,0.975}{-7.5,0.975}
    \MonoCut{-5.5,1.025}{-7.5,1.025}
    \MonoCut{-6.5,1.075}{-7.5,1.075}
    \MonoCut{6.5,1.05}{7.5,1.05}
    \MonoCut{5.5,0.95}{7.5,0.95}
    \MonoCut{5.5,1.0}{7.5,1.0}
    \SevenB{-3.5,2}
    \SevenB{5.5,2}
    \SevenB{-3.5,1}
    \SevenB{-4.5,1}
    \SevenB{-5.5,1}
    \SevenB{-6.5,1}
    \SevenB{4.5,1}
    \SevenB{5.5,1}
    \SevenB{6.5,1}
    \draw[decoration={brace,mirror,raise=10pt},decorate,thick]
  (-6.75,1) -- node[below=10pt] {$\scriptstyle{N_L}$ } (-3.25,1);
  \draw[decoration={brace,mirror,raise=10pt},decorate,thick]
  (4.25,1) -- node[below=10pt] {$\scriptstyle{N_R}$ } (6.75,1);
     \begin{scope}[yscale=-1,xscale=1]
    \Dbrane{-1.5,0}{-1,0.5}
    \Dbrane{3,0}{2,0.5}
    \Dbrane{-0.925,0.55}{1.924,0.55}
    \Dbrane{-1,0.5}{2,0.5}
    \Dbrane{-1.075,0.55}{2.075,0.55}
    \Dbrane{-1,0.5}{-2.5,1}
    \Dbrane{2,0.5}{3.5,1}
    \Dbrane{-2.55,1.075}{-3.5,1.075}
    \Dbrane{-2.5,1.025}{-3.5,1.025}
    \Dbrane{-2.42,0.975}{-3.5,0.975}
    \Dbrane{-2.34,0.925}{-3.5,0.925}
    \Dbrane{-3.5,0.975}{-4.5,0.975}
    \Dbrane{-3.5,1.025}{-4.5,1.025}
    \Dbrane{-3.5,1.075}{-4.5,1.075}
    \Dbrane{-5.5,1.075}{-6.5,1.075}
    \draw (-5,1) node {$\cdots$};
    \Dbrane{3.5,1}{4.5,1}
    \Dbrane{3.6,1.05}{4.5,1.05}
    \Dbrane{3.4,0.95}{4.5,0.95}
    \Dbrane{5.5,1.05}{6.5,1.05}
    \draw (5,1) node {$\cdots$};
    \Dbrane{3.5,1}{5.5,2}
    \Dbrane{-2.5,1}{-3.5,2}
    \MonoCut{-3.5,0.925}{-4.5,0.925}
    \MonoCut{-5.5,0.925}{-7.5,0.925}
    \MonoCut{-5.5,0.975}{-7.5,0.975}
    \MonoCut{-5.5,1.025}{-7.5,1.025}
    \MonoCut{-6.5,1.075}{-7.5,1.075}
    \MonoCut{6.5,1.05}{7.5,1.05}
    \MonoCut{5.5,0.95}{7.5,0.95}
    \MonoCut{5.5,1.0}{7.5,1.0}
    \SevenB{-3.5,2}
    \SevenB{5.5,2}
    \SevenB{-3.5,1}
    \SevenB{-4.5,1}
    \SevenB{-5.5,1}
    \SevenB{-6.5,1}
    \SevenB{4.5,1}
    \SevenB{5.5,1}
    \SevenB{6.5,1}
    \end{scope}
    \end{tikzpicture}
}}
       \\ \midrule
    $N_f$ even &
     \raisebox{-.5\height}{
 	\begin{tikzpicture}
 	\tikzset{node distance = 1cm};
	\tikzstyle{gauge} = [circle, draw,inner sep=2.5pt];
	\tikzstyle{flavour} = [regular polygon,regular polygon sides=4,inner 
sep=2.5pt, draw];
	\node (g1) [gauge,label=below:{\bb{1}}] {};
	\node (g2) [gauge,right of=g1,label=below:{\cc{1}}] {};
	\node (g3) [gauge,right of=g2,label=below:{\bb{2}}] {};
	\node (g4) [right of=g3] {$\cdots$};
	\node (g5) [gauge,right of=g4,label=below:{\cc{ \frac{N_f}{2} {-1} }}] {};
	\node (g6) [gauge,right of=g5,label=below:{\bb{ \frac{N_f }{2} }}] {};
	\node (g7) [gauge,right of=g6,label=below:{\cc{ \frac{N_f }{2} }}] {};
	\node (g8) [gauge,right of=g7,label=below:{\bb{ \frac{N_f }{2} }}] {};
	\node (g9) [gauge,right of=g8,label=below:{\cc{ \frac{N_f}{2} {-1} }}] {};
	\node (g10) [right of=g9] {$\cdots$};
	\node (g11) [gauge,right of=g10,label=below:{\bb{2}}] {};
	\node (g12) [gauge,right of=g11,label=below:{\cc{1}}] {};
	\node (g13) [gauge,right of=g12,label=below:{\bb{1}}] {};
	\node (u1) [flavour,above of=g7,label=right:{\dd{1}}] {};
	\draw (g1)--(g2) (g2)--(g3) (g3)--(g4) (g4)--(g5) (g5)--(g6) (g6)--(g7) (g7)--(g8) (g8)--(g9) (g9)--(g10) (g10)--(g11) (g11)--(g12) (g12)--(g13) (g7)--(u1);
	\end{tikzpicture}
	}  \\
 $N_f $ odd &
  \raisebox{-.5\height}{
 	\begin{tikzpicture}
 	\tikzset{node distance = 1cm};
	\tikzstyle{gauge} = [circle, draw,inner sep=2.5pt];
	\tikzstyle{flavour} = [regular polygon,regular polygon sides=4,inner 
sep=2.5pt, draw];
	\node (g1) [gauge,label=below:{\bb{1}}] {};
	\node (g2) [gauge,right of=g1,label=below:{\cc{1}}] {};
	\node (g3) [gauge,right of=g2,label=below:{\bb{2}}] {};
	\node (g4) [right of=g3] {$\cdots$};
	\node (g5) [gauge,right of=g4,label=below:{\bb{ \frac{N_f {-1} }{2} }}] {};
	\node (g6) [gauge,right of=g5,label=below:{\cc{ \frac{N_f {-1} }{2} }}] {};
	\node (g7) [gauge,right of=g6,label=below:{\bb{ \frac{N_f {+1} }{2} }}] {};
	\node (g8) [gauge,right of=g7,label=below:{\cc{ \frac{N_f {-1} }{2} }}] {};
	\node (g9) [gauge,right of=g8,label=below:{\bb{ \frac{N_f {-1} }{2} }}] {};
	\node (g10) [right of=g9] {$\cdots$};
	\node (g11) [gauge,right of=g10,label=below:{\bb{2}}] {};
	\node (g12) [gauge,right of=g11,label=below:{\cc{1}}] {};
	\node (g13) [gauge,right of=g12,label=below:{\bb{1}}] {};
	\node (u1) [flavour,above of=g7,label=right:{\cc{1}}] {};
	\draw (g1)--(g2) (g2)--(g3) (g3)--(g4) (g4)--(g5) (g5)--(g6) (g6)--(g7) (g7)--(g8) (g8)--(g9) (g9)--(g10) (g10)--(g11) (g11)--(g12) (g12)--(g13) (g7)--(u1);
	\end{tikzpicture}
	}  \\ \bottomrule
    \end{tabular}
    \caption{Brane web for $\orm(2k+1)$ with $N_f=N_L+N_R$ flavours. On a generic point of the Higgs branch $\orm(2k+1)$ with $N_f$ flavours is broken to pure $\orm(2k+1-N_f)$ gauge theory. The magnetic quivers describe then a moduli space of quaternionic dimension $\tfrac{1}{2} N_f(N_f+1)$.}
    \label{tab:O2k+1_finite}
\end{table}

\begin{table}[t]
    \centering
    \begin{tabular}{c|c}
    \toprule
        $N_f{=}2k{-}3$ & 
        \raisebox{-.5\height}{
 	\begin{tikzpicture}
 	\tikzset{node distance = 1cm};
	\tikzstyle{gauge} = [circle, draw,inner sep=2.5pt];
	\tikzstyle{flavour} = [regular polygon,regular polygon sides=4,inner 
sep=2.5pt, draw];
	\node (g1) [gauge,label=below:{\bb{0}}] {};
	\node (g2) [gauge,right of=g1,label=below:{\cc{1}}] {};
	\node (g3) [gauge,right of=g2,label=below:{\bb{1}}] {};
	\node (g4) [right of=g3] {$\cdots$};
	\node (g5) [gauge,right of=g4,label=below:{\bb{ \frac{N_f {-1} }{2} }}] {};
	\node (g6) [gauge,right of=g5,label=below:{\cc{ \frac{N_f {+1} }{2} }}] {};
	\node (g7) [gauge,right of=g6,label=below:{\bb{ \frac{N_f {+1} }{2} }}] {};
	\node (g8) [gauge,right of=g7,label=below:{\cc{ \frac{N_f {+1} }{2} }}] {};
	\node (g9) [gauge,right of=g8,label=below:{\bb{ \frac{N_f {-1} }{2} }}] {};
	\node (g10) [right of=g9] {$\cdots$};
	\node (g11) [gauge,right of=g10,label=below:{\bb{1}}] {};
	\node (g12) [gauge,right of=g11,label=below:{\cc{1}}] {};
	\node (g13) [gauge,right of=g12,label=below:{\bb{0}}] {};
	\node (u1) [gauge,above of=g7,label=right:{\cc{1}}] {};
	\draw (g1)--(g2) (g2)--(g3) (g3)--(g4) (g4)--(g5) (g5)--(g6) (g6)--(g7) (g7)--(g8) (g8)--(g9) (g9)--(g10) (g10)--(g11) (g11)--(g12) (g12)--(g13) (g7)--(u1);
	\end{tikzpicture}
	} \\ 
        $N_f{=}2k{-}4$ & 
        \raisebox{-.5\height}{
 	\begin{tikzpicture}
 	\tikzset{node distance = 1cm};
	\tikzstyle{gauge} = [circle, draw,inner sep=2.5pt];
	\tikzstyle{flavour} = [regular polygon,regular polygon sides=4,inner 
sep=2.5pt, draw];
	\node (g1) [gauge,label=below:{\bb{0}}] {};
	\node (g2) [gauge,right of=g1,label=below:{\cc{1}}] {};
	\node (g3) [gauge,right of=g2,label=below:{\bb{1}}] {};
	\node (g4) [right of=g3] {$\cdots$};
	\node (g5) [gauge,right of=g4,label=below:{\bb{ \frac{N_f}{2} {-1} }}] {};
	\node (g6) [gauge,right of=g5,label=below:{\cc{ \frac{N_f}{2} }}] {};
	\node (g7) [gauge,right of=g6,label=below:{\bb{ \frac{N_f}{2} }}] {};
	\node (g8) [gauge,right of=g7,label=below:{\cc{ \frac{N_f}{2} }}] {};
	\node (g9) [gauge,right of=g8,label=below:{\bb{ \frac{N_f}{2} {-1} }}] {};
	\node (g10) [right of=g9] {$\cdots$};
	\node (g11) [gauge,right of=g10,label=below:{\bb{1}}] {};
	\node (g12) [gauge,right of=g11,label=below:{\cc{1}}] {};
	\node (g13) [gauge,right of=g12,label=below:{\bb{0}}] {};
	\node (u1) [gauge,above of=g7,label=right:{\cc{1}}] {};
	\node (u2) [gauge,above of=u1,label=right:{\uu{1}}] {};
	\draw (g1)--(g2) (g2)--(g3) (g3)--(g4) (g4)--(g5) (g5)--(g6) (g6)--(g7) (g7)--(g8) (g8)--(g9) (g9)--(g10) (g10)--(g11) (g11)--(g12) (g12)--(g13) (g7)--(u1) (u1)--(u2);
	\end{tikzpicture}
	} \\  
        $\begin{matrix} N_f {<}2k{-}3 \\ \text{ $N_f$ odd} \end{matrix}$ &  
        \raisebox{-.5\height}{
 	\begin{tikzpicture}
 	\tikzset{node distance = 1cm};
	\tikzstyle{gauge} = [circle, draw,inner sep=2.5pt];
	\tikzstyle{flavour} = [regular polygon,regular polygon sides=4,inner 
sep=2.5pt, draw];
	\node (g1) [gauge,label=below:{\bb{0}}] {};
	\node (g2) [gauge,right of=g1,label=below:{\cc{1}}] {};
	\node (g3) [gauge,right of=g2,label=below:{\bb{1}}] {};
	\node (g4) [right of=g3] {$\cdots$};
	\node (g5) [gauge,right of=g4,label=below:{\cc{ \frac{N_f {-1} }{2} }}] {};
	\node (g6) [gauge,right of=g5,label=below:{\bb{ \frac{N_f {-1} }{2} }}] {};
	\node (g7) [gauge,right of=g6,label=below:{\cc{ \frac{N_f {+1} }{2} }}] {};
	\node (g8) [gauge,right of=g7,label=below:{\bb{ \frac{N_f {-1} }{2} }}] {};
	\node (g9) [gauge,right of=g8,label=below:{\cc{ \frac{N_f {-1} }{2} }}] {};
	\node (g10) [right of=g9] {$\cdots$};
	\node (g11) [gauge,right of=g10,label=below:{\bb{1}}] {};
	\node (g12) [gauge,right of=g11,label=below:{\cc{1}}] {};
	\node (g13) [gauge,right of=g12,label=below:{\bb{0}}] {};
	\node (f1) [gauge,above of=g7,label=right:{\uu{1}}] {};
	\node (f2) [flavour,above of =f1, label=right:{$\scriptstyle{k-\tfrac{N_f+1}{2}-2}$}] {};
	\draw (g1)--(g2) (g2)--(g3) (g3)--(g4) (g4)--(g5) (g5)--(g6) (g6)--(g7) (g7)--(g8) (g8)--(g9) (g9)--(g10) (g10)--(g11) (g11)--(g12) (g12)--(g13) (g7)--(f1);
	\draw [line join=round,decorate, decoration={zigzag, segment length=4,amplitude=.9,post=lineto,post length=2pt}]  (f1) -- (f2);
	\end{tikzpicture}
	} \\ 
       $\begin{matrix} {N_f {<}2k{-}4} \\ \text{$N_f$ even} \end{matrix} $  & \raisebox{-.5\height}{
 	\begin{tikzpicture}
 	\tikzset{node distance = 1cm};
	\tikzstyle{gauge} = [circle, draw,inner sep=2.5pt];
	\tikzstyle{flavour} = [regular polygon,regular polygon sides=4,inner 
sep=2.5pt, draw];
	\node (g1) [gauge,label=below:{\bb{0}}] {};
	\node (g2) [gauge,right of=g1,label=below:{\cc{1}}] {};
	\node (g3) [gauge,right of=g2,label=below:{\bb{1}}] {};
	\node (g4) [right of=g3] {$\cdots$};
	\node (g5) [gauge,right of=g4,label=below:{\bb{ \frac{N_f}{2} {-1} }}] {};
	\node (g6) [gauge,right of=g5,label=below:{\cc{ \frac{N_f}{2} }}] {};
	\node (g7) [gauge,right of=g6,label=below:{\bb{ \frac{N_f}{2} }}] {};
	\node (g8) [gauge,right of=g7,label=below:{\cc{ \frac{N_f}{2} }}] {};
	\node (g9) [gauge,right of=g8,label=below:{\bb{ \frac{N_f}{2} {-1} }}] {};
	\node (g10) [right of=g9] {$\cdots$};
	\node (g11) [gauge,right of=g10,label=below:{\bb{1}}] {};
	\node (g12) [gauge,right of=g11,label=below:{\cc{1}}] {};
	\node (g13) [gauge,right of=g12,label=below:{\bb{0}}] {};
	\node (f1) [gauge,above of=g7,label=right:{\uu{1}}] {};
	\node (f2) [flavour,above of =f1, label=right:{$\scriptstyle{k-\tfrac{N_f}{2}-3}$}] {};
	\draw (g1)--(g2) (g2)--(g3) (g3)--(g4) (g4)--(g5) (g5)--(g6) (g6)--(g7) (g7)--(g8) (g8)--(g9) (g9)--(g10) (g10)--(g11) (g11)--(g12) (g12)--(g13) (g7)--(f1);
	\draw [line join=round,decorate, decoration={zigzag, segment length=4,amplitude=.9,post=lineto,post length=2pt}]  (f1) -- (f2);
	\end{tikzpicture}
	}  \\ \bottomrule
    \end{tabular}
    \caption{Infinite coupling Higgs branches for $\orm(2k)$ theories with $N_f$ flavours. The moduli space dimensions correctly match the increase due to the new Higgs branch directions at infinite coupling \cite{Zafrir:2015ftn} compared to finite coupling of Table \ref{tab:O2k_finite}.}
    \label{tab:O2k_infinite}
\end{table}

\begin{table}[t]
    \centering
    \begin{tabular}{c|c}
    \toprule
    $N_f{=}2k{-}2$     &  
    \raisebox{-.5\height}{
 	\begin{tikzpicture}
 	\tikzset{node distance = 1cm};
	\tikzstyle{gauge} = [circle, draw,inner sep=2.5pt];
	\tikzstyle{flavour} = [regular polygon,regular polygon sides=4,inner 
sep=2.5pt, draw];
	\node (g1) [gauge,label=below:{\bb{1}}] {};
	\node (g2) [gauge,right of=g1,label=below:{\cc{1}}] {};
	\node (g3) [gauge,right of=g2,label=below:{\bb{2}}] {};
	\node (g4) [right of=g3] {$\cdots$};
	\node (g5) [gauge,right of=g4,label=below:{\bb{ \frac{N_f  }{2} }}] {};
	\node (g6) [gauge,right of=g5,label=below:{\cc{ \frac{N_f }{2} }}] {};
	\node (g7) [gauge,right of=g6,label=below:{\bb{ \frac{N_f }{2} {+1} }}] {};
	\node (g8) [gauge,right of=g7,label=below:{\cc{ \frac{N_f }{2} }}] {};
	\node (g9) [gauge,right of=g8,label=below:{\bb{ \frac{N_f }{2} }}] {};
	\node (g10) [right of=g9] {$\cdots$};
	\node (g11) [gauge,right of=g10,label=below:{\bb{2}}] {};
	\node (g12) [gauge,right of=g11,label=below:{\cc{1}}] {};
	\node (g13) [gauge,right of=g12,label=below:{\bb{1}}] {};
	\node (u1) [gauge,above of=g7,label=right:{\cc{1}}] {};
	\draw (g1)--(g2) (g2)--(g3) (g3)--(g4) (g4)--(g5) (g5)--(g6) (g6)--(g7) (g7)--(g8) (g8)--(g9) (g9)--(g10) (g10)--(g11) (g11)--(g12) (g12)--(g13) (g7)--(u1);
	\end{tikzpicture}
	} 
    \\
   $\begin{matrix} N_f {\leq} 2k{-}3 \\ \text{$N_f$ odd}    \end{matrix}$     & 
    \raisebox{-.5\height}{
 	\begin{tikzpicture}
 	\tikzset{node distance = 1cm};
	\tikzstyle{gauge} = [circle, draw,inner sep=2.5pt];
	\tikzstyle{flavour} = [regular polygon,regular polygon sides=4,inner 
sep=2.5pt, draw];
	\node (g1) [gauge,label=below:{\bb{1}}] {};
	\node (g2) [gauge,right of=g1,label=below:{\cc{1}}] {};
	\node (g3) [gauge,right of=g2,label=below:{\bb{2}}] {};
	\node (g4) [right of=g3] {$\cdots$};
	\node (g5) [gauge,right of=g4,label=below:{\bb{ \frac{N_f {-1} }{2} }}] {};
	\node (g6) [gauge,right of=g5,label=below:{\cc{ \frac{N_f {-1} }{2} }}] {};
	\node (g7) [gauge,right of=g6,label=below:{\bb{ \frac{N_f {+1} }{2} }}] {};
	\node (g8) [gauge,right of=g7,label=below:{\cc{ \frac{N_f {-1} }{2} }}] {};
	\node (g9) [gauge,right of=g8,label=below:{\bb{ \frac{N_f {-1} }{2} }}] {};
	\node (g10) [right of=g9] {$\cdots$};
	\node (g11) [gauge,right of=g10,label=below:{\bb{2}}] {};
	\node (g12) [gauge,right of=g11,label=below:{\cc{1}}] {};
	\node (g13) [gauge,right of=g12,label=below:{\bb{1}}] {};
	\node (u1) [gauge,above of=g7,label=right:{\uu{1}}] {};
	\node (u2) [flavour,above of=u1,label=right:{$\scriptstyle{k-\tfrac{N_f+1}{2}-2}$}] {};
	\draw (g1)--(g2) (g2)--(g3) (g3)--(g4) (g4)--(g5) (g5)--(g6) (g6)--(g7) (g7)--(g8) (g8)--(g9) (g9)--(g10) (g10)--(g11) (g11)--(g12) (g12)--(g13) (g7)--(u1);
	\draw [line join=round,decorate, decoration={zigzag, segment length=4,amplitude=.9,post=lineto,post length=2pt}]  (u1) -- (u2);
	\end{tikzpicture}
	} 
    \\
    $\begin{matrix} N_f {<} 2k{-}2 \\ \text{$N_f$ even}    \end{matrix}$     & 
     \raisebox{-.5\height}{
 	\begin{tikzpicture}
 	\tikzset{node distance = 1cm};
	\tikzstyle{gauge} = [circle, draw,inner sep=2.5pt];
	\tikzstyle{flavour} = [regular polygon,regular polygon sides=4,inner 
sep=2.5pt, draw];
	\node (g1) [gauge,label=below:{\bb{1}}] {};
	\node (g2) [gauge,right of=g1,label=below:{\cc{1}}] {};
	\node (g3) [gauge,right of=g2,label=below:{\bb{2}}] {};
	\node (g4) [right of=g3] {$\cdots$};
	\node (g5) [gauge,right of=g4,label=below:{\cc{ \frac{N_f}{2} {-1} }}] {};
	\node (g6) [gauge,right of=g5,label=below:{\bb{ \frac{N_f }{2} }}] {};
	\node (g7) [gauge,right of=g6,label=below:{\cc{ \frac{N_f }{2} }}] {};
	\node (g8) [gauge,right of=g7,label=below:{\bb{ \frac{N_f }{2} }}] {};
	\node (g9) [gauge,right of=g8,label=below:{\cc{ \frac{N_f}{2} {-1} }}] {};
	\node (g10) [right of=g9] {$\cdots$};
	\node (g11) [gauge,right of=g10,label=below:{\bb{2}}] {};
	\node (g12) [gauge,right of=g11,label=below:{\cc{1}}] {};
	\node (g13) [gauge,right of=g12,label=below:{\bb{1}}] {};
	\node (u1) [gauge,above of=g7,label=right:{\uu{1}}] {};
	\node (u2) [flavour,above of=u1,label=right:{$\scriptstyle{k-\tfrac{N_f}{2}-2}$}] {};
	\draw (g1)--(g2) (g2)--(g3) (g3)--(g4) (g4)--(g5) (g5)--(g6) (g6)--(g7) (g7)--(g8) (g8)--(g9) (g9)--(g10) (g10)--(g11) (g11)--(g12) (g12)--(g13) (g7)--(u1);
	\draw [line join=round,decorate, decoration={zigzag, segment length=4,amplitude=.9,post=lineto,post length=2pt}]  (u1) -- (u2);
	\end{tikzpicture}
	} 
    \\ \bottomrule
    \end{tabular}
    \caption{Infinite coupling Higgs branches for $\orm(2k+1)$ theories with $N_f$ flavours. The moduli space dimensions match the increase due to the new Higgs branch directions at infinite coupling \cite{Zafrir:2015ftn} compared to finite coupling of Table \ref{tab:O2k+1_finite}.}
    \label{tab:O2k+1_infinite}
\end{table}

%
%
\paragraph{Further predictions.}
In addition to 5-brane webs on \Ofp\ planes, one may equally well study configurations involving \Ofm\ and \Ofmt\ plane, which low-energy descriptions in terms of $\orm(k)$ gauge theories. For \Ofm\ planes the constraints on the number of flavours $ N_f \leq 2k-4$ for a non-trivial interaction fixed point are well-known \cite{Intriligator:1997pq,Brunner:1997gk}. In addition, it has been argued in \cite[Sec.\ 3.1]{Bergman:2015dpa} that $N_f=2k-3$ also leads to an interacting 5d fixed point. The relevant brane configuration is depicted in Table \ref{tab:O2k_finite}. Analysing the Higgs branch via the Conjecture \ref{conj:rules} and a generalisation of Observation \ref{obs:matter}, leads to the magnetic quivers for finite coupling in Table \ref{tab:O2k_finite} and infinite coupling in Table \ref{tab:O2k_infinite}.
Likewise, brane webs with \Ofmt\ planes can be reached via Higgsing the brane web of $\orm(2k+2)$ with $N_f+1$ fundamental flavours to the web corresponding to $\orm(2k+1)$ with $N_f$ flavours, see Table \ref{tab:O2k+1_finite}. The brane web yields the constraint $N_f\leq 2k-3$, while \cite{Bergman:2015dpa} showed the existence of another non-trivial fixed point for $N_f=2k-2$. The finite coupling magnetic quivers are summarised in Table \ref{tab:O2k+1_finite}, while the  infinite coupling magnetic quivers are shown in Table \ref{tab:O2k+1_infinite}.

In contrast to the symplectic case, the orthogonal gauge theories do not admit 
complete Higgsing such that already the finite coupling Higgs branch represents 
a computational challenge. Here, the magnetic quivers derived provide a 
\emph{prediction} for the Higgs branches at all values of the coupling. So far, 
the magnetic quivers do consistently reproduce the expected dimensions. Similar 
to the open puzzles in $6$d theories with orthogonal gauge groups 
\cite{Cabrera:2019dob}, a more detailed analysis is left for future research.
%
%
\paragraph{Outlook.}
Having considered all allowed numbers of flavours for $\sprm(k)$ theories, one should have observed instances of multiple cones. As known from \cite{Ferlito:2016grh}, for $N_f =2k$ the Higgs branch is a union of two cones. In terms of the brane web, this should be manifest in two inequivalent maximal subdivisions, as for instance in \cite{Cabrera:2018jxt}. However, so far there is no sign of more than one maximal subdivision in any of the cases based on \Of\ planes. In contrast, the O$7^-$ construction does admit two inequivalent subdivision such that the Higgs branch is, in fact, recovered to be a union of two cones.

In addition, for cases with non-complete Higgsing such as $\sprm(k)$ with $N_f 
<2k$ flavours as well as $\orm(k)$ with $N_f \leq k-3$, one can expect the 
appearance of nilpotent operators, as discussed for $\surm(k)$ gauge theories 
in \cite{Bourget:2019rtl}. Further progress in this direction requires, in part, 
a detailed analysis of the Higgs branch using different methods. 
\paragraph{Acknowledgements.}
We are grateful to Santiago Cabrera, Hirotaka Hayashi, Rudolph Kalveks, 
Sung-Soo Kim, Kimyeong Lee, Dominik Miketa, Sakura Sch\"afer-Nameki, Futoshi 
Yagi, Gabi Zafrir, and Anton Zajac for helpful discussions.
The work of A.B., J.F.G., A.H. and Z.Z. was supported by STFC grant 
ST/P000762/1.  
The work of M.S. was supported by the National Thousand-Young-Talents Program of China, the National Natural Science Foundation of China (grant no.\ 11950410497), and the China Postdoctoral Science Foundation (grant no.\ 2019M650616).
We thank the Simons Center for Geometry and Physics, Stony Brook University for the hospitality and the partial support during an early stage of this work at the Simons Summer workshop 2019. 
We thank the 11th Joburg Workshop on String Theory from 8-13 December 2019 for hospitality, and the MIT-Imperial College London Seed Fund for support.
\clearpage
%
\appendix
\section{Background material}
\label{app:background}
\subsection{Type II brane configurations with 8 supercharges}
Consider Type II superstring theory with
\begin{compactitem}
\item \NS\ branes in $x^0,x^1,\ldots,x^5$ direction,
\item \Dp\ branes in $x^0,x^1,\ldots,x^{p-1},x^6$, direction,
\item \Dpp\ branes in $x^0,x^1,\ldots,x^{p-1},x^7,x^8,x^9$ direction.
\end{compactitem}
For $0\leq p\leq6$ the configuration preserves 8 supercharges and gives rise to a $p$ dimensional world-volume theory living on the \Dp\ branes suspended between \NS\ branes.
Since every such \Dp-\Dpp-\NS\ brane configuration is T-dual to the set-up of \cite{Hanany:1996ie},
a \Dp\ brane is \emph{created} or \emph{annihilated} whenever a \NS\ passes through a \Dpp\ brane. In the presence of \Op\ planes, brane creation and annihilation is modified due to the non-trivial brane charge of the orientifold. In units of the physical \Dp\ branes, the RR-charges of the \Op\ planes are as follows \cite{Hanany:1999sj,Feng:2000eq}: 
\begin{equation}
    \textrm{charge} (\Oppm )=\pm 2^{p-5} \, , \quad 
    \textrm{charge}(\Opmt )=\frac{1}{2}-2^{p-5} \, , \quad 
    \textrm{charge}(\Oppt )=2^{p-5} \,.
\end{equation}
As a remark, the NS-charges of the \Op\ planes are zero.
Moreover, the orientifolds change whenever passing through a half \NS\ or half \Dpp\ brane \cite{Evans:1997hk,Hanany:1999sj,Hanany:2000fq}:
on the one hand, an \Oppm\ passing through a half \NS\ turns into an \Opmp; likewise, an \Oppmt\  becomes an \Opmpt. On the other hand, an \Oppm\ passing through a half \Dpp\ turns into an \Oppmt, and vice versa.

If there are \Dp\ branes parallel to the \Op\ planes, the world volume gauge theory becomes ortho-symplectic. Similarly, effects are induced on the world volume of the \Dpp. Table \ref{tab:orientifold} summarises these known results and introduces the notation, which follows \cite{Gaiotto:2008ak}.
\subsection{5-branes webs with 7-branes and orientifold 5-planes}
For this paper, $(p,q)$ 5-branes webs in the presence of $[p,q]$ 7-branes and 
orientifold 5-planes are relevant. The integers $p$ and $q$ are assumed to be 
coprime throughout. The Type IIB brane construction is summarised in Table 
\ref{tab:brane_setup}.
\paragraph{Brane bending due to orientifold planes}
Since \Of\ planes change whenever crossing an half \NS\ brane, one finds that a 
$(0,1)$ 5-brane (i.e.\ an \NS\ brane) that separates a \Ofp\ and \Ofm\ plane is 
bent. Using charge conservation, the quantum corrected configuration becomes 
\cite{Zafrir:2015ftn}:
\begin{equation}
 \raisebox{-.5\height}{
    \begin{tikzpicture}
    \OPlus{0,0}{2,0}
    \draw (2,0)--(4,1) (2,0)--(4,-1);
    \node at (4.6,1) {$\scriptstyle{(2,1)}$};
    \node at (4.6,-1) {$\scriptstyle{(2,-1)}$};
    \end{tikzpicture}
    }
    \label{eq:bending_O5-_O5+}
\end{equation}
Attempting to repeat the analysis for a classical $(0,1)$ 5-brane separating a \Ofpt\ and \Ofmt\ plane is known to be troublesome due to arising fractional brane charges. A solution involving monodromy cuts has been proposed in \cite{Zafrir:2015ftn}.
\paragraph{Monodromy of the 7-brane.}
For the $[p,q]$ 7-brane, the associated monodromy matrix is given by
\begin{equation}
    M_{[p,q]}=\begin{pmatrix}
    1-pq & p^2\\
    -q^2 & 1+pq
    \end{pmatrix} 
    \label{eq:monodromy}
\end{equation}
such that the action is clockwise; in more detail, a schematic depiction looks like
\begin{equation}
 \raisebox{-.5\height}{
    \begin{tikzpicture}
        \draw (0,-1)--(0,0)--(1,1);
        \node at (0,-1.25) {$(r,s)$};
        \node at (1.7,1.2) {$M_{[p,q]}\cdot (r,s)$};
        \MonoCut{-1,0}{1,0}
        \SevenB{1,0}
        \node at (1,-0.5) {$[p,q]$};
        \draw[->] (2,0)--(3,0);
        \draw(9,-1)--(9,0)--(10,1);
        \node at (9,-1.25) {$(r,s)$};
        \node at (10.7,1.2) {$M_{[p,q]}\cdot(r,s)$};
        \draw (5,0.05)--(9.05,0.05);
        \draw (5,0)--(9,0);
        \draw (5,-0.05)--(9,-0.05);
        \node at (7,0.5) {$|ps-qr|\cdot(p,q)$};
        \MonoCut{3.5,0}{5,0}
        \SevenB{5,0}
        \node at (5,-0.5) {$[p,q]$};
    \end{tikzpicture}
    }
    \label{eq:brane_creation}
\end{equation}
where $M_{[p,q]}\cdot (r,s)$ denotes the matrix product with a vector.
\paragraph{\Ofpt\ and \Ofmt\ planes and bending.}
Following an argument presented in \cite{Zafrir:2015ftn}, one considers the well-defined set-up \eqref{eq:bending_O5-_O5+}, adds two $[1,0]$ 7-branes on the left-hand-side 
\begin{subequations}
\begin{align}
 \raisebox{-.5\height}{
    \begin{tikzpicture}
    \OPlus{-2,0}{-1,0}
    \OPlusTilde{-1,0}{0,0}
    \OPlus{0,0}{1,0}
    \MonoCut{-2,0.075}{-1,0.075}
    \MonoCut{-2,-0.075}{0,-0.075}
    \MonoCut{3,1}{3,1.5}
    \MonoCut{3,-1}{3,-1.5}
    \SevenB{-1,0}
    \SevenB{0,0}
    \Dbrane{1,0}{3,1}
    \Dbrane{1,0}{3,-1}
    \SevenB{3,1}
    \SevenB{3,-1}
    \draw (1.2,0.5) node {$\scriptstyle{(2,1)}$};
    \draw (3.5,1) node {$\scriptstyle{[2,1]}$};
    \draw (-1.0,0.35) node {$\scriptstyle{[1,0]}$};
    \end{tikzpicture}
    }
\end{align}
and moves them successively through the 5-branes, accounting for brane creation and annihilation. The first transitions is not accompanied by brane creation
\begin{align}
 \raisebox{-.5\height}{
    \begin{tikzpicture}
    \OPlus{-2,0}{-1,0}
    \OPlusTilde{-1,0}{0,0}
    \OMinusTilde{0,0}{3,0}
    \MonoCut{-2,0.075}{-1,0.075}
    \MonoCut{-2,-0.075}{3,-0.075}
    \MonoCut{2,1}{2,1.5}
    \MonoCut{2,-1}{2,-1.5}
    \SevenB{-1,0}
    \SevenB{3,0}
    \Dbrane{0,0}{2,1}
    \Dbrane{0,0}{2,-1}
    \SevenB{2,1}
    \SevenB{2,-1}
    \draw (0.5,0.5) node {$\scriptstyle{(2,1)}$};
    \draw (2.5,1) node {$\scriptstyle{[2,1]}$};
    \draw (-1.0,0.35) node {$\scriptstyle{[1,0]}$};
    \end{tikzpicture}
    }
\end{align}
while the second transition is indeed accompanied by brane creation
\begin{align}
 \raisebox{-.5\height}{
    \begin{tikzpicture}
    \OPlus{-2,0}{-1,0}
    \OMinusTilde{2,0}{3,0}
    \Dbrane{-0.7,0.15}{2,0.15}
    \Dbrane{-0.7,-0.15}{2,-0.15}
    \MonoCut{-2,0.075}{2,0.075}
    \MonoCut{-2,-0.075}{3,-0.075}
    \MonoCut{1,1}{1,1.5}
    \MonoCut{1,-1}{1,-1.5}
    \SevenB{2,0}
    \SevenB{3,0}
    \Dbrane{-1,0}{1,1}
    \Dbrane{-1,0}{1,-1}
    \SevenB{1,1}
    \SevenB{1,-1}
    \draw (-0.5,0.5) node {$\scriptstyle{(2,1)}$};
    \draw (1.5,1) node {$\scriptstyle{[2,1]}$};
    \draw (3.0,0.35) node {$\scriptstyle{[1,0]}$};
    \end{tikzpicture}
    }
\end{align}
Thereafter, one can move the monodromy cuts, which then affects the $(2,1)$ 5-brane as described in \eqref{eq:monodromy}. Thus,
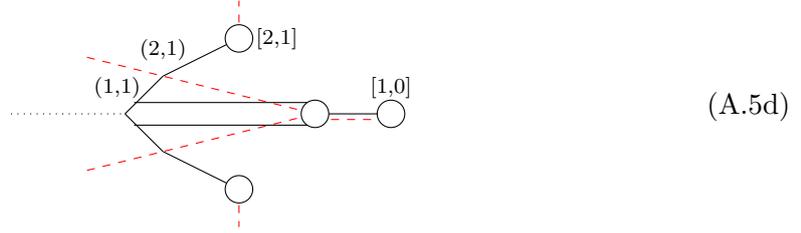
\begin{align}
 \raisebox{-.5\height}{
    \begin{tikzpicture}
    \OPlus{-2,0}{-0.5,0}
    \OMinusTilde{2,0}{3,0}
    \Dbrane{-0.38,0.15}{2,0.15}
    \Dbrane{-0.38,-0.15}{2,-0.15}
    \MonoCut{-1,0.75}{2,0}
    \MonoCut{-1,-0.75}{2,0}
    \MonoCut{2,-0.075}{3,-0.075}
    \MonoCut{1,1}{1,1.5}
    \MonoCut{1,-1}{1,-1.5}
    \SevenB{2,0}
    \SevenB{3,0}
    \Dbrane{-0.5,0}{0,0.5}
    \Dbrane{-0.5,0}{0,-0.5}
    \Dbrane{0,0.5}{1,1}
    \Dbrane{0,-0.5}{1,-1}
    \SevenB{1,1}
    \SevenB{1,-1}
    \draw (0,0.85) node {$\scriptstyle{(2,1)}$};
    \draw (-0.6,0.35) node {$\scriptstyle{(1,1)}$};
    \draw (1.5,1) node {$\scriptstyle{[2,1]}$};
    \draw (3.0,0.35) node {$\scriptstyle{[1,0]}$};
    \end{tikzpicture}
    }
\end{align}
and finally one ends up with
\begin{align}
 \raisebox{-.5\height}{
    \begin{tikzpicture}
    \OPlus{-2,0}{-0.5,0}
    \OMinusTilde{2,0}{3,0}
    \Dbrane{-0.38,0.15}{2,0.15}
    \Dbrane{-0.38,-0.15}{2,-0.15}
    \MonoCut{2,1}{2,0}
    \MonoCut{2,-1}{2,0}
    \MonoCut{2,-0.075}{3,-0.075}
    \MonoCut{0.5,1}{0.5,1.5}
    \MonoCut{0.5,-1}{0.5,-1.5}
    \SevenB{2,0}
    \SevenB{3,0}
    \Dbrane{-0.5,0}{0.5,1}
    \Dbrane{-0.5,0}{0.5,-1}
    \SevenB{0.5,1}
    \SevenB{0.5,-1}
    \draw (-0.6,0.35) node {$\scriptstyle{(1,1)}$};
    \draw (1,1) node {$\scriptstyle{[1,1]}$};
    \draw (3.0,0.35) node {$\scriptstyle{[1,0]}$};
    \end{tikzpicture}
    }
\end{align}
\end{subequations}
Hence, the understanding of \Ofpt\ and \Ofmt\ becomes
\begin{compactitem}
\item \Ofmt\ equals an \Ofm\ together with half a stuck \Dfive\ and a (half-) monodromy cut.
\item \Ofpt\ contains a (half-) monodromy cut as well.
\end{compactitem}
The (half-)monodromy cut is associated to a $[1,0]$ 7-brane and coincides with the orientifold drawn $(x^5,x^6)$-plane.

\begin{table}[t]
\centering
 \begin{tabular}{c|c}
 \toprule
Orientifold & Splitting pattern \\ \midrule
\Ofm & 
\raisebox{-.5\height}{
    \begin{tikzpicture}
    \Dbrane{0,0.5}{1,0.5}
    \Dbrane{0,-0.5}{1,-0.5}
    \draw[white] (-0.5,0) -- (1.5,0);
    \draw[red,dashed,->] (0,-0.5)--(0,-1);
    \draw[red,dashed,->] (0,0.5)--(0,1);
    \SevenB{0,0.5}
    \SevenB{0,-0.5}
    \draw[thick,->] (2,0)--(3,0);
    \Dbrane{4,0}{5,0}
    \Dbrane{5,0.1}{6,0.1}
    \Dbrane{5,-0.1}{6,-0.1}
    \draw[red,dashed,->] (4,0)--(4,0.5);
    \draw[red,dashed,->] (4,0)--(4,-0.5);
    \begin{scope}[red,dashed,decoration={
    markings,mark=at position 0.5 with {\arrow{>}}}
    ] 
    \draw[postaction={decorate}] (5,0.05)--(4,0.05);
    \end{scope}
    \SevenB{4,0}
    \SevenB{5,0}
     \end{tikzpicture}
    }
\\[1.5cm]
\Ofmt & 
\raisebox{-.5\height}{
    \begin{tikzpicture}
    \Dbrane{0,0.5}{1,0.5}
    \Dbrane{0,-0.5}{1,-0.5}
    \OMinusTilde{-0.5,0}{1.5,0}
    \begin{scope}[red,dashed,decoration={
    markings,mark=at position 0.5 with {\arrow{>}}}
    ] 
    \draw[postaction={decorate}] (1.5,0.05)--(-0.5,0.05);
    \end{scope}
    \draw[red,dashed,->] (0,-0.5)--(0,-1);
    \draw[red,dashed,->] (0,0.5)--(0,1);
    \SevenB{0,0.5}
    \SevenB{0,-0.5}
    \draw[thick,->] (2,0)--(3,0);
    \OMinusTilde{3.5,0}{4,0}
    \draw[red,dashed,->] (4,0.05)--(3.5,0.05);
    \Dbrane{4,0.1}{6,0.1}
    \Dbrane{4,-0.1}{6,-0.1}
    \OMinusTilde{5,0}{6,0}
    \begin{scope}[red,dashed,decoration={
    markings,mark=at position 0.5 with {\arrow{>}}}
    ] 
    \draw[postaction={decorate}] (6,0.05)--(5,0.05);
    \end{scope}
    \draw[red,dashed,->] (5,0)--(5,0.5);
    \draw[red,dashed,->] (5,0)--(5,-0.5);
    \SevenB{4,0}
    \SevenB{5,0}
     \end{tikzpicture}
    }
\\[1.5cm]
\Ofp &
\raisebox{-.5\height}{
    \begin{tikzpicture}
    \Dbrane{0,0.5}{1,0.5}
    \Dbrane{0,-0.5}{1,-0.5}
    \OPlus{-0.5,0}{1.5,0}
    \draw[red,dashed,->] (0,-0.5)--(0,-1);
    \draw[red,dashed,->] (0,0.5)--(0,1);
    \SevenB{0,0.5}
    \SevenB{0,-0.5}
    \draw[thick,->] (2,0)--(3,0);
    \OPlus{3.5,0}{4,0}
    \Dbrane{4,0.1}{6,0.1}
    \Dbrane{4,-0.1}{6,-0.1}
    \OPlusTilde{4,0}{5,0}
    \begin{scope}[red,dashed,decoration={
    markings,mark=at position 0.5 with {\arrow{>}}}
    ] 
    \draw[postaction={decorate}] (5,0.05)--(4,0.05);
    \end{scope}
    \OPlus{5,0}{6,0}
    \draw[red,dashed,->] (4,0)--(4,0.5);
    \draw[red,dashed,->] (4,0)--(4,-0.5);
    \SevenB{4,0}
    \SevenB{5,0}
     \end{tikzpicture}
    }
\\[1.5cm]
\Ofpt &
\raisebox{-.5\height}{
    \begin{tikzpicture}
    \Dbrane{0,0.5}{1,0.5}
    \Dbrane{0,-0.5}{1,-0.5}
    \OPlusTilde{-0.5,0}{1.5,0}
    \begin{scope}[red,dashed,decoration={
    markings,mark=at position 0.5 with {\arrow{>}}}
    ] 
    \draw[postaction={decorate}] (1.5,0.05)--(-0.5,0.05);
    \end{scope}
    \draw[red,dashed,->] (0,-0.5)--(0,-1);
    \draw[red,dashed,->] (0,0.5)--(0,1);
    \SevenB{0,0.5}
    \SevenB{0,-0.5}
    \draw[thick,->] (2,0)--(3,0);
    \OPlusTilde{3.5,0}{4,0}
    \draw[red,dashed,->] (4,0.05)--(3.5,0.05);
    \Dbrane{4,0.1}{6,0.1}
    \Dbrane{4,-0.1}{6,-0.1}
    \OPlus{4,0}{5,0}
    \OPlusTilde{5,0}{6,0}
    \draw[red,dashed,->] (5,0)--(5,0.5);
    \draw[red,dashed,->] (5,0)--(5,-0.5);
    \begin{scope}[red,dashed,decoration={
    markings,mark=at position 0.5 with {\arrow{>}}}
    ] 
    \draw[postaction={decorate}] (6,0.05)--(5,0.05);
    \end{scope}
    \SevenB{4,0}
    \SevenB{5,0}
     \end{tikzpicture}
    }
 \\ \bottomrule
 \end{tabular}
\caption{Splitting of \Dseven\ branes, with \Dfive\ branes ending on them, on 
various \Of\ planes.}
\label{tab:split_7-brane}
\end{table}
\paragraph{Splitting 7-branes on \Of\ planes.}
Having half 7-branes merging on a \Of\ plane and subsequently splitting along 
the orientifold, one needs to take care of how branes are created in the 
process. The different cases are summarised in Table \ref{tab:split_7-brane}.
\subsection{3d \texorpdfstring{$\Ncal=4$}{N=4} Coulomb branches}
\label{app:Coulomb_branch}
Associated to a magnetic quiver is a space of dressed monopole operators or, 
loosely speaking, a 3d $\Ncal=4$ Coulomb branch. In particular for linear quiver 
gauge theories of either unitary gauge groups or alternating ortho-symplectic 
gauge nodes, the IR global symmetry $G$ is usually well approximated by 
analysing the subset of balanced nodes.

\paragraph{Unitary quivers.}
A unitary gauge node $\urm(k)$  with $N_f$ flavours is called \cite[Sec.\ 2.4]{Gaiotto:2008ak}
\begin{align}
    \text{\emph{good} if } 
    \quad
    N_f \geq 2k
    \;, \qquad
    \text{and \emph{balanced} if }
    \quad
    N_f = 2k 
    \;.
\end{align}
Then for a quiver comprised of unitary nodes the following is expected to hold:
\begin{compactitem}
\item The subset of balanced gauge nodes forms the Dynkin diagram of the non-abelian part of the Coulomb branch global symmetry.
\item The number of unbalanced nodes minus one yields the number of $\uo$ factors inside the Coulomb branch global symmetry.
\end{compactitem}
Note that this procedure might only provide a subgroup of the global symmetry of the Coulomb branch.
\paragraph{Ortho-symplectic linear quivers.}
For orthogonal and symplectic 3d $\Ncal=4$ gauge theories, the conditions for \emph{good} and \emph{balanced} are as follows \cite[Sec.\ 5.1-5.2]{Gaiotto:2008ak}:
A $\orm(k)$ (or $\mathrm{SO}(k)$) gauge group with fundamental flavours, transforming under a $\usprm(2N_f)$ global symmetry, is called
\begin{align}
    \text{\emph{good} if } 
    \quad
    N_f \geq k-1
    \;, \qquad
    \text{and \emph{balanced} if }
    \quad
    N_f = k-1 
    \;.
\end{align}
Similarly, a $\usprm(2k)=\sprm(k)$ gauge group with fundamental flavours, transforming under a $\orm(2N_f)$ global symmetry, is called
\begin{align}
    \text{\emph{good} if } 
    \quad
    N_f \geq 2k+1
    \;, \qquad
    \text{and \emph{balanced} if }
    \quad
    N_f = 2k+1 
    \;.
\end{align}
Consider a \emph{linear} quiver with alternating orthogonal and symplectic gauge 
nodes, then a chain of $p$ balanced nodes is expected to give rise to the 
following enhanced Coulomb branch symmetry $G$:
\begin{compactitem}
\item An $\sorm(p+1)$ symmetry, if there are no $\sorm(2)$ (or $\orm(2)$) gauge nodes at the ends.
\item An $\sorm(p+2)$ symmetry, if there is an $\sorm(2)$ (or $\orm(2)$) gauge node at one of the ends.
\item An $\sorm(p+3)$ symmetry, if there is an $\sorm(2)$ (or $\orm(2)$) at each end. 
\end{compactitem}
\subsection{Summary companion paper}
\label{app:companion}
As indicated in the main text, the choice of gauge group for a 
unitary-orthosymplectic quiver needs to be specified when considering the 
Coulomb 
branch moduli space. In this appendix, the main idea of the companion paper 
\cite{Bourget:2020xdz} is summarised.
Based on the brane configurations, the magnetic gauge nodes are associated with 
the following groups
 \begin{align}
  \balg_n \to \sorm(2n+1) \, , \quad 
  \calg_n \to \usprm(2n) \, , \quad 
  \dalg_n \to \sorm(2n) \,.
 \end{align}
 However, the choice of overall magnetic gauge group of a given quiver remains 
and is resolved in \cite{Bourget:2020xdz} as follows:
\begin{compactitem}
 \item Framed orthosymplectic quiver: The magnetic gauge group of the quiver is 
the product group over all nodes, because the kernel of the representation 
defining the matter content is trivial.
\item Unframed (unitary-) orthosymplectic quiver without any 
$\sorm(\text{odd})$: The matter representation for the product gauge group has a 
nontrivial $\Z_2$ kernel. In order to reproduce the expected moduli spaces, the 
gauge group is chosen to be the product gauge group modulo this diagonal $\Z_2$.
\item Unframed (unitary-) orthosymplectic quiver with $\sorm(\text{odd})$: The 
matter representation for the product gauge group has a trivial kernel. 
Hence, the gauge group of the quiver is the product gauge group.
\end{compactitem}
Once the gauge group of a quiver is specified, the magnetic lattice is uniquely 
determined and the Coulomb branch Hilbert series can be evaluated via 
\cite{Cremonesi:2013lqa}.

\bibliographystyle{JHEP}
\bibliography{bibli.bib}

\end{document}